\documentclass[a4paper,11pt]{article}
\pdfoutput=1
\usepackage{jheppub}
\usepackage[utf8]{inputenc}
\usepackage{appendix}
\makeatletter
\gdef\@fpheader{}
\makeatother
\usepackage{pdflscape}
\usepackage{booktabs}
\usepackage{afterpage}
\usepackage{hyperref}
\usepackage[colorinlistoftodos]{todonotes}

\definecolor{goodgreen}{RGB}{55,169,49}

\newcommand{\verticalcenter}[1]{\raisebox{-0.5\height}{\begin{tikzpicture}
    \node at (0,0) {#1};
\end{tikzpicture}}}

\usepackage{pifont}
\newcommand{\cmark}{\text{\ding{51}}}
\newcommand{\xmark}{\text{\ding{55}}}

\colorlet{colors01}{red!10}
\colorlet{colors02}{yellow!10}
\colorlet{colors03}{green!10}
\colorlet{colors04}{blue!10}
\colorlet{colors05}{purple!10}

\usepackage[many]{tcolorbox}

\tcolorboxenvironment{test}{
  enhanced,
  borderline={0.8pt}{0pt}{black},
  borderline={0.4pt}{2pt}{black},
  boxrule=0.4pt,
  colback=white,
  coltitle=black,
  sharp corners,
}
\usepackage{placeins}
\usepackage{enumerate}
\usepackage{oubraces}
\usepackage{tikz}
\usetikzlibrary{shapes.geometric}
\tikzset{gauge1/.style={draw=none,minimum size=0.4cm,fill=white,circle, draw}}
\tikzset{gauge5/.style={draw=none,minimum size=0.6cm,fill=white,circle, draw}}
\tikzset{supergauge/.style={draw=none,minimum size=0.9cm,fill=white,circle, draw}}
\tikzset{bluenode/.style={draw=none,minimum size=0.4cm,fill=blue,circle, draw}}
\tikzset{rednode/.style={draw=none,minimum size=0.4cm,fill=red,circle, draw}}
\tikzset{gauge3/.style={draw=none,minimum size=0.4cm,fill=white,circle, draw}}
\tikzset{dotsize/.style={draw=none,minimum size=0.6pt,fill=black,circle,inner sep=1pt, draw}}
\tikzset{mini/.style={draw=none,minimum size=1pt,fill=white,circle,inner sep=3pt, draw}}
\tikzset{miniG/.style={draw=none,minimum size=1pt,fill=black,circle,inner sep=3pt, draw}}
\tikzset{cyane/.style={draw=none,minimum size=0.4cm,fill=cyan,circle, draw}}
\tikzset{pinklinet/.style={draw=none,minimum size=0.4cm,fill=magenta,circle, draw}}
\tikzset{greenlinet/.style={draw=none,minimum size=0.4cm,fill=green,circle, draw}}
\tikzset{blacknode/.style={draw=none,minimum size=0.4cm,fill=black,circle, draw}}
\tikzset{brownlinet/.style={draw=none,minimum size=0.4cm,fill=olive,circle, draw}}
\tikzset{magicmintlinet/.style={draw=none,minimum size=0.4cm,fill=red,circle, draw}}
\tikzset{orangeet/.style={draw=none,minimum size=0.4cm,fill=orange,circle, draw}}
\tikzset{grayet/.style={draw=none,minimum size=0.4cm,fill=gray,circle, draw}}
\tikzset{blueet/.style={draw=none,minimum size=0.4cm,fill=blue,circle, draw}}
\tikzset{flavour2/.style={draw=none,minimum size=0.8cm,fill=white, regular polygon,regular polygon sides=4,draw}}
\tikzset{flavour2/.style={draw=none,minimum size=0.6cm,fill=white, regular polygon,regular polygon sides=4,draw}}
\tikzset{redsquare/.style={draw=none,minimum size=0.6cm,fill=red, regular polygon,regular polygon sides=4,draw}}
\tikzset{bluesquare/.style={draw=none,minimum size=0.6cm,fill=blue, regular polygon,regular polygon sides=4,draw}}
\tikzset{greenflavor/.style={draw=none,minimum size=0.6cm,fill=green, regular polygon,regular polygon sides=4,draw}}
\tikzset{brownflavor/.style={draw=none,minimum size=0.6cm,fill=brown, regular polygon,regular polygon sides=4,draw}}
\tikzset{pinkflavor/.style={draw=none,minimum size=0.6cm,fill=magenta, regular polygon,regular polygon sides=4,draw}}
\tikzset{grayflavor/.style={draw=none,minimum size=0.6cm,fill=gray, regular polygon,regular polygon sides=4,draw}}
\tikzset{none/.style={draw=none}}
\tikzset{new edge style 1/.style={dashed}}
\tikzset{dashedline/.style={dashed}}
\tikzset{brace1/.style={decorate,decoration={brace,amplitude=5pt,mirror}}}
\tikzset{bluee/.style={line width=0.5mm,blue}}
\tikzset{orangee/.style={line width=0.5mm,orange}}
\tikzset{magentae/.style={line width=0.5mm,magenta}}
\tikzset{rede/.style={line width=0.5mm,red}}
\tikzset{thickred/.style={line width=5mm,red}}
\tikzset{greene/.style={line width=0.5mm,green}}
\tikzset{darke/.style={line width=0.5mm,black}}
\tikzset{cyaneX/.style={line width=0.5mm,cyan}}
\tikzset{new edge style 3/.style={dashed,red}}
\tikzset{magicmintline/.style={line width=0.5mm,gray}}
\tikzset{brownline/.style={line width=0.5mm,brown}}
\tikzset{greenline/.style={line width=0.5mm,green}}
\tikzset{oliveline/.style={line width=0.5mm,green}}
\tikzset{darkgreenline/.style={line width=0.5mm,olive}}
\tikzset{pinkline/.style={line width=0.5mm,magenta}}
\tikzset{dottedz/.style={line width=0.5mm,black,dotted}}
\tikzset{pinkline2/.style={line width=0.5mm,magenta,dotted}}
\tikzset{brace2/.style={decorate,decoration={brace,amplitude=5pt}}}
\tikzset{reddotted/.style={line width=0.5mm,red,dotted}}
\tikzset{bluedotted/.style={line width=0.5mm,blue,dotted}}
\tikzset{magicmintdotted/.style={line width=0.5mm,gray,dotted}}
\tikzset{greendotted/.style={line width=0.5mm,green,dotted}}
\tikzset{browndotted/.style={line width=0.5mm,brown,dotted}}
\tikzset{arrowed/.style={line width=0.5mm,->}}
\pgfdeclarelayer{edgelayer}
\pgfdeclarelayer{nodelayer}
\usetikzlibrary{decorations.pathreplacing}
\pgfsetlayers{edgelayer,nodelayer,main}

\usepackage{xcolor,colortbl}

\usepackage{ytableau}
\tikzstyle{brane}=[draw]
\tikzset{D7/.style={circle, draw=black, inner sep=0pt, fill=white, minimum size=3mm}}
\tikzset{hasse/.style={circle, fill,inner sep=2pt}}
\tikzset{flavor/.style={regular polygon,regular polygon sides=4,inner sep=2.5pt, draw}}
\tikzset{gauge/.style={circle, draw,inner sep=2.5pt}}
\tikzset{gaugeb/.style={circle, draw,fill=black,inner sep=2.5pt}}
\tikzset{gaugered/.style={circle, draw,fill=red,inner sep=2.5pt}}
\tikzset{gaugeblue/.style={circle, draw,fill=blue,inner sep=2.5pt}}
\tikzset{gaugegreen/.style={circle, draw,fill=green,inner sep=2.5pt}}
\tikzset{bd/.style={circle, draw=black, inner sep=0pt, fill=black, minimum size=2mm}}
\tikzset{wd/.style={circle, draw=black, inner sep=0pt, fill=white, minimum size=2mm}}
\tikzset{Dynkin/.style={circle, draw=black, inner sep=0pt, fill=white, minimum size=2mm}}
\tikzstyle{ligne}=[draw, thick] 
\tikzset{doublearrow/.style={ draw=black!75, color=black!75, thick, double distance=3pt, }} 
\tikzset{bluedouble/.style={ draw=blue!75, color=blue!75, thick, double distance=3pt, }} 
\newcommand{\midarrow}{\tikz 
\draw[-triangle 90] (0,0) -- +(.1,0);} 
\newcommand{\midarrowrev}{\tikz 
\draw[-triangle 90] +(.1,0) -- (0,0);}
\tikzset{reddashed/.style={ draw=red, color=red, thick }}
\tikzset{redarrow/.style={ draw=red, -> , thick }}

\newcommand{\DynkinA}{$ \small{\text{DynkinA}^{ \text{\tiny U/SU}}}\;$}
\newcommand{\DynkinB}{$\small{\text{DynkinB}^{ \text{\tiny U/SU}}}\;$}
\newcommand{\DynkinC}{$\small{\text{DynkinC}^{ \text{\tiny U/SU}}}\;$}
\newcommand{\DynkinD}{$\small{\text{DynkinD}^{ \text{\tiny U/SU}}}\;$}
\newcommand{\DynkinBC}{${\small\text{DynkinBC}^{ \text{\tiny U/SU}}}\;$}
\newcommand{\DynkinBCD}{${\small\text{DynkinBCD}^{ \text{\tiny U/SU}}}\;$}
\newcommand{\DynkinABCD}{$\small{\text{DynkinABCD}^{ \text{\tiny U/SU}}}\;$}
\newcommand{\DynkinASU}{$\small{\text{DynkinA}^{ \text{\tiny SU}}}\;$}
\newcommand{\DynkinDSU}{$\small{\text{DynkinD}^{ \text{\tiny SU}}}\;$}
\newcommand{\DynkinAmir}{$\small{{\text{DynkinA}^{ \text{\tiny U/SU}}_{\text{\tiny mirror}}}}\;$}
\newcommand{\DynkinBCmir}{$\small{\text{DynkinBC}^{ \text{\tiny U/SU}}_{\text{\tiny mirror}}}\;$}
\newcommand{\DynkinBCDmir}{$\small{\text{DynkinBCD}^{ \text{\tiny U/SU}}_{\text{\tiny mirror}}}\;$}
\newcommand{\DynkinABCDmir}{$\small{\text{DynkinABCD}^{ \text{\tiny U/SU}}_{\text{\tiny mirror}}}\;$}
\newcommand{\DynkinAmirbar}{$\small{\overline{\text{DynkinA}}^{ \text{\tiny U/SU}_{\text{\tiny mirror}}}}\;$}
\newcommand{\DynkinAbar}{$\small{\overline{\text{DynkinA\;}}^{ \text{\tiny U/SU}
}}$}
\newcommand{\DynkinBCDbar}{$\small{\overline{\text{DynkinBCD\;}}^{ \text{\tiny U/SU}
}}$}
\newcommand{\DynkinBCbar}{$\small{\overline{\text{DynkinBC\;}}^{ \text{\tiny U/SU}
}}$}
\newcommand{\DynkinABCDbar}{$\small{\overline{\text{DynkinABCD\;}}^{ \text{\tiny U/SU}
}}$}
\newcommand{\DynkinBCDmirbar}{$\small{\overline{\text{DynkinBCD\;}}^{ \text{\tiny U/SU}}_{\text{\tiny mirror}}}\;$}
\newcommand{\DynkinBCmirr}{$\small{\text{DynkinBC}_{\text{\tiny mirror}}}\;$}

\usetikzlibrary{calc}
\tikzset{gaugeJ/.style={inner sep=1mm,draw=none,fill=white,minimum size=2mm,circle, draw}}
\tikzset{flavourJ/.style={draw=none,minimum size=0.3mm,fill=white, regular polygon,regular polygon sides=4,draw}}
\tikzset{hasseJ/.style={circle, fill,inner sep=2pt}}

\usepackage{multirow}

\newcommand{\convexpath}[2]{
  [   
  create hullcoords/.code={
    \global\edef\namelist{#1}
    \foreach [count=\counter] \nodename in \namelist {
      \global\edef\numberofnodes{\counter}
      \coordinate (hullcoord\counter) at (\nodename);
    }
    \coordinate (hullcoord0) at (hullcoord\numberofnodes);
    \pgfmathtruncatemacro\lastnumber{\numberofnodes+1}
    \coordinate (hullcoord\lastnumber) at (hullcoord1);
  },
  create hullcoords
  ]
  ($(hullcoord1)!#2!-90:(hullcoord0)$)
  \foreach [
  evaluate=\currentnode as \previousnode using \currentnode-1,
  evaluate=\currentnode as \nextnode using \currentnode+1
  ] \currentnode in {1,...,\numberofnodes} {
    let \p1 = ($(hullcoord\currentnode) - (hullcoord\previousnode)$),
    \n1 = {atan2(\y1,\x1) + 90}, 
    \p2 = ($(hullcoord\nextnode) - (hullcoord\currentnode)$),
    \n2 = {atan2(\y2,\x2) + 90},
    \n{delta} = {Mod(\n2-\n1,360) - 360}
    in 
    {arc [start angle=\n1, delta angle=\n{delta}, radius=#2]}
    -- ($(hullcoord\nextnode)!#2!-90:(hullcoord\currentnode)$) 
  }
}

\title{A Bound on 3d Mirror Pairs}
\preprint{}

\author{Zhenghao Zhong}
\affiliation{Mathematical Institute, University of Oxford,\\
Andrew Wiles Building, Woodstock Road, Oxford, OX2 6GG, UK}
\emailAdd{zhenghao.zhong@maths.ox.ac.uk}

\abstract{A distinctive duality present in 3d $\mathcal{N}=4$ theories is the 3d mirror symmetry. Under this duality, the Coulomb (Higgs) branch of one theory corresponds to the Higgs (Coulomb) branch of its mirror dual. This paper is divided into two parts. In the first part, we examine quiver gauge theories constructed from unitary gauge groups arranged in the shape of ABCD-type Dynkin diagrams. This is arguably the largest family of quivers in the literature with known 3d mirror pairs. Using brane lockings and magnetic quivers, we show how this family can be vastly expanded by replacing any number of the unitary gauge groups with special unitary gauge groups and finding the mirror pairs. In the second part, we argue that in the landscape of 3d mirror pairs, most Lagrangian (quiver gauge theories) will not have 3d mirror duals that are also Lagrangian (quiver gauge theories). For unitary quiver gauge theories, we conjecture that any quiver with an exceptional affine Dynkin diagram as a subquiver cannot have a Lagrangian (quiver gauge theory) mirror.

}

\begin{document} 

\maketitle

\section{Introduction}
Supersymmetric gauge theories with 8 supercharges exhibit rich vacuum structures, making the study of the moduli space of vacua highly valuable. In $3d$ $\mathcal{N}=4$ gauge theories, unlike in other dimensions, both the Higgs branch and Coulomb branch moduli spaces can be hyperKähler cones. This gave rise to an IR duality known as $3d$ mirror symmetry \cite{Intriligator:1996ex}, where the Coulomb (Higgs) branch of one theory is equivalent to the Higgs (Coulomb) branch of its mirror dual. Over the last few decades, numerous 3d mirror pairs have been discovered in the literature.

The obvious advantage of this duality is that if the moduli space of one theory is difficult to study, one can instead analyze its mirror pair. For instance, in 3d $\mathcal{N}=4$ theories, the Coulomb branch receives quantum corrections, while the Higgs branch remains a classical object. Therefore, to study the Coulomb branch, one often examines the Higgs branch of its mirror. Over the past decade, there has been significant progress in understanding the Coulomb branch, allowing the reverse approach when the Higgs branch is more challenging to study than the Coulomb branch of the mirror.


In the literature, almost all quiver gauge theories whose 3d mirrors are also quivers (Lagrangian) theories fall into two main families: $T^\sigma_\rho[G]$ and DynkinG, where the classical algebra $G$ belongs to $A_n, B_n, C_n,$ or $D_n$ (these are the algebras of $SU(n+1)$, $SO(2n+1)$, $Sp(n)$ and $SO(2n)$ respectively). The former was introduced in \cite{Gaiotto:2008ak}, where for $G=A_n$, the theories are linear quivers (gauge groups connected in a linear chain) with unitary gauge groups. For $G = B_n, C_n, D_n$, these are orthosymplectic linear quivers with (special) orthogonal and symplectic gauge groups. The 3d mirrors, $T^\rho_\sigma[G^\vee]$, where $G^\vee$ is the Langlands dual, are also linear unitary quivers for $G=A_n$ and linear orthosymplectic quivers for $G = B_n, C_n, D_n$. The DynkinG family, which is the focus of this paper, consists of unitary gauge groups arranged in the shape of $G$-type Dynkin diagrams\footnote{These are also quivers whose Coulomb branches are slices of affine Grassmannian of $G$ \cite{Braverman:2016pwk,Bourget:2021siw}. }. The 3d mirrors, $\text{DynkinG}^{\text{mirror}}$, are linear unitary quivers for $G = A_n$ and linear quivers with mixed  unitary and orthosymplectic groups for $G = B_n, C_n, D_n$. In the case of $G = A_n$, the two families, $T^\sigma_\rho[A_n]$ and $\text{DynkinA}_{n}$, coincide. These represent almost all known 3d mirror pairs in the literature where both mirror pairs have a quiver (Lagrangian) description. Although some exceptions exist, they are very limited.

For the case of DynkinA (we will drop the subscript to avoid cluttering), any number of the unitary gauge groups can be replaced by special unitary gauge groups and the 3d mirror remains a quiver (Lagrangian) theory. This is carried out in \cite{USU} leading to the vastly larger family \DynkinA and its mirror \DynkinAmir. The proposed mirror pairs are obtained using a mechanism called \emph{brane locking}. The goal here is to then extend this locking mechanism for \DynkinBCD which requires the inclusion of orientifold planes to find their mirror. This leads to interesting features such as the dynamical branes locking together with orientifold planes. The proposed 3d mirrors \DynkinBCDmir are checked through Hilbert series computations. 

This approach vastly expands the landscape of quiver gauge theories with quiver (Lagrangian) 3d mirror duals. The next question is: how far can this be pushed? One of the main results of this paper is our argument that the vast majority of quiver gauge theories do not have 3d mirror duals that are also quiver (Lagrangian) gauge theories. We arrive at this conclusion using the recent \textbf{Decay and Fission} algorithm \cite{Decay, Fission}, which implements the Higgs mechanism on these theories. Based on this, we conjecture a simple rule for determining whether a quiver made solely of unitary gauge groups will have a Lagrangian 3d mirror: it cannot contain, as a subquiver, an affine or twisted affine Dynkin diagram of $G_2, F_4, E_6, E_7,$ $E_8$.



The summary of the paper is as follows.  \S\ref{sect2} takes the locking mechanism introduced in \cite{USU} and generalized it to \DynkinABCD quivers with mixed U and SU nodes in order to find their 3d mirrors. \S\ref{top}  discusses the feasibility of bootstrapping mirror pairs by gauging common topological symmetries. Finally, in  \S\ref{thelimit}, we argue that almost all non-linear quiver gauge theories will not have quiver (Lagrangian) mirror duals. Readers that are only interested in this can skip directly to this section. Our results are summarized in Figure \ref{ultimatetable}.
\begin{landscape}
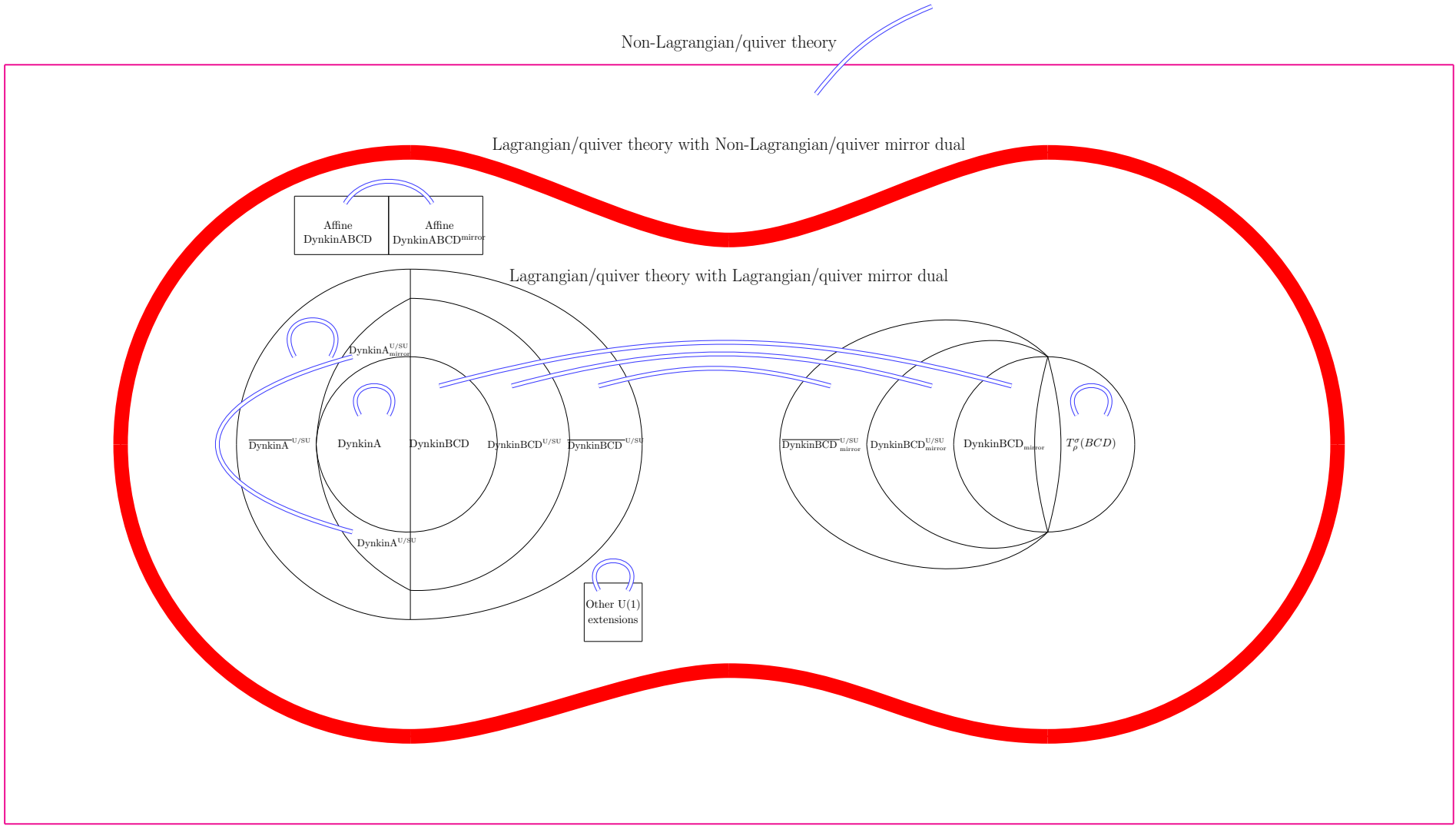
\begin{figure}[ht]
    \centering
\scalebox{0.5}{\begin{tikzpicture}
	\begin{pgfonlayer}{nodelayer}
		\node [style=none] (1) at (1, 2) {};
		\node [style=none] (2) at (1, -8) {};
		\node [style=none] (3) at (-2.25, -3) {};
		\node [style=none] (4) at (4, -3) {};
		\node [style=none] (5) at (6.5, -3) {};
		\node [style=none] (7) at (-5, -3) {};
		\node [style=none] (8) at (9, -3) {};
		\node [style=none] (9) at (1, 0) {};
		\node [style=none] (10) at (1, -6) {};
		\node [style=none] (11) at (1, 3) {};
		\node [style=none] (12) at (1, -9) {};
		\node [style=none] (13) at (-2.25, -3) {};
		\node [style=none] (14) at (23, 0) {};
		\node [style=none] (15) at (23, -6) {};
		\node [style=none] (16) at (19.75, -3) {};
		\node [style=none] (17) at (26, -3) {};
		\node [style=none] (19) at (13.75, -3) {};
		\node [style=none] (21) at (23, 0) {};
		\node [style=none] (22) at (23, -6) {};
		\node [style=none] (23) at (23, 0) {};
		\node [style=none] (24) at (23, -6) {};
		\node [style=none] (25) at (16.75, -3) {};
		\node [style=none] (28) at (7, -9.75) {};
		\node [style=none] (29) at (9, -7.75) {};
		\node [style=none] (30) at (7, -7.75) {};
		\node [style=none] (31) at (9, -9.75) {};
		\node [style=none] (32) at (-9, -3) {};
		\node [style=none] (33) at (1, 7) {};
		\node [style=none] (34) at (12, 4) {};
		\node [style=none] (35) at (23, 7) {};
		\node [style=none] (37) at (23, -13) {};
		\node [style=none] (38) at (12, -10.75) {};
		\node [style=none] (39) at (1, -13) {};
		\node [style=none] (52) at (33, -3) {};
		\node [style=none] (53) at (23, 7) {};
		\node [style=none] (54) at (12, 4) {};
		\node [style=none] (55) at (12, -10.75) {};
		\node [style=none] (56) at (23, -13) {};
		\node [style=none] (57) at (-13, 10) {};
		\node [style=none] (58) at (37, 10) {};
		\node [style=none] (59) at (-13, -16) {};
		\node [style=none] (60) at (37, -16) {};
		\node [style=none] (61) at (2, -3) {DynkinBCD};
		\node [style=none] (62) at (5, -3) {\DynkinBCD};
		\node [style=none] (63) at (7.75, -3) {\DynkinBCDbar};
		\node [style=none] (64) at (15.25, -3) {\DynkinBCDmirbar};
		\node [style=none] (65) at (18.25, -3) {\DynkinBCDmir};
		\node [style=none] (66) at (21.5, -3) {$\text{DynkinBCD}_{\text{\tiny mirror}}$};
		\node [style=none] (67) at (24.5, -3) {$T^\sigma_\rho(BCD)$};
		\node [style=none] (68) at (-0.75, -3) {DynkinA};
		\node [style=none] (69) at (0.25, -6.375) {\DynkinA};
		\node [style=none] (70) at (0, 0.25) {\DynkinAmir};
		\node [style=none] (71) at (-3.5, -3) {\DynkinAbar};
		\node [style=none] (72) at (8, -8.5) {Other U(1)};
		\node [style=none] (73) at (12, 2.75) {\LARGE{Lagrangian/quiver theory with Lagrangian/quiver mirror dual}};
		\node [style=none] (74) at (12, 7.25) {\LARGE{Lagrangian/quiver theory with Non-Lagrangian/quiver mirror dual}};
		\node [style=none] (75) at (12, 10.75) {\LARGE{Non-Lagrangian/quiver theory}};
		\node [style=none] (76) at (8, -9) {extensions};
		\node [style=none] (77) at (7.5, -1) {};
		\node [style=none] (78) at (15.5, -1) {};
		\node [style=none] (79) at (19, -1) {};
		\node [style=none] (80) at (4.5, -1) {};
		\node [style=none] (81) at (2, -1) {};
		\node [style=none] (82) at (21.75, -1) {};
		\node [style=none] (83) at (24, -2) {};
		\node [style=none] (84) at (25, -2) {};
		\node [style=none] (85) at (-0.75, -2) {};
		\node [style=none] (86) at (0.25, -2) {};
		\node [style=none] (87) at (-3, 0) {};
		\node [style=none] (88) at (-1.75, 0) {};
		\node [style=none] (89) at (-1, 0) {};
		\node [style=none] (90) at (-1, -6) {};
		\node [style=none] (91) at (7.5, -8) {};
		\node [style=none] (92) at (8.5, -8) {};
		\node [style=none] (93) at (15, 9) {};
		\node [style=none] (94) at (19, 12) {};
		\node [style=none] (95) at (-3, 3.5) {};
		\node [style=none] (96) at (0.25, 5.5) {};
		\node [style=none] (97) at (-3, 5.5) {};
		\node [style=none] (98) at (0.25, 3.5) {};
		\node [style=none] (99) at (-1.5, 4.5) {Affine};
		\node [style=none] (101) at (-1.25, 5.25) {};
		\node [style=none] (102) at (1.75, 5.25) {};
		\node [style=none] (103) at (0.25, 3.5) {};
		\node [style=none] (104) at (3.5, 5.5) {};
		\node [style=none] (105) at (0.25, 5.5) {};
		\node [style=none] (106) at (3.5, 3.5) {};
		\node [style=none] (108) at (2, 4.5) {Affine};
		\node [style=none] (109) at (-1.25, 5.25) {};
		\node [style=none] (110) at (1.75, 5.25) {};
		\node [style=none] (111) at (-1.5, 4) {DynkinABCD};
		\node [style=none] (112) at (2, 4) {$\text{DynkinABCD}^{\text{mirror}}$};
	\end{pgfonlayer}
	\begin{pgfonlayer}{edgelayer}
		\draw (1.center) to (2.center);
		\draw (1.center) to (11.center);
		\draw (2.center) to (12.center);
		\draw [bend left=45] (3.center) to (9.center);
		\draw [bend left=45] (9.center) to (4.center);
		\draw [bend right=45] (3.center) to (10.center);
		\draw [bend right=45] (10.center) to (4.center);
		\draw [in=90, out=0] (1.center) to (5.center);
		\draw [bend right=45] (2.center) to (5.center);
		\draw [bend right=45] (11.center) to (7.center);
		\draw [bend left=45] (12.center) to (7.center);
		\draw [in=90, out=0] (11.center) to (8.center);
		\draw [in=-90, out=0] (12.center) to (8.center);
		\draw [bend right] (13.center) to (2.center);
		\draw [bend right] (1.center) to (13.center);
		\draw (14.center) to (23.center);
		\draw (15.center) to (24.center);
		\draw [bend left=45] (16.center) to (21.center);
		\draw [bend left=45] (21.center) to (17.center);
		\draw [bend right=45] (16.center) to (22.center);
		\draw [bend right=45] (22.center) to (17.center);
		\draw [in=90, out=135] (23.center) to (19.center);
		\draw [in=-90, out=-135] (24.center) to (19.center);
		\draw [bend right=60] (25.center) to (15.center);
		\draw [bend right=60] (14.center) to (25.center);
		\draw (28.center) to (30.center);
		\draw (30.center) to (29.center);
		\draw (28.center) to (31.center);
		\draw (31.center) to (29.center);
		\draw [in=-180, out=90] (32.center) to (33.center);
		\draw [in=-180, out=0, looseness=0.75] (33.center) to (34.center);
		\draw [in=180, out=0] (38.center) to (37.center);
		\draw [in=180, out=-90] (32.center) to (39.center);
		\draw [in=-180, out=0, looseness=0.75] (39.center) to (38.center);
		\draw [in=0, out=90] (52.center) to (53.center);
		\draw [in=0, out=-180, looseness=0.75] (53.center) to (54.center);
		\draw [in=0, out=-90] (52.center) to (56.center);
		\draw [style=rede, in=-180, out=0, looseness=0.75] (54.center) to (53.center);
		\draw [style=rede, in=0, out=-180, looseness=0.75] (54.center) to (33.center);
		\draw [style=rede, bend left=315] (33.center) to (32.center);
		\draw [style=rede, bend left=45] (39.center) to (32.center);
		\draw [style=rede, in=180, out=0, looseness=0.75] (39.center) to (55.center);
		\draw [style=rede, in=180, out=0] (55.center) to (56.center);
		\draw [style=rede, bend right=45] (56.center) to (52.center);
		\draw [style=rede, bend right=45] (52.center) to (53.center);
		\draw [style=pinkline] (58.center) to (60.center);
		\draw [style=pinkline] (60.center) to (59.center);
		\draw [style=pinkline] (57.center) to (59.center);
		\draw [style=pinkline] (57.center) to (58.center);
		\draw [style=thickred, in=-180, out=0, looseness=0.75] (54.center) to (53.center);
		\draw [style=thickred, bend left=45] (53.center) to (52.center);
		\draw [style=thickred, bend left=45] (52.center) to (56.center);
		\draw [style=thickred, in=0, out=180] (56.center) to (55.center);
		\draw [style=thickred, in=0, out=-180, looseness=0.75] (55.center) to (39.center);
		\draw [style=thickred, bend left=45] (39.center) to (32.center);
		\draw [style=thickred, bend left=45] (32.center) to (33.center);
		\draw [style=thickred, in=180, out=0, looseness=0.75] (33.center) to (54.center);
		\draw [style=bluedouble, bend left=15] (77.center) to (78.center);
		\draw [style=bluedouble, bend left=15] (80.center) to (79.center);
		\draw [style=bluedouble, bend left=15] (81.center) to (82.center);
		\draw [style=bluedouble, bend left=120, looseness=4.00] (83.center) to (84.center);
		\draw [style=bluedouble, bend left=120, looseness=4.00] (85.center) to (86.center);
		\draw [style=bluedouble, bend left=120, looseness=4.00] (87.center) to (88.center);
		\draw [style=bluedouble, bend right=75, looseness=2.75] (89.center) to (90.center);
		\draw [style=bluedouble, bend left=120, looseness=4.00] (91.center) to (92.center);
		\draw [style=bluedouble, bend right=15] (94.center) to (93.center);
		\draw [bend left=15] (24.center) to (23.center);
		\draw [bend left=15] (23.center) to (24.center);
		\draw (95.center) to (97.center);
		\draw (97.center) to (96.center);
		\draw (95.center) to (98.center);
		\draw (98.center) to (96.center);
		\draw (103.center) to (105.center);
		\draw (105.center) to (104.center);
		\draw (103.center) to (106.center);
		\draw (106.center) to (104.center);
		\draw [style=bluedouble, bend left=60] (109.center) to (110.center);
	\end{pgfonlayer}
\end{tikzpicture}
}
    \caption{The landscape of 3d mirrors. The blue bridges represents 3d mirror pairs. The red line is the bound set by our  \hyperref[conjecture]{conjecture} where Lagrangian/quiver theories no longer have Lagrangian/quiver mirror duals. }
    \label{ultimatetable}
\end{figure}

\end{landscape}
\FloatBarrier
\paragraph{Notations}
Summary of notations and names of families we will be using. 
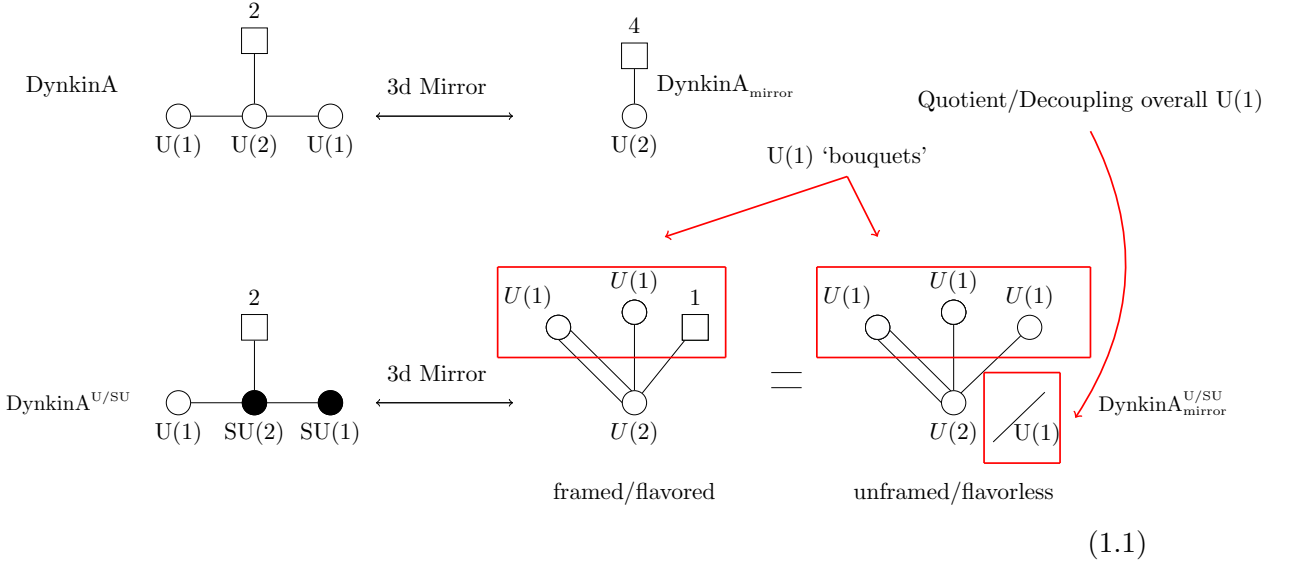
\begin{equation}
    \scalebox{0.8}{\begin{tikzpicture}
	\begin{pgfonlayer}{nodelayer}
		\node [style=gauge3] (0) at (-4, 0) {};
		\node [style=gauge3] (1) at (-2.75, 0) {};
		\node [style=gauge3] (2) at (-1.5, 0) {};
		\node [style=flavour2] (3) at (-2.75, 1.25) {};
		\node [style=none] (4) at (-2.75, 1.75) {2};
		\node [style=none] (5) at (-4, -0.5) {U(1)};
		\node [style=none] (6) at (-2.75, -0.5) {U(2)};
		\node [style=none] (7) at (-1.5, -0.5) {U(1)};
		\node [style=none] (8) at (-0.75, 0) {};
		\node [style=none] (9) at (1.5, 0) {};
		\node [style=gauge3] (10) at (3.5, 0) {};
		\node [style=flavour2] (11) at (3.5, 1) {};
		\node [style=none] (12) at (3.5, -0.5) {U(2)};
		\node [style=none] (13) at (3.5, 1.5) {4};
		\node [style=none] (14) at (0.25, 0.5) {3d Mirror};
		\node [style=gauge3] (15) at (-4, -4.75) {};
		\node [style=gauge3] (16) at (-2.75, -4.75) {};
		\node [style=gauge3] (17) at (-1.5, -4.75) {};
		\node [style=flavour2] (18) at (-2.75, -3.5) {};
		\node [style=none] (19) at (-2.75, -3) {2};
		\node [style=none] (20) at (-4, -5.25) {U(1)};
		\node [style=none] (21) at (-2.75, -5.25) {SU(2)};
		\node [style=none] (22) at (-1.5, -5.25) {SU(1)};
		\node [style=none] (23) at (-0.75, -4.75) {};
		\node [style=none] (24) at (1.5, -4.75) {};
		\node [style=none] (29) at (0.25, -4.25) {3d Mirror};
		\node [style=none] (30) at (-5.75, 0.5) {DynkinA};
		\node [style=none] (31) at (-5.75, -4.75) {\DynkinA};
		\node [style=blacknode] (32) at (-2.75, -4.75) {};
		\node [style=blacknode] (33) at (-1.5, -4.75) {};
		\node [style=gauge3] (34) at (3.5, -4.75) {};
		\node [style=gauge3] (35) at (2.25, -3.5) {};
		\node [style=gauge3] (36) at (3.5, -3.25) {};
		\node [style=none] (37) at (1.75, -3) {$U(1)$};
		\node [style=none] (38) at (3.5, -2.75) {$U(1)$};
		\node [style=none] (39) at (4.5, -3) {1};
		\node [style=none] (40) at (3.5, -5.25) {$U(2)$};
		\node [style=flavour2] (41) at (4.5, -3.5) {};
		\node [style=pinklinet] (42) at (3.5, -3.25) {};
		\node [style=blueet] (43) at (2.25, -3.5) {};
		\node [style=greenflavor] (44) at (4.5, -3.5) {};
		\node [style=brownlinet] (45) at (3.5, -3.25) {};
		\node [style=none] (46) at (2.125, -3.575) {};
		\node [style=none] (47) at (2.3, -3.4) {};
		\node [style=none] (48) at (3.375, -4.825) {};
		\node [style=none] (49) at (3.55, -4.65) {};
		\node [style=gauge3] (50) at (3.5, -3.25) {};
		\node [style=gauge3] (51) at (2.25, -3.5) {};
		\node [style=flavour2] (52) at (4.5, -3.5) {};
		\node [style=gauge3] (53) at (8.75, -4.75) {};
		\node [style=gauge3] (54) at (7.5, -3.5) {};
		\node [style=gauge3] (55) at (8.75, -3.25) {};
		\node [style=none] (56) at (7, -3) {$U(1)$};
		\node [style=none] (57) at (8.75, -2.75) {$U(1)$};
		\node [style=none] (58) at (10, -3) {$U(1)$};
		\node [style=none] (59) at (8.75, -5.25) {$U(2)$};
		\node [style=pinklinet] (61) at (8.75, -3.25) {};
		\node [style=blueet] (62) at (7.5, -3.5) {};
		\node [style=brownlinet] (64) at (8.75, -3.25) {};
		\node [style=none] (65) at (7.375, -3.575) {};
		\node [style=none] (66) at (7.55, -3.4) {};
		\node [style=none] (67) at (8.625, -4.825) {};
		\node [style=none] (68) at (8.8, -4.65) {};
		\node [style=gauge3] (69) at (8.75, -3.25) {};
		\node [style=gauge3] (70) at (7.5, -3.5) {};
		\node [style=gauge3] (71) at (10, -3.5) {};
		\node [style=none] (72) at (5.75, -4.25) {};
		\node [style=none] (73) at (6.25, -4.25) {};
		\node [style=none] (74) at (6.25, -4.5) {};
		\node [style=none] (75) at (5.75, -4.5) {};
		\node [style=none] (76) at (9.4, -5.4) {};
		\node [style=none] (77) at (10.25, -4.575) {};
		\node [style=none] (78) at (10.125, -5.275) {U(1)};
		\node [style=none] (79) at (5, 0.5) {$\text{DynkinA}_{\text{\tiny mirror}}$};
		\node [style=none] (80) at (12.25, -4.75) {\DynkinAmir};
		\node [style=none] (81) at (1.25, -2.5) {};
		\node [style=none] (82) at (1.25, -4) {};
		\node [style=none] (83) at (5, -4) {};
		\node [style=none] (84) at (5, -2.5) {};
		\node [style=none] (85) at (6.5, -2.5) {};
		\node [style=none] (86) at (6.5, -4) {};
		\node [style=none] (87) at (11, -4) {};
		\node [style=none] (88) at (11, -2.5) {};
		\node [style=none] (89) at (7, -1) {};
		\node [style=none] (90) at (4, -2) {};
		\node [style=none] (91) at (7.5, -2) {};
		\node [style=none] (92) at (4, -2) {};
		\node [style=none] (93) at (4, -2) {};
		\node [style=none] (94) at (7, -0.675) {U(1) `bouquets'};
		\node [style=none] (95) at (3.5, -6.25) {framed/flavored};
		\node [style=none] (96) at (8.75, -6.25) {unframed/flavorless};
		\node [style=none] (97) at (9.25, -4.25) {};
		\node [style=none] (98) at (9.25, -5.75) {};
		\node [style=none] (99) at (10.5, -5.75) {};
		\node [style=none] (100) at (10.5, -4.25) {};
		\node [style=none] (101) at (11, -0.25) {};
		\node [style=none] (102) at (10.75, -5) {};
		\node [style=none] (103) at (11, 0.25) {Quotient/Decoupling overall U(1)};
	\end{pgfonlayer}
	\begin{pgfonlayer}{edgelayer}
		\draw (0) to (2);
		\draw (1) to (3);
		\draw [style=->] (8.center) to (9.center);
		\draw [style=->] (9.center) to (8.center);
		\draw (11) to (10);
		\draw (15) to (17);
		\draw (16) to (18);
		\draw [style=->] (23.center) to (24.center);
		\draw [style=->] (24.center) to (23.center);
		\draw (34) to (36);
		\draw (34) to (41);
		\draw (46.center) to (48.center);
		\draw (49.center) to (47.center);
		\draw (53) to (55);
		\draw (65.center) to (67.center);
		\draw (68.center) to (66.center);
		\draw (71) to (67.center);
		\draw (74.center) to (75.center);
		\draw (72.center) to (73.center);
		\draw (77.center) to (76.center);
		\draw [style=reddashed] (81.center) to (84.center);
		\draw [style=reddashed] (84.center) to (83.center);
		\draw [style=reddashed] (83.center) to (82.center);
		\draw [style=reddashed] (82.center) to (81.center);
		\draw [style=reddashed] (85.center) to (86.center);
		\draw [style=reddashed] (86.center) to (87.center);
		\draw [style=reddashed] (87.center) to (88.center);
		\draw [style=reddashed] (88.center) to (85.center);
		\draw [style=redarrow] (89.center) to (93.center);
		\draw [style=redarrow] (89.center) to (91.center);
		\draw [style=reddashed] (97.center) to (98.center);
		\draw [style=reddashed] (98.center) to (99.center);
		\draw [style=reddashed] (99.center) to (100.center);
		\draw [style=reddashed] (100.center) to (97.center);
		\draw [style=redarrow, bend left] (101.center) to (102.center);
	\end{pgfonlayer}
\end{tikzpicture}
}
\end{equation}

\begin{itemize}
    \item \textbf{Black nodes}: Special unitary gauge groups. 
    \item \textbf{Linear quiver}: Gauge groups form a linear line. 
    \item \textbf{DynkinA}: Linear quiver (Shape of A-type finite Dynkin diagram). All gauge groups are unitary. 
    \item \textbf{$\text{DynkinA}_{\text{\tiny mirror}}$}: 3d mirror of DynkinA. 
    \item \textbf{$\text{DynkinA}^{\text{\tiny{SU}}}$}: Linear quiver. All gauge groups are special unitary. 
    \item \textbf{$\text{DynkinA}^{\text{\tiny{SU}}}_{\text{\tiny mirror}}$}: 3d mirror of $\text{DynkinA}^{\text{\tiny{SU}}}$.  
    \item \textbf{\DynkinA}: Linear quiver. Gauge groups are a mixture of unitary and special unitary gauge groups. 
    \item \textbf{\DynkinAmir}: 3d mirror of \DynkinA.
     \item \textbf{\DynkinABCD}: Quiver in the shape of finite ABCD Dynkin diagrams. Gauge groups are a mixture of unitary and special unitary gauge groups. 
    \item \textbf{\DynkinABCDmir}: 3d mirror of \DynkinABCD. 
    \item \textbf{U(1) Bouquets}: The changing of unitary to special unitary gauge groups causes the flavor nodes in the 3d mirror to become a `bouquet' of U(1)s. U(1) bouquets are gaugings of U(1) flavor symmetries due to the gauging of U(1) topological symmetries in the original theory. 
    \item \textbf{Framed/Flavored}: Quivers that contains explicit flavor nodes. 
    \item \textbf{Unframed/Flavorless}: Quivers that do not have an explicit flavor node. For quivers with only unitary gauge groups that are unframed/flavorless, we always quotient/decouple an overall U(1) (see above). 
\end{itemize}

\section{Expanding the landscape}\label{sect2}
We will now focus on  \DynkinABCD theories, which are made of unitary and special unitary gauge groups, and their 3d mirrors. For 3d mirror pairs to be well defined, we need to focus on quivers with gauge groups that are \textbf{good} in the sense of \cite{Gaiotto:2008ak} where each $U(N_i)$ or $SU(N_i)$ gauge group is connected to $N_f \geq 2N_i$ hypermultiplets. This is an important condition and if not satisfied, the Coulomb branch may no longer be a  hyperK\"ahler cone and therefore cannot be matched with any Higgs branches which are always hyperK\"ahler cones (or unions thereof)\footnote{For $SU(N_i)$ with $N_f=2N_i-1$, the Coulomb branch \emph{might} still be a hyperK\"ahler cone. But as it is less clear if it is always the case, we will not look at such cases. Furthermore, a \emph{good} (special) orthogonal or symplectic gauge node is less well defined but since \DynkinABCD  quivers are all made of unitary and special unitary gauge nodes, we do not have to worry about those.}. 

When all gauge groups are unitary,  $\text{DynkinA}$  quivers can be constructed using D3-D5-NS5 brane systems \cite{Hanany:1996ie} whereas DynkinBCD  quivers also require inserting an ON orientifold plane with appropriate charges \cite{Hanany:1999sj,Cremonesi:2014xha}. The 3d mirror duals can then be found after taking the S-duality. For $\text{DynkinA}$ quivers, the 3d mirror will also be an $\text{DynkinA}$ quiver with unitary gauge groups. For DynkinBCD, the 3d mirror will be a linear quiver with mixed unitary and (special) orthogonal/symplectic gauge groups. For DynkinAD, the duality can be checked through explicit computation of Higgs branch and Coulomb branch Hilbert series for the pair of quivers. On the otherhand, for DynkinBC, the quivers are non-simply laced and it is not known how to compute partition functions such as the Hilbert series for their Higgs branch. Nevertheless, their Coulomb branches can be studied using the monopole formula \cite{Cremonesi:2013lqa} and compared to the Higgs branch of the unitary-orthosymplectic mirror dual \DynkinBCmirr. Everything so far are known results in the literature.  One of the two goals of this paper is to greatly expand the landscape of 3d mirror pairs by taking DynkinG theories and replacing any number of the unitary gauge groups with special unitary gauge groups ($\text{DynkinG}^{\text{\tiny U/SU}}$) and finding their mirror pairs ($\text{DynkinG}^{\text{\tiny U/SU}}_{\text{\tiny mirror}}$). 

In \cite{USU}, a novel mechanism called \emph{brane locking} was introduced that allows us to study the moduli spaces of quivers with mixed unitary and special unitary gauge groups: \DynkinA. Starting with a $5d$ $\mathcal{N}=1$ gauge theory with only special unitary gauge groups \DynkinASU, a brane set up describing this theory can be constructed using $(p,q)$5-branes and $[p,q]$7-branes. Changing SU gauge groups to U gauge groups in the 5d theory then translates to \emph{locking} different 5-branes/subwebs by forcing them to move together. This is because each locking reduces the  dynamical degrees of freedom by one. From this brane web, a \emph{magnetic} quiver \cite{Cabrera:2018jxt} can then be read off which is defined as a $3d$ $\mathcal{N}=4$ theory whose Coulomb branch is the same as the Higgs branch of the $5d$ $\mathcal{N}=1$ \DynkinA quiver . 

For the goal of this paper, rather than studying the relation between a 5d theory and its magnetic quiver, we are interested in 3d mirror pairs. From non-renormalization theorem \cite{Argyres:1996eh} it is well-know that the classical Higgs branch\footnote{At the UV superconformal fixed point, it is well known that the 5d Higgs branch will receive quantum corrections. However, in the brane web, we never go to the fixed point and always maintains a finite gauge coupling. Therefore, the Higgs branch we see in our brane web set up is exactly the classical Higgs branch of the theory.}  of $5d$ $\mathcal{N}=1$ and $3d$ $\mathcal{N}=4$ theories are the same. Therefore, whatever conclusion we drew upon the Higgs branch of the \DynkinA 5d theory will be the same as if the quiver (with the same gauge and flavor groups) were a 3d theory. From this perspective, we view \DynkinA as a $3d$ $\mathcal{N}=4$ theory, making both \DynkinA and \DynkinAmir $3d$ $\mathcal{N}=4$ theories. Since \DynkinA are \emph{good} quivers, we further conjecture that the $3d$ Coulomb branch of \DynkinA coincides with the $3d$ Higgs branch of \DynkinAmir, thereby forming 3d mirror pairs. This non-trivial conjecture has been checked in a slew of examples in \cite{USU} through explicit Hilbert series computations.\newline
\\
In summary, we are still looking at a pair of 3d $\mathcal{N}=4$ theories (\DynkinA and \DynkinAmir), and the usage of 5d brane webs is just a detour to obtain the correct pairs. 
\\

In this paper, we extend the above mechanism for DynkinBCD quivers with mixed U \& SU gauge groups: \DynkinBCD. The new addition to the brane set up is an ON orientifold plane \cite{Hanany:1999sj} with the appropriate charges which can also be \emph{locked} together with the branes. By only focusing on good quivers, the brane web set up takes a particularly simple form with many features parallel to the D3-D5-NS5 set up of the same quiver. This similarity is expanded upon in Appendix \ref{similarbranes}. Where possible, the Coulomb branch and Higgs branch Hilbert series on both side are computed and are found to be consistent with them being mirror pairs. 

\subsection{Brane locking and \texorpdfstring{$A$}{} type Dynkin quivers: (\texorpdfstring{\DynkinA}{}) }\label{sec2}
A $\text{DynkinA}$ quiver is a quiver where the gauge groups connect to form a finite-A type Dynkin diagram, i.e a linear chain. In this section, we briefly summarize the brane locking mechanism introduced for $\text{DynkinA}$ quivers in \cite{USU}. 

\subsubsection{Brane webs -- \texorpdfstring{$(p,q)$}{}5-brane and \texorpdfstring{$[p,q]$}{}7-brane set up}
To begin, we take a look at the following $\text{DynkinA}$ quiver and its 3d mirror:
\begin{equation}
\scalebox{0.8}{\begin{tikzpicture}
	\begin{pgfonlayer}{nodelayer}
		\node [style=gauge3] (0) at (-4, 0) {};
		\node [style=gauge3] (1) at (-2.75, 0) {};
		\node [style=gauge3] (2) at (-1.5, 0) {};
		\node [style=flavour2] (3) at (-2.75, 1.25) {};
		\node [style=none] (4) at (-2.75, 1.75) {2};
		\node [style=none] (5) at (-4, -0.5) {U(1)};
		\node [style=none] (6) at (-2.75, -0.5) {U(2)};
		\node [style=none] (7) at (-1.5, -0.5) {U(1)};
		\node [style=none] (8) at (-0.75, 0) {};
		\node [style=none] (9) at (1.5, 0) {};
		\node [style=gauge3] (10) at (2.75, 0) {};
		\node [style=flavour2] (11) at (2.75, 1) {};
		\node [style=none] (12) at (2.75, -0.5) {U(2)};
		\node [style=none] (13) at (2.75, 1.5) {4};
		\node [style=none] (14) at (0.25, 0.5) {3d Mirror};
	\end{pgfonlayer}
	\begin{pgfonlayer}{edgelayer}
		\draw (0) to (2);
		\draw (1) to (3);
		\draw [style=->] (8.center) to (9.center);
		\draw [style=->] (9.center) to (8.center);
		\draw (11) to (10);
	\end{pgfonlayer}
\end{tikzpicture}
}
\label{firstmirror}
\end{equation}
Following the same steps as in \cite{USU}, we first study this quiver as a 5d $\mathcal{N}=1$ theory using the brane web set up pioneered in \cite{Aharony:1997bh}. However, a vanilla brane web set up cannot describe unitary gauge groups as U(1) degrees of freedom decouples from each of them, leading to a gauge theory with only special unitary gauge groups (from now on colored as black nodes):
\begin{equation}
\scalebox{0.8}{\begin{tikzpicture}
	\begin{pgfonlayer}{nodelayer}
		\node [style=gauge3] (0) at (-3.75, 0) {};
		\node [style=gauge3] (1) at (-2.5, 0) {};
		\node [style=gauge3] (2) at (-1.25, 0) {};
		\node [style=flavour2] (3) at (-2.5, 1.25) {};
		\node [style=none] (4) at (-2.5, 1.75) {2};
		\node [style=none] (5) at (-3.75, -0.5) {SU(1)};
		\node [style=none] (6) at (-2.5, -0.5) {SU(2)};
		\node [style=none] (7) at (-1.25, -0.5) {SU(1)};
		\node [style=blacknode] (8) at (-3.75, 0) {};
		\node [style=blacknode] (9) at (-2.5, 0) {};
		\node [style=blacknode] (10) at (-1.25, 0) {};
	\end{pgfonlayer}
	\begin{pgfonlayer}{edgelayer}
		\draw (0) to (2);
		\draw (1) to (3);
	\end{pgfonlayer}
\end{tikzpicture}
}
\end{equation}
The SU(1) gauge groups just contributes as a flavor U(1), nevertheless, we write it as a gauge group just to keep track of which gauge nodes are turned from U to SU. Since all the gauge nodes are SU, it is part of the  \DynkinASU family. The brane web is:
\begin{equation}
\scalebox{0.8}{\begin{tikzpicture}
	\begin{pgfonlayer}{nodelayer}
		\node [style=gauge3] (0) at (2.75, 1.25) {};
		\node [style=gauge3] (1) at (1.75, -1) {};
		\node [style=none] (2) at (5, 0) {};
		\node [style=none] (3) at (5, 1) {};
		\node [style=none] (4) at (6.5, -1) {};
		\node [style=none] (5) at (8.25, -1) {};
		\node [style=none] (6) at (8.25, 0) {};
		\node [style=none] (7) at (7, 1) {};
		\node [style=gauge3] (8) at (4, 2) {};
		\node [style=gauge3] (9) at (7, 2) {};
		\node [style=gauge3] (10) at (9.25, -2) {};
		\node [style=gauge3] (11) at (6.5, -2) {};
		\node [style=none] (12) at (10, 0) {};
		\node [style=gauge3] (13) at (10, 1) {};
		\node [style=gauge3] (14) at (11, -1) {};
		\node [style=gauge3] (15) at (5.5, 1.5) {};
		\node [style=gauge3] (16) at (6.5, 1.5) {};
		\node [style=none] (17) at (0.5, -0.5) {};
		\node [style=none] (18) at (0.5, -2.5) {};
		\node [style=none] (19) at (2.5, -2.5) {};
		\node [style=gauge1] (20) at (0.5, -2.5) {};
		\node [style=none] (21) at (2.925, -2.5) {$x^6$};
		\node [style=none] (22) at (0.5, -3.1) {$x^7,x^8,x^9$};
		\node [style=none] (23) at (0.5, -2.5) {{\Huge $\times$}};
		\node [style=none] (24) at (0.5, 0) {$x^5$};
		\node [style=none] (25) at (10.5, 0.5) {NS5};
		\node [style=none] (26) at (3.75, -0.5) {D5};
		\node [style=none] (27) at (8.25, 1) {(1,1)5-brane};
		\node [style=none] (28) at (10, 1.5) {[0,1]7-brane};
		\node [style=none] (29) at (5.5, 2) {D7};
		\node [style=none] (30) at (12.25, -1) {[1,1]7-brane};
		\node [style=none] (31) at (2.75, 0) {};
	\end{pgfonlayer}
	\begin{pgfonlayer}{edgelayer}
		\draw (5.center) to (10);
		\draw (5.center) to (6.center);
		\draw (6.center) to (12.center);
		\draw (12.center) to (13);
		\draw (12.center) to (14);
		\draw (7.center) to (6.center);
		\draw (7.center) to (9);
		\draw (7.center) to (3.center);
		\draw (3.center) to (8);
		\draw (2.center) to (4.center);
		\draw (2.center) to (3.center);
		\draw (4.center) to (5.center);
		\draw (4.center) to (11);
		\draw (18.center) to (20.center);
		\draw [style=->] (18.center) to (17.center);
		\draw [style=->] (18.center) to (19.center);
		\draw (0) to (31.center);
		\draw (31.center) to (2.center);
		\draw (31.center) to (1);
	\end{pgfonlayer}
\end{tikzpicture}}
\end{equation}
where horizontal lines are D5 branes, vertical lines are NS5 branes, sloped lines are (1,1)5-branes and nodes are 7-branes as labeled. This brane web describes the Coulomb branch phase of the 5d theory which has one real dimension. The stretching and contracting of the six-sided polygon parameterizes the VEV of the scalar field in the vector multiplet. The next step is to enter the Higgs branch phase by setting the masses of the hypermultiplets to zero:
\begin{equation}
\scalebox{0.8}{\begin{tikzpicture}
	\begin{pgfonlayer}{nodelayer}
		\node [style=none] (20) at (0.25, -0.75) {};
		\node [style=none] (21) at (0.25, -2.75) {};
		\node [style=none] (22) at (2.25, -2.75) {};
		\node [style=gauge1] (23) at (0.25, -2.75) {};
		\node [style=none] (24) at (2.675, -2.75) {$x^6$};
		\node [style=none] (25) at (0.25, -3.35) {$x^7,x^8,x^9$};
		\node [style=none] (26) at (0.25, -2.75) {{\Huge $\times$}};
		\node [style=none] (27) at (0.25, -0.25) {$x^5$};
		\node [style=gauge3] (28) at (3, 0) {};
		\node [style=gauge3] (29) at (6.75, 0) {};
		\node [style=none] (30) at (3, 0.125) {};
		\node [style=none] (31) at (3, -0.125) {};
		\node [style=none] (32) at (6.75, 0.125) {};
		\node [style=none] (33) at (6.75, -0.125) {};
		\node [style=gauge3] (34) at (3.75, 1.5) {};
		\node [style=gauge3] (35) at (3.75, -1.5) {};
		\node [style=gauge3] (36) at (4.5, 1.5) {};
		\node [style=gauge3] (37) at (4.5, -1.5) {};
		\node [style=gauge3] (38) at (5.25, 1.5) {};
		\node [style=gauge3] (39) at (5.25, -1.5) {};
		\node [style=gauge3] (40) at (6, 1.5) {};
		\node [style=gauge3] (41) at (6, -1.5) {};
		\node [style=none] (42) at (8, 0) {};
		\node [style=none] (43) at (10.25, 0) {};
		\node [style=none] (44) at (9, 0.5) {Magnetic Quiver};
		\node [style=gauge3] (45) at (12.75, -1) {};
		\node [style=gauge3] (46) at (11.5, 0.25) {};
		\node [style=gauge3] (47) at (12.25, 0.75) {};
		\node [style=gauge3] (48) at (13.25, 0.75) {};
		\node [style=none] (50) at (11, 0.75) {$U(1)$};
		\node [style=none] (51) at (12.25, 1.25) {$U(1)$};
		\node [style=none] (52) at (13.25, 1.25) {$U(1)$};
		\node [style=none] (53) at (14.5, 0.75) {$U(1)$};
		\node [style=none] (54) at (12.75, -1.5) {$U(2)$};
		\node [style=flavour2] (55) at (14, 0.25) {};
	\end{pgfonlayer}
	\begin{pgfonlayer}{edgelayer}
		\draw (21.center) to (23.center);
		\draw [style=->] (21.center) to (20.center);
		\draw [style=->] (21.center) to (22.center);
		\draw (32.center) to (30.center);
		\draw (31.center) to (33.center);
		\draw (34) to (35);
		\draw (37) to (36);
		\draw (38) to (39);
		\draw (41) to (40);
		\draw [style=->] (42.center) to (43.center);
		\draw (45) to (46);
		\draw (47) to (45);
		\draw (45) to (48);
		\draw (45) to (55);
	\end{pgfonlayer}
\end{tikzpicture}}
\label{firstweb}
\end{equation}
where the magnetic quiver can be obtained following the algorithm outlined in \cite{Cabrera:2018jxt}. Here, we settle on some conventions for the magnetic quiver that is different from other papers such as \cite{USU} where the magnetic quivers are always left \emph{unframed/flavorless}, i.e without any explicit flavor node. For quivers with only unitary gauge groups, it is always implied that an overall freely acting U(1) decouples which corresponds to fixing the center of mass of the brane system. However, later on we will be studying  unitary-orthosymplectic quivers where such automatic decoupling of U(1) does not occur for unframed/flavourless quivers. Hence, for clarity and consistency, we will always frame the magnetic quivers (by decoupling any freely acting U(1)) whenever possible in this section. 

Every set of branes/subwebs that move independently from others in the brane system will contribute to a gauge group in the magnetic quiver \cite{Cabrera:2018jxt}. In 5d, the NS5s carry dynamical degrees of freedom and therefore each of them correspond to one of the U(1) gauge nodes in the magnetic quiver. To make this clear, we choose a different color to represent the NS5s that are free to move independently from each other and the corresponding gauge nodes in the magnetic quiver: 
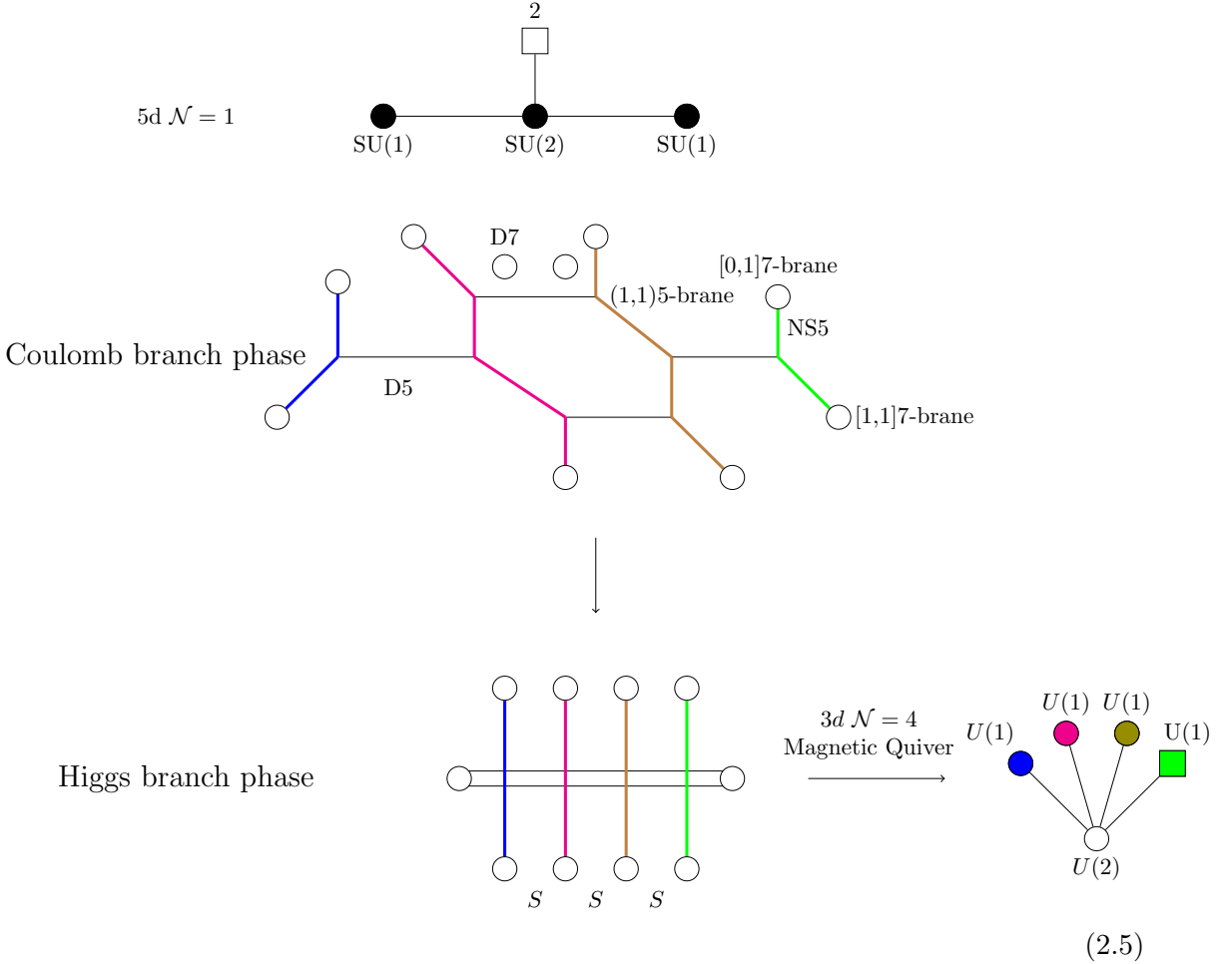
\begin{equation}
\scalebox{0.8}{\begin{tikzpicture}
	\begin{pgfonlayer}{nodelayer}
		\node [style=gauge3] (0) at (-2.5, 2) {};
		\node [style=gauge3] (2) at (-3.5, -0.25) {};
		\node [style=none] (3) at (-0.25, 0.75) {};
		\node [style=none] (4) at (-0.25, 1.75) {};
		\node [style=none] (6) at (1.25, -0.25) {};
		\node [style=none] (8) at (3, -0.25) {};
		\node [style=none] (9) at (3, 0.75) {};
		\node [style=none] (10) at (1.75, 1.75) {};
		\node [style=gauge3] (11) at (-1.25, 2.75) {};
		\node [style=gauge3] (12) at (1.75, 2.75) {};
		\node [style=gauge3] (13) at (4, -1.25) {};
		\node [style=gauge3] (14) at (1.25, -1.25) {};
		\node [style=none] (15) at (4.75, 0.75) {};
		\node [style=gauge3] (16) at (4.75, 1.75) {};
		\node [style=gauge3] (17) at (5.75, -0.25) {};
		\node [style=gauge3] (18) at (0.25, 2.25) {};
		\node [style=gauge3] (19) at (1.25, 2.25) {};
		\node [style=none] (28) at (5.25, 1.25) {NS5};
		\node [style=none] (29) at (-1.5, 0.25) {D5};
		\node [style=none] (30) at (3, 1.75) {(1,1)5-brane};
		\node [style=none] (31) at (4.75, 2.25) {[0,1]7-brane};
		\node [style=none] (32) at (0.25, 2.75) {D7};
		\node [style=none] (33) at (7, -0.25) {[1,1]7-brane};
		\node [style=none] (34) at (-2.5, 0.75) {};
		\node [style=gauge3] (35) at (-0.5, -6.25) {};
		\node [style=gauge3] (36) at (4, -6.25) {};
		\node [style=none] (37) at (-0.5, -6.125) {};
		\node [style=none] (38) at (-0.5, -6.375) {};
		\node [style=none] (39) at (4, -6.125) {};
		\node [style=none] (40) at (4, -6.375) {};
		\node [style=gauge3] (41) at (0.25, -4.75) {};
		\node [style=gauge3] (42) at (0.25, -7.75) {};
		\node [style=gauge3] (43) at (1.25, -4.75) {};
		\node [style=gauge3] (44) at (1.25, -7.75) {};
		\node [style=gauge3] (45) at (2.25, -4.75) {};
		\node [style=gauge3] (46) at (2.25, -7.75) {};
		\node [style=gauge3] (47) at (3.25, -4.75) {};
		\node [style=gauge3] (48) at (3.25, -7.75) {};
		\node [style=none] (49) at (5.25, -6.25) {};
		\node [style=none] (50) at (7.5, -6.25) {};
		\node [style=none] (51) at (6.25, -5.75) {Magnetic Quiver};
		\node [style=gauge3] (52) at (10, -7.25) {};
		\node [style=gauge3] (53) at (8.75, -6) {};
		\node [style=gauge3] (54) at (9.5, -5.5) {};
		\node [style=gauge3] (55) at (10.5, -5.5) {};
		\node [style=none] (56) at (8.25, -5.5) {$U(1)$};
		\node [style=none] (57) at (9.5, -5) {$U(1)$};
		\node [style=none] (58) at (10.5, -5) {$U(1)$};
		\node [style=none] (59) at (11.5, -5.5) {U(1)};
		\node [style=none] (60) at (10, -7.75) {$U(2)$};
		\node [style=flavour2] (61) at (11.25, -6) {};
		\node [style=none] (62) at (0.75, -8.25) {$S$};
		\node [style=none] (63) at (1.75, -8.25) {$S$};
		\node [style=none] (64) at (2.75, -8.25) {$S$};
		\node [style=pinklinet] (65) at (10.5, -5.5) {};
		\node [style=pinklinet] (66) at (9.5, -5.5) {};
		\node [style=blueet] (67) at (8.75, -6) {};
		\node [style=greenflavor] (68) at (11.25, -6) {};
		\node [style=brownlinet] (69) at (10.5, -5.5) {};
		\node [style=none] (70) at (-5.5, 0.75) {\Large{Coulomb branch phase}};
		\node [style=none] (71) at (-5, -6.25) {\Large{Higgs  branch phase}};
		\node [style=none] (72) at (1.75, -2.25) {};
		\node [style=none] (73) at (1.75, -3.5) {};
		\node [style=gauge3] (74) at (-1.75, 4.75) {};
		\node [style=gauge3] (75) at (0.75, 4.75) {};
		\node [style=gauge3] (76) at (3.25, 4.75) {};
		\node [style=flavour2] (77) at (0.75, 6) {};
		\node [style=none] (78) at (0.75, 6.5) {2};
		\node [style=none] (79) at (-1.75, 4.25) {SU(1)};
		\node [style=none] (80) at (0.75, 4.25) {SU(2)};
		\node [style=none] (81) at (3.25, 4.25) {SU(1)};
		\node [style=none] (82) at (-5, 4.75) {5d $\mathcal{N}=1$};
		\node [style=none] (83) at (6.25, -5.25) {$3d$ $\mathcal{N}=4$};
		\node [style=blacknode] (84) at (-1.75, 4.75) {};
		\node [style=blacknode] (85) at (0.75, 4.75) {};
		\node [style=blacknode] (86) at (3.25, 4.75) {};
	\end{pgfonlayer}
	\begin{pgfonlayer}{edgelayer}
		\draw (9.center) to (15.center);
		\draw (10.center) to (4.center);
		\draw (6.center) to (8.center);
		\draw (34.center) to (3.center);
		\draw (39.center) to (37.center);
		\draw (38.center) to (40.center);
		\draw [style=->] (49.center) to (50.center);
		\draw (52) to (53);
		\draw (54) to (52);
		\draw (52) to (55);
		\draw (52) to (61);
		\draw [style=pinkline] (43) to (44);
		\draw [style=bluee] (41) to (42);
		\draw [style=greenline] (47) to (48);
		\draw [style=brownline] (45) to (46);
		\draw [style=->] (72.center) to (73.center);
		\draw [style=bluee] (0) to (34.center);
		\draw [style=bluee] (34.center) to (2);
		\draw [style=pinkline] (11) to (4.center);
		\draw [style=pinkline] (4.center) to (3.center);
		\draw [style=pinkline] (3.center) to (6.center);
		\draw [style=pinkline] (6.center) to (14);
		\draw [style=brownline] (12) to (10.center);
		\draw [style=brownline] (10.center) to (9.center);
		\draw [style=brownline] (9.center) to (8.center);
		\draw [style=brownline] (8.center) to (13);
		\draw [style=greenline] (16) to (15.center);
		\draw [style=greenline] (15.center) to (17);
		\draw (74) to (76);
		\draw (75) to (77);
	\end{pgfonlayer}
\end{tikzpicture}
}
\label{braneweb2}
\end{equation}
Here, we also show the origin of the different colored NS5s in the Coulomb branch phase. Above the Coulomb branch phase is the  SU quiver that is stretched so that one can see which gauge group corresponds to which part of the brane set up. The D5 between the blue and pink lines and the D5 between the brown and green lines correspond to the two $SU(1)$ gauge groups. The polygon between the pink and brown lines corresponds to the $SU(2)$ gauge group. The labeling of unitary (U) and special unitary (S) below makes clear the position of the U/SU nodes in the \DynkinASU quiver once we enter the Higgs branch phase. Using this labeling system allows us to immediately see what happens to the magnetic quiver when we turn SU to U nodes in the brane system. 

Following the locking prescription \cite{USU}, turning SU nodes to U nodes in the  \DynkinASU quiver translates to locking NS5 branes so that they move together. Since the otherwise two independent NS5s are forced together, a U(1) factor   decouples. The resulting magnetic quiver takes the form:

\begin{equation}
  \scalebox{0.8}{  \begin{tikzpicture}
	\begin{pgfonlayer}{nodelayer}
		\node [style=gauge3] (28) at (2.25, 0) {};
		\node [style=gauge3] (29) at (6.75, 0) {};
		\node [style=none] (30) at (2.25, 0.125) {};
		\node [style=none] (31) at (2.25, -0.125) {};
		\node [style=none] (32) at (6.75, 0.125) {};
		\node [style=none] (33) at (6.75, -0.125) {};
		\node [style=gauge3] (34) at (3, 1.5) {};
		\node [style=gauge3] (35) at (3, -1.5) {};
		\node [style=gauge3] (36) at (4, 1.5) {};
		\node [style=gauge3] (37) at (4, -1.5) {};
		\node [style=gauge3] (38) at (5, 1.5) {};
		\node [style=gauge3] (39) at (5, -1.5) {};
		\node [style=gauge3] (40) at (6, 1.5) {};
		\node [style=gauge3] (41) at (6, -1.5) {};
		\node [style=none] (42) at (8, 0) {};
		\node [style=none] (43) at (10.25, 0) {};
		\node [style=none] (44) at (9, 0.5) {Magnetic Quiver};
		\node [style=gauge3] (45) at (12.75, -1) {};
		\node [style=gauge3] (46) at (11.5, 0.25) {};
		\node [style=gauge3] (48) at (12.75, 0.5) {};
		\node [style=none] (50) at (11, 0.75) {$U(1)$};
		\node [style=none] (52) at (12.75, 1) {$U(1)$};
		\node [style=none] (53) at (14.5, 0.75) {$U(1)$};
		\node [style=none] (54) at (12.75, -1.5) {$U(2)$};
		\node [style=flavour2] (55) at (14, 0.25) {};
		\node [style=none] (56) at (3.5, -2) {$U$};
		\node [style=none] (57) at (4.5, -2) {$S$};
		\node [style=none] (58) at (5.5, -2) {$S$};
		\node [style=pinklinet] (59) at (12.75, 0.5) {};
		\node [style=blueet] (61) at (11.5, 0.25) {};
		\node [style=greenflavor] (62) at (14, 0.25) {};
		\node [style=brownlinet] (63) at (12.75, 0.5) {};
		\node [style=none] (64) at (11.375, 0.175) {};
		\node [style=none] (65) at (11.55, 0.35) {};
		\node [style=none] (66) at (12.625, -1.075) {};
		\node [style=none] (67) at (12.8, -0.9) {};
	\end{pgfonlayer}
	\begin{pgfonlayer}{edgelayer}
		\draw (32.center) to (30.center);
		\draw (31.center) to (33.center);
		\draw [style=->] (42.center) to (43.center);
		\draw (45) to (48);
		\draw (45) to (55);
		\draw [style=bluee] (34) to (35);
		\draw [style=greenline] (40) to (41);
		\draw [style=brownline] (38) to (39);
		\draw [style=bluee] (36) to (37);
		\draw (64.center) to (66.center);
		\draw (67.center) to (65.center);
	\end{pgfonlayer}
\end{tikzpicture}}
\label{braneweb3}
\end{equation}
where the bouquet of two U(1)s in the magnetic quiver in \eqref{braneweb2} becomes a single U(1) but with a multiplicity two link. The multiplicity of the links is the stable intersection number between the blue NS5s and the D5s when viewed as tropical curves \cite{Cabrera:2018jxt}. For the brane set ups in this section, the intersection number is simply how many times the NS5s intersect the D-branes. A multiplicity $n$ link means $n$ copies of bifundamental hypers in between. 

Massless instanton operators in 5d are known to enhance the Higgs branch, giving it a quantum correction. The distance between two adjacent NS5 branes gives $\frac{1}{g^2}$ where $g$ is the gauge coupling. Since the distances between our NS5s are non-zero, the gauge couplings are all finite. Since the masses of instanton operators that appear in the 5d theory are proportional to  $\frac{1}{g^2}$, they remain massive and do not contribute to enhancing the Higgs branch. Therefore, in our set up, the Higgs branch remains classical. Following the non-renormalization theorem, we know that the classical Higgs branch is the same in $5d$ $\mathcal{N}=1$ and $3d$ $\mathcal{N}=4$. This allows us to view both  \DynkinA  and its magnetic quiver as $3d$ $\mathcal{N}=4$ theories and conjecture that they actually form a 3d mirror pair:
\begin{equation}
  \scalebox{0.8}{ \begin{tikzpicture}
	\begin{pgfonlayer}{nodelayer}
		\node [style=gauge3] (0) at (0, 0) {};
		\node [style=gauge3] (1) at (1.25, 0) {};
		\node [style=gauge3] (2) at (2.5, 0) {};
		\node [style=flavour2] (3) at (1.25, 1.25) {};
		\node [style=none] (4) at (1.25, 1.75) {2};
		\node [style=none] (5) at (0, -0.5) {$U(1)$};
		\node [style=none] (6) at (1.25, -0.5) {$SU(2)$};
		\node [style=none] (7) at (2.5, -0.5) {$SU(1)$};
		\node [style=none] (8) at (3.25, 0) {};
		\node [style=none] (9) at (5.25, 0) {};
		\node [style=none] (10) at (4.25, 0.5) {3d Mirror};
		\node [style=gauge3] (11) at (7.5, 0) {};
		\node [style=gauge3] (12) at (6.25, 1.25) {};
		\node [style=gauge3] (13) at (7.5, 1.5) {};
		\node [style=none] (14) at (5.75, 1.75) {$U(1)$};
		\node [style=none] (15) at (7.5, 2) {$U(1)$};
		\node [style=none] (16) at (9.25, 1.75) {$U(1)$};
		\node [style=none] (17) at (7.5, -0.5) {$U(2)$};
		\node [style=flavour2] (18) at (8.75, 1.25) {};
		\node [style=pinklinet] (19) at (7.5, 1.5) {};
		\node [style=blueet] (20) at (6.25, 1.25) {};
		\node [style=greenflavor] (21) at (8.75, 1.25) {};
		\node [style=brownlinet] (22) at (7.5, 1.5) {};
		\node [style=none] (23) at (6.125, 1.175) {};
		\node [style=none] (24) at (6.3, 1.35) {};
		\node [style=none] (25) at (7.375, -0.075) {};
		\node [style=none] (26) at (7.55, 0.1) {};
		\node [style=gauge3] (27) at (7.5, 1.5) {};
		\node [style=gauge3] (28) at (6.25, 1.25) {};
		\node [style=flavour2] (29) at (8.75, 1.25) {};
		\node [style=blacknode] (30) at (1.25, 0) {};
		\node [style=blacknode] (31) at (2.5, 0) {};
	\end{pgfonlayer}
	\begin{pgfonlayer}{edgelayer}
		\draw (0) to (2);
		\draw (1) to (3);
		\draw [style=->] (8.center) to (9.center);
		\draw [style=->] (9.center) to (8.center);
		\draw (11) to (13);
		\draw (11) to (18);
		\draw (23.center) to (25.center);
		\draw (26.center) to (24.center);
	\end{pgfonlayer}
\end{tikzpicture}}
\end{equation}
Finally, setting all the gauge groups in the \DynkinA quiver to unitary means locking all the NS5s together:
\begin{equation}
   \scalebox{0.8}{ \begin{tikzpicture}
	\begin{pgfonlayer}{nodelayer}
		\node [style=gauge3] (0) at (2.25, 0) {};
		\node [style=gauge3] (1) at (6.75, 0) {};
		\node [style=none] (2) at (2.25, 0.125) {};
		\node [style=none] (3) at (2.25, -0.125) {};
		\node [style=none] (4) at (6.75, 0.125) {};
		\node [style=none] (5) at (6.75, -0.125) {};
		\node [style=gauge3] (6) at (3, 1.5) {};
		\node [style=gauge3] (7) at (3, -1.5) {};
		\node [style=gauge3] (8) at (4, 1.5) {};
		\node [style=gauge3] (9) at (4, -1.5) {};
		\node [style=gauge3] (10) at (5, 1.5) {};
		\node [style=gauge3] (11) at (5, -1.5) {};
		\node [style=gauge3] (12) at (6, 1.5) {};
		\node [style=gauge3] (13) at (6, -1.5) {};
		\node [style=none] (14) at (8, 0) {};
		\node [style=none] (15) at (10, 0) {};
		\node [style=none] (16) at (9, 0.5) {Magnetic Quiver};
		\node [style=gauge3] (17) at (12, -0.75) {};
		\node [style=none] (18) at (12, 1.25) {4};
		\node [style=none] (19) at (12, -1.25) {$U(2)$};
		\node [style=flavour2] (20) at (12, 0.75) {};
		\node [style=none] (21) at (3.5, -2) {$U$};
		\node [style=none] (22) at (4.5, -2) {$U$};
		\node [style=none] (23) at (5.5, -2) {$U$};
		\node [style=greenflavor] (24) at (12, 0.75) {};
		\node [style=none] (25) at (11.875, -0.825) {};
		\node [style=none] (26) at (12.05, -0.65) {};
	\end{pgfonlayer}
	\begin{pgfonlayer}{edgelayer}
		\draw (4.center) to (2.center);
		\draw (3.center) to (5.center);
		\draw [style=->] (14.center) to (15.center);
		\draw (17) to (20);
		\draw [style=greenline] (12) to (13);
		\draw [style=greenline] (11) to (10);
		\draw [style=greenline] (8) to (9);
		\draw [style=greenline] (7) to (6);
	\end{pgfonlayer}
\end{tikzpicture}}
\label{allUSU4}
\end{equation}
which reproduces the 3d mirror in \eqref{firstmirror}. All possible U/SU combinations are given in Table \ref{SU4}. This locking procedure can be generalized to any  \DynkinA quivers as shown in \cite{USU}. All the 3d mirror pairs are also verified through explicit Hilbert series computations.

\paragraph{Broken global symmetry}
Starting with  DynkinA in \eqref{firstmirror}, we see that all the unitary gauge nodes are balanced. The balanced nodes form an $A_3$ finite Dynkin diagram which tells us the Coulomb branch global symmetry is SU(4) \cite{Gaiotto:2008ak}. Turning U nodes to SU breaks the symmetry into a subgroup of SU(4) as the SU node is not balanced. In Table \ref{symmetrybreaking},  we see how the Coulomb branch global symmetry can be determined by the arrangement of U/SU nodes in the quiver. 
\begin{table}
    \centering
    \begin{tabular}{c|c} \hline
    Dynkin diagram & Coulomb Branch Global Symmetry \\ \hline
  \scalebox{0.7}{     \begin{tikzpicture}
	\begin{pgfonlayer}{nodelayer}
		\node [style=gauge3] (0) at (5, 0) {};
		\node [style=gauge3] (1) at (6, 0) {};
		\node [style=gauge3] (2) at (7, 0) {};
		\node [style=none] (3) at (5, -0.5) {U};
		\node [style=none] (4) at (6, -0.5) {U};
		\node [style=none] (5) at (7, -0.5) {U};
	\end{pgfonlayer}
	\begin{pgfonlayer}{edgelayer}
		\draw (0) to (1);
		\draw (1) to (2);
	\end{pgfonlayer}
\end{tikzpicture}}
  & $SU(4)$ \\ \hline 
\scalebox{0.7}{\begin{tikzpicture}
	\begin{pgfonlayer}{nodelayer}
		\node [style=gauge3] (0) at (5, 0) {};
		\node [style=gauge3] (1) at (6, 0) {};
		\node [style=gauge3] (2) at (7, 0) {};
		\node [style=none] (3) at (5, -0.5) {S};
		\node [style=none] (4) at (6, -0.5) {U};
		\node [style=none] (5) at (7, -0.5) {U};
		\node [style=blacknode] (6) at (5, 0) {};
	\end{pgfonlayer}
	\begin{pgfonlayer}{edgelayer}
		\draw (0) to (1);
		\draw (1) to (2);
	\end{pgfonlayer}
\end{tikzpicture}}
       & $SU(1) \times SU(3)$ \\ \hline 
\scalebox{0.7}{\begin{tikzpicture}
	\begin{pgfonlayer}{nodelayer}
		\node [style=gauge3] (0) at (5, 0) {};
		\node [style=gauge3] (1) at (6, 0) {};
		\node [style=gauge3] (2) at (7, 0) {};
		\node [style=none] (3) at (5, -0.5) {U};
		\node [style=none] (4) at (6, -0.5) {S};
		\node [style=none] (5) at (7, -0.5) {U};
		\node [style=blacknode] (6) at (6, 0) {};
	\end{pgfonlayer}
	\begin{pgfonlayer}{edgelayer}
		\draw (0) to (1);
		\draw (1) to (2);
	\end{pgfonlayer}
\end{tikzpicture}}
 & $SU(2)\times SU(1) \times SU(2)$ \\ \hline \scalebox{0.7}{\begin{tikzpicture}
	\begin{pgfonlayer}{nodelayer}
		\node [style=gauge3] (0) at (5, 0) {};
		\node [style=gauge3] (1) at (6, 0) {};
		\node [style=gauge3] (2) at (7, 0) {};
		\node [style=none] (3) at (5, -0.5) {U};
		\node [style=none] (4) at (7, -0.5) {S};
		\node [style=none] (5) at (6, -0.5) {U};
		\node [style=blacknode] (6) at (7, 0) {};
	\end{pgfonlayer}
	\begin{pgfonlayer}{edgelayer}
		\draw (0) to (1);
		\draw (1) to (2);
	\end{pgfonlayer}
\end{tikzpicture}}
& $SU(3) \times SU(1)$ \\ \hline 
\scalebox{0.7}{\begin{tikzpicture}
	\begin{pgfonlayer}{nodelayer}
		\node [style=gauge3] (0) at (5, 0) {};
		\node [style=gauge3] (1) at (6, 0) {};
		\node [style=gauge3] (2) at (7, 0) {};
		\node [style=none] (3) at (5, -0.5) {S};
		\node [style=none] (4) at (6, -0.5) {S};
		\node [style=none] (5) at (7, -0.5) {U};
		\node [style=blacknode] (6) at (6, 0) {};
		\node [style=blacknode] (7) at (5, 0) {};
	\end{pgfonlayer}
	\begin{pgfonlayer}{edgelayer}
		\draw (0) to (1);
		\draw (1) to (2);
	\end{pgfonlayer}
\end{tikzpicture}}
 & $SU(1) \times SU(1) \times SU(2)$
\\ \hline 
\scalebox{0.7}{\begin{tikzpicture}
	\begin{pgfonlayer}{nodelayer}
		\node [style=gauge3] (0) at (5, 0) {};
		\node [style=gauge3] (1) at (6, 0) {};
		\node [style=gauge3] (2) at (7, 0) {};
		\node [style=none] (3) at (5, -0.5) {S};
		\node [style=none] (4) at (7, -0.5) {S};
		\node [style=none] (5) at (6, -0.5) {U};
		\node [style=blacknode] (6) at (5, 0) {};
		\node [style=blacknode] (7) at (7, 0) {};
	\end{pgfonlayer}
	\begin{pgfonlayer}{edgelayer}
		\draw (0) to (1);
		\draw (1) to (2);
	\end{pgfonlayer}
\end{tikzpicture}}
 & $SU(1)\times SU(2) \times SU(1)$ 
 \\ \hline 
\scalebox{0.7}{\begin{tikzpicture}
	\begin{pgfonlayer}{nodelayer}
		\node [style=gauge3] (0) at (5, 0) {};
		\node [style=gauge3] (1) at (6, 0) {};
		\node [style=gauge3] (2) at (7, 0) {};
		\node [style=none] (3) at (6, -0.5) {S};
		\node [style=none] (4) at (7, -0.5) {S};
		\node [style=none] (5) at (5, -0.5) {U};
		\node [style=blacknode] (6) at (6, 0) {};
		\node [style=blacknode] (7) at (7, 0) {};
	\end{pgfonlayer}
	\begin{pgfonlayer}{edgelayer}
		\draw (0) to (1);
		\draw (1) to (2);
	\end{pgfonlayer}
\end{tikzpicture}}
& $SU(2) \times SU(1) \times SU(1)$
\\ \hline 
\scalebox{0.7}{\begin{tikzpicture}
	\begin{pgfonlayer}{nodelayer}
		\node [style=gauge3] (0) at (5, 0) {};
		\node [style=gauge3] (1) at (6, 0) {};
		\node [style=gauge3] (2) at (7, 0) {};
		\node [style=none] (3) at (6, -0.5) {S};
		\node [style=none] (4) at (7, -0.5) {S};
		\node [style=none] (5) at (5, -0.5) {S};
		\node [style=blacknode] (6) at (6, 0) {};
		\node [style=blacknode] (7) at (5, 0) {};
		\node [style=blacknode] (8) at (7, 0) {};
	\end{pgfonlayer}
	\begin{pgfonlayer}{edgelayer}
		\draw (0) to (1);
		\draw (1) to (2);
	\end{pgfonlayer}
\end{tikzpicture}}
& $SU(1)\times SU(1) \times SU(1)\times SU(1)$
\\ \hline
\end{tabular}
    \caption{On the left, we take the gauge nodes of the DynkinA in \eqref{firstmirror} which takes the form of the finite Dynkin diagram of $A_3$. $k$ unitary nodes that are connected to each other form a sub Dynkin diagram which contributes $SU(k)$ factor to the Coulomb branch global symmetry. Special unitary gauge nodes does not contribute to the global symmetry (which we denote as contributing $SU(1)$ which is trivial), hence breaking up the $SU(4)$ global symmetry into its subgroups given on the right column.}
    \label{symmetrybreaking}
\end{table}

\subsection{BCD-type Dynkin quivers (\texorpdfstring{\DynkinBCD}{})}\label{dtype}
A brane set up describing  $\text{DynkinBCD}$ quivers will involve an ON orientifold plane \cite{Hanany:1999sj}. An $\mathrm{ON}^-$ produces   $\text{DynkinD}$ quiver, $\mathrm{\widetilde{ON}}^-$ produces a $\text{DynkinB}$ quiver and $\mathrm{ON}^+$ produces a $\text{DynkinC}$ quiver. Examples in the literature include, but are not limited to, D3-F1 brane systems in \cite{Hanany:2001iy}, $3d$ $\mathcal{N}=4$ gauge theories on D3-D5-NS5 brane systems in \cite{Hanany:1999sj,Cremonesi:2014xha} and $5d$ $\mathcal{N}=1$ gauge theories on brane webs in \cite{Hayashi:2015vhy,Kimura:2019gon,Kim:2022dbr,Wei:2022hjx}. 

\subsubsection{D-type Dynkin quivers (\texorpdfstring{\DynkinD}{})}\label{dynkinD}
Equipped with brane locking, we proceed to DynkinD quivers. Let us begin with the following DynkinD quiver, which has all gauge nodes balanced, and its 3d mirror:
\begin{equation}
  \scalebox{0.8}{\begin{tikzpicture}
	\begin{pgfonlayer}{nodelayer}
		\node [style=gauge3] (0) at (-1.75, 0) {};
		\node [style=gauge3] (1) at (-0.5, 0) {};
		\node [style=gauge3] (2) at (-3, 0) {};
		\node [style=gauge3] (3) at (0.5, 1) {};
		\node [style=gauge3] (4) at (0.5, -1) {};
		\node [style=none] (7) at (0.5, 1.5) {U(2)};
		\node [style=none] (8) at (0.5, -1.5) {U(2)};
		\node [style=flavour2] (9) at (1.75, 1) {};
		\node [style=flavour2] (10) at (1.75, -1) {};
		\node [style=none] (11) at (-3, -0.5) {U(1)};
		\node [style=none] (12) at (-1.75, -0.5) {U(2)};
		\node [style=none] (13) at (-0.5, -0.5) {U(3)};
		\node [style=none] (14) at (1.75, 1.5) {1};
		\node [style=none] (15) at (1.75, -1.5) {1};
		\node [style=none] (16) at (3, 0) {};
		\node [style=none] (17) at (5, 0) {};
		\node [style=gauge3] (18) at (7, -0.75) {};
		\node [style=flavour2] (19) at (7, 0.75) {};
		\node [style=none] (20) at (7, -1.25) {USp(4)};
		\node [style=none] (21) at (7, 1.25) {SO(10)};
		\node [style=none] (22) at (4, 0.5) {3d mirror};
	\end{pgfonlayer}
	\begin{pgfonlayer}{edgelayer}
		\draw (1) to (3);
		\draw (1) to (4);
		\draw (3) to (9);
		\draw (10) to (4);
		\draw (1) to (0);
		\draw (0) to (2);
		\draw [style=->] (16.center) to (17.center);
		\draw [style=->] (17.center) to (16.center);
		\draw (19) to (18);
	\end{pgfonlayer}
\end{tikzpicture}}
\label{DtypeU}
\end{equation}

Following the steps in \S\ref{sec2}, we first turn all gauge nodes in \eqref{DtypeU} into special unitary, giving us the \DynkinDSU quiver:
\begin{equation}
  \scalebox{0.8}{\begin{tikzpicture}
	\begin{pgfonlayer}{nodelayer}
		\node [style=gauge3] (0) at (-1.75, 0) {};
		\node [style=gauge3] (1) at (-0.5, 0) {};
		\node [style=gauge3] (2) at (-3, 0) {};
		\node [style=gauge3] (3) at (0.5, 1) {};
		\node [style=gauge3] (4) at (0.5, -1) {};
		\node [style=none] (7) at (0.5, 1.5) {SU(2)};
		\node [style=none] (8) at (0.5, -1.5) {SU(2)};
		\node [style=flavour2] (9) at (1.75, 1) {};
		\node [style=flavour2] (10) at (1.75, -1) {};
		\node [style=none] (11) at (-3, -0.5) {SU(1)};
		\node [style=none] (12) at (-1.75, -0.5) {SU(2)};
		\node [style=none] (13) at (-0.5, -0.5) {SU(3)};
		\node [style=none] (14) at (1.75, 1.5) {1};
		\node [style=none] (15) at (1.75, -1.5) {1};
		\node [style=blacknode] (16) at (-3, 0) {};
		\node [style=blacknode] (17) at (-1.75, 0) {};
		\node [style=blacknode] (18) at (-0.5, 0) {};
		\node [style=blacknode] (19) at (0.5, 1) {};
		\node [style=blacknode] (20) at (0.5, -1) {};
	\end{pgfonlayer}
	\begin{pgfonlayer}{edgelayer}
		\draw (1) to (3);
		\draw (1) to (4);
		\draw (3) to (9);
		\draw (10) to (4);
		\draw (1) to (0);
		\draw (0) to (2);
	\end{pgfonlayer}
\end{tikzpicture}}
\label{DtypeSU}
\end{equation}

The Coulomb branch phase of the 5d brane web takes the following form: 
\begin{equation}
\scalebox{0.8}{\begin{tikzpicture}
	\begin{pgfonlayer}{nodelayer}
		\node [style=none] (0) at (-3, 2) {};
		\node [style=none] (1) at (-3, -2) {};
		\node [style=none] (2) at (-2, 2) {};
		\node [style=none] (3) at (-2, -2) {};
		\node [style=none] (4) at (-1, 2) {};
		\node [style=none] (5) at (-1, -2) {};
		\node [style=none] (6) at (0, 2) {};
		\node [style=none] (7) at (0, -2) {};
		\node [style=none] (8) at (1, 2) {};
		\node [style=none] (9) at (1, -2) {};
		\node [style=none] (16) at (-3, 0.25) {};
		\node [style=none] (17) at (-2, 0.25) {};
		\node [style=none] (18) at (-2, 0.5) {};
		\node [style=none] (19) at (-2, 0) {};
		\node [style=none] (20) at (-1, 0) {};
		\node [style=none] (21) at (-1, 0.5) {};
		\node [style=none] (22) at (-1, 0.75) {};
		\node [style=none] (23) at (-1, 0.25) {};
		\node [style=none] (24) at (-1, -0.25) {};
		\node [style=none] (25) at (0, -0.25) {};
		\node [style=none] (26) at (0, 0.25) {};
		\node [style=none] (27) at (0, 0.75) {};
		\node [style=none] (28) at (2, 2) {};
		\node [style=none] (29) at (2, -2) {};
		\node [style=none] (30) at (0, 1.25) {};
		\node [style=none] (31) at (0, 1) {};
		\node [style=none] (32) at (1, 1.25) {};
		\node [style=none] (33) at (1, 1) {};
		\node [style=none] (34) at (7, 2) {};
		\node [style=none] (35) at (7, -2) {};
		\node [style=none] (36) at (6, 2) {};
		\node [style=none] (37) at (6, -2) {};
		\node [style=none] (38) at (5, 2) {};
		\node [style=none] (39) at (5, -2) {};
		\node [style=none] (40) at (4, 2) {};
		\node [style=none] (41) at (4, -2) {};
		\node [style=none] (42) at (3, 2) {};
		\node [style=none] (43) at (3, -2) {};
		\node [style=none] (44) at (7, 0.25) {};
		\node [style=none] (45) at (6, 0.25) {};
		\node [style=none] (46) at (6, 0.5) {};
		\node [style=none] (47) at (6, 0) {};
		\node [style=none] (48) at (5, 0) {};
		\node [style=none] (49) at (5, 0.5) {};
		\node [style=none] (50) at (5, 0.75) {};
		\node [style=none] (51) at (5, 0.25) {};
		\node [style=none] (52) at (5, -0.25) {};
		\node [style=none] (53) at (4, -0.25) {};
		\node [style=none] (54) at (4, 0.25) {};
		\node [style=none] (55) at (4, 0.75) {};
		\node [style=none] (56) at (4, 1.25) {};
		\node [style=none] (57) at (4, 1) {};
		\node [style=none] (58) at (3, 1.25) {};
		\node [style=none] (59) at (3, 1) {};
		\node [style=none] (60) at (1, -1.25) {};
		\node [style=none] (61) at (1, -1.5) {};
		\node [style=none] (62) at (3, -0.75) {};
		\node [style=none] (63) at (3, -1) {};
		\node [style=none] (64) at (0, -0.75) {};
		\node [style=none] (65) at (0, -0.75) {};
		\node [style=none] (66) at (4, -1.25) {};
		\node [style=none] (67) at (4, -1.5) {};
		\node [style=none] (68) at (0, -1) {};
		\node [style=none] (77) at (2, 2.5) {\LARGE${ON^-}$};
		\node [style=gauge3] (78) at (0.5, 1.75) {};
		\node [style=gauge3] (79) at (3.5, 1.75) {};
		\node [style=gauge3] (80) at (4, 2) {};
		\node [style=gauge3] (81) at (3, 2) {};
		\node [style=gauge3] (82) at (3, -2) {};
		\node [style=gauge3] (83) at (4, -2) {};
		\node [style=gauge3] (84) at (5, -2) {};
		\node [style=gauge3] (85) at (5, 2) {};
		\node [style=gauge3] (86) at (6, 2) {};
		\node [style=gauge3] (87) at (6, -2) {};
		\node [style=gauge3] (88) at (7, -2) {};
		\node [style=gauge3] (89) at (7, 2) {};
		\node [style=gauge3] (90) at (-1, 2) {};
		\node [style=gauge3] (91) at (-1, -2) {};
		\node [style=gauge3] (92) at (0, -2) {};
		\node [style=gauge3] (93) at (0, 2) {};
		\node [style=gauge3] (94) at (-2, 2) {};
		\node [style=gauge3] (95) at (-2, -2) {};
		\node [style=gauge3] (96) at (-3, -2) {};
		\node [style=gauge3] (97) at (-3, 2) {};
	\end{pgfonlayer}
	\begin{pgfonlayer}{edgelayer}
		\draw (0.center) to (1.center);
		\draw (2.center) to (3.center);
		\draw (4.center) to (5.center);
		\draw (6.center) to (7.center);
		\draw (8.center) to (9.center);
		\draw (16.center) to (17.center);
		\draw (21.center) to (18.center);
		\draw (20.center) to (19.center);
		\draw (27.center) to (22.center);
		\draw (26.center) to (23.center);
		\draw (25.center) to (24.center);
		\draw (30.center) to (32.center);
		\draw (31.center) to (33.center);
		\draw (34.center) to (35.center);
		\draw (36.center) to (37.center);
		\draw (38.center) to (39.center);
		\draw (40.center) to (41.center);
		\draw (42.center) to (43.center);
		\draw (44.center) to (45.center);
		\draw (49.center) to (46.center);
		\draw (48.center) to (47.center);
		\draw (55.center) to (50.center);
		\draw (54.center) to (51.center);
		\draw (53.center) to (52.center);
		\draw (56.center) to (58.center);
		\draw (57.center) to (59.center);
		\draw (62.center) to (65.center);
		\draw (63.center) to (68.center);
		\draw (60.center) to (66.center);
		\draw (67.center) to (61.center);
		\draw [style=brownline] (0.center) to (1.center);
		\draw [style=brownline] (34.center) to (35.center);
		\draw [style=greenline] (2.center) to (3.center);
		\draw [style=greenline] (36.center) to (37.center);
		\draw [style=magicmintline] (38.center) to (39.center);
		\draw [style=magicmintline] (4.center) to (5.center);
		\draw [style=rede] (40.center) to (41.center);
		\draw [style=rede] (6.center) to (7.center);
		\draw [style=bluee] (8.center) to (9.center);
		\draw [style=bluee] (42.center) to (43.center);
		\draw [style=dottedz] (28.center) to (29.center);
	\end{pgfonlayer}
\end{tikzpicture}}
\label{branefullbalanceSU}
\end{equation}
Here, to simplify the diagram, we will use a similar convention adopted in  \cite{Hayashi:2015vhy,Kimura:2019gon} where $(p,1)$5-branes are drawn with vertical lines. The presence of the $\text{ON}^-$ plane means the NS5 branes on each side are half-NS5 branes. To help keep track of which gauge nodes are turned form SU to U, we color the NS5 branes in the Coulomb branch phase as well. 
Going into Higgs branch phase gives us:
\begin{equation}
\scalebox{0.8}{\begin{tikzpicture}
	\begin{pgfonlayer}{nodelayer}
		\node [style=none] (0) at (0.5, 2) {};
		\node [style=none] (1) at (0.5, -2) {};
		\node [style=none] (2) at (1.5, 2) {};
		\node [style=none] (3) at (1.5, -2) {};
		\node [style=none] (4) at (2.5, 2) {};
		\node [style=none] (5) at (2.5, -2) {};
		\node [style=none] (6) at (3.5, 2) {};
		\node [style=none] (7) at (3.5, -2) {};
		\node [style=none] (8) at (4.5, 2) {};
		\node [style=none] (9) at (4.5, -2) {};
		\node [style=none] (22) at (5.5, 2) {};
		\node [style=none] (23) at (5.5, -2) {};
		\node [style=none] (28) at (10.5, 2) {};
		\node [style=none] (29) at (10.5, -2) {};
		\node [style=none] (30) at (9.5, 2) {};
		\node [style=none] (31) at (9.5, -2) {};
		\node [style=none] (32) at (8.5, 2) {};
		\node [style=none] (33) at (8.5, -2) {};
		\node [style=none] (34) at (7.5, 2) {};
		\node [style=none] (35) at (7.5, -2) {};
		\node [style=none] (36) at (6.5, 2) {};
		\node [style=none] (37) at (6.5, -2) {};
		\node [style=none] (71) at (5.5, 2.5) {\LARGE${ON^-}$};
		\node [style=none] (72) at (-0.4, 0.5) {};
		\node [style=none] (73) at (-0.65, 0) {};
		\node [style=none] (74) at (11.125, 0.5) {};
		\node [style=none] (75) at (11.375, 0) {};
		\node [style=none] (76) at (-0.65, 0.25) {};
		\node [style=none] (77) at (-0.4, -0.25) {};
		\node [style=none] (78) at (11.375, 0.25) {};
		\node [style=none] (79) at (11.125, -0.25) {};
		\node [style=none] (84) at (0.5, 2) {};
		\node [style=none] (85) at (0.5, -2) {};
		\node [style=none] (86) at (1.5, 2) {};
		\node [style=none] (87) at (1.5, -2) {};
		\node [style=none] (88) at (2.5, 2) {};
		\node [style=none] (89) at (2.5, -2) {};
		\node [style=none] (90) at (3.5, 2) {};
		\node [style=none] (91) at (3.5, -2) {};
		\node [style=none] (92) at (4.5, 2) {};
		\node [style=none] (93) at (4.5, -2) {};
		\node [style=none] (94) at (10.5, 2) {};
		\node [style=none] (95) at (10.5, -2) {};
		\node [style=none] (96) at (9.5, 2) {};
		\node [style=none] (97) at (9.5, -2) {};
		\node [style=none] (98) at (8.5, 2) {};
		\node [style=none] (99) at (8.5, -2) {};
		\node [style=none] (100) at (7.5, 2) {};
		\node [style=none] (101) at (7.5, -2) {};
		\node [style=none] (102) at (6.5, 2) {};
		\node [style=none] (103) at (6.5, -2) {};
		\node [style=supergauge] (104) at (11.5, 0.125) {};
		\node [style=supergauge] (105) at (-0.75, 0.125) {};
		\node [style=gauge3] (106) at (0.5, 2) {};
		\node [style=gauge3] (107) at (0.5, -2) {};
		\node [style=gauge3] (108) at (1.5, -2) {};
		\node [style=gauge3] (109) at (1.5, 2) {};
		\node [style=gauge3] (110) at (2.5, 2) {};
		\node [style=gauge3] (111) at (2.5, -2) {};
		\node [style=gauge3] (112) at (3.5, -2) {};
		\node [style=gauge3] (113) at (3.5, 2) {};
		\node [style=gauge3] (114) at (4.5, 2) {};
		\node [style=gauge3] (115) at (4.5, -2) {};
		\node [style=gauge3] (116) at (6.5, -2) {};
		\node [style=gauge3] (117) at (6.5, 2) {};
		\node [style=gauge3] (118) at (7.5, 2) {};
		\node [style=gauge3] (119) at (7.5, -2) {};
		\node [style=gauge3] (120) at (8.5, -2) {};
		\node [style=gauge3] (121) at (8.5, 2) {};
		\node [style=gauge3] (122) at (9.5, 2) {};
		\node [style=gauge3] (123) at (9.5, -2) {};
		\node [style=gauge3] (124) at (10.5, -2) {};
		\node [style=gauge3] (125) at (10.5, 2) {};
	\end{pgfonlayer}
	\begin{pgfonlayer}{edgelayer}
		\draw (0.center) to (1.center);
		\draw (2.center) to (3.center);
		\draw (4.center) to (5.center);
		\draw (6.center) to (7.center);
		\draw (8.center) to (9.center);
		\draw (28.center) to (29.center);
		\draw (30.center) to (31.center);
		\draw (32.center) to (33.center);
		\draw (34.center) to (35.center);
		\draw (36.center) to (37.center);
		\draw (72.center) to (74.center);
		\draw (75.center) to (73.center);
		\draw (76.center) to (78.center);
		\draw (79.center) to (77.center);
		\draw (84.center) to (85.center);
		\draw (86.center) to (87.center);
		\draw (88.center) to (89.center);
		\draw (90.center) to (91.center);
		\draw (92.center) to (93.center);
		\draw (94.center) to (95.center);
		\draw (96.center) to (97.center);
		\draw (98.center) to (99.center);
		\draw (100.center) to (101.center);
		\draw (102.center) to (103.center);
		\draw [style=brownline] (84.center) to (85.center);
		\draw [style=brownline] (94.center) to (95.center);
		\draw [style=greenline] (86.center) to (87.center);
		\draw [style=greenline] (96.center) to (97.center);
		\draw [style=magicmintline] (98.center) to (99.center);
		\draw [style=magicmintline] (88.center) to (89.center);
		\draw [style=rede] (100.center) to (101.center);
		\draw [style=rede] (90.center) to (91.center);
		\draw [style=bluee] (92.center) to (93.center);
		\draw [style=bluee] (102.center) to (103.center);
		\draw [style=dottedz] (22.center) to (23.center);
	\end{pgfonlayer}
\end{tikzpicture}
}
\label{mirrorD5balanceSU}
\end{equation}
where we performed a set of Hanany-Witten transitions such that the vertical lines are now truly NS5s rather than some $(p,q)$ bound state. With all the gauge groups being SU, the NS5 branes all move independently from each other. However, since the left half and right half of the brane set up are reflected by the $\text{ON}^-$ plane, the half-NS5 and their reflections must move together and have the same color. Reading off the magnetic quiver:
\begin{equation}
  \scalebox{0.8}{ \begin{tikzpicture}
	\begin{pgfonlayer}{nodelayer}
		\node [style=bluenode] (0) at (0, 0) {};
		\node [style=none] (2) at (0, -0.5) {USp(4)};
		\node [style=gauge3] (4) at (0, 0) {};
		\node [style=brownlinet] (5) at (-1.5, 0.5) {};
		\node [style=greenlinet] (6) at (-1, 1.25) {};
		\node [style=cyane] (7) at (0, 1.75) {};
		\node [style=rednode] (8) at (1, 1.25) {};
		\node [style=bluenode] (9) at (1.5, 0.5) {};
		\node [style=none] (10) at (-2.25, 0.5) {U(1)};
		\node [style=none] (11) at (-1.25, 1.75) {U(1)};
		\node [style=none] (12) at (0, 2.25) {U(1)};
		\node [style=none] (13) at (1.25, 1.75) {U(1)};
		\node [style=none] (14) at (2.25, 0.5) {U(1)};
		\node [style=grayet] (15) at (0, 1.75) {};
	\end{pgfonlayer}
	\begin{pgfonlayer}{edgelayer}
		\draw (5) to (4);
		\draw (6) to (4);
		\draw (4) to (7);
		\draw (8) to (4);
		\draw (4) to (9);
	\end{pgfonlayer}
\end{tikzpicture}}
\label{DtypeSUmirror}
\end{equation}
which is a flavorless/unframed unitary-orthosymplectic quiver with a bouquet of U(1)s. Contrary to the unitary case, the presence of an orthosymplectic gauge node means there is no need to decouple an overall U(1). From the brane webs point of view, this is  because the orientifold plane already fixes the center of mass of the brane system. As with DynkinA quivers, we further conjecture that \eqref{DtypeSU} and \eqref{DtypeSUmirror} form a 3d mirror pair. This is verified through explicit Coulomb branch and Higgs branch Hilbert series computations. \textbf{Important:} since  the quiver is an unframed unitary-orthosymplectic quiver, there is an overall diagonal $\mathbb{Z}_2$ that one can choose to quotient out (this is the one-form symmetry of the theory), and this choice changes the magnetic lattice that one sums over the magnetic charges in the monopole formula \cite{Bourget:2020xdz}. In this paper, we always choose to quotient out this diagonal $\mathbb{Z}_2$  from all unframed/flavorless orthosymplectic quivers. 

Next, we turn some of the gauge groups from special unitary to unitary. One important thing to notice is that the quiver in \eqref{DtypeSU} behaves just like a linear DynkinA away from the bifurcation. We can see this in the brane system as well, where away from the $\text{ON}^-$ plane in \eqref{branefullbalanceSU}, the brane system is identical to the brane system of the linear DynkinA quiver. Thus, locking the four half-NS5 branes on the left, we expect a similar effect on the U(1) bouquet in the mirror quiver. Let us demonstrate by first turning the three gauge nodes forming a linear chain in \eqref{DtypeSU} into unitary gauge groups:
\begin{equation}
  \scalebox{0.8}{

}
\label{DtypeSULockA}
\end{equation}
In the Coulomb branch phase with the branes we wish to lock colored:
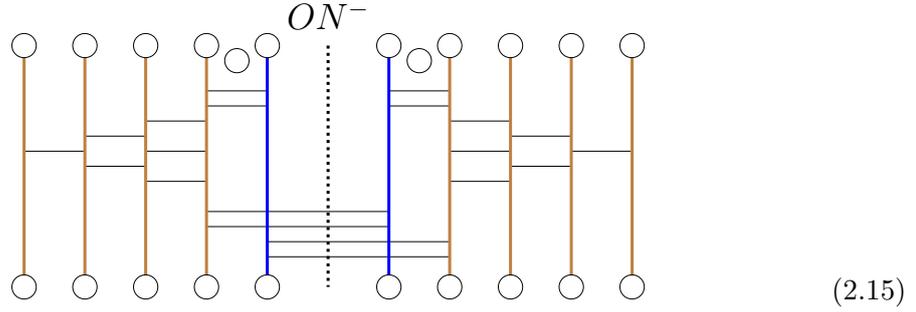
\begin{equation}
\scalebox{0.8}{%
}
\label{branefullbalanceSULockA}
\end{equation}
going to the Higgs phase:
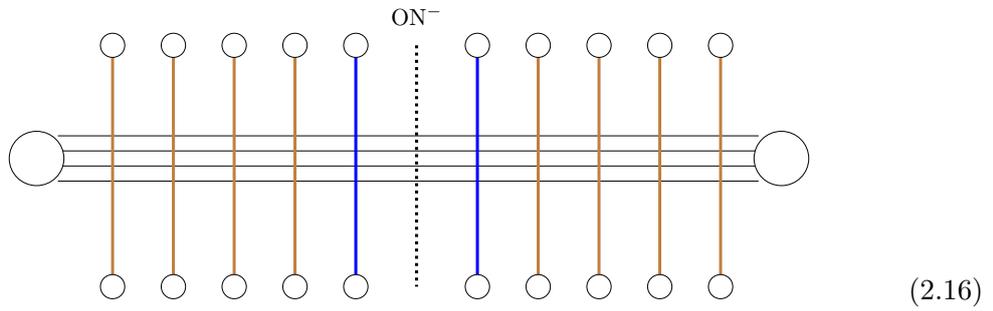
\begin{equation}
\scalebox{0.8}{%
}
\label{DtypeSUmirrorbraneLockA}
\end{equation}
and reading off the magnetic quiver:
\begin{equation}
    \scalebox{0.8}{ %
}
\label{DtypeSUMirrorLockA}
\end{equation}
Explicit Hilbert series computations shows \eqref{DtypeSULockA} and \eqref{DtypeSUMirrorLockA} are indeed 3d mirror pairs. This shows the locking mechanism for \DynkinA quivers can be straightforwardly applied to \DynkinD quivers when focusing on the gauge nodes away from the bifurcation.  Later on, we will see the same holds for \DynkinBC quivers as well. 

The next step is to push further by turning one of the bifurcated nodes into unitary as well:
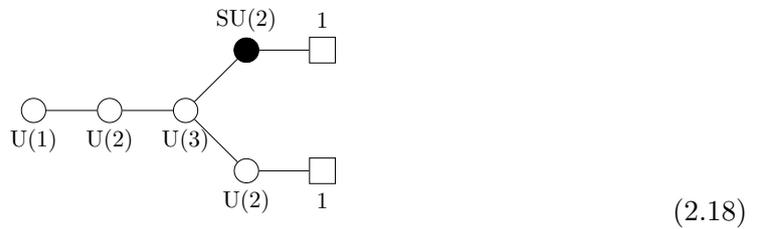
\begin{equation}
  \scalebox{0.8}{%
}
\label{DtypeSULockB}
\end{equation}
where the four unitary quivers form a linear chain. The brane set up now have all NS5 brane fixed:
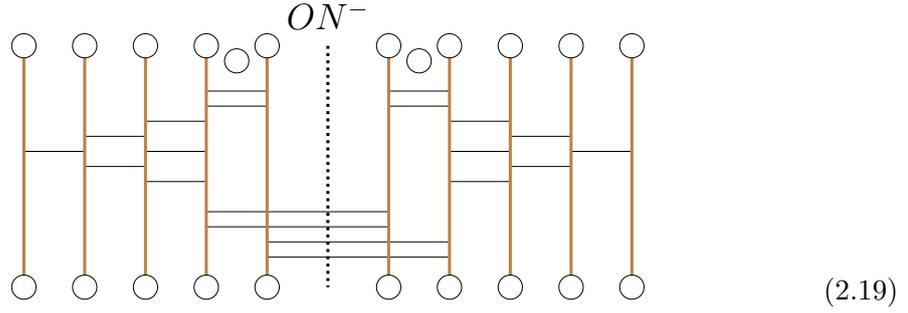
\begin{equation}
\scalebox{0.8}{
%
}
\label{branesetupDtypeLockB}
\end{equation}
The Higgs branch phase is:
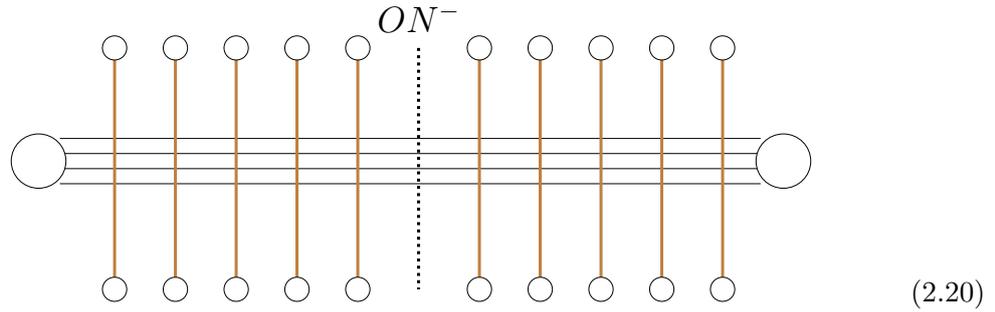
\begin{equation}
\scalebox{0.8}{%
}
\label{DtypeSUmirrorbraneLockAB}
\end{equation}
The five NS5s now moves together, contributing to a single U(1) in the magnetic quiver:
\begin{equation}
     \scalebox{0.8}{   %
   }
\label{DtypeSUMirrorLockB}
\end{equation}
Again, we conjecture \eqref{DtypeSULockB} and \eqref{DtypeSUMirrorLockB} form a 3d mirror pair which is verified by explicit Hilbert series computations. The Higgs branch global symmetry of of the magnetic quiver is SU(5). This is the same as the Coulomb branch global symmetry of \eqref{DtypeSULockB} obtained by looking at  the set of balanced nodes.


Next, we turn one of the nodes in the bifurcation into unitary whilst the rest of the gauge nodes remain special unitary:
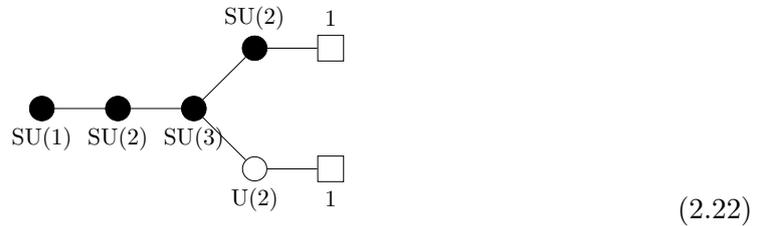
\begin{equation}
  \scalebox{0.8}{

}
\label{SULockC}
\end{equation}
The Coulomb branch phase gives:
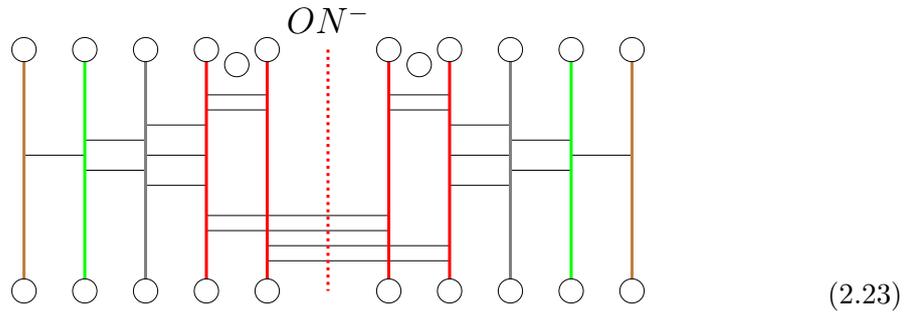
\begin{equation}
\scalebox{0.8}{%
}
\label{braneSULockC}
\end{equation}
The Higgs branch phase gives:
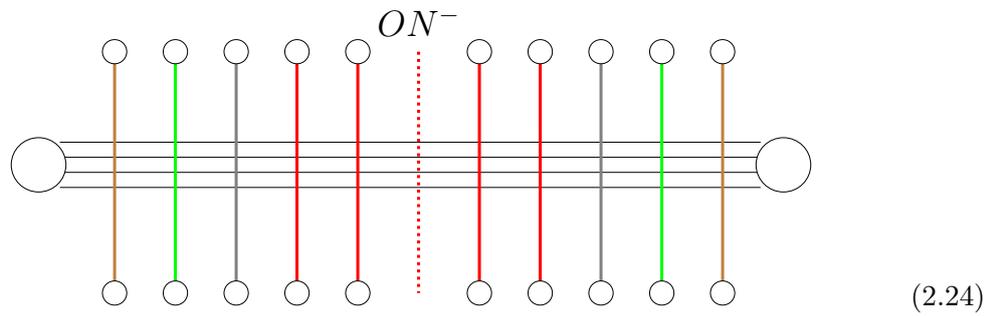
\begin{equation}
\scalebox{0.8}{%
}
\label{branesetupmirrorLockC}
\end{equation}
Since the $\text{ON}^-$ plane is always fixed, the NS5s that are locked with the orientifold plane (now in red) will be fixed as well and no longer contribute as dynamical degrees of freedom (they are stuck!). As a result, the magnetic quiver will have a flavor node with SO global symmetry:
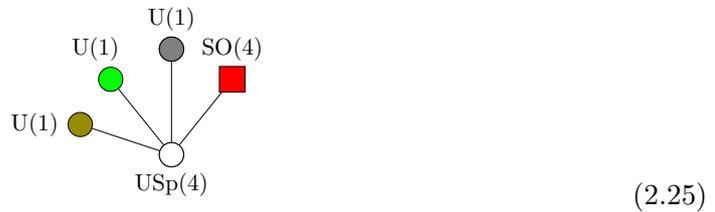
\begin{equation}
   \scalebox{0.8}{  %
}
\label{mirrorSULockC}  
\end{equation}
Explicit Hilbert series computations is consistent with  \eqref{SULockC} and  \eqref{mirrorSULockC} being 3d mirror pairs. 

Finally, turning all SU nodes to U gives the following brane set up:
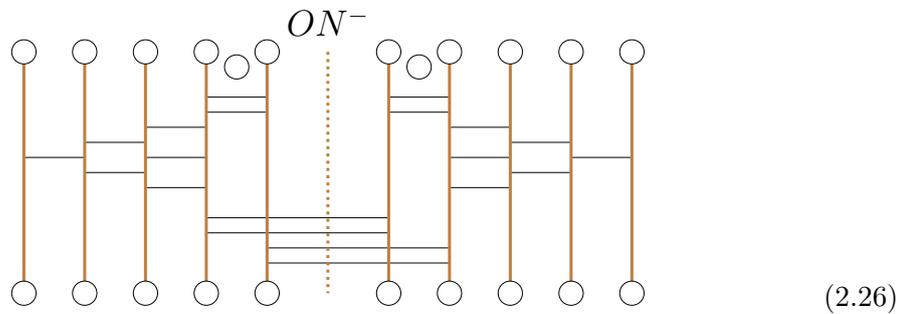
\begin{equation}
\scalebox{0.8}{
%
}
\label{DtypefullOn}
\end{equation}
and going to the Higgs phase:
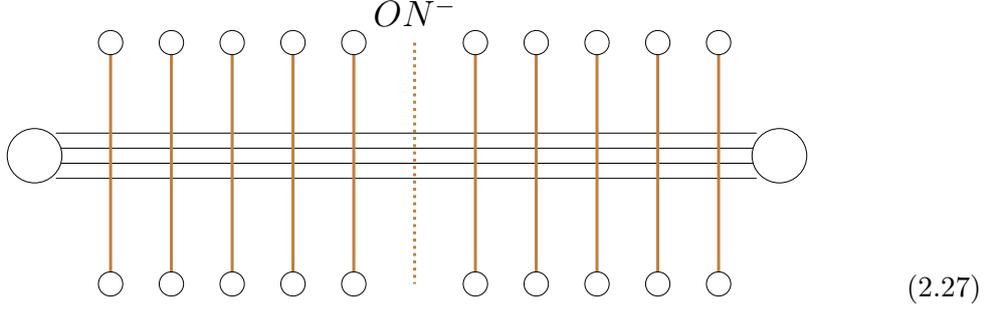
\begin{equation}
\scalebox{0.8}{%
}
\label{DtypefullOnmirror}
\end{equation}
All the NS5 branes are now locked with the $\text{ON}^-$ and become non-dynamical. The bouquet in the magnetic quiver is now only a single flavor node:
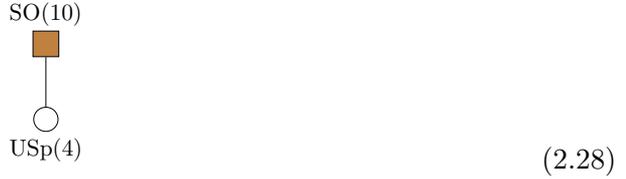
\begin{equation}
   \scalebox{0.8}{    %
}
\end{equation}
This reproduces the 3d mirror pair in \eqref{DtypeU}. All possible U \& SU combinations are given in Table \ref{DtypeUSU1} and \ref{DtypeUSU2} along with their 3d mirrors. Hilbert series computations on both sides have been computed to verify they are indeed 3d mirror pairs.

The locking mechanism can be easily generalized to the following family of quiver and their mirrors:
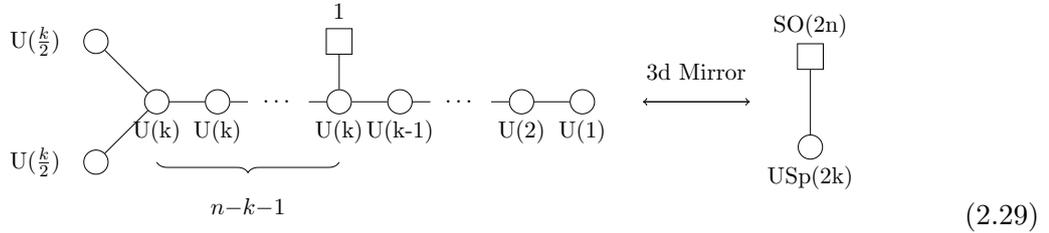
\begin{equation}
  \scalebox{0.8}{ %
}
\label{generalDA}
\end{equation}
We divide generic U \& SU combinations  into two cases. The first case is when both the bifurcated nodes are unitary. 
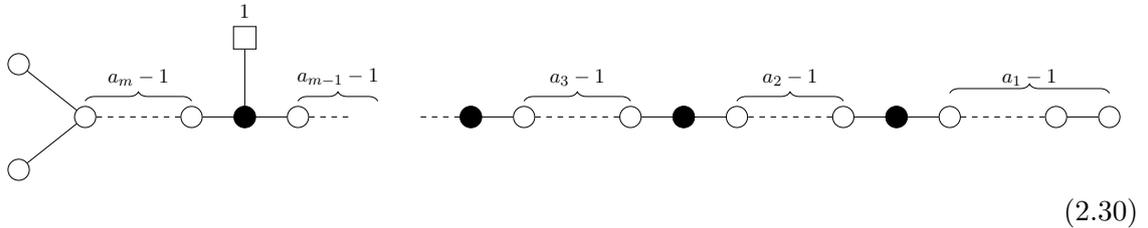
\begin{equation}
  \scalebox{0.7
  }{  %
}
\end{equation}
where
\begin{equation}
    \sum_{i=1}^{m}a_i+1=n
\end{equation}
The magnetic quiver contains a bouquet of U(1)s along with an SO flavor node:
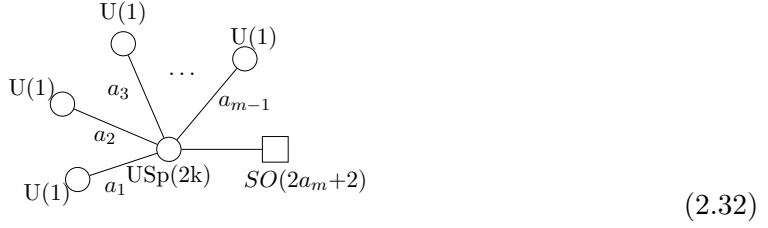
\begin{equation}
  \scalebox{0.8}{ %
}
\label{generalDAmirror}
\end{equation}
where the multiplicity of the links are labeled. Similar to the DynkinA cases, the arrangements of U/SU nodes in the \DynkinD quiver fully determines the bouquet in the magnetic quiver, i.e the number of U(1)s and the multiplicities of their links. The only difference is that \DynkinD quivers also gives an SO flavor node if the bifurcated nodes are both unitary. From the brane perspective, this is because two or more NS5 branes are locked with the $\text{ON}^-$ plane and become non-dynamical degrees of freedom. This can also be seen from the balance of gauge nodes: since the bifurcated nodes are both unitary, a subset of the balanced nodes in the quiver form a $D_{a_m+1}$ Dynkin diagram, thus indicating the Coulomb branch global symmetry contains an $SO(2a_m+2)$ subgroup. The magnetic quiver will then have a $SO(2a_m+2)$ flavor group. As usual, we conjecture that \eqref{generalDA} and \eqref{generalDAmirror} are 3d mirror pairs. If either or both of the bifurcated nodes are special unitary, then the bouquet in the magnetic quiver will not contain any SO flavor nodes. This brings us to the second case:
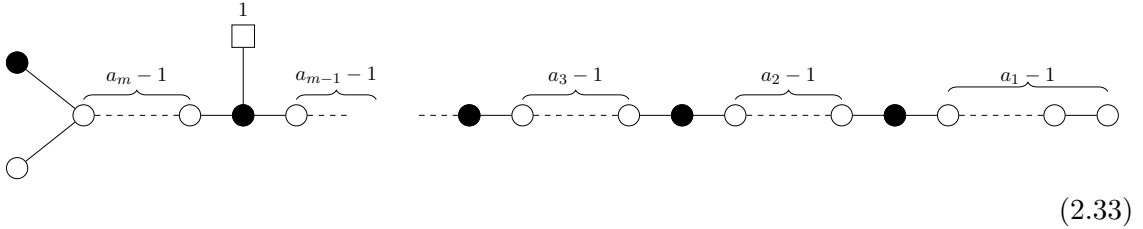
\begin{equation}
\scalebox{0.7}{
  \begin{tikzpicture}
	\begin{pgfonlayer}{nodelayer}
		\node [style=none] (8) at (7, 0) {};
		\node [style=none] (15) at (8.25, 0) {};
		\node [style=none] (16) at (18.25, 0.45) {};
		\node [style=none] (17) at (21.25, 0.45) {};
		\node [style=none] (18) at (14.25, 0.3) {};
		\node [style=none] (19) at (16.25, 0.3) {};
		\node [style=none] (20) at (6, 0.3) {};
		\node [style=none] (21) at (7.5, 0.3) {};
		\node [style=none] (22) at (19.75, 0.75) {$a_1-1$};
		\node [style=none] (23) at (15.25, 0.75) {$a_2-1$};
		\node [style=none] (24) at (6.75, 0.75) {$a_{m-1}-1$};
		\node [style=none] (43) at (10.25, 0.3) {};
		\node [style=none] (44) at (12.25, 0.3) {};
		\node [style=none] (45) at (11.25, 0.75) {$a_3-1$};
		\node [style=none] (46) at (2, 0.3) {};
		\node [style=none] (47) at (4, 0.3) {};
		\node [style=none] (48) at (3, 0.75) {$a_{m}-1$};
		\node [style=gauge3] (51) at (21.25, 0) {};
		\node [style=gauge3] (52) at (20.25, 0) {};
		\node [style=gauge3] (53) at (18.25, 0) {};
		\node [style=gauge3] (54) at (17.25, 0) {};
		\node [style=gauge3] (55) at (16.25, 0) {};
		\node [style=gauge3] (56) at (14.25, 0) {};
		\node [style=gauge3] (57) at (13.25, 0) {};
		\node [style=gauge3] (58) at (12.25, 0) {};
		\node [style=gauge3] (59) at (10.25, 0) {};
		\node [style=gauge3] (60) at (9.25, 0) {};
		\node [style=gauge3] (61) at (6, 0) {};
		\node [style=gauge3] (62) at (5, 0) {};
		\node [style=gauge3] (63) at (4, 0) {};
		\node [style=gauge3] (64) at (2, 0) {};
		\node [style=gauge3] (65) at (0.75, 1) {};
		\node [style=gauge3] (66) at (0.75, -1) {};
		\node [style=blacknode] (67) at (17.25, 0) {};
		\node [style=blacknode] (68) at (13.25, 0) {};
		\node [style=blacknode] (69) at (9.25, 0) {};
		\node [style=blacknode] (70) at (5, 0) {};
		\node [style=flavour2] (71) at (5, 1.5) {};
		\node [style=none] (72) at (5, 2) {1};
		\node [style=blacknode] (73) at (0.75, 1) {};
	\end{pgfonlayer}
	\begin{pgfonlayer}{edgelayer}
		\draw [style=brace2] (18.center) to (19.center);
		\draw [style=brace2] (20.center) to (21.center);
		\draw [style=brace2] (16.center) to (17.center);
		\draw [style=brace2] (43.center) to (44.center);
		\draw [style=brace2] (46.center) to (47.center);
		\draw (65) to (64);
		\draw (64) to (66);
		\draw [style=dashedline] (63) to (64);
		\draw (63) to (62);
		\draw (62) to (61);
		\draw (60) to (59);
		\draw [style=dashedline] (59) to (58);
		\draw (58) to (56);
		\draw [style=dashedline] (55) to (56);
		\draw (55) to (53);
		\draw [style=dashedline] (53) to (52);
		\draw (51) to (52);
		\draw (71) to (70);
		\draw [style=dashedline] (69) to (15.center);
		\draw [style=dashedline] (61) to (8.center);
	\end{pgfonlayer}
\end{tikzpicture}}
\label{generalDBX}
\end{equation}
where the 3d mirror is now:
\begin{equation}
  \scalebox{0.8}{ \begin{tikzpicture}
	\begin{pgfonlayer}{nodelayer}
		\node [style=none] (0) at (2.75, -0.45) {USp(2k)};
		\node [style=none] (1) at (1.875, -0.65) {$a_1$};
		\node [style=none] (2) at (1.7, 0.2) {$a_2$};
		\node [style=none] (3) at (1.925, 0.95) {$a_3$};
		\node [style=none] (4) at (4, 0.75) {$a_{k-1}$};
		\node [style=none] (5) at (0.75, -0.75) {U(1)};
		\node [style=none] (6) at (0.5, 1) {U(1)};
		\node [style=none] (7) at (2, 2.25) {U(1)};
		\node [style=none] (8) at (4.35, 2) {U(1)};
		\node [style=none] (9) at (4, -0.275) {$a_k{+}1$};
		\node [style=none] (10) at (3, 1.25) {$\dots$};
		\node [style=gauge3] (11) at (1.25, -0.5) {};
		\node [style=gauge3] (12) at (1, 0.75) {};
		\node [style=gauge3] (13) at (2, 1.75) {};
		\node [style=gauge3] (14) at (4, 1.5) {};
		\node [style=gauge3] (15) at (2.75, 0) {};
		\node [style=gauge3] (16) at (4.75, 0) {};
		\node [style=none] (17) at (5.5, 0) {U(1)};
	\end{pgfonlayer}
	\begin{pgfonlayer}{edgelayer}
		\draw (11) to (15);
		\draw (15) to (12);
		\draw (13) to (15);
		\draw (15) to (14);
		\draw (15) to (16);
	\end{pgfonlayer}
\end{tikzpicture}}
\label{generalDBmirrorX}
\end{equation}
As usual, we conjecture that \eqref{generalDBX} and \eqref{generalDBmirrorX} are 3d mirror pairs.

\subsubsection{B-type quivers (\texorpdfstring{\DynkinB}{})}\label{BCtype}
Now, we move onto DynkinBC quivers which have non-simply laced edges (represented by an arrow). This is in the same sense as the non-simply laced edge in Dynkin diagrams of BC type Lie algebra that distinguishes between long and short roots. It is currently unclear how to study the Higgs branches of non-simply laced quivers as we do not yet have a hyperK\"ahler construction for them. Nevertheless, the brane web set up can still be constructed using $\widetilde{\text{ON}}^-$ for DynkinB and $\text{ON}^+$ for DynkinC.  DynkinBC quivers in D3-D5-NS5 systems with ON planes have been studied in \cite{Hanany:1999sj, Cremonesi:2014xha} whereas brane webs of DynkinBC have also been explored in \cite{Kimura:2019gon}. Once the brane web is constructed, the magnetic quiver can then be readily read off. However, since we have no tool to compute the Higgs branch of DynkinBC quivers, we cannot verify the equivalence between the Higgs branch of DynkinBC quivers and the Coulomb branch of the mirror. On the other hand, we can view the DynkinBC quiver as a $3d$ $\mathcal{N}=4$ gauge theory, as we had done so far, and ask if its Coulomb branch is the same as the Higgs branch of the mirror. The monopole formula can be used to compute the Coulomb branch Hilbert series for $3d$ $\mathcal{N}=4$ non-simply laced quivers as shown in \cite{Cremonesi:2014xha}.   Therefore, we are  able to test one side of the 3d mirror: whether the Coulomb branch of the DynkinBC  quivers is the same as the Higgs branch of the `mirror dual'. 

Consider a fully balanced $B_4$ Dynkin quiver and its `mirror':
\begin{equation}
  \scalebox{0.8}{ \begin{tikzpicture}
	\begin{pgfonlayer}{nodelayer}
		\node [style=gauge3] (0) at (-0.75, 0) {};
		\node [style=gauge3] (1) at (0.5, 0) {};
		\node [style=gauge3] (2) at (1.75, 0) {};
		\node [style=gauge3] (3) at (3, 0) {};
		\node [style=none] (4) at (1.925, 0.125) {};
		\node [style=none] (5) at (1.925, -0.125) {};
		\node [style=none] (6) at (2.925, 0.125) {};
		\node [style=none] (7) at (2.925, -0.125) {};
		\node [style=none] (8) at (2.225, 0.375) {};
		\node [style=none] (9) at (2.6, 0) {};
		\node [style=none] (10) at (2.225, -0.375) {};
		\node [style=none] (11) at (-0.75, -0.5) {U(1)};
		\node [style=none] (12) at (0.5, -0.5) {U(2)};
		\node [style=none] (13) at (1.75, -0.5) {U(3)};
		\node [style=none] (14) at (3, -0.5) {U(2)};
		\node [style=flavour2] (15) at (3, 1.25) {};
		\node [style=none] (16) at (3, 1.75) {1};
		\node [style=none] (17) at (6.75, -0.5) {USp(4)};
		\node [style=none] (18) at (6.75, 1.75) {SO(9)};
		\node [style=gauge3] (19) at (6.75, 0) {};
		\node [style=brownflavor] (20) at (6.75, 1.25) {};
		\node [style=flavour2] (21) at (6.75, 1.25) {};
		\node [style=none] (22) at (4, 0) {};
		\node [style=none] (23) at (5.5, 0) {};
		\node [style=none] (24) at (4.75, 0.5) {3d ``Mirror''};
	\end{pgfonlayer}
	\begin{pgfonlayer}{edgelayer}
		\draw (4.center) to (6.center);
		\draw (5.center) to (7.center);
		\draw (8.center) to (9.center);
		\draw (9.center) to (10.center);
		\draw (1) to (2);
		\draw (1) to (0);
		\draw (15) to (3);
		\draw (20) to (19);
		\draw [style=->] (22.center) to (23.center);
		\draw [style=->] (23.center) to (22.center);
	\end{pgfonlayer}
\end{tikzpicture}}
\label{B4fistone}
\end{equation}
The theory has an USp gauge group connected to an odd number of hypermultiplets and thus suffers from Witten's global anomaly \cite{WITTEN1982324}. To avoid this, the gauge group also comes with a suitable Chern-Simons term. Since the classical Higgs branch is insensitive to Chern-Simons levels, we don't have to worry about it when computing Higgs branch Hilbert series. The Higgs branch is shown to be  closure of the next to minimal nilpotent orbit of $B_4$ \cite{Hanany:2016gbz} which is the same as the Coulomb branch of the $B_4$ quiver on the left (checked with explicit Hilbert series computations). 

Now, we replace all the gauge nodes with SU:
\begin{equation}
    \scalebox{0.8}{\begin{tikzpicture}
	\begin{pgfonlayer}{nodelayer}
		\node [style=gauge3] (0) at (-0.75, 0) {};
		\node [style=gauge3] (1) at (0.5, 0) {};
		\node [style=gauge3] (2) at (1.75, 0) {};
		\node [style=gauge3] (3) at (3, 0) {};
		\node [style=none] (4) at (1.925, 0.125) {};
		\node [style=none] (5) at (1.925, -0.125) {};
		\node [style=none] (6) at (2.925, 0.125) {};
		\node [style=none] (7) at (2.925, -0.125) {};
		\node [style=none] (8) at (2.225, 0.375) {};
		\node [style=none] (9) at (2.6, 0) {};
		\node [style=none] (10) at (2.225, -0.375) {};
		\node [style=none] (11) at (-0.75, -0.5) {SU(1)};
		\node [style=none] (12) at (0.5, -0.5) {SU(2)};
		\node [style=none] (13) at (1.75, -0.5) {SU(3)};
		\node [style=none] (14) at (3, -0.5) {SU(2)};
		\node [style=flavour2] (15) at (3, 1.25) {};
		\node [style=none] (16) at (3, 1.75) {1};
		\node [style=blacknode] (17) at (-0.75, 0) {};
		\node [style=blacknode] (18) at (0.5, 0) {};
		\node [style=blacknode] (19) at (1.75, 0) {};
		\node [style=blacknode] (20) at (3, 0) {};
	\end{pgfonlayer}
	\begin{pgfonlayer}{edgelayer}
		\draw (4.center) to (6.center);
		\draw (5.center) to (7.center);
		\draw (8.center) to (9.center);
		\draw (9.center) to (10.center);
		\draw (1) to (2);
		\draw (1) to (0);
		\draw (15) to (3);
	\end{pgfonlayer}
\end{tikzpicture}}
\end{equation}and draw the brane set up. The Coulomb branch phase gives:
\begin{equation}
\scalebox{0.8}{\begin{tikzpicture}
	\begin{pgfonlayer}{nodelayer}
		\node [style=none] (2) at (-2, 2) {};
		\node [style=none] (3) at (-2, -2) {};
		\node [style=none] (4) at (-1, 2) {};
		\node [style=none] (5) at (-1, -2) {};
		\node [style=none] (6) at (0, 2) {};
		\node [style=none] (7) at (0, -2) {};
		\node [style=none] (8) at (1, 2) {};
		\node [style=none] (9) at (1, -2) {};
		\node [style=none] (17) at (-2, 0.25) {};
		\node [style=none] (19) at (-2, 0) {};
		\node [style=none] (20) at (-1, 0) {};
		\node [style=none] (23) at (-1, 0.25) {};
		\node [style=none] (24) at (-1, -0.25) {};
		\node [style=none] (25) at (0, -0.25) {};
		\node [style=none] (26) at (0, 0.25) {};
		\node [style=none] (28) at (2, 2) {};
		\node [style=none] (29) at (2, -2) {};
		\node [style=none] (30) at (0, 0.5) {};
		\node [style=none] (31) at (0, 0) {};
		\node [style=none] (32) at (1, 0.5) {};
		\node [style=none] (33) at (1, 0) {};
		\node [style=none] (36) at (6, 2) {};
		\node [style=none] (37) at (6, -2) {};
		\node [style=none] (38) at (5, 2) {};
		\node [style=none] (39) at (5, -2) {};
		\node [style=none] (40) at (4, 2) {};
		\node [style=none] (41) at (4, -2) {};
		\node [style=none] (42) at (3, 2) {};
		\node [style=none] (43) at (3, -2) {};
		\node [style=none] (60) at (1, -1.25) {};
		\node [style=none] (61) at (1, -1.5) {};
		\node [style=none] (62) at (3, -0.75) {};
		\node [style=none] (63) at (3, -1) {};
		\node [style=none] (64) at (0, -0.75) {};
		\node [style=none] (65) at (1, -0.75) {};
		\node [style=none] (66) at (3, -1.25) {};
		\node [style=none] (67) at (3, -1.5) {};
		\node [style=none] (68) at (1, -1) {};
		\node [style=none] (77) at (2, 2.5) {\LARGE${\widetilde{ON}^-}$};
		\node [style=none] (78) at (0, -0.5) {};
		\node [style=none] (79) at (1, -0.5) {};
		\node [style=none] (81) at (6, 0) {};
		\node [style=none] (82) at (5, 0) {};
		\node [style=none] (83) at (5, 0.25) {};
		\node [style=none] (84) at (5, -0.25) {};
		\node [style=none] (85) at (4, -0.25) {};
		\node [style=none] (86) at (4, 0.25) {};
		\node [style=none] (87) at (4, 0.5) {};
		\node [style=none] (88) at (4, 0) {};
		\node [style=none] (89) at (3, 0.5) {};
		\node [style=none] (90) at (3, 0) {};
		\node [style=none] (91) at (4, -0.5) {};
		\node [style=none] (92) at (3, -0.5) {};
		\node [style=gauge3] (93) at (-2, 2) {};
		\node [style=gauge3] (94) at (-1, 2) {};
		\node [style=gauge3] (95) at (0.5, 1.5) {};
		\node [style=gauge3] (96) at (3.5, 1.5) {};
		\node [style=gauge3] (97) at (0, 2) {};
		\node [style=gauge3] (98) at (1, 2) {};
		\node [style=gauge3] (99) at (3, 2) {};
		\node [style=gauge3] (100) at (4, 2) {};
		\node [style=gauge3] (101) at (5, 2) {};
		\node [style=gauge3] (102) at (6, 2) {};
		\node [style=gauge3] (103) at (6, -2) {};
		\node [style=gauge3] (104) at (5, -2) {};
		\node [style=gauge3] (105) at (4, -2) {};
		\node [style=gauge3] (106) at (3, -2) {};
		\node [style=gauge3] (107) at (1, -2) {};
		\node [style=gauge3] (108) at (0, -2) {};
		\node [style=gauge3] (109) at (-1, -2) {};
		\node [style=gauge3] (110) at (-2, -2) {};
	\end{pgfonlayer}
	\begin{pgfonlayer}{edgelayer}
		\draw (20.center) to (19.center);
		\draw (26.center) to (23.center);
		\draw (25.center) to (24.center);
		\draw (30.center) to (32.center);
		\draw (31.center) to (33.center);
		\draw (62.center) to (65.center);
		\draw (63.center) to (68.center);
		\draw (60.center) to (66.center);
		\draw (67.center) to (61.center);
		\draw (79.center) to (78.center);
		\draw (82.center) to (81.center);
		\draw (86.center) to (83.center);
		\draw (85.center) to (84.center);
		\draw (87.center) to (89.center);
		\draw (88.center) to (90.center);
		\draw (92.center) to (91.center);
		\draw [style=brownline] (2.center) to (3.center);
		\draw [style=brownline] (37.center) to (36.center);
		\draw [style=greenline] (4.center) to (5.center);
		\draw [style=greenline] (39.center) to (38.center);
		\draw [style=rede] (6.center) to (7.center);
		\draw [style=rede] (40.center) to (41.center);
		\draw [style=bluee] (8.center) to (9.center);
		\draw [style=bluee] (42.center) to (43.center);
		\draw [style=dottedz] (28.center) to (29.center);
	\end{pgfonlayer}
\end{tikzpicture}}
\end{equation}
Going to the Higgs phase:
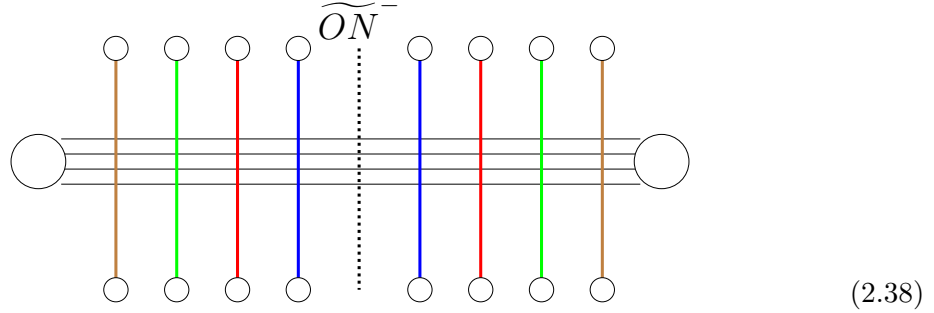
\begin{equation}
\scalebox{0.8}{\begin{tikzpicture}
	\begin{pgfonlayer}{nodelayer}
		\node [style=none] (0) at (3.25, 2) {};
		\node [style=none] (1) at (3.25, -2) {};
		\node [style=none] (2) at (4.25, 2) {};
		\node [style=none] (3) at (4.25, -2) {};
		\node [style=none] (6) at (5.25, 2) {};
		\node [style=none] (7) at (5.25, -2) {};
		\node [style=none] (8) at (6.25, 2) {};
		\node [style=none] (9) at (6.25, -2) {};
		\node [style=none] (10) at (7.25, 2) {};
		\node [style=none] (11) at (7.25, -2) {};
		\node [style=none] (12) at (11.25, 2) {};
		\node [style=none] (13) at (11.25, -2) {};
		\node [style=none] (14) at (10.25, 2) {};
		\node [style=none] (15) at (10.25, -2) {};
		\node [style=none] (18) at (9.25, 2) {};
		\node [style=none] (19) at (9.25, -2) {};
		\node [style=none] (20) at (8.25, 2) {};
		\node [style=none] (21) at (8.25, -2) {};
		\node [style=none] (26) at (7.25, 2.5) {\LARGE${\widetilde{ON}^-}$};
		\node [style=none] (27) at (2.35, 0.5) {};
		\node [style=none] (28) at (2.1, 0) {};
		\node [style=none] (29) at (11.875, 0.5) {};
		\node [style=none] (30) at (12.125, 0) {};
		\node [style=none] (31) at (2.1, 0.25) {};
		\node [style=none] (32) at (2.35, -0.25) {};
		\node [style=none] (33) at (12.125, 0.25) {};
		\node [style=none] (34) at (11.875, -0.25) {};
		\node [style=none] (39) at (3.25, 2) {};
		\node [style=none] (40) at (3.25, -2) {};
		\node [style=none] (41) at (4.25, 2) {};
		\node [style=none] (42) at (4.25, -2) {};
		\node [style=none] (45) at (5.25, 2) {};
		\node [style=none] (46) at (5.25, -2) {};
		\node [style=none] (47) at (6.25, 2) {};
		\node [style=none] (48) at (6.25, -2) {};
		\node [style=none] (49) at (11.25, 2) {};
		\node [style=none] (50) at (11.25, -2) {};
		\node [style=none] (51) at (10.25, 2) {};
		\node [style=none] (52) at (10.25, -2) {};
		\node [style=none] (55) at (9.25, 2) {};
		\node [style=none] (56) at (9.25, -2) {};
		\node [style=none] (57) at (8.25, 2) {};
		\node [style=none] (58) at (8.25, -2) {};
		\node [style=supergauge] (59) at (12.225, 0.125) {};
		\node [style=supergauge] (60) at (1.975, 0.125) {};
		\node [style=gauge3] (61) at (3.25, 2) {};
		\node [style=gauge3] (62) at (4.25, 2) {};
		\node [style=gauge3] (63) at (5.25, 2) {};
		\node [style=gauge3] (64) at (6.25, 2) {};
		\node [style=gauge3] (65) at (8.25, 2) {};
		\node [style=gauge3] (66) at (9.25, 2) {};
		\node [style=gauge3] (67) at (10.25, 2) {};
		\node [style=gauge3] (68) at (11.25, 2) {};
		\node [style=gauge3] (69) at (11.25, -2) {};
		\node [style=gauge3] (70) at (10.25, -2) {};
		\node [style=gauge3] (71) at (9.25, -2) {};
		\node [style=gauge3] (72) at (8.25, -2) {};
		\node [style=gauge3] (73) at (6.25, -2) {};
		\node [style=gauge3] (74) at (5.25, -2) {};
		\node [style=gauge3] (75) at (4.25, -2) {};
		\node [style=gauge3] (76) at (3.25, -2) {};
	\end{pgfonlayer}
	\begin{pgfonlayer}{edgelayer}
		\draw (0.center) to (1.center);
		\draw (2.center) to (3.center);
		\draw (6.center) to (7.center);
		\draw (8.center) to (9.center);
		\draw (12.center) to (13.center);
		\draw (14.center) to (15.center);
		\draw (18.center) to (19.center);
		\draw (20.center) to (21.center);
		\draw (27.center) to (29.center);
		\draw (30.center) to (28.center);
		\draw (31.center) to (33.center);
		\draw (34.center) to (32.center);
		\draw (39.center) to (40.center);
		\draw (41.center) to (42.center);
		\draw (45.center) to (46.center);
		\draw (47.center) to (48.center);
		\draw (49.center) to (50.center);
		\draw (51.center) to (52.center);
		\draw (55.center) to (56.center);
		\draw (57.center) to (58.center);
		\draw [style=brownline] (39.center) to (40.center);
		\draw [style=brownline] (49.center) to (50.center);
		\draw [style=greenline] (41.center) to (42.center);
		\draw [style=greenline] (51.center) to (52.center);
		\draw [style=rede] (55.center) to (56.center);
		\draw [style=rede] (45.center) to (46.center);
		\draw [style=bluee] (47.center) to (48.center);
		\draw [style=bluee] (57.center) to (58.center);
		\draw [style=dottedz] (10.center) to (11.center);
	\end{pgfonlayer}
\end{tikzpicture}}
\end{equation}
The 3d mirror is then a bouquet of U(1)s but with an additional half-hyper which we label as an SO(1):
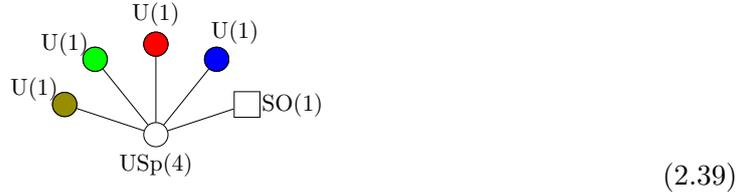
\begin{equation}
  \scalebox{0.8}{ \begin{tikzpicture}
	\begin{pgfonlayer}{nodelayer}
		\node [style=gauge3] (0) at (0, 0) {};
		\node [style=none] (1) at (0, -0.5) {USp(4)};
		\node [style=gauge3] (2) at (-1.5, 0.5) {};
		\node [style=gauge3] (3) at (-1, 1.25) {};
		\node [style=gauge3] (4) at (0, 1.5) {};
		\node [style=flavour2] (5) at (1.5, 0.5) {};
		\node [style=none] (6) at (-2, 0.75) {U(1)};
		\node [style=none] (7) at (-1.5, 1.5) {U(1)};
		\node [style=none] (8) at (0, 2) {U(1)};
		\node [style=gauge3] (9) at (1, 1.25) {};
		\node [style=none] (10) at (1.3, 1.7) {U(1)};
		\node [style=none] (11) at (2.25, 0.5) {SO(1)};
		\node [style=brownlinet] (12) at (-1.5, 0.5) {};
		\node [style=greenlinet] (13) at (-1, 1.25) {};
		\node [style=magicmintlinet] (14) at (0, 1.5) {};
		\node [style=blueet] (15) at (1, 1.25) {};
	\end{pgfonlayer}
	\begin{pgfonlayer}{edgelayer}
		\draw (2) to (0);
		\draw (0) to (3);
		\draw (4) to (0);
		\draw (0) to (5);
		\draw (0) to (9);
	\end{pgfonlayer}
\end{tikzpicture}}
\end{equation}
Since the number of half-hypers are odd, it cannot be divided into an integer number of $U(1)\cong SO(2)$ gauge nodes and there  will always be a half-hyper left over. In terms of the brane system this is because a $\widetilde{\text{ON}}^-$ is equivalent to a $\text{ON}^-$ plane with a half-NS5 brane stuck in the middle which contributes the extra half-hyper. 

The U(2) in the DynkinB quiver in \eqref{B4fistone}(with the flavor node attached) is a \emph{short node} (in the sense of long and short roots of the B-type Lie algebra). If all the gauge groups are unitary, then the Coulomb branch global symmetry is SO(9). The 3d mirror will then have a SO(9) flavor symmetry as we see in \eqref{B4fistone}. If this U(2) is turned into SU(2), then the global symmetry is broken to SU(4). The 3d mirror will then have a single U(1) gauge group connected with a multiplicity 4 link and a half hyper:
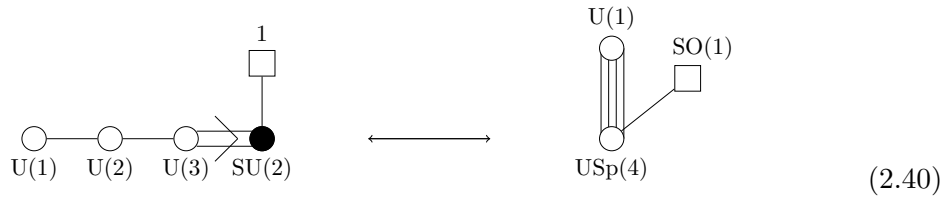
\begin{equation}
     \scalebox{0.8}{  
}
\end{equation}
Just like in \DynkinD quivers, if the short node is a U(2), then the $\widetilde{\text{ON}}^-$ will be `locked' with the NS5 branes and the SO(odd) flavor symmetry in the mirror will be enhanced. For instance:
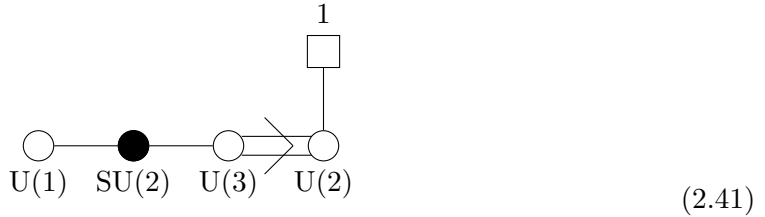
\begin{equation}
    %
\end{equation}
Drawing the brane set up and going to the Higgs phase with the appropriate lockings give:
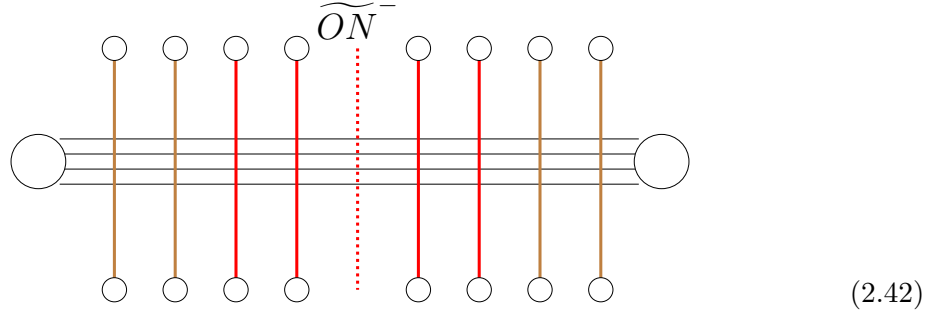
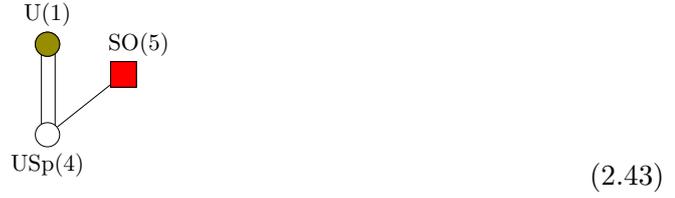
\begin{equation}
\scalebox{0.8}{%
}
\end{equation}
Remember that we started with nine half-hypers. The double bond to the U(1) consists of four half-hypers (two full hypers) and the SO(5) contributes five half-hypers so the overall number of half hypers are conserved as expected. Hilbert series computation also confirmed these relationships. 

\paragraph{\DynkinB \; balanced (General case)}
The above case can be generalized to the following family and its mirror:
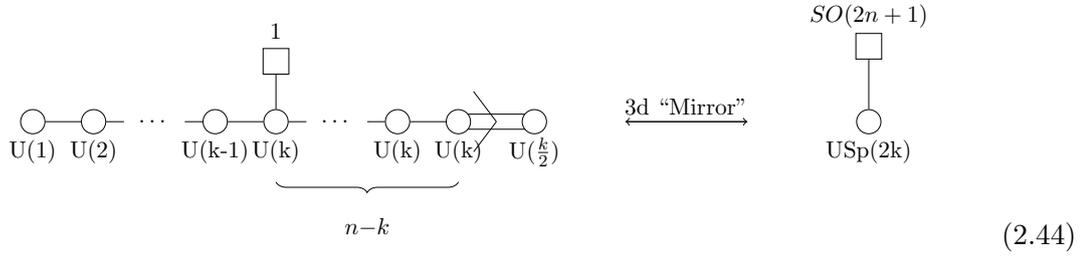
\begin{equation}
  \scalebox{0.8}{ %
}
\end{equation}
As with the D-type case, we divide general combinations of U \& SU nodes into two cases. In the first case, the short node is unitary :
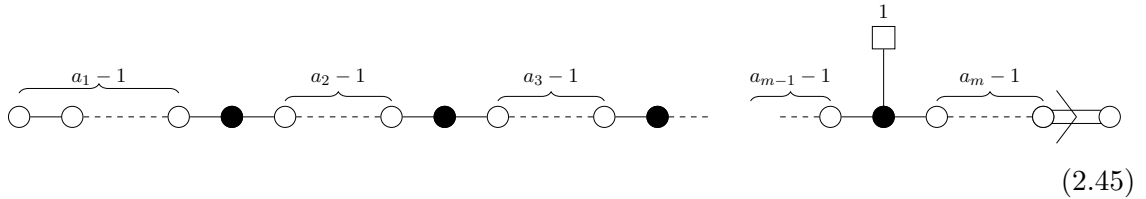
\begin{equation}
   \scalebox{0.7}{ %
}
\end{equation}
where
\begin{equation}
    \sum_{i=1}^{m}a_i=n
\end{equation}
the 3d mirror is then:
\begin{equation}
    \scalebox{0.8}{   %
}
\end{equation}
If the short node is special unitary:
\begin{equation}
  \scalebox{0.7}{ %
}
\end{equation}
and the $SO(2n+1)$ Coulomb branch global symmetry will break up into one of its subgroups. The 3d mirror is then:
\begin{equation}
    \scalebox{0.8}{   %
}
\end{equation}

\subsubsection{C-type quivers (\texorpdfstring{\DynkinC}{})}
Finally, we look at DynkinC quivers where the brane web includes an $\text{ON}^+$ plane. Consider the following $C_4$ Dynkin quiver and its 3d mirror:
\begin{equation}
  \scalebox{0.8}{%
}
\label{3dmirrorC}
\end{equation}
Notice the gauge group in the mirror is O(2) rather than SO(2) \cite{Hanany:2016gbz}. In general, the  mirror of DynkinC quivers will always have orthogonal rather than special orthogonal gauge groups. 
Turning all the gauge groups to SU gives:
\begin{equation}
  \scalebox{0.8}{ %
}
\end{equation}
whose brane set gives:
\begin{equation}
\scalebox{0.7}{   %
}
\end{equation}
and going into the Higgs phase:
\begin{equation}
   \scalebox{0.8}{%
}
\end{equation}
The U(2) gauge group on the right side of the \DynkinC quiver is a long node which corresponds to the interval containing the $\text{ON}^+$ plane in the brane set up. If this gauge node is an SU(2) instead, then the 3d mirror will not have any flavor nodes:
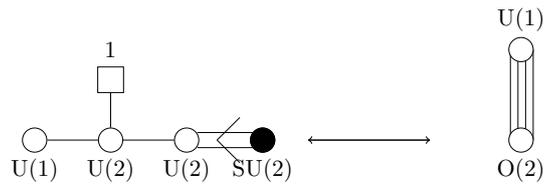
\begin{equation}
  \scalebox{0.8}{ %
}
\end{equation}
where the USp(8) global symmetry is broken to SU(4). On the otherhand, if the long node is U(2), then there will be USp flavor node in the mirror quiver:
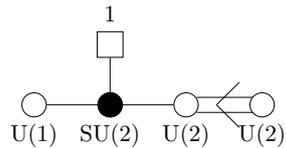
\begin{equation}
  \scalebox{0.8}{%
}
\end{equation}
where the brane set up after going to the Higgs phase gives:
\begin{equation}
 \scalebox{0.8}{  %
}
\end{equation}

\paragraph{\DynkinC\; balanced (General case)}
The above case can be generalized to the following family:
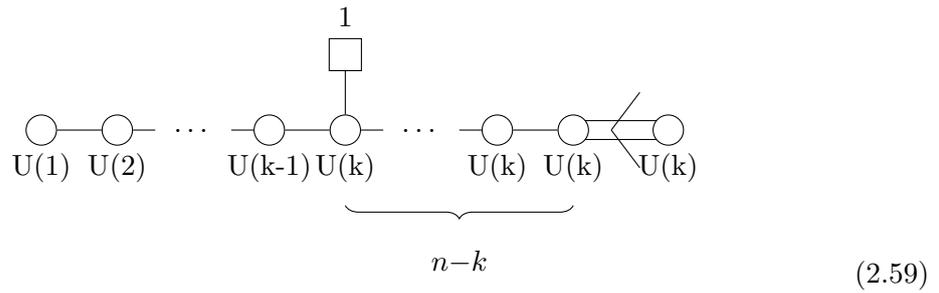
\begin{equation}
%
\end{equation}
Case 1: The long node is U(k):
\begin{equation}
   \scalebox{0.7}{%
}
\end{equation}
Case 2: The long node is SU(k):
\begin{equation}
  \scalebox{0.7}{  %
}
\end{equation}
and the 3d mirror no longer display any explicit flavor symmetry:
\begin{equation}
   \scalebox{0.8}{    %
}
\end{equation}

\paragraph{Are they really 3d mirrors?}
Since we are only able to test the Coulomb branch of the non-simply laced quiver with the Higgs branch of the orthosymplectic quiver, how sure are we that the other side of the duality matches as well? As explained above, there is currently no method in computing the Higgs branch of non-simply laced quivers so we can't match the Hilbert series to the Coulomb branch of the orthosymplectic quiver. This issue was addressed, in the case where all  gauge groups are unitary, in \cite{Cremonesi:2014xha}. Such a brane system can be constructed using a D3-D5-NS5 set up with various $\text{ON}$ planes. In this set up we take an S-dual which turns $\text{ON}$ planes into  $\text{O5}$ planes and read off the 3d mirror. The 3d mirrors there are consistent with our results here (once all the NS5s are locked together). From the point of view of S-duality, the argument for 3d mirror is clearer. Therefore, despite the lack of Hilbert series computation for the other side of the duality, we argue our proposed mirror pairs for \DynkinBC quivers are correct.

\paragraph{Overbalanced finite Dynkin quivers}
So far, we only looked at  Dynkin quivers where all the gauge nodes are balanced (in the sense that for each $U(N_i)$ or $SU(N_i)$ gauge group there are $N_f=2N_i$ hypers). If all the gauge groups in the balanced quiver are unitary, then the 3d mirror only has a single flavor node. This is because the topological symmetry of the Dynkin quivers will be a single non-Abelian group which translates to the flavor symmetry in the mirror being a single group. The situation changes when some or all of the gauge nodes are no longer balanced. Details of the more general constructions are given in Appendix \ref{overbalanced}. 

\subsection{Beyond Dynkin type quivers}\label{beyond}
In the previous section, we examined the 3d mirror of \DynkinABCD quivers with mixed U/SU gauge groups. A natural question arises: if we take \DynkinAmir and start turning U gauge groups into SU gauge groups, what will the mirror of this theory look like? Clearly, this process will not return us to \DynkinABCD quivers but will lead to something new. To make this clear, consider the following set of 3d mirror pairs:
\begin{equation}
  \scalebox{0.8}{\begin{tikzpicture}
	\begin{pgfonlayer}{nodelayer}
		\node [style=gauge3] (0) at (-4, 1.25) {};
		\node [style=gauge3] (1) at (-2.5, 1.25) {};
		\node [style=flavour2] (2) at (-4, 2.75) {};
		\node [style=flavour2] (3) at (-2.5, 2.75) {};
		\node [style=none] (4) at (-4, 0.75) {U(2)};
		\node [style=none] (5) at (-2.5, 0.75) {SU(2)};
		\node [style=none] (6) at (-4, 3.25) {4};
		\node [style=none] (7) at (-2.5, 3.25) {4};
		\node [style=blacknode] (8) at (-2.5, 1.25) {};
		\node [style=none] (9) at (-1.5, 2) {};
		\node [style=none] (10) at (0, 2) {};
		\node [style=gauge3] (11) at (0.75, 1.25) {};
		\node [style=gauge3] (12) at (2.25, 1.25) {};
		\node [style=flavour2] (14) at (2.25, 2.75) {};
		\node [style=none] (15) at (0.75, 0.75) {U(1)};
		\node [style=none] (16) at (2.25, 0.75) {U(2)};
		\node [style=none] (18) at (2.25, 3.25) {1};
		\node [style=gauge3] (20) at (3.75, 1.25) {};
		\node [style=gauge3] (21) at (5.25, 1.25) {};
		\node [style=none] (22) at (3.75, 0.75) {U(2)};
		\node [style=none] (23) at (5.25, 0.75) {U(2)};
		\node [style=gauge3] (25) at (6.75, 1.25) {};
		\node [style=gauge3] (26) at (8.25, 1.25) {};
		\node [style=none] (27) at (6.75, 0.75) {U(2)};
		\node [style=none] (28) at (8.25, 0.75) {U(2)};
		\node [style=gauge3] (30) at (9.75, 1.25) {};
		\node [style=none] (31) at (9.75, 0.75) {U(1)};
		\node [style=flavour2] (32) at (5.25, 2.75) {};
		\node [style=blacknode] (33) at (8.25, 2.75) {};
		\node [style=none] (34) at (5.25, 3.25) {1};
		\node [style=none] (35) at (8.25, 3.25) {U(1)};
		\node [style=none] (36) at (5.75, 0.75) {};
		\node [style=none] (37) at (5.75, -3.5) {};
		\node [style=none] (38) at (6.75, -1.5) {/U(1)};
		\node [style=gauge3] (39) at (0.75, -4) {};
		\node [style=gauge3] (40) at (2.25, -4) {};
		\node [style=flavour2] (41) at (2.25, -2.5) {};
		\node [style=none] (42) at (0.75, -4.5) {U(1)};
		\node [style=none] (43) at (2.25, -4.5) {U(2)};
		\node [style=none] (44) at (2.25, -2) {1};
		\node [style=gauge3] (45) at (3.75, -4) {};
		\node [style=gauge3] (46) at (5.25, -4) {};
		\node [style=none] (47) at (3.75, -4.5) {U(2)};
		\node [style=none] (48) at (5.25, -4.5) {SU(2)};
		\node [style=gauge3] (49) at (6.75, -4) {};
		\node [style=gauge3] (50) at (8.25, -4) {};
		\node [style=none] (51) at (6.75, -4.5) {U(2)};
		\node [style=none] (52) at (8.25, -4.5) {U(2)};
		\node [style=gauge3] (53) at (9.75, -4) {};
		\node [style=none] (54) at (9.75, -4.5) {U(1)};
		\node [style=flavour2] (55) at (5.25, -2.5) {};
		\node [style=blacknode] (56) at (8.25, -2.5) {};
		\node [style=none] (57) at (5.25, -2) {1};
		\node [style=none] (58) at (8.25, -2) {U(1)};
		\node [style=blacknode] (59) at (5.25, -4) {};
		\node [style=none] (60) at (-1.5, -4) {};
		\node [style=none] (61) at (0, -4) {};
		\node [style=gauge3] (62) at (-4, -4) {};
		\node [style=gauge3] (63) at (-2.5, -4) {};
		\node [style=none] (66) at (-4, -4.5) {U(2)};
		\node [style=none] (67) at (-2.5, -4.5) {SU(2)};
		\node [style=none] (68) at (-4, -2) {U(1)};
		\node [style=none] (69) at (-2.5, -2) {4};
		\node [style=blacknode] (70) at (-2.5, -4) {};
		\node [style=gauge3] (72) at (8.25, 2.75) {};
		\node [style=gauge3] (73) at (8.25, -2.5) {};
		\node [style=none] (74) at (-4.175, -2.5) {};
		\node [style=none] (75) at (-4.05, -2.5) {};
		\node [style=none] (76) at (-3.925, -2.5) {};
		\node [style=none] (77) at (-3.8, -2.5) {};
		\node [style=none] (78) at (-4.175, -4) {};
		\node [style=none] (79) at (-4.05, -4) {};
		\node [style=none] (80) at (-3.925, -4) {};
		\node [style=none] (81) at (-3.8, -4) {};
		\node [style=none] (82) at (-0.75, 2.25) {3d mirror};
		\node [style=none] (83) at (-0.75, -3.75) {3d mirror};
		\node [style=gauge3] (84) at (-4, -2.5) {};
		\node [style=flavour2] (85) at (-2.5, -2.5) {};
		\node [style=gauge3] (86) at (-4, 6.5) {};
		\node [style=gauge3] (87) at (-2.5, 6.5) {};
		\node [style=flavour2] (88) at (-4, 8) {};
		\node [style=flavour2] (89) at (-2.5, 8) {};
		\node [style=none] (90) at (-4, 6) {U(2)};
		\node [style=none] (91) at (-2.5, 6) {U(2)};
		\node [style=none] (92) at (-4, 8.5) {4};
		\node [style=none] (93) at (-2.5, 8.5) {4};
		\node [style=none] (95) at (-1.5, 7.25) {};
		\node [style=none] (96) at (0, 7.25) {};
		\node [style=gauge3] (97) at (0.75, 6.5) {};
		\node [style=gauge3] (98) at (2.25, 6.5) {};
		\node [style=flavour2] (99) at (2.25, 8) {};
		\node [style=none] (100) at (0.75, 6) {U(1)};
		\node [style=none] (101) at (2.25, 6) {U(2)};
		\node [style=none] (102) at (2.25, 8.5) {1};
		\node [style=gauge3] (103) at (3.75, 6.5) {};
		\node [style=gauge3] (104) at (5.25, 6.5) {};
		\node [style=none] (105) at (3.75, 6) {U(2)};
		\node [style=none] (106) at (5.25, 6) {U(2)};
		\node [style=gauge3] (107) at (6.75, 6.5) {};
		\node [style=gauge3] (108) at (8.25, 6.5) {};
		\node [style=none] (109) at (6.75, 6) {U(2)};
		\node [style=none] (110) at (8.25, 6) {U(2)};
		\node [style=gauge3] (111) at (9.75, 6.5) {};
		\node [style=none] (112) at (9.75, 6) {U(1)};
		\node [style=flavour2] (113) at (5.25, 8) {};
		\node [style=blacknode] (114) at (8.25, 8) {};
		\node [style=none] (115) at (5.25, 8.5) {1};
		\node [style=none] (116) at (8.25, 8.5) {1};
		\node [style=none] (117) at (-2.5, 5.5) {};
		\node [style=none] (118) at (-2, 1.75) {};
		\node [style=none] (119) at (-1.25, 3.75) {/U(1)};
		\node [style=gauge3] (120) at (8.25, 8) {};
		\node [style=none] (121) at (-0.75, 7.5) {3d mirror};
		\node [style=flavour2] (122) at (8.25, 8) {};
		\node [style=none] (123) at (-5.5, 7.25) {DynkinA};
		\node [style=none] (124) at (11.75, 7.25) {$\text{DynkinA}_{\text{\tiny mirror}}$};
		\node [style=none] (125) at (-5.5, 2) {\DynkinA};
		\node [style=none] (126) at (11.75, 2) {\DynkinAmir};
		\node [style=none] (127) at (-5.5, -3) {\DynkinAbar};
		\node [style=none] (128) at (11.75, -3) {\DynkinAbar};
	\end{pgfonlayer}
	\begin{pgfonlayer}{edgelayer}
		\draw (2) to (0);
		\draw (0) to (1);
		\draw (1) to (3);
		\draw [style=arrowed] (9.center) to (10.center);
		\draw [style=arrowed] (10.center) to (9.center);
		\draw (11) to (12);
		\draw (12) to (14);
		\draw (20) to (21);
		\draw (25) to (26);
		\draw (12) to (30);
		\draw (32) to (21);
		\draw (33) to (26);
		\draw [style=arrowed, bend left, looseness=0.75] (36.center) to (37.center);
		\draw (39) to (40);
		\draw (40) to (41);
		\draw (45) to (46);
		\draw (49) to (50);
		\draw (40) to (53);
		\draw (55) to (46);
		\draw (56) to (50);
		\draw [style=arrowed] (60.center) to (61.center);
		\draw [style=arrowed] (61.center) to (60.center);
		\draw (62) to (63);
		\draw (77.center) to (81.center);
		\draw (80.center) to (76.center);
		\draw (75.center) to (79.center);
		\draw (78.center) to (74.center);
		\draw (85) to (70);
		\draw (88) to (86);
		\draw (86) to (87);
		\draw (87) to (89);
		\draw [style=arrowed] (95.center) to (96.center);
		\draw [style=arrowed] (96.center) to (95.center);
		\draw (97) to (98);
		\draw (98) to (99);
		\draw (103) to (104);
		\draw (107) to (108);
		\draw (98) to (111);
		\draw (113) to (104);
		\draw (114) to (108);
		\draw [style=arrowed, bend left=15] (117.center) to (118.center);
	\end{pgfonlayer}
\end{tikzpicture}
}
\label{bigeq}
\end{equation}
Let’s review how we constructed these mirror pairs. We start with the two-gauge-group quiver on the top left (DynkinA), which has a known 3d mirror on the top right ($\text{DynkinA}_{\text{\tiny mirror}}$). In the next row, we turn the right U(2) into an SU(2) after ungauging a U(1) in the DynkinA, making the theory a \DynkinA. Following our usual U/SU algorithm, we then obtain the mirror quiver on the central right (\DynkinAmir). Next, focusing on this \DynkinAmir quiver, we ungauge a U(1), so the central U(2) becomes an SU(2). At this point, we introduce something new.  Because of that U(1) bouquet on the right, the quiver is non-linear—and thus outside the family \DynkinA or \DynkinAmir \;\footnote{By accident, this does belong to the family of D-type Dynkin quivers we studied earlier. But in general this will not be the case.}. We label this quiver as belonging to the new family \DynkinAbar. Next, we conjecture that we can simply ignore this U(1) bouquet and proceed with the U/SU algorithm. This leads to the gauging of a U(1) flavor node in the mirror on the bottom left. Indeed, upon computing the Hilbert series on both sides, it is confirmed to be correct. 

Unlike previous cases, both quivers in the bottom belong to the all encompassing \DynkinAbar\; family. In fact, a sequence of \DynkinA $\rightarrow$ take 3d mirror $\rightarrow$ take combination of U/SU $\rightarrow$ take 3d mirror $\rightarrow$ \dots will never take you outside of the \DynkinAbar\; family. 

When discussing \DynkinA quivers in \cite{USU},  we always have a framed/flavored quiver on one side (\DynkinA), while the mirror  (\DynkinAmir)  contains only unitary gauge groups, which are left unframed/flavorless. We left the unitary quiver flavorless because when all the gauge groups are unitary, there is always an overall U(1) subgroup that decouples, and it \emph{does not matter} from which gauge group(s) it is decoupled. Thus, this ambiguity does not affect the result. For example, in the central row of \eqref{bigeq}, we could have alternatively choose the following U(1) ungauging to get the flavored unitary quiver on the right:
\begin{equation}
   \scalebox{0.8}{  \begin{tikzpicture}
	\begin{pgfonlayer}{nodelayer}
		\node [style=gauge3] (0) at (-4, 1.25) {};
		\node [style=gauge3] (1) at (-2.5, 1.25) {};
		\node [style=flavour2] (2) at (-4, 2.75) {};
		\node [style=flavour2] (3) at (-2.5, 2.75) {};
		\node [style=none] (4) at (-4, 0.75) {U(2)};
		\node [style=none] (5) at (-2.5, 0.75) {SU(2)};
		\node [style=none] (6) at (-4, 3.25) {4};
		\node [style=none] (7) at (-2.5, 3.25) {4};
		\node [style=blacknode] (8) at (-2.5, 1.25) {};
		\node [style=none] (9) at (-1.5, 2) {};
		\node [style=none] (10) at (0, 2) {};
		\node [style=gauge3] (11) at (0.75, 1.25) {};
		\node [style=gauge3] (12) at (2.25, 1.25) {};
		\node [style=none] (15) at (0.75, 0.75) {U(1)};
		\node [style=none] (16) at (2.25, 0.75) {U(2)};
		\node [style=none] (18) at (8.25, 3.25) {1};
		\node [style=gauge3] (20) at (3.75, 1.25) {};
		\node [style=gauge3] (21) at (5.25, 1.25) {};
		\node [style=none] (22) at (3.75, 0.75) {U(2)};
		\node [style=none] (23) at (5.25, 0.75) {U(2)};
		\node [style=gauge3] (25) at (6.75, 1.25) {};
		\node [style=gauge3] (26) at (8.25, 1.25) {};
		\node [style=none] (27) at (6.75, 0.75) {U(2)};
		\node [style=none] (28) at (8.25, 0.75) {U(2)};
		\node [style=gauge3] (30) at (9.75, 1.25) {};
		\node [style=none] (31) at (9.75, 0.75) {U(1)};
		\node [style=blacknode] (33) at (8.25, 2.75) {};
		\node [style=none] (35) at (3.75, 3.25) {U(1)};
		\node [style=none] (36) at (5.75, 0.75) {};
		\node [style=none] (37) at (5.5, -3.5) {};
		\node [style=none] (38) at (6.5, -1.5) {/U(1)};
		\node [style=none] (60) at (-1.5, -4) {};
		\node [style=none] (61) at (0, -4) {};
		\node [style=gauge3] (62) at (-4, -4) {};
		\node [style=gauge3] (63) at (-2.5, -4) {};
		\node [style=none] (66) at (-4, -4.5) {U(2)};
		\node [style=none] (67) at (-2.5, -4.5) {SU(2)};
		\node [style=none] (68) at (-4, -2) {U(1)};
		\node [style=none] (69) at (-2.5, -2) {4};
		\node [style=blacknode] (70) at (-2.5, -4) {};
		\node [style=none] (74) at (-4.175, -2.5) {};
		\node [style=none] (75) at (-4.05, -2.5) {};
		\node [style=none] (76) at (-3.925, -2.5) {};
		\node [style=none] (77) at (-3.8, -2.5) {};
		\node [style=none] (78) at (-4.175, -4) {};
		\node [style=none] (79) at (-4.05, -4) {};
		\node [style=none] (80) at (-3.925, -4) {};
		\node [style=none] (81) at (-3.8, -4) {};
		\node [style=none] (82) at (-0.75, 2.25) {3d mirror};
		\node [style=none] (83) at (-0.75, -3) {3d mirror};
		\node [style=gauge3] (84) at (-4, -2.5) {};
		\node [style=flavour2] (85) at (-2.5, -2.5) {};
		\node [style=gauge3] (86) at (3.75, 2.75) {};
		\node [style=flavour2] (87) at (8.25, 2.75) {};
		\node [style=gauge3] (89) at (0.75, -4) {};
		\node [style=gauge3] (90) at (2.25, -4) {};
		\node [style=none] (91) at (0.75, -4.5) {U(1)};
		\node [style=none] (92) at (2.25, -4.5) {U(2)};
		\node [style=none] (93) at (8.25, -2) {1};
		\node [style=gauge3] (94) at (3.75, -4) {};
		\node [style=gauge3] (95) at (5.25, -4) {};
		\node [style=none] (96) at (3.75, -4.5) {U(2)};
		\node [style=none] (97) at (5.25, -4.5) {SU(2)};
		\node [style=gauge3] (98) at (6.75, -4) {};
		\node [style=gauge3] (99) at (8.25, -4) {};
		\node [style=none] (100) at (6.75, -4.5) {U(2)};
		\node [style=none] (101) at (8.25, -4.5) {U(2)};
		\node [style=gauge3] (102) at (9.75, -4) {};
		\node [style=none] (103) at (9.75, -4.5) {U(1)};
		\node [style=blacknode] (104) at (8.25, -2.5) {};
		\node [style=none] (105) at (3.75, -2) {U(1)};
		\node [style=gauge3] (106) at (3.75, -2.5) {};
		\node [style=flavour2] (107) at (8.25, -2.5) {};
		\node [style=blacknode] (108) at (5.25, -4) {};
		\node [style=none] (109) at (-0.25, -3.5) {};
		\node [style=none] (110) at (-1.25, -4.5) {};
		\node [style=none] (111) at (-1.25, -3.5) {};
		\node [style=none] (112) at (-0.25, -4.5) {};
		\node [style=none] (113) at (-1.575, -7.25) {};
		\node [style=none] (114) at (0.425, -5.5) {};
		\node [style=gauge3] (115) at (-4.075, -7.25) {};
		\node [style=gauge3] (116) at (-2.575, -7.25) {};
		\node [style=none] (117) at (-4.075, -7.75) {U(2)};
		\node [style=none] (118) at (-2.575, -7.75) {SU(2)};
		\node [style=none] (119) at (-2.575, -5.25) {U(1)};
		\node [style=none] (120) at (-4.075, -5.25) {4};
		\node [style=blacknode] (121) at (-2.575, -7.25) {};
		\node [style=none] (122) at (-2.75, -5.75) {};
		\node [style=none] (123) at (-2.625, -5.75) {};
		\node [style=none] (124) at (-2.5, -5.75) {};
		\node [style=none] (125) at (-2.375, -5.75) {};
		\node [style=none] (126) at (-2.75, -7.25) {};
		\node [style=none] (127) at (-2.625, -7.25) {};
		\node [style=none] (128) at (-2.5, -7.25) {};
		\node [style=none] (129) at (-2.375, -7.25) {};
		\node [style=none] (130) at (-0.075, -7) {3d mirror};
		\node [style=gauge3] (131) at (-2.575, -5.75) {};
		\node [style=flavour2] (132) at (-4.075, -5.75) {};
	\end{pgfonlayer}
	\begin{pgfonlayer}{edgelayer}
		\draw (2) to (0);
		\draw (0) to (1);
		\draw (1) to (3);
		\draw [style=arrowed] (9.center) to (10.center);
		\draw [style=arrowed] (10.center) to (9.center);
		\draw (11) to (12);
		\draw (20) to (21);
		\draw (25) to (26);
		\draw (12) to (30);
		\draw (33) to (26);
		\draw [style=arrowed, bend left=15] (36.center) to (37.center);
		\draw [style=arrowed] (60.center) to (61.center);
		\draw [style=arrowed] (61.center) to (60.center);
		\draw (62) to (63);
		\draw (77.center) to (81.center);
		\draw (80.center) to (76.center);
		\draw (75.center) to (79.center);
		\draw (78.center) to (74.center);
		\draw (85) to (70);
		\draw (12) to (86);
		\draw (86) to (21);
		\draw (89) to (90);
		\draw (94) to (95);
		\draw (98) to (99);
		\draw (90) to (102);
		\draw (104) to (99);
		\draw (90) to (106);
		\draw (106) to (95);
		\draw (111.center) to (112.center);
		\draw (110.center) to (109.center);
		\draw [style=arrowed] (113.center) to (114.center);
		\draw [style=arrowed] (114.center) to (113.center);
		\draw (115) to (116);
		\draw (125.center) to (129.center);
		\draw (128.center) to (124.center);
		\draw (123.center) to (127.center);
		\draw (126.center) to (122.center);
		\draw (132) to (115);
	\end{pgfonlayer}
\end{tikzpicture}}
\end{equation}
The top right quiver here is the same theory in every way as the central right quiver in \eqref{bigeq}. However, once we start converting gauge groups from unitary to special unitary, the choice \emph{does} matter. The reasoning is because for a quiver with only unitary gauge groups, we are free to move the decoupled U(1) anywhere around the quiver, but in the presence of a non-unitary gauge group (e.g special unitary), this freedom is lost. For instance, compare the bottom right quiver in \eqref{bigeq} with the bottom right quiver here—while in both cases we turned the central U(2) into an SU(2), the resulting theories are now different. Consequently, their 3d mirrors differ as well. 

Turning our attention to the mirror on the bottom left, our algorithm does not clearly indicate whether one should gauge the flavor group on the SU(2) or the U(2) gauge group. Only one choice is correct and one can lift this degeneracy in the following way: gauging the flavor group on top of the U(2) results in a global symmetry of $SU(4) \times SO(8)$, whereas gauging it on the SU(2) gives $SU(4) \times U(4)$. Comparing this to the Coulomb branch global symmetry on the mirror quiver, which has a balanced $D_4$ sub diagram, it becomes clear the choice containing $SO(8)$ is correct. However, the difference may not always be apparent at the level of global symmetry. If that is the case, the next step is to explicitly compute  Hilbert series for the different gauging choices. From examples we've done, one of the choices is bound to be correct.  We leave the development of a simpler algorithm that resolves this ambiguity for future work.

One can extend this to the BCD types. The advantage here is that the ambiguity of decoupling U(1)s that we had above will not occur since the center of mass is always fixed. However, a sequence of taking 3d mirrors and U/SU combinations will not stay within a single family. Instead, it will give us the \DynkinBCDbar family and its 3d mirror \DynkinBCDmirbar. This is clear because \DynkinBCDbar will never include orthosymplectic gauge groups, whereas \DynkinBCDmirbar always will.

\section{Bootstrapping 3d mirrors by gauging topological symmetries}\label{top}
A crucial feature of all the examples we've constructed so far is that both theories in the mirror pairs are Lagrangian (quiver gauge theories)\footnote{In this paper, we considered \DynkinBCbar quivers as Lagrangian. See Section \ref{conc} for a more detailed discussion on this.}. The importance of constructing Lagrangian mirror pairs is that both theories are accessible through the wealth of tools available in the literature, such as partition functions, Higgsing algorithms, and more. Depending on various features of the theory—such as unitary vs orthosymplectic gauge groups or linear vs non-linear quivers—one of the two theories is often easier to study. If a quiver gauge theory proves difficult to analyze directly, it can be useful to study it through its mirror dual—but this is not possible if the mirror turns out to be non-Lagrangian. In this section we discuss the possibility of bootstrapping 3d mirror pairs by gauging topological symmetries, but also discussing whether the resulting mirrors are Lagrangian. 

So far, our exercises involves the art of gauging and ungauging U(1) symmetries in finding new 3d mirrors.  The reason why such a process is effective in finding 3d mirrors is because for $3d$ $\mathcal{N}=4$ theories, each unitary gauge group  carries a U(1) topological global symmetry on the Coulomb branch. The act of ungauging a U(1) gauge symmetry, as we've done so far, to turn a unitary gauge group to a special unitary gauge group is the same as the act of gauging this topological U(1) global symmetry. In the following, rather than saying ungauging gauge symmetries, it is more instructive to talk about the equivalent process of gauging topological global symmetries instead. On the 3d mirror, this means gauging a U(1) flavor symmetry. Gauging Abelian flavor symmetries is straightforward, with the primary challenge being the identification of the specific U(1) flavor symmetry to gauge—this is where our algorithms come into play\footnote{The gauging of U(1) flavor and U(1) topological symmetries to finding 3d mirror pairs also plays a prominent role in \cite{Closset:2020afy,Closset:2020scj}. }. However, gauging non-Abelian topological and flavor symmetries is a different story. 

It is well known that these U(1) topological symmetries may enhance to some non-Abelian topological symmetry $G$ in the IR. The question is: if $\mathcal{T}$ has a mirror theory $\mathcal{T}^{\text{mirror}}$, can we gauge a non-Abelian topological symmetry $G$ in $\mathcal{T}$ to obtain a new mirror, which is simply $\mathcal{T}^{\text{mirror}}$ with a subgroup $G \subset F$ gauged, where $F$ is a flavor symmetry? Consider the following quiver:
\begin{equation}
\scalebox{0.7}{
\begin{tikzpicture}
	\begin{pgfonlayer}{nodelayer}
		\node [style=gauge3] (1) at (-2.5, 1.25) {};
		\node [style=flavour2] (3) at (-2.5, 2.75) {};
		\node [style=none] (5) at (-2.5, 0.75) {Sp(4)};
		\node [style=none] (7) at (-2.5, 3.25) {16};
		\node [style=none] (9) at (-1, 2) {};
		\node [style=none] (10) at (0.5, 2) {};
		\node [style=gauge3] (11) at (1.25, 1.25) {};
		\node [style=gauge3] (12) at (2.75, 1.25) {};
		\node [style=none] (15) at (1.25, 0.75) {U(1)};
		\node [style=none] (16) at (2.75, 0.75) {U(2)};
		\node [style=gauge3] (20) at (4.25, 1.25) {};
		\node [style=gauge3] (21) at (5.75, 1.25) {};
		\node [style=none] (22) at (4.25, 0.75) {U(3)};
		\node [style=none] (23) at (5.75, 0.75) {U(4)};
		\node [style=gauge3] (25) at (7.25, 1.25) {};
		\node [style=gauge3] (26) at (8.75, 1.25) {};
		\node [style=none] (27) at (7.25, 0.75) {U(5)};
		\node [style=none] (28) at (8.75, 0.75) {U(6)};
		\node [style=gauge3] (30) at (10.25, 1.25) {};
		\node [style=none] (31) at (10.25, 0.75) {U(4)};
		\node [style=none] (82) at (-0.25, 2.25) {3d mirror};
		\node [style=flavour2] (83) at (11.5, 1.25) {};
		\node [style=none] (84) at (11.5, 0.75) {2};
		\node [style=gauge3] (85) at (8.75, 2.5) {};
		\node [style=none] (86) at (8.75, 3) {U(3)};
		\node [style=none] (87) at (6.5, 0) {};
		\node [style=none] (88) at (6.5, -2.75) {};
		\node [style=none] (90) at (-2.5, 0) {};
		\node [style=none] (91) at (-2.5, -2.75) {};
		\node [style=none] (93) at (0.75, -1) {(Gauge SU(2) topological symmetry)};
		\node [style=none] (94) at (9.25, -1) {(Gauge SU(2) flavor symmetry)};
		\node [style=gauge3] (95) at (1.25, -4) {};
		\node [style=gauge3] (96) at (2.75, -4) {};
		\node [style=none] (97) at (1.25, -4.5) {U(1)};
		\node [style=none] (98) at (2.75, -4.5) {U(2)};
		\node [style=gauge3] (99) at (4.25, -4) {};
		\node [style=gauge3] (100) at (5.75, -4) {};
		\node [style=none] (101) at (4.25, -4.5) {U(3)};
		\node [style=none] (102) at (5.75, -4.5) {U(4)};
		\node [style=gauge3] (103) at (7.25, -4) {};
		\node [style=gauge3] (104) at (8.75, -4) {};
		\node [style=none] (105) at (7.25, -4.5) {U(5)};
		\node [style=none] (106) at (8.75, -4.5) {U(6)};
		\node [style=gauge3] (107) at (10.25, -4) {};
		\node [style=none] (108) at (10.25, -4.5) {U(4)};
		\node [style=none] (110) at (11.5, -4.5) {SU(2)};
		\node [style=gauge3] (111) at (8.75, -2.75) {};
		\node [style=none] (112) at (8.75, -2.25) {U(3)};
		\node [style=gauge3] (113) at (11.5, -4) {};
		\node [style=none] (114) at (-1, -4) {};
		\node [style=none] (115) at (0.5, -4) {};
		\node [style=none] (116) at (-0.25, -3.75) {3d mirror};
		\node [style=none] (117) at (-2.5, -3.75) {Non-Lagrangian};
		\node [style=none] (118) at (-2.5, -4.25) {$E_8$ theory};
		\node [style=none] (119) at (12.5, -4) {};
		\node [style=none] (120) at (11.75, -5.25) {};
		\node [style=none] (121) at (12.5, -5) {$\mathbb{Z}_2$};
		\node [style=none] (122) at (7, -5.5) {};
		\node [style=none] (123) at (7.25, -5.5) {};
		\node [style=none] (124) at (7, -6) {};
		\node [style=none] (125) at (7.25, -6) {};
		\node [style=gauge3] (126) at (1.25, -8.25) {};
		\node [style=gauge3] (127) at (2.75, -8.25) {};
		\node [style=none] (128) at (1.25, -8.75) {U(1)};
		\node [style=none] (129) at (2.75, -8.75) {U(2)};
		\node [style=gauge3] (130) at (4.25, -8.25) {};
		\node [style=gauge3] (131) at (5.75, -8.25) {};
		\node [style=none] (132) at (4.25, -8.75) {U(3)};
		\node [style=none] (133) at (5.75, -8.75) {U(4)};
		\node [style=gauge3] (134) at (7.25, -8.25) {};
		\node [style=gauge3] (135) at (8.75, -8.25) {};
		\node [style=none] (136) at (7.25, -8.75) {U(5)};
		\node [style=none] (137) at (8.75, -8.75) {U(6)};
		\node [style=gauge3] (138) at (10.25, -8.25) {};
		\node [style=none] (139) at (10.25, -8.75) {U(4)};
		\node [style=none] (140) at (11.5, -8.75) {U(2)};
		\node [style=gauge3] (141) at (8.75, -7) {};
		\node [style=none] (142) at (8.75, -6.5) {U(3)};
		\node [style=gauge3] (143) at (11.5, -8.25) {};
		\node [style=none] (144) at (12.5, -8.25) {};
		\node [style=none] (145) at (11.75, -9.5) {};
		\node [style=none] (146) at (12.5, -9.25) {U(1)};
	\end{pgfonlayer}
	\begin{pgfonlayer}{edgelayer}
		\draw (1) to (3);
		\draw [style=arrowed] (9.center) to (10.center);
		\draw [style=arrowed] (10.center) to (9.center);
		\draw (11) to (12);
		\draw (20) to (21);
		\draw (25) to (26);
		\draw (12) to (30);
		\draw (26) to (85);
		\draw (30) to (83);
		\draw [style=arrowed] (87.center) to (88.center);
		\draw [style=arrowed] (90.center) to (91.center);
		\draw (95) to (96);
		\draw (99) to (100);
		\draw (103) to (104);
		\draw (96) to (107);
		\draw (104) to (111);
		\draw [style=arrowed] (114.center) to (115.center);
		\draw [style=arrowed] (115.center) to (114.center);
		\draw (107) to (113);
		\draw (120.center) to (119.center);
		\draw (123.center) to (125.center);
		\draw (124.center) to (122.center);
		\draw (126) to (127);
		\draw (130) to (131);
		\draw (134) to (135);
		\draw (127) to (138);
		\draw (135) to (141);
		\draw (138) to (143);
		\draw (145.center) to (144.center);
	\end{pgfonlayer}
\end{tikzpicture}}
\end{equation}
The $Sp(4)$ gauge group on the top left side has a $-1$ balance according to the definition in \cite{Gaiotto:2008ak}. Nevertheless, it has been argued in \cite{exo3} that the Coulomb branch is well defined and has a topological global symmetry of $SU(2)$. We indeed see this in the 3d mirror where there is a $SU(2)$ flavor node. After gauging the $SU(2)$, there is an overall $\mathbb{Z}_2$ that is also quotiented out (see \cite{Bourget:2020xdz} for more discussion on this topic) and the resulting quiver is the same as the affine $E_8$ unitary quiver\footnote{Since it is a unitary quiver, there is always a U(1) that automatically decouples. We will draw the U(1) quotient rather than explicitly flavoring/framing the quiver in this section for clarity.}. The proposal at the bottom is that $Sp(4)$ gauge theory with 16 half-hypers with a $SU(2)$ topological symmetry gauged is 3d mirror to the $E_8$ affine Dynkin quiver.   This relation can be checked\footnote{ We are grateful of Simone Giacomelli  and Noppadol Mekareeya  who carried out this computation and verified the result.} following a similar computation involving $Sp(2)$ gauge group with 10 half-hypers with a $U(1)$ topological symmetry gauged being 3d mirror to the $E_6$ affine Dynkin quiver  \cite{Carta:2021whq}.    
\\ \newline
If we adopt this approach, gauging non-Abelian topological symmetries become an efficient, almost trivial, way of bootstrapping 3d mirror pairs. 
\\ \newline
Any non-linear quiver gauge theory with unitary gauge groups can be obtained by gluing/gauging common flavor nodes $G_i$ of linear quivers from the $T_\rho (G_i)$ family. These linear quivers have   3d mirrors $T^\rho (G_i)$. These $G_i$ flavor symmetries becomes the IR $G_i$ topological symmetries in the 3d mirror, thus the mirror gluing process is the gauging of these non-Abelian topological symmetries. Let's see this in more detail for the 4d $\mathcal{N}=2$  class $\mathcal{S}$ theory  known as $T_4$ compactified on a circle and its 3d mirror. 
\begin{equation}
\scalebox{0.8}{\begin{tikzpicture}
	\begin{pgfonlayer}{nodelayer}
		\node [style=gauge3] (0) at (-0.75, -3.75) {};
		\node [style=gauge3] (1) at (0.25, -3.75) {};
		\node [style=gauge3] (2) at (1.25, -3.75) {};
		\node [style=flavour2] (3) at (2.25, -3.75) {};
		\node [style=none] (7) at (-0.75, -4.25) {U(1)};
		\node [style=none] (8) at (0.25, -4.25) {U(2)};
		\node [style=none] (9) at (1.25, -4.25) {U(3)};
		\node [style=none] (10) at (2.25, -4.25) {4};
		\node [style=gauge3] (11) at (6.25, -3.75) {};
		\node [style=gauge3] (12) at (5.25, -3.75) {};
		\node [style=gauge3] (13) at (4.25, -3.75) {};
		\node [style=flavour2] (14) at (3.25, -3.75) {};
		\node [style=none] (18) at (6.25, -4.25) {U(1)};
		\node [style=none] (19) at (5.25, -4.25) {U(2)};
		\node [style=none] (20) at (4.25, -4.25) {U(3)};
		\node [style=none] (21) at (3.25, -4.25) {4};
		\node [style=gauge3] (22) at (2.75, 0) {};
		\node [style=gauge3] (23) at (2.75, -1) {};
		\node [style=gauge3] (24) at (2.75, -2) {};
		\node [style=flavour2] (25) at (2.75, -3) {};
		\node [style=none] (29) at (3.25, 0) {U(1)};
		\node [style=none] (30) at (3.25, -1) {U(2)};
		\node [style=none] (31) at (3.25, -2) {U(3)};
		\node [style=none] (32) at (3.25, -3) {4};
		\node [style=none] (33) at (2.75, -2.5) {};
		\node [style=none] (34) at (1.75, -3.5) {};
		\node [style=none] (35) at (3.75, -3.5) {};
		\node [style=none] (36) at (2.75, -4.5) {};
		\node [style=none] (37) at (2.75, -5) {\textcolor{brown}{Gauge $SU(4)$ flavor symmetry}};
		\node [style=none] (38) at (-2.75, -2.25) {};
		\node [style=none] (39) at (-1.25, -2.25) {};
		\node [style=none] (41) at (-2, -1.75) {3d mirror};
		\node [style=none] (132) at (2.625, -5.75) {};
		\node [style=none] (133) at (2.625, -6.25) {};
		\node [style=none] (134) at (2.875, -6.25) {};
		\node [style=none] (135) at (2.875, -5.75) {};
		\node [style=none] (150) at (-3.25, -3.5) {};
		\node [style=none] (151) at (-4.5, -5) {};
		\node [style=none] (152) at (-3, -4.75) {\textcolor{red}{Gauge $SU(4)$ }};
		\node [style=none] (153) at (-3.5, -5.25) {\textcolor{red}{ topological symmetry}};
		\node [style=none] (154) at (-6, -5.75) {};
		\node [style=none] (155) at (-6, -6.25) {};
		\node [style=none] (156) at (-5.75, -6.25) {};
		\node [style=none] (157) at (-5.75, -5.75) {};
		\node [style=none] (158) at (-5.75, -8.75) {Non-Lagrangian $T_4$ theory on $S^1$};
		\node [style=gauge3] (159) at (-0.25, -9.75) {};
		\node [style=gauge3] (160) at (0.75, -9.75) {};
		\node [style=gauge3] (161) at (1.75, -9.75) {};
		\node [style=none] (163) at (-0.25, -10.25) {U(1)};
		\node [style=none] (164) at (0.75, -10.25) {U(2)};
		\node [style=none] (165) at (1.75, -10.25) {U(3)};
		\node [style=none] (166) at (2.75, -10.25) {U(4)};
		\node [style=gauge3] (167) at (5.75, -9.75) {};
		\node [style=gauge3] (168) at (4.75, -9.75) {};
		\node [style=gauge3] (169) at (3.75, -9.75) {};
		\node [style=none] (171) at (5.75, -10.25) {U(1)};
		\node [style=none] (172) at (4.75, -10.25) {U(2)};
		\node [style=none] (173) at (3.75, -10.25) {U(3)};
		\node [style=gauge3] (175) at (2.75, -6.75) {};
		\node [style=gauge3] (176) at (2.75, -7.75) {};
		\node [style=gauge3] (177) at (2.75, -8.75) {};
		\node [style=none] (179) at (3.5, -6.75) {U(1)};
		\node [style=none] (180) at (3.5, -7.75) {U(2)};
		\node [style=none] (181) at (3.5, -8.75) {U(3)};
		\node [style=gauge3] (183) at (2.75, -9.75) {};
		\node [style=none] (184) at (6.75, -9.75) {};
		\node [style=none] (185) at (5.5, -11.25) {};
		\node [style=none] (186) at (6.5, -10.75) {U(1)};
		\node [style=none] (187) at (-2.75, -8.75) {};
		\node [style=none] (188) at (-1.25, -8.75) {};
		\node [style=none] (189) at (-2, -8.25) {3d mirror};
		\node [style=gauge3] (190) at (-7.5, -0.5) {};
		\node [style=gauge3] (191) at (-6.5, -0.5) {};
		\node [style=gauge3] (192) at (-5.5, -0.5) {};
		\node [style=flavour2] (193) at (-4.5, -0.5) {};
		\node [style=none] (194) at (-7.5, -1) {\textcolor{red}{U(1)}};
		\node [style=none] (195) at (-6.5, -1) {\textcolor{red}{U(2)}};
		\node [style=none] (196) at (-5.5, -1) {\textcolor{red}{U(3)}};
		\node [style=none] (197) at (-4.5, -1) {4};
		\node [style=gauge3] (198) at (-7.5, -2) {};
		\node [style=gauge3] (199) at (-6.5, -2) {};
		\node [style=gauge3] (200) at (-5.5, -2) {};
		\node [style=flavour2] (201) at (-4.5, -2) {};
		\node [style=none] (205) at (-4.5, -2.5) {4};
		\node [style=gauge3] (206) at (-7.5, -3.5) {};
		\node [style=gauge3] (207) at (-6.5, -3.5) {};
		\node [style=gauge3] (208) at (-5.5, -3.5) {};
		\node [style=flavour2] (209) at (-4.5, -3.5) {};
		\node [style=none] (213) at (-4.5, -4) {4};
		\node [style=none] (214) at (-7.5, -2.5) {\textcolor{red}{U(1)}};
		\node [style=none] (215) at (-6.5, -2.5) {\textcolor{red}{U(2)}};
		\node [style=none] (216) at (-5.5, -2.5) {\textcolor{red}{U(3)}};
		\node [style=none] (217) at (-7.5, -4) {\textcolor{red}{U(1)}};
		\node [style=none] (218) at (-6.5, -4) {\textcolor{red}{U(2)}};
		\node [style=none] (219) at (-5.5, -4) {\textcolor{red}{U(3)}};
		\node [style=gauge3] (220) at (-7.25, 5) {};
		\node [style=gauge3] (221) at (-6.25, 5) {};
		\node [style=gauge3] (222) at (-5.25, 5) {};
		\node [style=flavour2] (223) at (-4.25, 5) {};
		\node [style=none] (224) at (-7.25, 4.5) {U(1)};
		\node [style=none] (225) at (-6.25, 4.5) {U(2)};
		\node [style=none] (226) at (-5.25, 4.5) {U(3)};
		\node [style=none] (227) at (-4.25, 4.5) {4};
		\node [style=gauge3] (228) at (-7.25, 4) {};
		\node [style=gauge3] (229) at (-6.25, 4) {};
		\node [style=gauge3] (230) at (-5.25, 4) {};
		\node [style=flavour2] (231) at (-4.25, 4) {};
		\node [style=none] (232) at (-7.25, 3.5) {U(1)};
		\node [style=none] (233) at (-6.25, 3.5) {U(2)};
		\node [style=none] (234) at (-5.25, 3.5) {U(3)};
		\node [style=none] (235) at (-4.25, 3.5) {4};
		\node [style=gauge3] (236) at (-7.25, 2.75) {};
		\node [style=gauge3] (237) at (-6.25, 2.75) {};
		\node [style=gauge3] (238) at (-5.25, 2.75) {};
		\node [style=flavour2] (239) at (-4.25, 2.75) {};
		\node [style=none] (240) at (-7.25, 2.25) {U(1)};
		\node [style=none] (241) at (-6.25, 2.25) {U(2)};
		\node [style=none] (242) at (-5.25, 2.25) {U(3)};
		\node [style=none] (243) at (-4.25, 2.25) {4};
		\node [style=gauge3] (244) at (4.25, 5) {};
		\node [style=gauge3] (245) at (3.25, 5) {};
		\node [style=gauge3] (246) at (2.25, 5) {};
		\node [style=flavour2] (247) at (1.25, 5) {};
		\node [style=none] (248) at (4.25, 4.5) {U(1)};
		\node [style=none] (249) at (3.25, 4.5) {U(2)};
		\node [style=none] (250) at (2.25, 4.5) {U(3)};
		\node [style=none] (251) at (1.25, 4.5) {4};
		\node [style=gauge3] (252) at (4.25, 4) {};
		\node [style=gauge3] (253) at (3.25, 4) {};
		\node [style=gauge3] (254) at (2.25, 4) {};
		\node [style=flavour2] (255) at (1.25, 4) {};
		\node [style=none] (256) at (4.25, 3.5) {U(1)};
		\node [style=none] (257) at (3.25, 3.5) {U(2)};
		\node [style=none] (258) at (2.25, 3.5) {U(3)};
		\node [style=none] (259) at (1.25, 3.5) {4};
		\node [style=gauge3] (260) at (4.25, 2.75) {};
		\node [style=gauge3] (261) at (3.25, 2.75) {};
		\node [style=gauge3] (262) at (2.25, 2.75) {};
		\node [style=flavour2] (263) at (1.25, 2.75) {};
		\node [style=none] (264) at (4.25, 2.25) {U(1)};
		\node [style=none] (265) at (3.25, 2.25) {U(2)};
		\node [style=none] (266) at (2.25, 2.25) {U(3)};
		\node [style=none] (267) at (1.25, 2.25) {4};
		\node [style=none] (268) at (-2.75, 3.5) {};
		\node [style=none] (269) at (-1.25, 3.5) {};
		\node [style=none] (270) at (-2, 4) {3d mirror};
		\node [style=none] (271) at (2.75, 1.75) {};
		\node [style=none] (272) at (2.75, 0.75) {};
		\node [style=none] (273) at (4.75, 1.5) {Gauge common};
		\node [style=none] (274) at (4.75, 1) {SU(4) flavor symmetry};
		\node [style=none] (275) at (-6, 1.75) {};
		\node [style=none] (276) at (-6, 0.75) {};
		\node [style=none] (277) at (-4, 1.5) {Gauge common};
		\node [style=none] (278) at (-3, 1) {SU(4) topological symmetry};
	\end{pgfonlayer}
	\begin{pgfonlayer}{edgelayer}
		\draw (0) to (1);
		\draw (11) to (12);
		\draw (22) to (23);
		\draw [style=browndotted, bend right=45] (34.center) to (36.center);
		\draw [style=browndotted, bend right=45] (36.center) to (35.center);
		\draw [style=browndotted, bend right=45] (35.center) to (33.center);
		\draw [style=browndotted, bend left=315] (33.center) to (34.center);
		\draw [style=arrowed] (38.center) to (39.center);
		\draw [style=arrowed] (39.center) to (38.center);
		\draw (135.center) to (134.center);
		\draw (133.center) to (132.center);
		\draw [style=rede] (150.center) to (151.center);
		\draw (157.center) to (156.center);
		\draw (155.center) to (154.center);
		\draw (1) to (3);
		\draw (25) to (23);
		\draw (13) to (12);
		\draw (14) to (13);
		\draw (159) to (160);
		\draw (167) to (168);
		\draw (175) to (176);
		\draw (169) to (168);
		\draw (176) to (183);
		\draw (183) to (169);
		\draw (183) to (160);
		\draw (185.center) to (184.center);
		\draw [style=arrowed] (187.center) to (188.center);
		\draw [style=arrowed] (188.center) to (187.center);
		\draw (190) to (191);
		\draw (191) to (193);
		\draw (198) to (199);
		\draw (199) to (201);
		\draw (206) to (207);
		\draw (207) to (209);
		\draw (220) to (221);
		\draw (221) to (223);
		\draw (228) to (229);
		\draw (229) to (231);
		\draw (236) to (237);
		\draw (237) to (239);
		\draw (244) to (245);
		\draw (245) to (247);
		\draw (252) to (253);
		\draw (253) to (255);
		\draw (260) to (261);
		\draw (261) to (263);
		\draw [style=arrowed] (268.center) to (269.center);
		\draw [style=arrowed] (269.center) to (268.center);
		\draw [style=arrowed] (271.center) to (272.center);
		\draw [style=arrowed] (275.center) to (276.center);
	\end{pgfonlayer}
\end{tikzpicture}}
\end{equation}
Thus, we can view the $T_4$ theory on a circle as three sets of quivers with their common topological symmetry gauged \footnote{This gluing example has also been discussed in the context of implosion and contraction \cite{Dancer:2024lra}.}. On the Coulomb branch, one can also check this gauging directly by refining the Coulomb branch Hilbert series of the three pairs of $T[SU(4)]$ with fugacities valued in the common $SU(4)$ topological symmetry and taking the hyperK\"ahler quotient with respect to $SU(4)$.  

This is a crucial result, as in the literature, both the  $E_8$ theory and 4d $T_4$ theory compactified on a circle were argued not to be Lagrangian (quiver gauge theory). Nevertheless, we can present them as quiver theories with some of their topological symmetries gauged. However, this raises the question: is this considered a Lagrangian (quiver gauge theory)? We would argue they are \emph{not} Lagrangian because the non-Abelian topological symmetries are enhanced in the IR and do not manifest in the UV Lagrangian. In the UV Lagrangian, only the $U(1)$ topological symmetries manifest which is why we can gauge it and still argue the resulting theory is a Lagrangian (quiver gauge theory), as we've done earlier on. In the $U(1)$ case, this simply turns unitary into special unitary gauge groups. However, gauging non-Abelian topological symmetry cannot be done directly in the UV Lagrangian theory. 

Let us take a step back and consider the following relation:
\begin{equation}
 \scalebox{0.8}{\begin{tikzpicture}
	\begin{pgfonlayer}{nodelayer}
		\node [style=gauge3] (0) at (-0.75, -4.75) {};
		\node [style=gauge3] (1) at (0.25, -4.75) {};
		\node [style=gauge3] (2) at (1.25, -4.75) {};
		\node [style=flavour2] (3) at (2.25, -4.75) {};
		\node [style=none] (7) at (-0.75, -5.25) {U(1)};
		\node [style=none] (8) at (0.25, -5.25) {U(2)};
		\node [style=none] (9) at (1.25, -5.25) {U(3)};
		\node [style=none] (10) at (2.25, -5.25) {4};
		\node [style=gauge3] (11) at (6.25, -4.75) {};
		\node [style=gauge3] (12) at (5.25, -4.75) {};
		\node [style=gauge3] (13) at (4.25, -4.75) {};
		\node [style=flavour2] (14) at (3.25, -4.75) {};
		\node [style=none] (18) at (6.25, -5.25) {U(1)};
		\node [style=none] (19) at (5.25, -5.25) {U(2)};
		\node [style=none] (20) at (4.25, -5.25) {U(3)};
		\node [style=none] (21) at (3.25, -5.25) {4};
		\node [style=none] (33) at (2.75, -3.5) {};
		\node [style=none] (34) at (1.75, -4.5) {};
		\node [style=none] (35) at (3.75, -4.5) {};
		\node [style=none] (36) at (2.75, -5.5) {};
		\node [style=none] (37) at (2.75, -6) {\textcolor{brown}{Gauge $SU(4)$ flavor symmetry}};
		\node [style=none] (38) at (-2.75, -3.25) {};
		\node [style=none] (39) at (-1.25, -3.25) {};
		\node [style=none] (41) at (-2, -2.75) {3d mirror};
		\node [style=none] (132) at (2.625, -6.75) {};
		\node [style=none] (133) at (2.625, -7.25) {};
		\node [style=none] (134) at (2.875, -7.25) {};
		\node [style=none] (135) at (2.875, -6.75) {};
		\node [style=none] (150) at (-3.25, -4.5) {};
		\node [style=none] (151) at (-4.5, -6) {};
		\node [style=none] (152) at (-3, -5.75) {\textcolor{red}{Gauge $SU(4)$ }};
		\node [style=none] (153) at (-3.5, -6.25) {\textcolor{red}{ topological symmetry}};
		\node [style=none] (154) at (-6, -6.75) {};
		\node [style=none] (155) at (-6, -7.25) {};
		\node [style=none] (156) at (-5.75, -7.25) {};
		\node [style=none] (157) at (-5.75, -6.75) {};
		\node [style=none] (187) at (-2.75, -9.25) {};
		\node [style=none] (188) at (-1.25, -9.25) {};
		\node [style=none] (189) at (-2, -9) {3d mirror};
		\node [style=gauge3] (198) at (-7.5, -3.75) {};
		\node [style=gauge3] (199) at (-6.5, -3.75) {};
		\node [style=gauge3] (200) at (-5.5, -3.75) {};
		\node [style=flavour2] (201) at (-4.5, -3.75) {};
		\node [style=none] (205) at (-4.5, -4.25) {4};
		\node [style=gauge3] (206) at (-7.5, -4.75) {};
		\node [style=gauge3] (207) at (-6.5, -4.75) {};
		\node [style=gauge3] (208) at (-5.5, -4.75) {};
		\node [style=flavour2] (209) at (-4.5, -4.75) {};
		\node [style=none] (213) at (-4.5, -5.25) {4};
		\node [style=none] (214) at (-7.5, -4.25) {\textcolor{red}{U(1)}};
		\node [style=none] (215) at (-6.5, -4.25) {\textcolor{red}{U(2)}};
		\node [style=none] (216) at (-5.5, -4.25) {\textcolor{red}{U(3)}};
		\node [style=none] (217) at (-7.5, -5.25) {\textcolor{red}{U(1)}};
		\node [style=none] (218) at (-6.5, -5.25) {\textcolor{red}{U(2)}};
		\node [style=none] (219) at (-5.5, -5.25) {\textcolor{red}{U(3)}};
		\node [style=gauge3] (220) at (5.5, 3.75) {};
		\node [style=gauge3] (221) at (4.5, 3.75) {};
		\node [style=gauge3] (222) at (3.5, 3.75) {};
		\node [style=flavour2] (223) at (2.5, 3.75) {};
		\node [style=none] (224) at (5.5, 3.25) {U(1)};
		\node [style=none] (225) at (4.5, 3.25) {U(2)};
		\node [style=none] (226) at (3.5, 3.25) {U(3)};
		\node [style=none] (227) at (2.5, 3.25) {4};
		\node [style=gauge3] (228) at (5.5, 2.75) {};
		\node [style=gauge3] (229) at (4.5, 2.75) {};
		\node [style=gauge3] (230) at (3.5, 2.75) {};
		\node [style=flavour2] (231) at (2.5, 2.75) {};
		\node [style=none] (232) at (5.5, 2.25) {U(1)};
		\node [style=none] (233) at (4.5, 2.25) {U(2)};
		\node [style=none] (234) at (3.5, 2.25) {U(3)};
		\node [style=none] (235) at (2.5, 2.25) {4};
		\node [style=gauge3] (238) at (5.5, 6) {};
		\node [style=flavour2] (239) at (4.5, 6) {};
		\node [style=none] (242) at (5.5, 5.5) {U(1)};
		\node [style=none] (243) at (4.5, 5.5) {4};
		\node [style=gauge3] (244) at (-7.5, 3.75) {};
		\node [style=gauge3] (245) at (-6.5, 3.75) {};
		\node [style=gauge3] (246) at (-5.5, 3.75) {};
		\node [style=flavour2] (247) at (-4.5, 3.75) {};
		\node [style=none] (248) at (-7.5, 3.25) {U(1)};
		\node [style=none] (249) at (-6.5, 3.25) {U(2)};
		\node [style=none] (250) at (-5.5, 3.25) {U(3)};
		\node [style=none] (251) at (-4.5, 3.25) {4};
		\node [style=gauge3] (252) at (-7.5, 2.75) {};
		\node [style=gauge3] (253) at (-6.5, 2.75) {};
		\node [style=gauge3] (254) at (-5.5, 2.75) {};
		\node [style=flavour2] (255) at (-4.5, 2.75) {};
		\node [style=none] (256) at (-7.5, 2.25) {U(1)};
		\node [style=none] (257) at (-6.5, 2.25) {U(2)};
		\node [style=none] (258) at (-5.5, 2.25) {U(3)};
		\node [style=none] (259) at (-4.5, 2.25) {4};
		\node [style=none] (268) at (-2.75, 3.5) {};
		\node [style=none] (269) at (-1.25, 3.5) {};
		\node [style=none] (270) at (-2, 4) {3d mirror};
		\node [style=none] (271) at (2.75, 1.75) {};
		\node [style=none] (272) at (2.75, 0.75) {};
		\node [style=none] (273) at (4.75, 1.5) {Gauge common};
		\node [style=none] (274) at (4.75, 1) {SU(4) flavor symmetry};
		\node [style=none] (275) at (-6, 1.75) {};
		\node [style=none] (276) at (-6, 0.75) {};
		\node [style=none] (277) at (-4, 1.5) {Gauge common};
		\node [style=none] (278) at (-3, 1) {SU(4) topological symmetry};
		\node [style=gauge3] (279) at (5.5, 4.75) {};
		\node [style=flavour2] (280) at (4.5, 4.75) {};
		\node [style=none] (281) at (5.5, 4.25) {U(1)};
		\node [style=none] (282) at (4.5, 4.25) {4};
		\node [style=gauge3] (296) at (-7.5, 7.25) {};
		\node [style=gauge3] (297) at (-6.5, 7.25) {};
		\node [style=gauge3] (298) at (-6.5, 7.25) {};
		\node [style=none] (299) at (-7.5, 6.75) {U(1)};
		\node [style=none] (300) at (-6.5, 6.75) {U(1)};
		\node [style=gauge3] (301) at (-5.5, 7.25) {};
		\node [style=none] (302) at (-5.5, 6.75) {U(1)};
		\node [style=gauge3] (312) at (2.25, -3) {};
		\node [style=flavour2] (313) at (2.25, -4) {};
		\node [style=none] (314) at (2.25, -2.5) {U(1)};
		\node [style=none] (315) at (1.75, -4) {4};
		\node [style=gauge3] (316) at (3.25, -3) {};
		\node [style=flavour2] (317) at (3.25, -4) {};
		\node [style=none] (318) at (3.25, -2.5) {U(1)};
		\node [style=none] (319) at (3.75, -4) {4};
		\node [style=gauge3] (336) at (-7.5, -0.75) {};
		\node [style=gauge3] (337) at (-6.5, -0.75) {};
		\node [style=gauge3] (338) at (-6.5, -0.75) {};
		\node [style=gauge3] (339) at (-5.5, -0.75) {};
		\node [style=none] (342) at (-7.5, -1.25) {\textcolor{red}{U(1)}};
		\node [style=none] (343) at (-6.5, -1.25) {\textcolor{red}{U(1)}};
		\node [style=none] (344) at (-5.5, -1.25) {\textcolor{red}{U(1)}};
		\node [style=gauge3] (350) at (-5.875, -9.5) {};
		\node [style=flavour2] (351) at (-5.875, -8.5) {};
		\node [style=none] (352) at (-5.875, -10) {SU(4)};
		\node [style=none] (353) at (-5.875, -8) {8};
		\node [style=gauge3] (354) at (-0.25, -9.75) {};
		\node [style=gauge3] (355) at (0.75, -9.75) {};
		\node [style=gauge3] (356) at (1.75, -9.75) {};
		\node [style=none] (358) at (-0.25, -10.25) {U(1)};
		\node [style=none] (359) at (0.75, -10.25) {U(2)};
		\node [style=none] (360) at (1.75, -10.25) {U(3)};
		\node [style=gauge3] (362) at (5.75, -9.75) {};
		\node [style=gauge3] (363) at (4.75, -9.75) {};
		\node [style=gauge3] (364) at (3.75, -9.75) {};
		\node [style=none] (366) at (5.75, -10.25) {U(1)};
		\node [style=none] (367) at (4.75, -10.25) {U(2)};
		\node [style=none] (368) at (3.75, -10.25) {U(3)};
		\node [style=gauge3] (374) at (2.25, -8.75) {};
		\node [style=none] (376) at (2.25, -8.25) {U(1)};
		\node [style=gauge3] (378) at (3.25, -8.75) {};
		\node [style=none] (380) at (3.25, -8.25) {U(1)};
		\node [style=gauge3] (381) at (2.75, -9.75) {};
		\node [style=none] (382) at (2.75, -10.25) {U(4)};
		\node [style=flavour2] (383) at (-7.5, 8.25) {};
		\node [style=flavour2] (384) at (-5.5, 8.25) {};
		\node [style=none] (385) at (-5.5, 8.75) {1};
		\node [style=none] (386) at (-7.5, 8.75) {1};
		\node [style=gauge3] (387) at (-7.5, 4.75) {};
		\node [style=gauge3] (388) at (-6.5, 4.75) {};
		\node [style=gauge3] (389) at (-6.5, 4.75) {};
		\node [style=none] (390) at (-7.5, 4.25) {U(1)};
		\node [style=none] (391) at (-6.5, 4.25) {U(1)};
		\node [style=gauge3] (392) at (-5.5, 4.75) {};
		\node [style=none] (393) at (-5.5, 4.25) {U(1)};
		\node [style=flavour2] (395) at (-7.5, 5.75) {};
		\node [style=flavour2] (396) at (-5.5, 5.75) {};
		\node [style=none] (397) at (-5.5, 6.25) {1};
		\node [style=none] (398) at (-7.5, 6.25) {1};
		\node [style=flavour2] (399) at (-7.5, 0.25) {};
		\node [style=flavour2] (400) at (-5.5, 0.25) {};
		\node [style=none] (401) at (-5, 0.25) {1};
		\node [style=none] (402) at (-8, 0.25) {1};
		\node [style=gauge3] (403) at (-7.5, -2.75) {};
		\node [style=gauge3] (404) at (-6.5, -2.75) {};
		\node [style=gauge3] (405) at (-6.5, -2.75) {};
		\node [style=gauge3] (406) at (-5.5, -2.75) {};
		\node [style=none] (407) at (-7.5, -3.25) {\textcolor{red}{U(1)}};
		\node [style=none] (408) at (-6.5, -3.25) {\textcolor{red}{U(1)}};
		\node [style=flavour2] (411) at (-7.5, -1.75) {};
		\node [style=flavour2] (412) at (-5.5, -1.75) {};
		\node [style=none] (413) at (-5, -1.75) {1};
		\node [style=none] (414) at (-8, -1.75) {1};
		\node [style=none] (415) at (7, -9.5) {};
		\node [style=none] (416) at (5.75, -11) {};
		\node [style=none] (417) at (6.75, -10.5) {U(1)};
		\node [style=none] (418) at (-5.5, -3.25) {\textcolor{red}{U(1)}};
	\end{pgfonlayer}
	\begin{pgfonlayer}{edgelayer}
		\draw (0) to (1);
		\draw (11) to (12);
		\draw [style=browndotted, bend right=45] (34.center) to (36.center);
		\draw [style=browndotted, bend right=45] (36.center) to (35.center);
		\draw [style=browndotted, bend right=45] (35.center) to (33.center);
		\draw [style=browndotted, bend left=315] (33.center) to (34.center);
		\draw [style=arrowed] (38.center) to (39.center);
		\draw [style=arrowed] (39.center) to (38.center);
		\draw (135.center) to (134.center);
		\draw (133.center) to (132.center);
		\draw [style=rede] (150.center) to (151.center);
		\draw (157.center) to (156.center);
		\draw (155.center) to (154.center);
		\draw (1) to (3);
		\draw (13) to (12);
		\draw (14) to (13);
		\draw [style=arrowed] (187.center) to (188.center);
		\draw [style=arrowed] (188.center) to (187.center);
		\draw (198) to (199);
		\draw (199) to (201);
		\draw (206) to (207);
		\draw (207) to (209);
		\draw (220) to (221);
		\draw (221) to (223);
		\draw (228) to (229);
		\draw (229) to (231);
		\draw (244) to (245);
		\draw (245) to (247);
		\draw (252) to (253);
		\draw (253) to (255);
		\draw [style=arrowed] (268.center) to (269.center);
		\draw [style=arrowed] (269.center) to (268.center);
		\draw [style=arrowed] (271.center) to (272.center);
		\draw [style=arrowed] (275.center) to (276.center);
		\draw (238) to (239);
		\draw (279) to (280);
		\draw (296) to (297);
		\draw (301) to (297);
		\draw (312) to (313);
		\draw (316) to (317);
		\draw (336) to (337);
		\draw (339) to (337);
		\draw (350) to (351);
		\draw (354) to (355);
		\draw (362) to (363);
		\draw (364) to (363);
		\draw (355) to (364);
		\draw (381) to (374);
		\draw (378) to (381);
		\draw (301) to (384);
		\draw (296) to (383);
		\draw (387) to (388);
		\draw (392) to (388);
		\draw (392) to (396);
		\draw (387) to (395);
		\draw (339) to (400);
		\draw (336) to (399);
		\draw (403) to (404);
		\draw (406) to (404);
		\draw (406) to (412);
		\draw (403) to (411);
		\draw (416.center) to (415.center);
	\end{pgfonlayer}
\end{tikzpicture}}
\end{equation}
On the right, we glue the common $SU(4)$ flavor symmetry. If we follow straight from our previous line of argument, the left hand side consists of gluing the common non-Abelian $SU(4)$ topological symmetry, an operation that robs the theory of its Lagrangian description. However, it is well known that the mirror is simply $SU(4)$ gauge theory with 8 flavors. Such coincidences can occur where, on one side, we have multiple quivers with their common topological symmetries gauged in the IR, while on the other, we have a UV Lagrangian quiver gauge theory that flows to the same IR theory.  

We reached the following conclusion for this section. It is well known that finding 3d mirrors of non-linear unitary quivers is a challenging problem, with very few examples in the literature. However, any non-linear unitary quiver can be constructed from building blocks of linear unitary quivers, which are then connected through a gluing process. These linear quivers have known 3d mirrors. In this section, we demonstrated that gauging flavor symmetries serves as the gluing method to construct non-linear quivers. Consequently, the 3d mirror of a non-linear quiver can be constructed by first finding the mirrors of the linear unitary quivers and then gluing them together through the gauging of topological symmetries. However, the general process of gauging non-Abelian topological symmetries is a non-Lagrangian construction. As a result, this method of bootstrapping 3d mirrors for non-linear quivers leads to non-Lagrangian mirrors, which are not really useful. While the bootstrapping method always leads to non-Lagrangian mirrors for non-linear quivers, it does not rule out the possibility that some of these mirrors might be the IR limit of an alternative UV Lagrangian theory. 

The rarity of Lagrangian mirror pairs for non-linear quivers in the literature points to a different phenomenon, which we will explore in the next section.


\paragraph{What is a Lagrangian theory?}
Before proceeding further, we should clarify what we mean by a Lagrangian theory. Generally, we consider a theory to be Lagrangian if symmetries such as gauge symmetries and global symmetries are all manifest in the Lagrangian. In $3d$ $\mathcal{N}=4$, a theory is Lagrangian if it has a UV Lagrangian description, since enhancements from monopole operators in the IR  always renders the description non-Lagrangian. Whenever a $3d$ $\mathcal{N}=4$ theory is identified in the literature using a quiver gauge theory, the quiver describes its UV Lagrangian.

However, there are tools, such as the monopole formula and the 3d superconformal index, that take the UV Lagrangian as input and extract information hidden from the Lagrangian, such as IR enhancement of topological symmetries. Moreover, enhancements in topological symmetries can be easily derived from the UV Lagrangian (quiver gauge theory) by computing the balance of gauge groups \cite{Gaiotto:2008ak}. Therefore, one may be inclined to overlook the fact that such information cannot be extracted from the UV Lagrangian through conventional means and still consider the theory Lagrangian. If adopting this philosophy, then gauging non-Abelian topological symmetries could serve as the method for bootstrapping any 3d mirror from building blocks of linear quivers.

We do not adopt this philosophy here for a valid reason: it renders the 3d mirror pair irrelevant. 3d mirror symmetry is meant to describe a duality between two distinct theories that flow to the same theory in the IR, under the exchange of Coulomb and Higgs branches. The main goal of finding a theory's mirror is to provide an alternative description that might make the theory’s moduli spaces more approachable through the mirror. Since topological and flavor symmetries are known to be exchanged under this duality in the IR, constructing 3d mirrors by simply gauging common flavor and topological symmetries becomes trivial, offering no new insights beyond the original theory. Therefore, constructing a mirror pair in this way  does not aid in studying the moduli spaces.
\\ \\
So for our purpose, we adopt the conventional definition that a theory is Lagrangian if all the symmetries in the theory are manifest in the UV Lagrangian. This also means \emph{a Lagrangian $3d$ $\mathcal{N}=4$ theory is also a quiver gauge theory. }

\section{The Bound}\label{thelimit}
In this section, we give a simple argument why almost all  quivers made of unitary gauge groups beyond \DynkinAbar\; family will \textbf{not} have a Lagrangian (quiver gauge theory) mirror. In fact, even beyond unitary quivers it is true that:
\begin{center}
\textbf{\textcolor{red}{Most non-linear quivers don't have Lagrangian mirror duals}}
\end{center}

The \textbf{Decay and Fission} algorithm introduced in \cite{Decay,Fission} gives a tool to perform Higgs mechanism on supersymmetric theories with eight supercharges. The algorithm relies on the use of magnetic quivers (MQ) \cite{Cabrera:2019izd}. As a reminder, given a theory $\mathcal{T}$, which can be a supersymmetric gauge theory or SCFT, with eight supercharges in dimensions $d=3,4,5,6$, the corresponding magnetic quiver is a $3d$ $\mathcal{N}=4$ theory such that:
\begin{equation}
    \text{Higgs}^{d=3,4,5,6}(\mathcal{T})=\text{Coulomb}^{3d\; \mathcal{N}=4}(\text{Magnetic Quiver})
\end{equation}
As we are only dealing with 3d theories in this paper, $\mathcal{T}$ will always be a $3d$ $\mathcal{N}=4$ theory (can be a higher dimensional theory compactified to 3d) and the relation boils down to 3d mirror symmetry\footnote{There is a few caveats here such as the Higgs branch  of $\mathcal{T}$ need to be a single hyperK\"ahler cone  (see \cite{USU} for more detail on magnetic quiver vs 3d mirror symmetry). These caveats will \emph{not} affect our conjectures in this section as the conclusion is the same even if the Higgs branch is made of multiple cones. }.

A quick review of the Decay and Fission algorithm: 
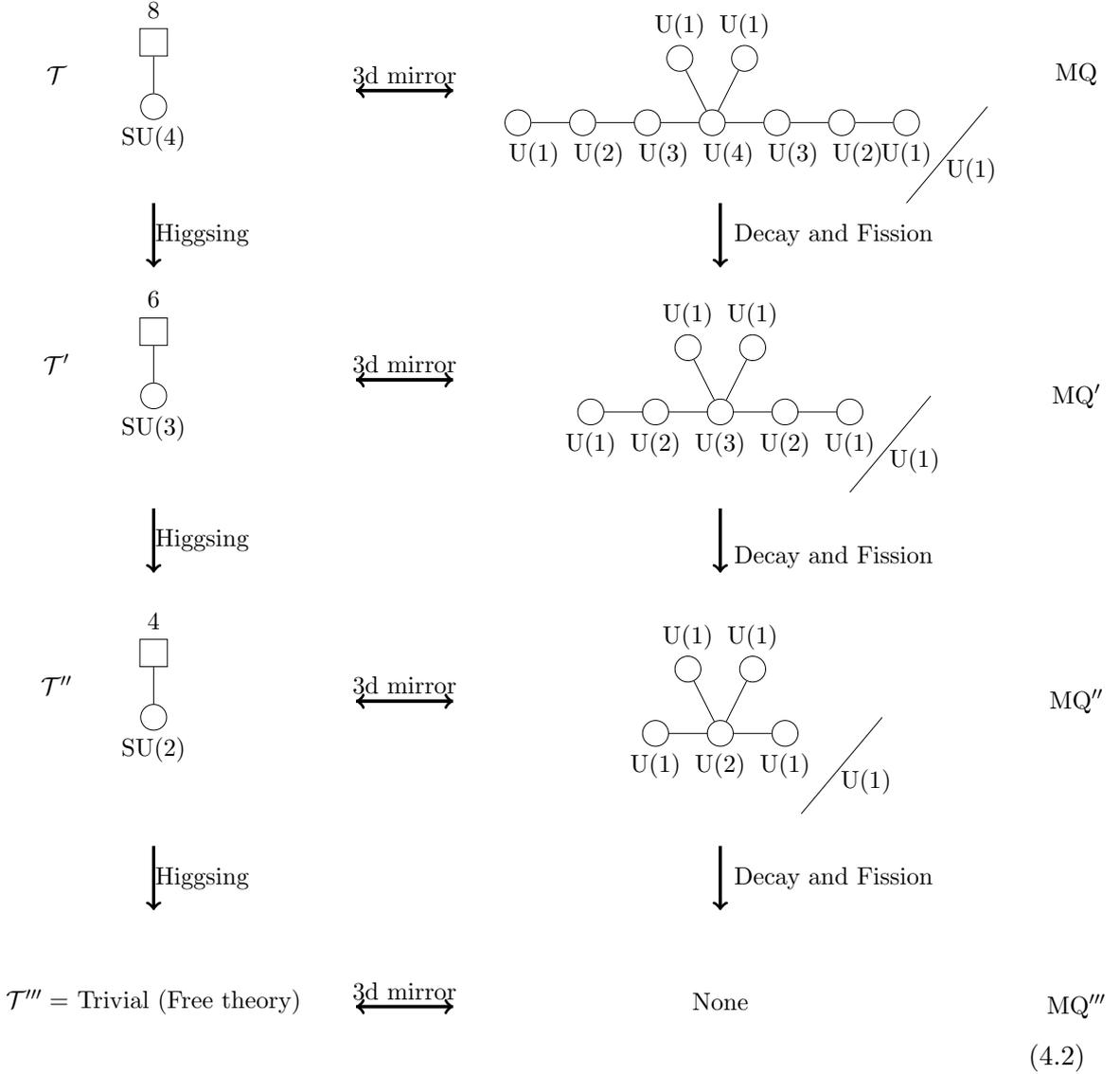
\begin{equation}
    \scalebox{0.9}{\begin{tikzpicture}
	\begin{pgfonlayer}{nodelayer}
		\node [style=none] (187) at (-2.375, -9.25) {};
		\node [style=none] (188) at (-0.875, -9.25) {};
		\node [style=none] (189) at (-1.625, -9) {3d mirror};
		\node [style=gauge3] (350) at (-5.5, -9.5) {};
		\node [style=flavour2] (351) at (-5.5, -8.5) {};
		\node [style=none] (352) at (-5.5, -10) {SU(4)};
		\node [style=none] (353) at (-5.5, -8) {8};
		\node [style=gauge3] (354) at (0.125, -9.75) {};
		\node [style=gauge3] (355) at (1.125, -9.75) {};
		\node [style=gauge3] (356) at (2.125, -9.75) {};
		\node [style=none] (358) at (0.375, -10.25) {U(1)};
		\node [style=none] (359) at (1.375, -10.25) {U(2)};
		\node [style=none] (360) at (2.375, -10.25) {U(3)};
		\node [style=gauge3] (362) at (6.125, -9.75) {};
		\node [style=gauge3] (363) at (5.125, -9.75) {};
		\node [style=gauge3] (364) at (4.125, -9.75) {};
		\node [style=none] (366) at (6.125, -10.25) {U(1)};
		\node [style=none] (367) at (5.375, -10.25) {U(2)};
		\node [style=none] (368) at (4.375, -10.25) {U(3)};
		\node [style=gauge3] (374) at (2.625, -8.75) {};
		\node [style=none] (376) at (2.625, -8.25) {U(1)};
		\node [style=gauge3] (378) at (3.625, -8.75) {};
		\node [style=none] (380) at (3.625, -8.25) {U(1)};
		\node [style=gauge3] (381) at (3.125, -9.75) {};
		\node [style=none] (382) at (3.375, -10.25) {U(4)};
		\node [style=none] (415) at (7.375, -9.5) {};
		\node [style=none] (416) at (6.125, -11) {};
		\node [style=none] (417) at (7.125, -10.5) {U(1)};
		\node [style=none] (418) at (-2.375, -13.75) {};
		\node [style=none] (419) at (-0.875, -13.75) {};
		\node [style=none] (420) at (-1.625, -13.5) {3d mirror};
		\node [style=gauge3] (421) at (-5.5, -14) {};
		\node [style=flavour2] (422) at (-5.5, -13) {};
		\node [style=none] (423) at (-5.5, -14.5) {SU(3)};
		\node [style=none] (424) at (-5.5, -12.5) {6};
		\node [style=gauge3] (426) at (1.25, -14.25) {};
		\node [style=gauge3] (427) at (2.25, -14.25) {};
		\node [style=none] (429) at (1.25, -14.75) {U(1)};
		\node [style=none] (430) at (2.25, -14.75) {U(2)};
		\node [style=gauge3] (432) at (5.25, -14.25) {};
		\node [style=gauge3] (433) at (4.25, -14.25) {};
		\node [style=none] (435) at (5.25, -14.75) {U(1)};
		\node [style=none] (436) at (4.25, -14.75) {U(2)};
		\node [style=gauge3] (437) at (2.75, -13.25) {};
		\node [style=none] (438) at (2.75, -12.75) {U(1)};
		\node [style=gauge3] (439) at (3.75, -13.25) {};
		\node [style=none] (440) at (3.75, -12.75) {U(1)};
		\node [style=gauge3] (441) at (3.25, -14.25) {};
		\node [style=none] (442) at (3.25, -14.75) {U(3)};
		\node [style=none] (443) at (6.5, -14) {};
		\node [style=none] (444) at (5.25, -15.5) {};
		\node [style=none] (445) at (6.25, -15) {U(1)};
		\node [style=none] (446) at (-2.375, -18.75) {};
		\node [style=none] (447) at (-0.875, -18.75) {};
		\node [style=none] (448) at (-1.625, -18.5) {3d mirror};
		\node [style=gauge3] (449) at (-5.5, -19) {};
		\node [style=flavour2] (450) at (-5.5, -18) {};
		\node [style=none] (451) at (-5.5, -19.5) {SU(2)};
		\node [style=none] (452) at (-5.5, -17.5) {4};
		\node [style=gauge3] (455) at (2.25, -19.25) {};
		\node [style=none] (458) at (2.25, -19.75) {U(1)};
		\node [style=gauge3] (461) at (4.25, -19.25) {};
		\node [style=none] (464) at (4.25, -19.75) {U(1)};
		\node [style=gauge3] (465) at (2.75, -18.25) {};
		\node [style=none] (466) at (2.75, -17.75) {U(1)};
		\node [style=gauge3] (467) at (3.75, -18.25) {};
		\node [style=none] (468) at (3.75, -17.75) {U(1)};
		\node [style=gauge3] (469) at (3.25, -19.25) {};
		\node [style=none] (470) at (3.25, -19.75) {U(2)};
		\node [style=none] (471) at (5.75, -19) {};
		\node [style=none] (472) at (4.5, -20.5) {};
		\node [style=none] (473) at (5.5, -20) {U(1)};
		\node [style=none] (474) at (-2.375, -23.5) {};
		\node [style=none] (475) at (-0.875, -23.5) {};
		\node [style=none] (476) at (-1.625, -23.25) {3d mirror};
		\node [style=none] (502) at (-5.5, -11) {};
		\node [style=none] (503) at (-5.5, -12) {};
		\node [style=none] (504) at (3.25, -11) {};
		\node [style=none] (505) at (3.25, -12) {};
		\node [style=none] (506) at (-4.75, -11.5) {Higgsing};
		\node [style=none] (507) at (5, -11.5) {Decay and Fission};
		\node [style=none] (508) at (-5.5, -15.75) {};
		\node [style=none] (509) at (-5.5, -16.75) {};
		\node [style=none] (510) at (3.25, -15.75) {};
		\node [style=none] (511) at (3.25, -16.75) {};
		\node [style=none] (512) at (-4.75, -16.25) {Higgsing};
		\node [style=none] (513) at (5, -16.5) {Decay and Fission};
		\node [style=none] (514) at (-5.5, -21) {};
		\node [style=none] (515) at (-5.5, -22) {};
		\node [style=none] (516) at (3.25, -21) {};
		\node [style=none] (517) at (3.25, -22) {};
		\node [style=none] (518) at (-4.75, -21.5) {Higgsing};
		\node [style=none] (519) at (5, -21.5) {Decay and Fission};
		\node [style=none] (520) at (-5.5, -23.425) {$\mathcal{T}'''=$ Trivial (Free theory)};
		\node [style=none] (521) at (3.25, -23.425) {None};
		\node [style=none] (522) at (-7, -9) {$\mathcal{T}$};
		\node [style=none] (523) at (-7, -13.5) {$\mathcal{T}'$};
		\node [style=none] (524) at (-7, -18.5) {$\mathcal{T}''$};
		\node [style=none] (525) at (8.75, -9) {$\text{MQ}$};
		\node [style=none] (526) at (8.75, -14) {$\text{MQ}'$};
		\node [style=none] (527) at (8.75, -18.75) {$\text{MQ}''$};
		\node [style=none] (528) at (8.75, -23.5) {$\text{MQ}'''$};
	\end{pgfonlayer}
	\begin{pgfonlayer}{edgelayer}
		\draw [style=arrowed] (187.center) to (188.center);
		\draw [style=arrowed] (188.center) to (187.center);
		\draw (350) to (351);
		\draw (354) to (355);
		\draw (362) to (363);
		\draw (364) to (363);
		\draw (355) to (364);
		\draw (381) to (374);
		\draw (378) to (381);
		\draw (416.center) to (415.center);
		\draw [style=arrowed] (418.center) to (419.center);
		\draw [style=arrowed] (419.center) to (418.center);
		\draw (421) to (422);
		\draw (433) to (432);
		\draw (426) to (433);
		\draw (441) to (437);
		\draw (439) to (441);
		\draw (444.center) to (443.center);
		\draw [style=arrowed] (446.center) to (447.center);
		\draw [style=arrowed] (447.center) to (446.center);
		\draw (449) to (450);
		\draw (469) to (465);
		\draw (467) to (469);
		\draw (472.center) to (471.center);
		\draw [style=arrowed] (474.center) to (475.center);
		\draw [style=arrowed] (475.center) to (474.center);
		\draw [style=arrowed] (502.center) to (503.center);
		\draw [style=arrowed] (504.center) to (505.center);
		\draw [style=arrowed] (508.center) to (509.center);
		\draw [style=arrowed] (510.center) to (511.center);
		\draw [style=arrowed] (514.center) to (515.center);
		\draw [style=arrowed] (516.center) to (517.center);
		\draw (455) to (461);
	\end{pgfonlayer}
\end{tikzpicture}
}
\end{equation}
Let us stick with the notations of $\mathcal{T}$ and its 3d mirror/magnetic quiver (MQ) to avoid confusion when using the algorithm. Every time $\mathcal{T}$ undergoes a minimal Higgsing (a single Higgsing that cannot be decomposed into multiple Higgsings), its 3d mirror is given by the magnetic quiver (MQ) undergoing decay and fission. Consequently, $\mathcal{T}'$ is the 3d mirror of $\text{MQ}'$, and so on.

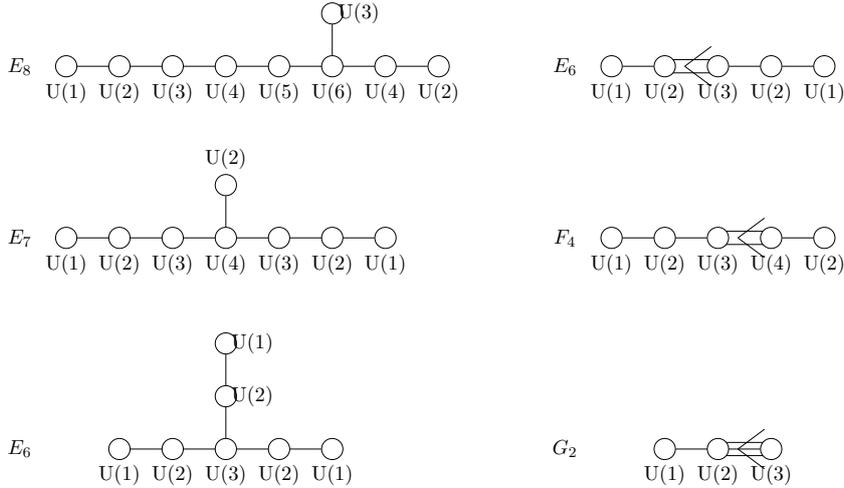
\begin{figure}[ht]
\centering
\scalebox{0.7}{
   \begin{tikzpicture}
	\begin{pgfonlayer}{nodelayer}
		\node [style=gauge3] (590) at (0.375, -18.75) {};
		\node [style=gauge3] (591) at (1.375, -18.75) {};
		\node [style=gauge3] (592) at (2.375, -18.75) {};
		\node [style=gauge3] (593) at (3.375, -18.75) {};
		\node [style=gauge3] (594) at (4.375, -18.75) {};
		\node [style=gauge3] (595) at (5.375, -18.75) {};
		\node [style=gauge3] (596) at (6.375, -18.75) {};
		\node [style=gauge3] (597) at (7.375, -18.75) {};
		\node [style=gauge3] (598) at (5.375, -17.75) {};
		\node [style=none] (599) at (0.375, -19.25) {U(1)};
		\node [style=none] (600) at (1.375, -19.25) {U(2)};
		\node [style=none] (601) at (2.375, -19.25) {U(3)};
		\node [style=none] (602) at (3.375, -19.25) {U(4)};
		\node [style=none] (603) at (4.375, -19.25) {U(5)};
		\node [style=none] (604) at (5.375, -19.25) {U(6)};
		\node [style=none] (607) at (6.375, -19.25) {U(4)};
		\node [style=none] (608) at (7.375, -19.25) {U(2)};
		\node [style=none] (609) at (5.875, -17.75) {U(3)};
		\node [style=gauge3] (610) at (0.375, -22) {};
		\node [style=gauge3] (611) at (1.375, -22) {};
		\node [style=gauge3] (612) at (2.375, -22) {};
		\node [style=gauge3] (613) at (3.375, -22) {};
		\node [style=gauge3] (614) at (4.375, -22) {};
		\node [style=gauge3] (615) at (5.375, -22) {};
		\node [style=gauge3] (616) at (6.375, -22) {};
		\node [style=none] (619) at (0.375, -22.5) {U(1)};
		\node [style=none] (620) at (1.375, -22.5) {U(2)};
		\node [style=none] (621) at (2.375, -22.5) {U(3)};
		\node [style=none] (622) at (3.375, -22.5) {U(4)};
		\node [style=none] (623) at (4.375, -22.5) {U(3)};
		\node [style=none] (624) at (5.375, -22.5) {U(2)};
		\node [style=none] (625) at (6.375, -22.5) {U(1)};
		\node [style=gauge3] (626) at (3.375, -21) {};
		\node [style=none] (627) at (3.375, -20.5) {U(2)};
		\node [style=gauge3] (629) at (1.375, -26) {};
		\node [style=gauge3] (630) at (2.375, -26) {};
		\node [style=gauge3] (631) at (3.375, -26) {};
		\node [style=gauge3] (632) at (4.375, -26) {};
		\node [style=gauge3] (633) at (5.375, -26) {};
		\node [style=none] (636) at (1.375, -26.5) {U(1)};
		\node [style=none] (637) at (2.375, -26.5) {U(2)};
		\node [style=none] (638) at (3.375, -26.5) {U(3)};
		\node [style=none] (639) at (4.375, -26.5) {U(2)};
		\node [style=none] (640) at (5.375, -26.5) {U(1)};
		\node [style=gauge3] (642) at (3.375, -25) {};
		\node [style=none] (643) at (3.875, -25) {U(2)};
		\node [style=gauge3] (644) at (3.375, -24) {};
		\node [style=none] (645) at (3.875, -24) {U(1)};
		\node [style=gauge3] (646) at (10.625, -18.75) {};
		\node [style=gauge3] (647) at (11.625, -18.75) {};
		\node [style=gauge3] (648) at (12.625, -18.75) {};
		\node [style=gauge3] (649) at (13.625, -18.75) {};
		\node [style=gauge3] (650) at (14.625, -18.75) {};
		\node [style=none] (651) at (10.625, -19.25) {U(1)};
		\node [style=none] (652) at (11.625, -19.25) {U(2)};
		\node [style=none] (653) at (12.625, -19.25) {U(3)};
		\node [style=none] (654) at (13.625, -19.25) {U(2)};
		\node [style=none] (655) at (14.625, -19.25) {U(1)};
		\node [style=none] (656) at (11.75, -18.625) {};
		\node [style=none] (657) at (12.5, -18.625) {};
		\node [style=none] (658) at (12.5, -18.875) {};
		\node [style=none] (659) at (11.75, -18.875) {};
		\node [style=none] (660) at (12.5, -18.375) {};
		\node [style=none] (661) at (12, -18.75) {};
		\node [style=none] (662) at (12.5, -19.125) {};
		\node [style=gauge3] (663) at (10.625, -22) {};
		\node [style=gauge3] (664) at (11.625, -22) {};
		\node [style=gauge3] (665) at (12.625, -22) {};
		\node [style=gauge3] (666) at (13.625, -22) {};
		\node [style=gauge3] (667) at (14.625, -22) {};
		\node [style=none] (668) at (10.625, -22.5) {U(1)};
		\node [style=none] (669) at (11.625, -22.5) {U(2)};
		\node [style=none] (670) at (12.625, -22.5) {U(3)};
		\node [style=none] (671) at (13.625, -22.5) {U(4)};
		\node [style=none] (672) at (14.625, -22.5) {U(2)};
		\node [style=none] (673) at (12.75, -21.875) {};
		\node [style=none] (674) at (13.5, -21.875) {};
		\node [style=none] (675) at (13.5, -22.125) {};
		\node [style=none] (676) at (12.75, -22.125) {};
		\node [style=none] (677) at (13.5, -21.625) {};
		\node [style=none] (678) at (13, -22) {};
		\node [style=none] (679) at (13.5, -22.375) {};
		\node [style=gauge3] (680) at (11.625, -26) {};
		\node [style=gauge3] (681) at (12.625, -26) {};
		\node [style=gauge3] (682) at (13.625, -26) {};
		\node [style=none] (683) at (11.625, -26.5) {U(1)};
		\node [style=none] (684) at (12.625, -26.5) {U(2)};
		\node [style=none] (685) at (13.625, -26.5) {U(3)};
		\node [style=none] (686) at (12.75, -25.875) {};
		\node [style=none] (687) at (13.5, -25.875) {};
		\node [style=none] (688) at (13.5, -26.125) {};
		\node [style=none] (689) at (12.75, -26.125) {};
		\node [style=none] (690) at (13.5, -25.625) {};
		\node [style=none] (691) at (13, -26) {};
		\node [style=none] (692) at (13.5, -26.375) {};
		\node [style=none] (693) at (-0.5, -18.75) {$E_8$};
		\node [style=none] (694) at (-0.5, -22) {$E_7$};
		\node [style=none] (695) at (-0.5, -26) {$E_6$};
		\node [style=none] (696) at (9.75, -18.75) {$E_6$};
		\node [style=none] (697) at (9.75, -22) {$F_4$};
		\node [style=none] (698) at (9.75, -26) {$G_2$};
	\end{pgfonlayer}
	\begin{pgfonlayer}{edgelayer}
		\draw (598) to (595);
		\draw (590) to (597);
		\draw (610) to (616);
		\draw (613) to (626);
		\draw (631) to (642);
		\draw (629) to (633);
		\draw (642) to (644);
		\draw (662.center) to (661.center);
		\draw (661.center) to (660.center);
		\draw (656.center) to (657.center);
		\draw (658.center) to (659.center);
		\draw (650) to (649);
		\draw (679.center) to (678.center);
		\draw (678.center) to (677.center);
		\draw (673.center) to (674.center);
		\draw (675.center) to (676.center);
		\draw (663) to (665);
		\draw (667) to (666);
		\draw (649) to (648);
		\draw (646) to (647);
		\draw (692.center) to (691.center);
		\draw (691.center) to (690.center);
		\draw (686.center) to (687.center);
		\draw (688.center) to (689.center);
		\draw (682) to (680);
	\end{pgfonlayer}
\end{tikzpicture}}
\caption{Here is a list of $3d$ $\mathcal{N}=4$ theories that are made of unitary gauge groups and takes the form of affine Dynkin diagrams and twisted affine Dynkin diagrams of exceptional EFG algebras. Any theory undergoing decay and fission that reaches one of these quivers will have a non-Lagrangian/non-quiver gauge theory 3d mirror. }
\label{exceptionends}
\end{figure}

Now, on to the real problem. Consider the following theory $\mathcal{Z}$ with a non-linear and non-\DynkinAbar type mirror:
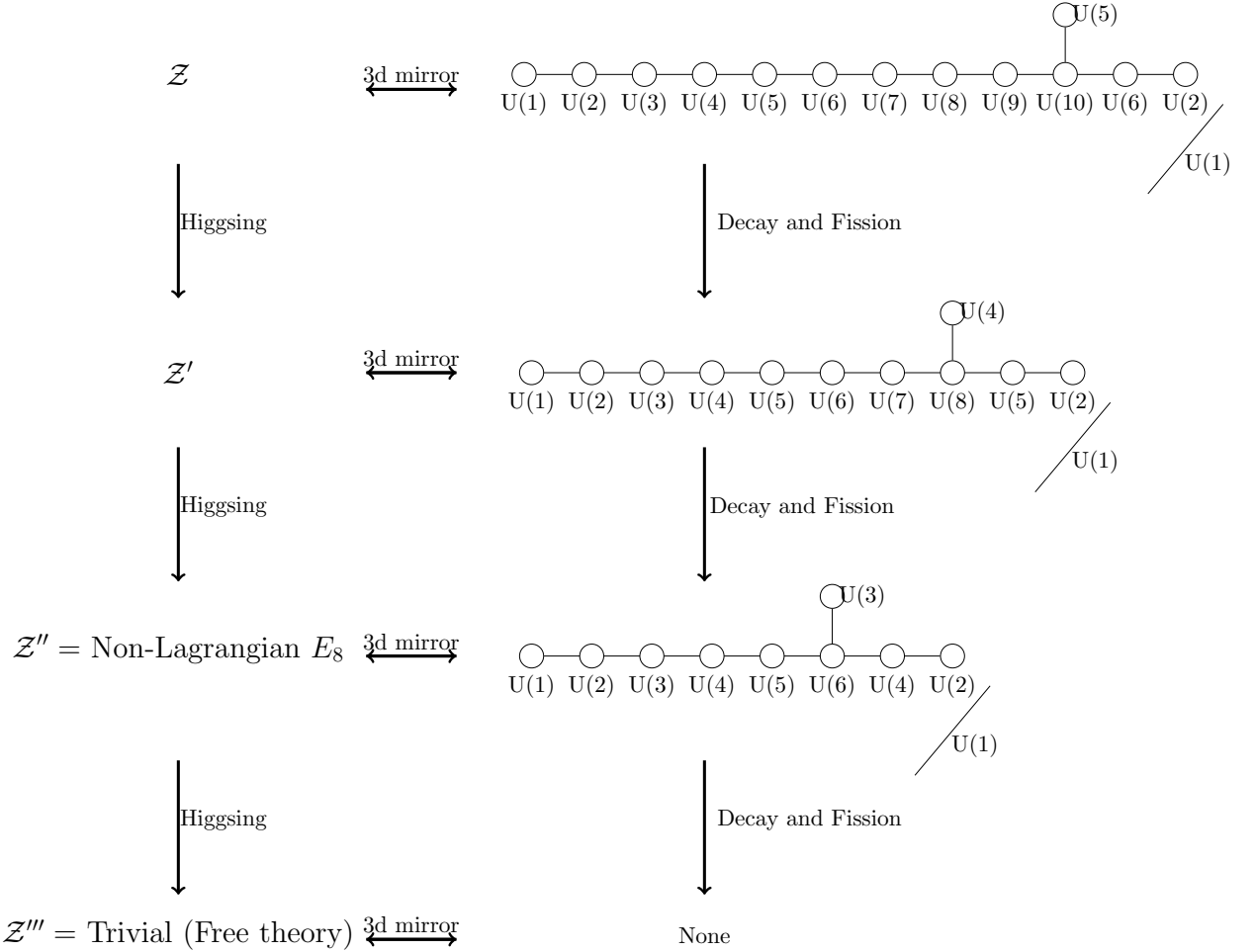
\begin{equation}
\scalebox{0.8}{\begin{tikzpicture}
	\begin{pgfonlayer}{nodelayer}
		\node [style=none] (187) at (-2.375, -9.25) {};
		\node [style=none] (188) at (-0.875, -9.25) {};
		\node [style=none] (189) at (-1.625, -9) {3d mirror};
		\node [style=none] (415) at (11.875, -9.5) {};
		\node [style=none] (416) at (10.625, -11) {};
		\node [style=none] (417) at (11.625, -10.5) {U(1)};
		\node [style=none] (446) at (-2.375, -18.75) {};
		\node [style=none] (447) at (-0.875, -18.75) {};
		\node [style=none] (448) at (-1.625, -18.5) {3d mirror};
		\node [style=none] (474) at (-2.375, -23.5) {};
		\node [style=none] (475) at (-0.875, -23.5) {};
		\node [style=none] (476) at (-1.625, -23.25) {3d mirror};
		\node [style=none] (502) at (-5.5, -10.5) {};
		\node [style=none] (503) at (-5.5, -12.75) {};
		\node [style=none] (504) at (3.25, -10.5) {};
		\node [style=none] (505) at (3.25, -12.75) {};
		\node [style=none] (506) at (-4.75, -11.5) {Higgsing};
		\node [style=none] (507) at (5, -11.5) {Decay and Fission};
		\node [style=none] (508) at (-5.5, -15.25) {};
		\node [style=none] (509) at (-5.5, -17.5) {};
		\node [style=none] (510) at (3.25, -15.25) {};
		\node [style=none] (511) at (3.25, -17.5) {};
		\node [style=none] (512) at (-4.75, -16.25) {Higgsing};
		\node [style=none] (513) at (4.875, -16.25) {Decay and Fission};
		\node [style=none] (514) at (-5.5, -20.5) {};
		\node [style=none] (515) at (-5.5, -22.75) {};
		\node [style=none] (516) at (3.25, -20.5) {};
		\node [style=none] (517) at (3.25, -22.75) {};
		\node [style=none] (518) at (-4.75, -21.5) {Higgsing};
		\node [style=none] (519) at (5, -21.5) {Decay and Fission};
		\node [style=none] (520) at (-5.5, -23.425) {\Large{$\mathcal{Z'''}=$} Trivial (Free theory)};
		\node [style=none] (521) at (3.25, -23.425) {None};
		\node [style=none] (522) at (-5.5, -9) {\Large{$\mathcal{Z}$}};
		\node [style=none] (523) at (-5.5, -14) {\Large{$\mathcal{Z}'$}};
		\node [style=none] (524) at (-5.5, -18.625) {\Large{$\mathcal{Z}''=$} Non-Lagrangian $E_8$};
		\node [style=gauge3] (530) at (0.25, -9) {};
		\node [style=gauge3] (531) at (1.25, -9) {};
		\node [style=gauge3] (532) at (2.25, -9) {};
		\node [style=gauge3] (533) at (3.25, -9) {};
		\node [style=gauge3] (534) at (4.25, -9) {};
		\node [style=gauge3] (535) at (5.25, -9) {};
		\node [style=gauge3] (536) at (6.25, -9) {};
		\node [style=gauge3] (537) at (7.25, -9) {};
		\node [style=gauge3] (538) at (8.25, -9) {};
		\node [style=gauge3] (539) at (9.25, -9) {};
		\node [style=gauge3] (540) at (10.25, -9) {};
		\node [style=gauge3] (541) at (11.25, -9) {};
		\node [style=gauge3] (542) at (9.25, -8) {};
		\node [style=none] (543) at (0.25, -9.5) {U(1)};
		\node [style=none] (544) at (1.25, -9.5) {U(2)};
		\node [style=none] (545) at (2.25, -9.5) {U(3)};
		\node [style=none] (546) at (3.25, -9.5) {U(4)};
		\node [style=none] (547) at (4.25, -9.5) {U(5)};
		\node [style=none] (548) at (5.25, -9.5) {U(6)};
		\node [style=none] (549) at (6.25, -9.5) {U(7)};
		\node [style=none] (550) at (7.25, -9.5) {U(8)};
		\node [style=none] (551) at (8.25, -9.5) {U(9)};
		\node [style=none] (552) at (9.25, -9.5) {U(10)};
		\node [style=none] (553) at (10.25, -9.5) {U(6)};
		\node [style=none] (554) at (11.25, -9.5) {U(2)};
		\node [style=none] (555) at (9.75, -8) {U(5)};
		\node [style=none] (556) at (10, -14.5) {};
		\node [style=none] (557) at (8.75, -16) {};
		\node [style=none] (558) at (9.75, -15.5) {U(1)};
		\node [style=gauge3] (561) at (0.375, -14) {};
		\node [style=gauge3] (562) at (1.375, -14) {};
		\node [style=gauge3] (563) at (2.375, -14) {};
		\node [style=gauge3] (564) at (3.375, -14) {};
		\node [style=gauge3] (565) at (4.375, -14) {};
		\node [style=gauge3] (566) at (5.375, -14) {};
		\node [style=gauge3] (567) at (6.375, -14) {};
		\node [style=gauge3] (568) at (7.375, -14) {};
		\node [style=gauge3] (569) at (8.375, -14) {};
		\node [style=gauge3] (570) at (9.375, -14) {};
		\node [style=gauge3] (571) at (7.375, -13) {};
		\node [style=none] (572) at (0.375, -14.5) {U(1)};
		\node [style=none] (573) at (1.375, -14.5) {U(2)};
		\node [style=none] (574) at (2.375, -14.5) {U(3)};
		\node [style=none] (575) at (3.375, -14.5) {U(4)};
		\node [style=none] (576) at (4.375, -14.5) {U(5)};
		\node [style=none] (577) at (5.375, -14.5) {U(6)};
		\node [style=none] (578) at (6.375, -14.5) {U(7)};
		\node [style=none] (581) at (7.375, -14.5) {U(8)};
		\node [style=none] (582) at (8.375, -14.5) {U(5)};
		\node [style=none] (583) at (9.375, -14.5) {U(2)};
		\node [style=none] (584) at (7.875, -13) {U(4)};
		\node [style=none] (585) at (8, -19.25) {};
		\node [style=none] (586) at (6.75, -20.75) {};
		\node [style=none] (587) at (7.75, -20.25) {U(1)};
		\node [style=gauge3] (590) at (0.375, -18.75) {};
		\node [style=gauge3] (591) at (1.375, -18.75) {};
		\node [style=gauge3] (592) at (2.375, -18.75) {};
		\node [style=gauge3] (593) at (3.375, -18.75) {};
		\node [style=gauge3] (594) at (4.375, -18.75) {};
		\node [style=gauge3] (595) at (5.375, -18.75) {};
		\node [style=gauge3] (596) at (6.375, -18.75) {};
		\node [style=gauge3] (597) at (7.375, -18.75) {};
		\node [style=gauge3] (598) at (5.375, -17.75) {};
		\node [style=none] (599) at (0.375, -19.25) {U(1)};
		\node [style=none] (600) at (1.375, -19.25) {U(2)};
		\node [style=none] (601) at (2.375, -19.25) {U(3)};
		\node [style=none] (602) at (3.375, -19.25) {U(4)};
		\node [style=none] (603) at (4.375, -19.25) {U(5)};
		\node [style=none] (604) at (5.375, -19.25) {U(6)};
		\node [style=none] (607) at (6.375, -19.25) {U(4)};
		\node [style=none] (608) at (7.375, -19.25) {U(2)};
		\node [style=none] (609) at (5.875, -17.75) {U(3)};
		\node [style=none] (610) at (-2.375, -14) {};
		\node [style=none] (611) at (-0.875, -14) {};
		\node [style=none] (612) at (-1.625, -13.75) {3d mirror};
	\end{pgfonlayer}
	\begin{pgfonlayer}{edgelayer}
		\draw [style=arrowed] (187.center) to (188.center);
		\draw [style=arrowed] (188.center) to (187.center);
		\draw (416.center) to (415.center);
		\draw [style=arrowed] (446.center) to (447.center);
		\draw [style=arrowed] (447.center) to (446.center);
		\draw [style=arrowed] (474.center) to (475.center);
		\draw [style=arrowed] (475.center) to (474.center);
		\draw [style=arrowed] (502.center) to (503.center);
		\draw [style=arrowed] (504.center) to (505.center);
		\draw [style=arrowed] (508.center) to (509.center);
		\draw [style=arrowed] (510.center) to (511.center);
		\draw [style=arrowed] (514.center) to (515.center);
		\draw [style=arrowed] (516.center) to (517.center);
		\draw (530) to (541);
		\draw (542) to (539);
		\draw (557.center) to (556.center);
		\draw (571) to (568);
		\draw (561) to (570);
		\draw (586.center) to (585.center);
		\draw (598) to (595);
		\draw [style=arrowed] (610.center) to (611.center);
		\draw [style=arrowed] (611.center) to (610.center);
		\draw (590) to (597);
	\end{pgfonlayer}
\end{tikzpicture}
}
\end{equation}
The theory $\mathcal{Z}$ is the $6d$ $\mathcal{N}=(1,0)$ SCFT of $Sp(3)$ gauge theory with 12 fundamental flavors at the infinite coupling limit, compactified on $T^3$ to $3d$ $\mathcal{N}=4$. The magnetic quiver is given in \cite{Hanany:2018uhm}. We observe that, after undergoing two decays, the magnetic quiver/3d mirror ultimately transforms into the $E_8$ affine Dynkin quiver. This implies that performing two (minimal) Higgsings on $\mathcal{Z}$ leads to $\mathcal{Z''}$, which is the 3d mirror of the $E_8$ affine Dynkin quiver. The critical point here is that the $E_8$ affine Dynkin quiver \textbf{does not} possess a Lagrangian (quiver gauge theory) as its 3d mirror. Consequently, $\mathcal{Z''}$ is not Lagrangian. Since $\mathcal{Z}$, after Higgsing, gives rise to a non-Lagrangian theory, it is reasonable to argue that $\mathcal{Z}$ itself  is also non-Lagrangian. This phenomenon is not confined to the $E_8$ affine Dynkin quiver; any magnetic quiver/3d mirror that decays into one of the quivers shown in Figure \ref{exceptionends} will yield a non-Lagrangian mirror theory. This leads to the following conjecture:

\begin{tcolorbox}[enhanced,
  borderline={0.8pt}{0pt}{black},
  borderline={0.4pt}{2pt}{black},
  boxrule=0.4pt,
  colback=white,
  coltitle=black,
  sharp corners]
\textbf{3d mirror conjecture:} \\ 
Any $3d$ $\mathcal{N}=4$ quiver that reaches one of the quivers in Figure \ref{exceptionends} after undergoing Decay and Fission, does not have a Lagrangian (quiver gauge theory) 3d mirror dual.
\label{conjecture}
\end{tcolorbox}
The relation between Decay and Fission and Quiver Subtraction  of \cite{Bourget:2019aer} tells us that any affine Dynkin quiver that appears as a subquiver will also be one of the penultimate quivers in decay and fission (quivers that decays to nothing). Therefore, we can equivalently make the statement: 
\begin{tcolorbox}[enhanced,
  borderline={0.8pt}{0pt}{black},
  borderline={0.4pt}{2pt}{black},
  boxrule=0.4pt,
  colback=white,
  coltitle=black,
  sharp corners]
\textbf{3d mirror corollary:} \\ 
Any $3d$ $\mathcal{N}=4$  quiver that contains a quiver in Figure \ref{exceptionends} as a subquiver, does not have a Lagrangian (quiver gauge theory) 3d mirror.
\end{tcolorbox}
By subquiver, we mean a subset of the gauge nodes forms one of the quivers in Figure \ref{exceptionends} and the gauge group ranks are greater or equal to the ranks there. In fact, the corollary can be extended to \emph{any} subquiver that, by itself, is known to have a non-Lagrangian 3d mirror. Notice we didn't constrain this to quivers made of only unitary gauge groups as it is true even for unitary subquivers of unitary-orthosymplectic quivers. 

In general, consider  \S\;  and its highly non-linear 3d mirror: 
\begin{equation}
    \scalebox{0.5}{\begin{tikzpicture}
	\begin{pgfonlayer}{nodelayer}
		\node [style=none] (187) at (-2.375, -9) {};
		\node [style=none] (188) at (-0.875, -9) {};
		\node [style=none] (189) at (-1.625, -8.5) {\huge{3d mirror}};
		\node [style=none] (415) at (8.375, -9) {};
		\node [style=none] (416) at (7.125, -10.5) {};
		\node [style=none] (417) at (8.125, -10) {U(1)};
		\node [style=none] (446) at (-2.375, -19) {};
		\node [style=none] (447) at (-0.875, -19) {};
		\node [style=none] (448) at (-1.625, -18.5) {\huge{3d mirror}};
		\node [style=none] (474) at (-2.375, -23.5) {};
		\node [style=none] (475) at (-0.875, -23.5) {};
		\node [style=none] (476) at (-1.3, -23) {\huge{3d mirror}};
		\node [style=none] (502) at (-5.5, -10.5) {};
		\node [style=none] (503) at (-5.5, -12.75) {};
		\node [style=none] (504) at (3.25, -10.5) {};
		\node [style=none] (505) at (3.25, -12.75) {};
		\node [style=none] (506) at (-4.25, -11.5) {\huge{Higgsing}};
		\node [style=none] (507) at (5.5, -11.5) {\LARGE{Decay and Fission}};
		\node [style=none] (514) at (-5.5, -20.5) {};
		\node [style=none] (515) at (-5.5, -22.75) {};
		\node [style=none] (516) at (3.25, -20.5) {};
		\node [style=none] (517) at (3.25, -22.75) {};
		\node [style=none] (518) at (-4.25, -21.5) {\huge{Higgsing}};
		\node [style=none] (519) at (5, -21.5) {Decay and Fission};
		\node [style=none] (520) at (-5.5, -23.425) {\huge{Trivial (Free theory)}};
		\node [style=none] (521) at (3.25, -23.425) {\huge{None}};
		\node [style=none] (522) at (-5.5, -9) {\Huge{\S}};
		\node [style=none] (524) at (-5.5, -18.875) {\LARGE{Non-Lagrangian $E_8$}};
		\node [style=gauge3] (530) at (0.25, -9) {};
		\node [style=gauge3] (531) at (1.25, -9) {};
		\node [style=gauge3] (532) at (2.25, -9) {};
		\node [style=gauge3] (533) at (3.25, -9) {};
		\node [style=gauge3] (534) at (4.25, -9) {};
		\node [style=gauge3] (535) at (5.25, -9) {};
		\node [style=gauge3] (536) at (6.25, -9) {};
		\node [style=gauge3] (537) at (7.25, -9) {};
		\node [style=none] (543) at (0.25, -9.5) {U(2)};
		\node [style=none] (544) at (1.25, -9.5) {U(4)};
		\node [style=none] (545) at (2.25, -9.5) {U(7)};
		\node [style=none] (546) at (3.25, -9.5) {U(10)};
		\node [style=none] (547) at (4.25, -9.5) {U(12)};
		\node [style=none] (548) at (5.25, -9.5) {U(10)};
		\node [style=none] (549) at (6.25, -9.5) {U(7)};
		\node [style=none] (550) at (7.25, -9.5) {U(3)};
		\node [style=none] (585) at (8, -19.25) {};
		\node [style=none] (586) at (6.75, -20.75) {};
		\node [style=none] (587) at (7.75, -20.25) {U(1)};
		\node [style=gauge3] (590) at (0.375, -18.75) {};
		\node [style=gauge3] (591) at (1.375, -18.75) {};
		\node [style=gauge3] (592) at (2.375, -18.75) {};
		\node [style=gauge3] (593) at (3.375, -18.75) {};
		\node [style=gauge3] (594) at (4.375, -18.75) {};
		\node [style=gauge3] (595) at (5.375, -18.75) {};
		\node [style=gauge3] (596) at (6.375, -18.75) {};
		\node [style=gauge3] (597) at (7.375, -18.75) {};
		\node [style=gauge3] (598) at (5.375, -17.75) {};
		\node [style=none] (599) at (0.375, -19.25) {U(1)};
		\node [style=none] (600) at (1.375, -19.25) {U(2)};
		\node [style=none] (601) at (2.375, -19.25) {U(3)};
		\node [style=none] (602) at (3.375, -19.25) {U(4)};
		\node [style=none] (603) at (4.375, -19.25) {U(5)};
		\node [style=none] (604) at (5.375, -19.25) {U(6)};
		\node [style=none] (607) at (6.375, -19.25) {U(4)};
		\node [style=none] (608) at (7.375, -19.25) {U(2)};
		\node [style=none] (609) at (5.875, -17.75) {U(3)};
		\node [style=gauge3] (613) at (3.25, -8) {};
		\node [style=gauge3] (614) at (5, -7) {};
		\node [style=gauge3] (615) at (6.25, -8) {};
		\node [style=gauge3] (616) at (3.25, -6.75) {};
		\node [style=gauge3] (617) at (4.5, -8) {};
		\node [style=none] (618) at (3.25, -6.25) {U(2)};
		\node [style=none] (619) at (5, -6.5) {U(2)};
		\node [style=none] (620) at (6.75, -8) {U(3)};
		\node [style=none] (621) at (5, -8) {U(2)};
		\node [style=none] (622) at (2.5, -8) {U(4)};
		\node [style=none] (623) at (-5.5, -14.25) {\vdots};
		\node [style=none] (624) at (3.25, -14.25) {\vdots};
		\node [style=none] (625) at (-5.5, -15.75) {};
		\node [style=none] (626) at (-5.5, -18) {};
		\node [style=none] (627) at (3.25, -15.75) {};
		\node [style=none] (628) at (3.25, -18) {};
		\node [style=none] (629) at (-4.75, -16.75) {Higgsing};
		\node [style=none] (630) at (5, -16.75) {Decay and Fission};
		\node [style=none] (631) at (-8.625, -15.75) {};
		\node [style=none] (632) at (-9.875, -17.25) {};
		\node [style=none] (633) at (-8.375, -16.75) {Higgsing};
		\node [style=gauge3] (634) at (-11, -19) {};
		\node [style=flavour2] (635) at (-11, -17.75) {};
		\node [style=none] (636) at (-11, -17.25) {3};
		\node [style=none] (637) at (-11, -19.5) {U(1)};
		\node [style=none] (638) at (-10.625, -15) {};
		\node [style=none] (639) at (-11.875, -16.5) {};
		\node [style=none] (640) at (-11.625, -15.25) {Higgsing};
		\node [style=none] (641) at (-10.75, -16) {$\dots$};
		\node [style=none] (642) at (-10.875, -20.75) {};
		\node [style=none] (643) at (-6.875, -22.75) {};
		\node [style=none] (644) at (-10.625, -22.25) {Higgsing};
		\node [style=none] (645) at (9.625, -15.75) {};
		\node [style=none] (646) at (10.875, -17.25) {};
		\node [style=gauge3] (648) at (10.25, -19.25) {};
		\node [style=none] (651) at (10.25, -19.75) {U(1)};
		\node [style=none] (652) at (11.625, -15) {};
		\node [style=none] (653) at (12.875, -16.5) {};
		\node [style=none] (655) at (11.25, -16) {$\dots$};
		\node [style=none] (656) at (8.625, -16.75) {Decay and Fission};
		\node [style=none] (657) at (13.875, -15.5) {Decay and Fission};
		\node [style=gauge3] (658) at (11.25, -18.25) {};
		\node [style=gauge3] (659) at (12.25, -19.25) {};
		\node [style=none] (660) at (12.25, -19.75) {U(1)};
		\node [style=none] (661) at (11.25, -17.75) {U(1)};
		\node [style=none] (662) at (13.25, -19.25) {};
		\node [style=none] (663) at (12, -20.75) {};
		\node [style=none] (664) at (13, -20.25) {U(1)};
		\node [style=none] (665) at (10.5, -20.5) {};
		\node [style=none] (666) at (4.75, -22.75) {};
		\node [style=none] (667) at (8.625, -22) {Decay and Fission};
		\node [style=none] (668) at (-5.5, -13.5) {\vdots};
		\node [style=none] (669) at (3.25, -13.5) {\vdots};
		\node [style=none] (670) at (-5.5, -15) {\vdots};
		\node [style=none] (671) at (3.25, -15) {\vdots};
	\end{pgfonlayer}
	\begin{pgfonlayer}{edgelayer}
		\draw [style=arrowed] (187.center) to (188.center);
		\draw [style=arrowed] (188.center) to (187.center);
		\draw (416.center) to (415.center);
		\draw [style=arrowed] (446.center) to (447.center);
		\draw [style=arrowed] (447.center) to (446.center);
		\draw [style=arrowed] (474.center) to (475.center);
		\draw [style=arrowed] (475.center) to (474.center);
		\draw [style=arrowed] (502.center) to (503.center);
		\draw [style=arrowed] (504.center) to (505.center);
		\draw [style=arrowed] (514.center) to (515.center);
		\draw [style=arrowed] (516.center) to (517.center);
		\draw (586.center) to (585.center);
		\draw (598) to (595);
		\draw (590) to (597);
		\draw (532) to (613);
		\draw (530) to (537);
		\draw (536) to (615);
		\draw (614) to (617);
		\draw (617) to (616);
		\draw (613) to (617);
		\draw (617) to (536);
		\draw (617) to (534);
		\draw (615) to (614);
		\draw (616) to (613);
		\draw (531) to (616);
		\draw (533) to (613);
		\draw (534) to (616);
		\draw (535) to (615);
		\draw [style=arrowed] (625.center) to (626.center);
		\draw [style=arrowed] (627.center) to (628.center);
		\draw [style=arrowed] (631.center) to (632.center);
		\draw (634) to (635);
		\draw [style=arrowed] (638.center) to (639.center);
		\draw [style=arrowed] (642.center) to (643.center);
		\draw [style=arrowed] (645.center) to (646.center);
		\draw [style=arrowed] (652.center) to (653.center);
		\draw (648) to (659);
		\draw (659) to (658);
		\draw (658) to (648);
		\draw (663.center) to (662.center);
		\draw [style=arrowed] (665.center) to (666.center);
	\end{pgfonlayer}
\end{tikzpicture}}
\end{equation}
Depending on the path of Higgsings, \S\;  may end up on many different penultimate theories $\S_i$ that are then minimally Higgsed to a  trivial/free theory. Some of the theories $\S_i$  may have a mirror that does not belong in Figure \ref{exceptionends} and therefore making $\S_i$ a Lagrangian theory. One of them $\S_1$ mirrors to the affine $A_2$ quiver, which means that $\S_1$ is the Lagrangian theory of  $U(1)$ gauge theory with three flavors. This means there is a path of Higgsings that Higgses \S\; into $U(1)$ with three flavors. However, we also see there is an affine $E_8$ subquiver embedded in the 3d mirror of  \S, and thus concludes that one of the penultimate quivers is the $E_8$ affine quiver, which has a non-Lagrangian 3d mirror. This means there is a different path of Higgsings that Higgses \S\; into this non-Lagrangian $E_8$ theory. In the end, as long as \emph{one of} these paths of Higgsings ends up in a non-Lagrangian theory, the starting theory \S\; is non-Lagrangian. This is why our bound is very powerful as regardless of how complicated a quiver is, if there is one subquiver that belongs to Figure \ref{exceptionends}, the mirror will be non-Lagrangian. This conjecture utilized the very reasonable assumption that a Lagrangian theory cannot be Higgsed into a non-Lagrangian theory\footnote{One way to see this is that given a Lagrangian (quiver gauge theory), one can always use group-theoretic methods to find all possible minimal Higgsings (see \cite{Bennett:2024loi} for detailed examples). Since a Lagrangian theory exhibits classical flavor symmetries, the Higgsed theories will be quiver gauge theories  with classical flavor symmetries as well. The converse, of course, is not true. A non-Lagrangian theory can be Higgsed into a Lagrangian theory. And of course, for other kinds of RG flows like mass deformation or tuning gauge couplings it is possible for a Lagrangian theory to flow into a non-Lagrangian theory.}. 

A similar statement can be made for unitary-orthosymplectic quivers. Some of the quivers in Figure \ref{exceptionends} have unitary-orthosymplectic analogues in \cite{O5,Bourget:2020xdz,product} so if any unitary-orthosymplectic quiver contains them as a subquiver, it will not have a Lagrangian (quiver gauge theory) 3d mirror by the same reasoning. These lists are not exhaustive\footnote{The examples in Figure \ref{exceptionends} are referred to as \emph{stable} quivers in \cite{Decay,Fission}. A way to construct an exhaustive list is to find all stable quivers and see whether they have Lagrangian 3d mirrors.}, so there could be more unitary-orthosymplectic subquivers that makes the mirror non-Lagrangian. This is also a call back to the statement at the beginning of the section as non-linearity gives the opportunity for exceptional unitary or orthosymplectic subquivers to appear. This is why the odds are against a generic non-linear quiver of classical gauge groups from having a Lagrangian 3d mirror.  

A non-linear quiver with exceptional Coulomb branch global symmetry is already an indication that the 3d mirror is non-Lagrangian. However, from analyzing the balance of the gauge nodes in the mirror of \S, it is clear that the Coulomb branch global symmetry group is classical, consisting of products  of unitary and special unitary groups. Thus, the conjecture shifts the argument from identifying an exceptional global symmetry (sub)group to identifying an exceptional Dynkin subquiver. It fact, in general, quivers with non-Lagrangian mirrors only possess classical global symmetry groups, as the balancing conditions required to exhibit exceptional global symmetry are a rarity.

\paragraph{Higher dimensional SCFTs} This is also why most class $\mathcal{S}$ theories with three or more punctures, when compactified on a circle, will not be Lagrangian: their 3d mirrors are highly non-linear, star-shaped quivers \cite{Benini:2010uu}. Similarly, most 4d $\mathcal{N}=2$ S-fold theories, when compactified on a circle, are 3d mirrors to quivers containing twisted affine EFG Dynkin subdiagrams \cite{Giacomelli:2020gee,Bourget:2020mez,Giacomelli:2024dbd}, and thus lack a Lagrangian description as well. In contrast, many Argyres-Douglas theories (such as $D_p$ theories constructed with one regular and one irregular puncture), when compactified on a circle, are 3d mirrors \cite{Nanopoulos:2010bv,Closset:2020afy,Giacomelli:2020ryy} to theories without any subdiagram in Figure \ref{exceptionends} and are indeed known to have Lagrangian descriptions. This further shows that our conjectures, in combination with magnetic quivers, can help diagnose whether a higher-dimensional SCFT, when compactified to 3d, admits a Lagrangian description. The compactified theory being (Non-)Lagrangian is also a good indication of whether  the higher-dimensional theory itself is (Non-)Lagrangian.

\section{Conclusion and Future outlook}\label{conc}
In this paper, we studied a large class of 3d $\mathcal{N}=4$ quiver gauge theories that have Lagrangian (quiver gauge theory) 3d mirrors, known as the DynkinABCD family. Extending the work of \cite{USU}, which expanded the DynkinA family to the \DynkinA family by including mixtures of unitary and special unitary gauge groups, we similarly extended the DynkinBCD family to the \DynkinBCD family. We also introduced the new families \DynkinAbar\; and \DynkinBCDbar as further extensions. These additions significantly broaden the landscape of known 3d mirror pairs.

In the second part of the paper, we conjectured a sharp boundary in the landscape of Lagrangian (quiver gauge theory) mirror pairs. This boundary helps explain why most non-linear quivers do not have a Lagrangian (quiver gauge theory) 3d mirror. These results are summarized in Figure \ref{ultimatetable}.  

\paragraph{Future outlook and some Comments} 
In Figure \ref{ultimatetable}, we observe a region between the two blobs representing the now systematically studied \DynkinABCDbar and $T^\sigma_\rho(G)$ families, and the boundary set by our conjecture. What kind of theories exist within this gap? Let's consider a highly non-linear unitary quiver with only simply-laced edges and constructed solely from U(2) and U(1) gauge groups. Such a theory is inside our bound, yet it remains beyond any known family. Exploring these examples can help answer whether every quiver gauge theory within our bound has a Lagrangian (quiver gauge theory) 3d mirror. 

Let us also discuss some other families with known 3d mirrors in the literature, some of which lives in this gap. 

\paragraph{Affine Dynkin type quivers}
In Lie algebra theory, Dynkin diagrams can be either \emph{finite} or \emph{affine}. So far, in our discussion of Dynkin-type quivers, we have only considered the finite cases, where the gauge groups form finite Dynkin diagrams. Now, what happens when the gauge groups form affine Dynkin diagrams.

The 3d mirrors of affine Dynkin quivers with only unitary gauge groups are provided in \cite{Cremonesi:2014xha, Mekareeya:2015bla}. Affine DynkinA quivers takes the form of a circular quiver with flavors nodes on one or multiple number of gauge groups. The 3d mirror of an affine DykinA quiver is also an affine DynkinA quiver. However, naively applying our U/SU algorithm does not yield the correct mirror pair for affine-\DynkinA. This issue becomes apparent in the limiting case where one side of the mirror reduces to the ADHM quiver of U(k) with an adjoint and $n$ flavors. The 3d mirror in this case is known to be a circular quiver consisting of  $n-1$ U(k) gauge groups with a single fundamental flavor attached to one of them. If we instead consider  SU(k) with an adjoint with $n$ flavors and apply the U/SU algorithm, we would expect the U(1)  flavor node in the mirror to be gauged. However, since all the gauge nodes are unitary, this gauged U(1) symmetry decouples automatically and we return to the original circular quiver. The underlying reason is that in a circular brane setup, there is a $\mathrm{U}(1)$ degree of freedom that is lost. A linear brane setup with $n-1$ intervals corresponds to $n-1$ lockings, whereas a circular setup has only $n-2$ lockings. This missing $\mathrm{U}(1)$ affects the U/SU algorithm in a non-trivial way and prevents us from obtaining the correct 3d mirror in generic cases. 

In the case of affine-DynkinBCD type quivers, the quivers are modified to include two ON orientifold planes. These examples do not involve an additional loss of U(1) as the circular quivers, so it might be that our algorithm for finite Dynkin cases applies here more straightforwardly. We leave it for the future to resolve and explore these new theories and their 3d mirrors, hoping for a similar algorithm for what we have here but with slight modifications. 

\paragraph{Limit for orthosymplectic quiver}
The elephant in the room is to generalize our 3d mirror conjectures in Section \ref{thelimit} to orthosymplectic quivers as well. The arguments of our conjectures for unitary quivers are based on the usage of Decay and Fission as a Higgsing algorithm. An analogous and complete statement for orthosymplectic quivers will require a orthosymplectic quiver decay and fission algorithm. Such an algorithm is still in the works \cite{orthosymplecticdecay} although had already been explored to some degree in \cite{Bao:2024eoq}. 

\paragraph{Bad quivers}
So far, we discussed \emph{good} quivers gauge theories and their 3d mirrors. For \emph{bad} quivers, the definition of 3d mirror becomes problematic as we discussed in detail in \cite{USU}. However,  one may still wishes to treat them as 3d mirror pairs, perhaps as a local definition, which leads to interesting results (see \cite{Xie:2012hs,Giacomelli:2020ryy,Benvenuti:2017bpg,Beem:2023ofp}). These bad-\DynkinA and their `mirror' can still be derived using the U/SU algorithm as shown in \cite{USU}. The main difference in the `mirror' is that there are edges connecting the U(1) gauge groups in the bouquet, a feature that is absent for good 3d mirrors. An extension can be easily done for bad-\DynkinBCD theories as well, and one would expect a similar feature where the unitary-orthosymplectic nodes in the bouquet connect to each other by edges. Some of these quivers are mirrors of $(D_m,D_n)$ Argyres-Douglas theories and the mirror pairs are given in   \cite{Carta:2021dyx} which can be reproduced from our locking mechanism. 

Although the algorithm we have in this paper is very simple, it will still be nice, especially for bad-\DynkinBCD quivers, to have a \texttt{Mathematica} code similar to the one done for \DynkinA in \cite{USU} that generate the mirrors. 

\paragraph{Matter with more exotic representations}
The theories studied in \cite{O5,Bourget:2020xdz,exo2,exo3,product,exo1} are some examples of quiver theories where matter with representations beyond (bi)fundamental and adjoint representations made an appearance and their 3d mirrors were argued. This can certainly create a much more diverse patch in the 3d mirror landscape. However, many of these theories are actually the same (dual) to theories in (\DynkinABCDbar and $T^\sigma_\rho(G)$), and hence their 3d mirrors are something we already documented in Figure \ref{ultimatetable}. It will be more interesting to focus on the ones (if there is any) that does not have a known dual theory with ordinary matter representations but nevertheless has a Lagrangian (quiver) mirror. There are also rank-0 theories where the Higgs and/or Coulomb branches are trivial \cite{zero,zero11,zero10,zero9,zero8,zero7,zero6,zero5,zero4,zero3,zero2,zero1,Zhong:2024nff} and some of these have conjectured 3d mirror duals. However, at least one of the mirror pairs (whenever the Coulomb branch is trivial) is a non-Lagrangian theory. 

\paragraph{Are non-simply laced quivers Lagrangian?}
In Section \ref{sect2}, we examined non-simply laced quivers, \DynkinBCD, and their mirrors. Although a Lagrangian description for these quivers is not yet known, we treated them as such as they both possess a quiver description and a brane setup description involving orientifold planes. If one prefers to view these as non-Lagrangian theories, they can simply refine our 3d mirror conjecture by excluding any non-simply laced quiver gauge theory, resulting in an even more restrictive bound.

\paragraph{Other theories}
There are also \emph{good} quiver gauge theories with ordinary matter representations that lives between the red wall in Figure \ref{ultimatetable} and the theories discussed thus far. Although such theories are dearth, they do exist: see \cite{Dey:2020hfe,Dey:2021rxw}. These usually includes more U(1) extensions (in some highly non-linear way.) extended from a single linear chain of non-Abelian gauge groups. Since these are just a bunch of U(1)s, they do not violate our conjectures and can have Lagrangian mirror pairs. Perhaps the method introduced in \cite{Dey:2020hfe,Dey:2021rxw} and other methods such as \cite{Hwang:2021ulb,Giacomelli:2023zkk,Giacomelli:2024laq} can shed some light for such theories.  

\paragraph{Discrete gaugings}
As shown in \cite{Nawata:2023rdx,Bhardwaj:2023zix}, it is possible to take a theory with known 3d mirror, then gauge a one-form symmetry (a discrete gauging) and find the new 3d mirror. Such discrete gaugings can be done for all the theories in the landscape, and the 3d mirror will have this discrete action acting on the flavor symmetry. Since it is just discrete gauging, we  adopt the convention that it is not a completely new 3d mirror pair. 

\paragraph{Are non-Lagrangian theories always non-Lagrangian?}
When discussing non-Lagrangian theories, the question arises: are they truly non-Lagrangian, or do they have a Lagrangian description in the UV that we have yet to discover? In 4d $\mathcal{N}=2$ SCFTs, this phenomenon has been studied extensively (see \cite{Razamat:2022gpm} for a review of recent results). For instance, the non-Lagrangian $E_{6,7,8}$ Minahan-Nemeschansky SCFTs actually flow from Lagrangian 4d $\mathcal{N}=1$ theories \cite{Zafrir:2019hps, Razamat:2020gcc}. It would be interesting to explore whether 3d $\mathcal{N}=4$ $E_{6,7,8}$ theories can similarly flow from 3d $\mathcal{N}=2$ Lagrangian theories. However, generalizing such studies to find the 3d mirror for generic non-linear quivers is more challenging compared to, for example, constructing brane systems. Furthermore, even if these non-Lagrangian theories are found, it is unlikely they will have a 3d $\mathcal{N}=4$ UV Lagrangian (quiver gauge theory) description, which is the focus of our paper. This is a crucial point as finding 3d mirror pairs is not the end but merely the starting point, and it is the wealth of tools we use to study 3d $\mathcal{N}=4$ quiver gauge theories that makes finding mirror pairs worth it. 

\paragraph{Filling the gap}
A recent paper \cite{Bennett:2024loi} proposed a subtraction algorithm that performs Higgsings directly on 3d $\mathcal{N}=4$ Lagrangian quiver gauge theories. This serves as the dual algorithm to Decay and Fission. Therefore, if there is a theory within the gap where both 3d mirrors are Lagrangian, these two algorithms can be used together to obtain all the Higgsings of the theory and their respective 3d mirrors. Similarly, new set of proposed quotient and gauging algorithms in \cite{Bennett:2024llh,Hanany:2024fqf,Hanany:2023uzn,Hanany:2023tvn} carries potential of taking known Lagrangian pairs to finding new Lagrangian pairs as well. 


\section*{Acknowledgements}
We thank Francesco Benini,  Nick Dorey, Amihay Hanany,  Kenneth Intriligator, Chunhao Li, Noppadol Mekareeya, Sara Pasquetti,  Matteo Sacchi and Simone Giacomelli. We thank in particular Antoine Bourget and Marcus Sperling for giving precious comments on our draft.
I am supported by the ERC Consolidator Grant \# 864828 ``Algebraic Foundations of Supersymmetric Quantum Field Theory'' (SCFTAlg). I also thank the hospitality of New College, Oxford, where part of this project is completed. 
\clearpage

\appendix
\section{D3-D5-NS5 brane system}\label{similarbranes}
For DynkinABCD quivers, the information needed to arrive at the  3d mirror for any combinations of U \& SU  can be encoded in a  D3-D5-NS5 set up with ON planes. This is useful when it becomes cumbersome to draw the brane webs explicitly. 

Focusing on \eqref{firstmirror} but from a D3-D5-NS5 set up:
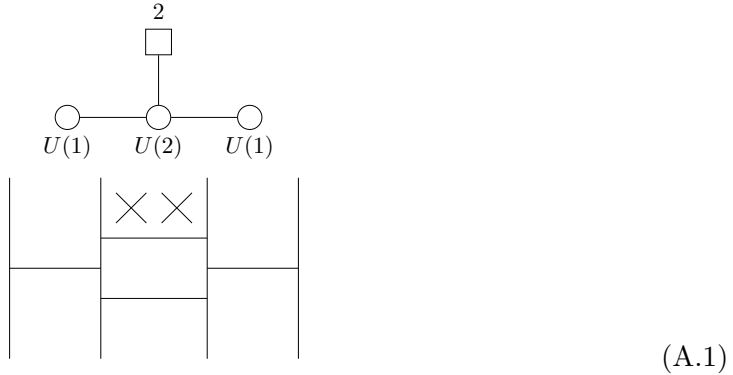
\begin{equation}
\scalebox{0.8}{\begin{tikzpicture}
	\begin{pgfonlayer}{nodelayer}
		\node [style=gauge3] (0) at (6.5, 0) {};
		\node [style=gauge3] (1) at (8, 0) {};
		\node [style=gauge3] (2) at (9.5, 0) {};
		\node [style=flavour2] (3) at (8, 1.25) {};
		\node [style=none] (4) at (8, 1.75) {2};
		\node [style=none] (5) at (6.5, -0.5) {$U(1)$};
		\node [style=none] (6) at (8, -0.5) {$U(2)$};
		\node [style=none] (7) at (9.5, -0.5) {$U(1)$};
		\node [style=none] (8) at (5.55, -1) {};
		\node [style=none] (9) at (5.55, -4) {};
		\node [style=none] (10) at (7.05, -1) {};
		\node [style=none] (11) at (7.05, -4) {};
		\node [style=none] (12) at (8.8, -1) {};
		\node [style=none] (13) at (8.8, -4) {};
		\node [style=none] (14) at (10.3, -1) {};
		\node [style=none] (15) at (10.3, -4) {};
		\node [style=none] (16) at (5.55, -2.5) {};
		\node [style=none] (17) at (7.05, -2.5) {};
		\node [style=none] (18) at (7.05, -2) {};
		\node [style=none] (19) at (7.05, -3) {};
		\node [style=none] (20) at (8.8, -3) {};
		\node [style=none] (21) at (8.8, -2.5) {};
		\node [style=none] (22) at (8.8, -2) {};
		\node [style=none] (23) at (10.3, -2.5) {};
		\node [style=none] (24) at (7.8, -1.25) {};
		\node [style=none] (25) at (7.3, -1.75) {};
		\node [style=none] (26) at (7.3, -1.25) {};
		\node [style=none] (27) at (7.8, -1.75) {};
		\node [style=none] (28) at (8.55, -1.25) {};
		\node [style=none] (29) at (8.05, -1.75) {};
		\node [style=none] (30) at (8.05, -1.25) {};
		\node [style=none] (31) at (8.55, -1.75) {};
	\end{pgfonlayer}
	\begin{pgfonlayer}{edgelayer}
		\draw (0) to (1);
		\draw (1) to (2);
		\draw (3) to (1);
		\draw (14.center) to (15.center);
		\draw (13.center) to (12.center);
		\draw (10.center) to (11.center);
		\draw (9.center) to (8.center);
		\draw (16.center) to (17.center);
		\draw (18.center) to (22.center);
		\draw (19.center) to (20.center);
		\draw (21.center) to (23.center);
		\draw (26.center) to (27.center);
		\draw (24.center) to (25.center);
		\draw (30.center) to (31.center);
		\draw (28.center) to (29.center);
	\end{pgfonlayer}
\end{tikzpicture}}
\end{equation}
which describes the Coulomb branch phase of the quiver where horizontal lines are D3 branes, vertical lines are NS5 branes and crosses are D5 branes. Going to the Higgs phase, we once again set the masses of the hypers to be zero and do some Hanany-Witten transitions:
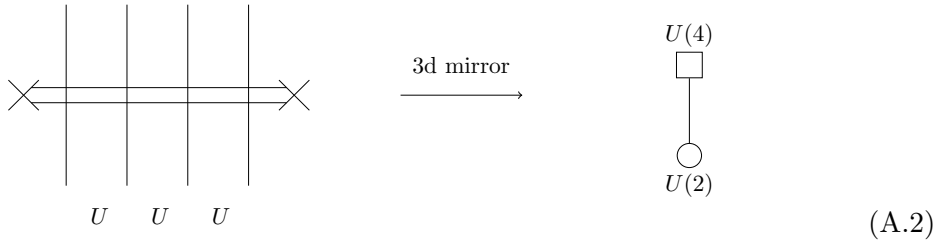
\begin{equation}
\scalebox{0.8}{\begin{tikzpicture}
	\begin{pgfonlayer}{nodelayer}
		\node [style=gauge3] (0) at (12.75, -8) {};
		\node [style=none] (1) at (12.75, -6) {$U(4)$};
		\node [style=none] (2) at (12.75, -8.5) {$U(2)$};
		\node [style=flavour2] (3) at (12.75, -6.5) {};
		\node [style=none] (4) at (12.625, -8.075) {};
		\node [style=none] (5) at (12.8, -7.9) {};
		\node [style=none] (6) at (1.925, -6.875) {};
		\node [style=none] (7) at (1.925, -7.125) {};
		\node [style=none] (8) at (6.125, -6.875) {};
		\node [style=none] (9) at (6.125, -7.125) {};
		\node [style=none] (10) at (3.05, -9) {$U$};
		\node [style=none] (11) at (4.05, -9) {$U$};
		\node [style=none] (12) at (5.05, -9) {$U$};
		\node [style=none] (13) at (6.5, -6.75) {};
		\node [style=none] (14) at (6, -7.25) {};
		\node [style=none] (15) at (6, -6.75) {};
		\node [style=none] (16) at (6.5, -7.25) {};
		\node [style=none] (17) at (2.05, -6.75) {};
		\node [style=none] (18) at (1.55, -7.25) {};
		\node [style=none] (19) at (1.55, -6.75) {};
		\node [style=none] (20) at (2.05, -7.25) {};
		\node [style=none] (21) at (2.5, -5.5) {};
		\node [style=none] (22) at (2.5, -8.5) {};
		\node [style=none] (23) at (3.5, -5.5) {};
		\node [style=none] (24) at (3.5, -8.5) {};
		\node [style=none] (25) at (4.5, -5.5) {};
		\node [style=none] (26) at (4.5, -8.5) {};
		\node [style=none] (27) at (5.5, -5.5) {};
		\node [style=none] (28) at (5.5, -8.5) {};
		\node [style=none] (29) at (10, -7) {};
		\node [style=none] (30) at (8, -7) {};
		\node [style=none] (31) at (9, -6.5) {3d mirror};
	\end{pgfonlayer}
	\begin{pgfonlayer}{edgelayer}
		\draw (0) to (3);
		\draw (8.center) to (6.center);
		\draw (7.center) to (9.center);
		\draw (15.center) to (16.center);
		\draw (13.center) to (14.center);
		\draw (19.center) to (20.center);
		\draw (17.center) to (18.center);
		\draw (27.center) to (28.center);
		\draw (26.center) to (25.center);
		\draw (23.center) to (24.center);
		\draw (22.center) to (21.center);
		\draw [style=->] (30.center) to (29.center);
	\end{pgfonlayer}
\end{tikzpicture}}
\end{equation}
Comparing this Higgs phase configuration with that in \eqref{allUSU4}, we observe that the setups are nearly identical, except that the D5-NS5-D7-NS7 brane system is replaced by a D3-NS5-D5 configuration. The key difference is that the [0,1]7-branes (NS7) do not seem to have an equivalent in the $3d$ brane system. This poses a challenge because, in the brane web setup, the NS5 branes are dynamical and can move along the [0,1]7-branes, contributing to the dynamical degrees of freedom. In contrast, in the D3-D5-NS5 configuration, the NS5 branes are non-dynamical and should not contribute to gauge groups in the $3d$ mirror. Nevertheless, we will still apply the “locking mechanism” on the NS5 branes, similar to how it was used in the brane web setups. 

\paragraph{``Locking'' in D3-D5-NS5}
Let us turn the all the unitary gauge nodes to special unitary by \emph{dictating} that each NS5s carry a U(1) degree of freedom. The 3d mirror is then:
\begin{equation}
  \scalebox{0.8}{  
}
\label{3dSU4}
\end{equation}
which reproduces the $3d$ mirror in \eqref{firstweb}. By locking the NS5 branes together and calculating the intersection number between the NS5 branes and the D3 branes, we obtain the exact same set of $3d$ mirrors for the various combinations of U and SU nodes shown in Table \ref{SU4}.

\begin{table}
    \centering
  \scalebox{0.9}{  \begin{tabular}{c|c|c|c}
    \scalebox{0.7}{ \begin{tikzpicture}
	\begin{pgfonlayer}{nodelayer}
		\node [style=gauge3] (0) at (0.5, 0) {};
		\node [style=gauge3] (1) at (1.75, 0) {};
		\node [style=gauge3] (2) at (3, 0) {};
		\node [style=flavour2] (3) at (1.75, 1.25) {};
		\node [style=none] (4) at (1.75, 1.75) {U(2)};
		\node [style=none] (5) at (0.5, -0.5) {$U(1)$};
		\node [style=none] (6) at (1.75, -0.5) {$U(2)$};
		\node [style=none] (7) at (3, -0.5) {$U(1)$};
	\end{pgfonlayer}
	\begin{pgfonlayer}{edgelayer}
		\draw (0) to (2);
		\draw (1) to (3);
	\end{pgfonlayer}
\end{tikzpicture}}
      & \scalebox{0.7}{\begin{tikzpicture}
	\begin{pgfonlayer}{nodelayer}
		\node [style=gauge3] (0) at (2.25, 0) {};
		\node [style=gauge3] (1) at (6.75, 0) {};
		\node [style=none] (2) at (2.25, 0.125) {};
		\node [style=none] (3) at (2.25, -0.125) {};
		\node [style=none] (4) at (6.75, 0.125) {};
		\node [style=none] (5) at (6.75, -0.125) {};
		\node [style=gauge3] (6) at (3, 1.5) {};
		\node [style=gauge3] (7) at (3, -1.5) {};
		\node [style=gauge3] (8) at (4, 1.5) {};
		\node [style=gauge3] (9) at (4, -1.5) {};
		\node [style=gauge3] (10) at (5, 1.5) {};
		\node [style=gauge3] (11) at (5, -1.5) {};
		\node [style=gauge3] (12) at (6, 1.5) {};
		\node [style=gauge3] (13) at (6, -1.5) {};
	\end{pgfonlayer}
	\begin{pgfonlayer}{edgelayer}
		\draw (4.center) to (2.center);
		\draw (3.center) to (5.center);
		\draw [style=pinkline] (8) to (9);
		\draw [style=bluee] (6) to (7);
		\draw [style=greenline] (12) to (13);
		\draw [style=brownline] (10) to (11);
		\draw [style=bluee] (8) to (9);
		\draw [style=greenline] (10) to (11);
		\draw [style=greenline] (8) to (9);
		\draw [style=greenline] (6) to (7);
	\end{pgfonlayer}
\end{tikzpicture}
} &\scalebox{0.7}{\begin{tikzpicture}
	\begin{pgfonlayer}{nodelayer}
		\node [style=none] (0) at (1.925, -6.875) {};
		\node [style=none] (1) at (1.925, -7.125) {};
		\node [style=none] (2) at (6.125, -6.875) {};
		\node [style=none] (3) at (6.125, -7.125) {};
		\node [style=none] (4) at (6.5, -6.75) {};
		\node [style=none] (5) at (6, -7.25) {};
		\node [style=none] (6) at (6, -6.75) {};
		\node [style=none] (7) at (6.5, -7.25) {};
		\node [style=none] (8) at (2.05, -6.75) {};
		\node [style=none] (9) at (1.55, -7.25) {};
		\node [style=none] (10) at (1.55, -6.75) {};
		\node [style=none] (11) at (2.05, -7.25) {};
		\node [style=none] (12) at (2.5, -5.5) {};
		\node [style=none] (13) at (2.5, -8.5) {};
		\node [style=none] (14) at (3.5, -5.5) {};
		\node [style=none] (15) at (3.5, -8.5) {};
		\node [style=none] (16) at (4.5, -5.5) {};
		\node [style=none] (17) at (4.5, -8.5) {};
		\node [style=none] (18) at (5.5, -5.5) {};
		\node [style=none] (19) at (5.5, -8.5) {};
	\end{pgfonlayer}
	\begin{pgfonlayer}{edgelayer}
		\draw (2.center) to (0.center);
		\draw (1.center) to (3.center);
		\draw (6.center) to (7.center);
		\draw (4.center) to (5.center);
		\draw (10.center) to (11.center);
		\draw (8.center) to (9.center);
		\draw [style=greenline] (18.center) to (19.center);
		\draw [style=greenline] (16.center) to (17.center);
		\draw [style=greenline] (14.center) to (15.center);
		\draw [style=greenline] (12.center) to (13.center);
	\end{pgfonlayer}
\end{tikzpicture}
} &\scalebox{0.7}{\begin{tikzpicture}
	\begin{pgfonlayer}{nodelayer}
		\node [style=gauge3] (0) at (4, -4) {};
		\node [style=flavour2] (1) at (4, -2.5) {};
		\node [style=none] (2) at (4, -2) {U(4)};
		\node [style=none] (3) at (4, -4.5) {$U(2)$};
	\end{pgfonlayer}
	\begin{pgfonlayer}{edgelayer}
		\draw (1) to (0);
	\end{pgfonlayer}
\end{tikzpicture}
}
\\
\end{tabular}}
    \caption{The first column on the left gives the A-type Dynkin quiver with different combiantion of U/SU gauge nodes. The second column gives the brane web describing the Higgs phase of the quiver theory where the different colored NS5 (vertical lines) can move independently. The circles are 7-branes and horizontal lines are D5 branes. The same locking mechanism is performed but on a D3-D5-NS5 system in the third column where the crosses are D5, vertical lines are NS5s and horizontal lines are D3s. The 3d mirror to the A-type Dynkin quiver is then given on the rightmost column. }
    \label{SU4}
\end{table}

Since DynkinA quivers are \emph{good} in the sense of \cite{Gaiotto:2008ak}, the brane web for the Higgs phase can always be arranged to consist  only of D5, NS5, D7 and $[0,1]$7-branes. In other words, after a suitable set of Hanany-Witten transitions \cite{Hanany:1996ie}, there will not be any bound states of 5-branes or 7-branes. Therefore, the same information can be encoded in a brane system using D3-D5-NS5 branes. Applying this trick, one no longer has to start from brane web configurations, but the D3-D5-NS5 set ups, which is much easier to draw, can be used immediately.  

\paragraph{Physical interpretation?}
Here, we must emphasize that the physical interpretation of the D3-D5-NS5 brane system with the locking mechanism is unclear as  NS5 branes in a 3d brane system are always non-dynamical. In this section, we use these brane systems merely as combinatorial tools to arrive at the correct 3d mirrors. The confidence that the predicted 3d mirrors are correct stems from explicit Hilbert series computations and agreement with those obtained from brane webs.

\section{\texorpdfstring{\DynkinD}{} Example}
In Table \ref{DtypeUSU1} and \ref{DtypeUSU2}, we give all the U/SU combinations for the \DynkinD quiver in Section \ref{dynkinD}. 
\begin{table}
    \centering
 \scalebox{0.8} {  
\begin{tabular}{c|c|c} 
    Electric Quiver & Brane system & Magnetic Quiver \\
\hline
    \end{tabular}}
    \caption{}
    \label{DtypeUSU1}
\end{table}

\begin{table}
    \centering
  \scalebox{0.8}{  \begin{tabular}{c|c|c} 
    Electric Quiver & Brane system & Magnetic Quiver \\
\hline

        \end{tabular}}
    \caption{}
    \label{DtypeUSU2}
\end{table}

\section{\texorpdfstring{\DynkinABCD}{} quiver with overbalanced node}\label{overbalanced}
In this  section, we investigate \DynkinABCD quivers that are \emph{overbalanced} in the sense that some or all of the  $U(N_i)$ or $SU(N_i)$ gauge groups have $N_f > 2N_i$ hypers. If there are $m$ overbalanced gauge nodes in the \DynkinABCD quiver, then $m+1$ gauge nodes in the mirror dual will have flavor nodes attached to them. Similarly, if the \DynkinABCD quiver contains a mixture of U \& SU nodes, then the mirror dual will have $m+1$ bouquets of U(1)s. 
\FloatBarrier 
In \cite{USU}, \DynkinA quivers with overbalanced gauge nodes and their 3d mirrors have already been studied. In essence, one starts with DynkinA with all unitary gauge nodes and find its mirror. The mirror will have one or more flavor nodes. Turning U to SU nodes in the DynkinA then results in some or all of the flavor nodes to explode into bouquets of U(1)s. Then, depending on the arrangements of U \& SU nodes, the bouquets of U(1)s  in the 3d mirror will be connected to different gauge groups with links of different multiplicity. This procedure, including a convenient algorithm involving `incomplete brane systems' is given in \cite{USU} which we will not repeat here. 

For \DynkinBCD quivers, a similar procedure follows. In fact, the 3d mirror will only have a single orthosymplectic gauge group and the rest will be unitary. This is obvious from the brane set up as there is only one ON plane. Another reasoning is that the set of balanced nodes in a BCD-type Dynkin quiver cannot yield more than one finite-Dynkin diagram of BCD type, so the Coulomb branch global symmetry cannot be product of two or more orthosymplectic groups. In fact, when some of the nodes are SU or are overbalanced, a $D_k$, $B_k$ or $C_k$ Dynkin quiver will have a  $SO(2k')\prod_i U(n_i) \subset SO(2k)$,  $SO(2k'+1)\prod_i U(n_i) \subset SO(2k+1)$  or $USp(2k')\prod_i U(n_i) \subset USp(2k)$ Coulomb branch global symmetry group respectively where $k'<k$. By virtue of mirror symmetry, this means the mirror pair can have at most a single orthosymplectic flavor node connected to a single  orthosymplectic gauge node, whereas there can be multiple unitary flavor nodes connected to unitary gauge groups. In summary,  general  DynkinBCD quivers with all unitary groups will have the following 3d mirror pairs:  
\begin{equation}
    \scalebox{0.8}{\begin{tikzpicture}
	\begin{pgfonlayer}{nodelayer}
		\node [style=gauge3] (0) at (-8, 0) {};
		\node [style=gauge3] (1) at (-7, 0) {};
		\node [style=gauge3] (2) at (-6, 0) {};
		\node [style=none] (3) at (-5, 0) {$\dots$};
		\node [style=gauge3] (4) at (-4, 0) {};
		\node [style=gauge3] (5) at (-3, 0) {};
		\node [style=gauge3] (6) at (-2.25, 0.75) {};
		\node [style=gauge3] (7) at (-2.25, -0.75) {};
		\node [style=gauge3] (8) at (-8, -8) {};
		\node [style=gauge3] (9) at (-7, -8) {};
		\node [style=gauge3] (10) at (-6, -8) {};
		\node [style=none] (11) at (-5, -8) {$\dots$};
		\node [style=gauge3] (12) at (-4, -8) {};
		\node [style=gauge3] (13) at (-3, -8) {};
		\node [style=gauge3] (15) at (-1.75, -8) {};
		\node [style=none] (16) at (-1.75, -7.875) {};
		\node [style=none] (17) at (-1.75, -8.125) {};
		\node [style=none] (18) at (-3, -7.875) {};
		\node [style=none] (19) at (-3, -8.125) {};
		\node [style=none] (20) at (-2, -7.5) {};
		\node [style=none] (21) at (-2, -8.5) {};
		\node [style=none] (22) at (-2.375, -8) {};
		\node [style=gauge3] (23) at (-8, -4) {};
		\node [style=gauge3] (24) at (-7, -4) {};
		\node [style=gauge3] (25) at (-6, -4) {};
		\node [style=none] (26) at (-5, -4) {$\dots$};
		\node [style=gauge3] (27) at (-4, -4) {};
		\node [style=gauge3] (28) at (-3, -4) {};
		\node [style=gauge3] (29) at (-1.75, -4) {};
		\node [style=none] (30) at (-1.75, -3.875) {};
		\node [style=none] (31) at (-1.75, -4.125) {};
		\node [style=none] (32) at (-3, -3.875) {};
		\node [style=none] (33) at (-3, -4.125) {};
		\node [style=none] (34) at (-2.625, -3.5) {};
		\node [style=none] (35) at (-2.625, -4.5) {};
		\node [style=none] (36) at (-2.25, -4) {};
		\node [style=none] (37) at (-4.5, 0) {};
		\node [style=none] (38) at (-5.5, 0) {};
		\node [style=none] (39) at (-5.5, -8) {};
		\node [style=none] (40) at (-4.5, -8) {};
		\node [style=none] (41) at (-4.5, -4) {};
		\node [style=none] (42) at (-5.5, -4) {};
		\node [style=flavour2] (43) at (-8, 1) {};
		\node [style=flavour2] (44) at (-7, 1) {};
		\node [style=flavour2] (45) at (-6, 1) {};
		\node [style=flavour2] (46) at (-4, 1) {};
		\node [style=flavour2] (47) at (-3, 1) {};
		\node [style=flavour2] (48) at (-1.5, 0.75) {};
		\node [style=flavour2] (49) at (-1.5, -0.75) {};
		\node [style=flavour2] (50) at (-3, -7) {};
		\node [style=flavour2] (51) at (-1.75, -7) {};
		\node [style=flavour2] (52) at (-4, -7) {};
		\node [style=flavour2] (53) at (-6, -7) {};
		\node [style=flavour2] (54) at (-7, -7) {};
		\node [style=flavour2] (55) at (-8, -7) {};
		\node [style=flavour2] (56) at (-8, -3) {};
		\node [style=flavour2] (57) at (-7, -3) {};
		\node [style=flavour2] (58) at (-6, -3) {};
		\node [style=flavour2] (59) at (-4, -3) {};
		\node [style=flavour2] (60) at (-3, -3) {};
		\node [style=flavour2] (61) at (-1.75, -3) {};
		\node [style=none] (62) at (-8, -0.5) {U};
		\node [style=none] (63) at (-7, -0.5) {U};
		\node [style=none] (64) at (-6, -0.5) {U};
		\node [style=none] (65) at (-4, -0.5) {U};
		\node [style=none] (66) at (-3, -0.5) {U};
		\node [style=none] (67) at (-2.25, 1.25) {U};
		\node [style=none] (68) at (-2.25, -1.25) {U};
		\node [style=none] (69) at (-8, -8.5) {U};
		\node [style=none] (70) at (-7, -8.5) {U};
		\node [style=none] (71) at (-6, -8.5) {U};
		\node [style=none] (72) at (-4, -8.5) {U};
		\node [style=none] (73) at (-3, -8.5) {U};
		\node [style=none] (74) at (-1.75, -8.5) {U};
		\node [style=none] (75) at (-8, -4.5) {U};
		\node [style=none] (76) at (-7, -4.5) {U};
		\node [style=none] (77) at (-6, -4.5) {U};
		\node [style=none] (78) at (-4, -4.5) {U};
		\node [style=none] (79) at (-3, -4.5) {U};
		\node [style=none] (80) at (-1.75, -4.5) {U};
		\node [style=none] (81) at (-10.25, 0) {D-type};
		\node [style=none] (82) at (-10, -4) {B-Type};
		\node [style=none] (83) at (-10, -8) {C-Type};
		\node [style=none] (84) at (0, 0) {};
		\node [style=none] (85) at (2, 0) {};
		\node [style=none] (86) at (0, -4) {};
		\node [style=none] (87) at (2, -4) {};
		\node [style=none] (88) at (0, -8) {};
		\node [style=none] (89) at (2, -8) {};
		\node [style=gauge3] (90) at (3, 0) {};
		\node [style=gauge3] (91) at (4, 0) {};
		\node [style=gauge3] (92) at (5, 0) {};
		\node [style=none] (93) at (6, 0) {$\dots$};
		\node [style=gauge3] (94) at (7, 0) {};
		\node [style=none] (95) at (6.5, 0) {};
		\node [style=none] (96) at (5.5, 0) {};
		\node [style=flavour2] (97) at (3, 1) {};
		\node [style=flavour2] (98) at (4, 1) {};
		\node [style=flavour2] (99) at (5, 1) {};
		\node [style=flavour2] (100) at (7, 1) {};
		\node [style=none] (101) at (3, -0.5) {U};
		\node [style=none] (102) at (4, -0.5) {U};
		\node [style=none] (103) at (5, -0.5) {U};
		\node [style=none] (104) at (7, -0.5) {U};
		\node [style=gauge3] (105) at (3, -4) {};
		\node [style=gauge3] (106) at (4, -4) {};
		\node [style=gauge3] (107) at (5, -4) {};
		\node [style=none] (108) at (6, -4) {$\dots$};
		\node [style=gauge3] (109) at (7, -4) {};
		\node [style=none] (110) at (6.5, -4) {};
		\node [style=none] (111) at (5.5, -4) {};
		\node [style=flavour2] (112) at (3, -3) {};
		\node [style=flavour2] (113) at (4, -3) {};
		\node [style=flavour2] (114) at (5, -3) {};
		\node [style=flavour2] (115) at (7, -3) {};
		\node [style=none] (116) at (3, -4.5) {U};
		\node [style=none] (117) at (4, -4.5) {U};
		\node [style=none] (118) at (5, -4.5) {U};
		\node [style=none] (119) at (7, -4.5) {U};
		\node [style=gauge3] (120) at (3, -8) {};
		\node [style=gauge3] (121) at (4, -8) {};
		\node [style=gauge3] (122) at (5, -8) {};
		\node [style=none] (123) at (6, -8) {$\dots$};
		\node [style=gauge3] (124) at (7, -8) {};
		\node [style=none] (125) at (6.5, -8) {};
		\node [style=none] (126) at (5.5, -8) {};
		\node [style=flavour2] (127) at (3, -7) {};
		\node [style=flavour2] (128) at (4, -7) {};
		\node [style=flavour2] (129) at (5, -7) {};
		\node [style=flavour2] (130) at (7, -7) {};
		\node [style=none] (131) at (3, -8.5) {U};
		\node [style=none] (132) at (4, -8.5) {U};
		\node [style=none] (133) at (5, -8.5) {U};
		\node [style=none] (134) at (7, -8.5) {U};
		\node [style=gauge3] (135) at (8.5, 0) {};
		\node [style=gauge3] (136) at (8.5, -4) {};
		\node [style=gauge3] (137) at (8.5, -8) {};
		\node [style=flavour2] (138) at (8.5, 1) {};
		\node [style=flavour2] (139) at (8.5, -3) {};
		\node [style=flavour2] (140) at (8.5, -7) {};
		\node [style=none] (141) at (8.5, -8.5) {O};
		\node [style=none] (142) at (8.5, -4.5) {USp(even)};
		\node [style=none] (143) at (8.5, -0.5) {USp(even)};
		\node [style=none] (144) at (8.5, 1.5) {SO(even)};
		\node [style=none] (145) at (8.5, -2.5) {SO(odd)};
		\node [style=none] (146) at (8.5, -6.5) {USp(even)};
	\end{pgfonlayer}
	\begin{pgfonlayer}{edgelayer}
		\draw (16.center) to (18.center);
		\draw (17.center) to (19.center);
		\draw (20.center) to (22.center);
		\draw (22.center) to (21.center);
		\draw (30.center) to (32.center);
		\draw (31.center) to (33.center);
		\draw (34.center) to (36.center);
		\draw (36.center) to (35.center);
		\draw (0) to (2);
		\draw (4) to (5);
		\draw (5) to (6);
		\draw (5) to (7);
		\draw (8) to (10);
		\draw (12) to (13);
		\draw (27) to (28);
		\draw (25) to (23);
		\draw (37.center) to (4);
		\draw (38.center) to (2);
		\draw (10) to (39.center);
		\draw (40.center) to (12);
		\draw (27) to (41.center);
		\draw (42.center) to (25);
		\draw (43) to (0);
		\draw (1) to (44);
		\draw (45) to (2);
		\draw (4) to (46);
		\draw (47) to (5);
		\draw (6) to (48);
		\draw (7) to (49);
		\draw (55) to (8);
		\draw (9) to (54);
		\draw (53) to (10);
		\draw (52) to (12);
		\draw (50) to (19.center);
		\draw (51) to (17.center);
		\draw (61) to (30.center);
		\draw (60) to (33.center);
		\draw (59) to (27);
		\draw (58) to (25);
		\draw (57) to (24);
		\draw (56) to (23);
		\draw [style=->] (84.center) to (85.center);
		\draw [style=->] (85.center) to (84.center);
		\draw [style=->] (86.center) to (87.center);
		\draw [style=->] (87.center) to (86.center);
		\draw [style=->] (88.center) to (89.center);
		\draw [style=->] (89.center) to (88.center);
		\draw (90) to (92);
		\draw (95.center) to (94);
		\draw (96.center) to (92);
		\draw (97) to (90);
		\draw (91) to (98);
		\draw (99) to (92);
		\draw (94) to (100);
		\draw (105) to (107);
		\draw (110.center) to (109);
		\draw (111.center) to (107);
		\draw (112) to (105);
		\draw (106) to (113);
		\draw (114) to (107);
		\draw (109) to (115);
		\draw (120) to (122);
		\draw (125.center) to (124);
		\draw (126.center) to (122);
		\draw (127) to (120);
		\draw (121) to (128);
		\draw (129) to (122);
		\draw (124) to (130);
		\draw (94) to (135);
		\draw (109) to (136);
		\draw (137) to (124);
		\draw (138) to (135);
		\draw (136) to (139);
		\draw (140) to (137);
	\end{pgfonlayer}
\end{tikzpicture}}
\end{equation}
where we leave the ranks of the groups arbitrary just to emphasize on the general structure of the 3d mirror: an orthosympletic gauge group followed by a sequence of unitary gauge groups. 

With some practice, one can instantly read off what the bouquets in the 3d mirror will be for any U \& SU arrangements. For \DynkinA quivers, this is written in detail in \cite{USU}. One thing to note is the importance of whether the gauge group in your Dynkin quiver is balanced U, overbalanced U, balanced SU or overbalanced SU. Let us demonstrate this with an $A_3$ Dynkin quiver with nodes or arbitrary ranks and their 3d mirror:
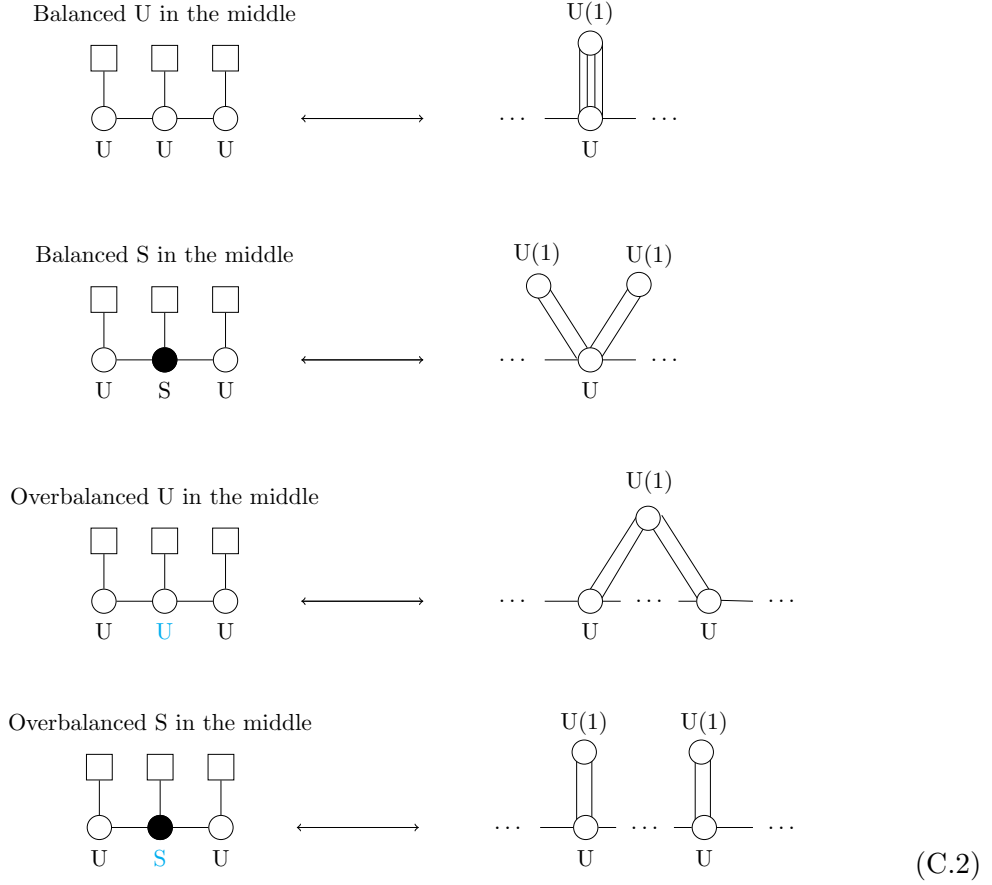
\begin{equation}
  \scalebox{0.8}{ \begin{tikzpicture}
	\begin{pgfonlayer}{nodelayer}
		\node [style=gauge3] (0) at (1.75, 0) {};
		\node [style=gauge3] (1) at (2.75, 0) {};
		\node [style=gauge3] (2) at (3.75, 0) {};
		\node [style=flavour2] (6) at (1.75, 1) {};
		\node [style=flavour2] (7) at (2.75, 1) {};
		\node [style=flavour2] (8) at (3.75, 1) {};
		\node [style=none] (9) at (1.75, -0.5) {U};
		\node [style=none] (10) at (2.75, -0.5) {U};
		\node [style=none] (11) at (3.75, -0.5) {U};
		\node [style=none] (15) at (5, 0) {};
		\node [style=none] (16) at (7, 0) {};
		\node [style=gauge3] (19) at (9.75, 0) {};
		\node [style=none] (20) at (11, 0) {$\dots$};
		\node [style=none] (21) at (11.5, 0) {};
		\node [style=none] (22) at (10.5, 0) {};
		\node [style=none] (28) at (9.75, -0.5) {U};
		\node [style=none] (29) at (8.5, 0) {$\dots$};
		\node [style=none] (30) at (9, 0) {};
		\node [style=none] (31) at (8, 0) {};
		\node [style=none] (32) at (2.75, 1.75) {Balanced U in the middle};
		\node [style=gauge3] (33) at (9.75, 1.25) {};
		\node [style=none] (34) at (9.75, 1.75) {U(1)};
		\node [style=none] (36) at (9.575, 1.25) {};
		\node [style=none] (37) at (9.7, 1.25) {};
		\node [style=none] (38) at (9.825, 1.25) {};
		\node [style=none] (39) at (9.95, 1.25) {};
		\node [style=none] (40) at (9.575, 0) {};
		\node [style=none] (41) at (9.7, 0) {};
		\node [style=none] (42) at (9.825, 0) {};
		\node [style=none] (43) at (9.95, 0) {};
		\node [style=gauge3] (44) at (1.75, -4) {};
		\node [style=gauge3] (45) at (2.75, -4) {};
		\node [style=gauge3] (46) at (3.75, -4) {};
		\node [style=flavour2] (47) at (1.75, -3) {};
		\node [style=flavour2] (48) at (2.75, -3) {};
		\node [style=flavour2] (49) at (3.75, -3) {};
		\node [style=none] (50) at (1.75, -4.5) {U};
		\node [style=none] (51) at (2.75, -4.5) {S};
		\node [style=none] (52) at (3.75, -4.5) {U};
		\node [style=none] (53) at (5, -4) {};
		\node [style=none] (54) at (7, -4) {};
		\node [style=gauge3] (55) at (9.75, -4) {};
		\node [style=none] (56) at (11, -4) {$\dots$};
		\node [style=none] (57) at (11.5, -4) {};
		\node [style=none] (58) at (10.5, -4) {};
		\node [style=none] (59) at (9.75, -4.5) {U};
		\node [style=none] (60) at (8.5, -4) {$\dots$};
		\node [style=none] (61) at (9, -4) {};
		\node [style=none] (62) at (8, -4) {};
		\node [style=none] (63) at (2.75, -2.25) {Balanced S in the middle};
		\node [style=none] (65) at (8.875, -2.25) {U(1)};
		\node [style=blacknode] (74) at (2.75, -4) {};
		\node [style=none] (75) at (8.75, -2.75) {};
		\node [style=none] (76) at (8.925, -2.575) {};
		\node [style=none] (77) at (9.525, -3.975) {};
		\node [style=none] (78) at (9.7, -3.8) {};
		\node [style=none] (79) at (10.7, -2.75) {};
		\node [style=none] (80) at (10.525, -2.575) {};
		\node [style=none] (81) at (9.925, -3.975) {};
		\node [style=none] (82) at (9.75, -3.8) {};
		\node [style=gauge3] (83) at (8.9, -2.775) {};
		\node [style=gauge3] (84) at (10.55, -2.75) {};
		\node [style=none] (85) at (10.725, -2.275) {U(1)};
		\node [style=gauge3] (86) at (1.75, -8) {};
		\node [style=gauge3] (87) at (2.75, -8) {};
		\node [style=gauge3] (88) at (3.75, -8) {};
		\node [style=flavour2] (89) at (1.75, -7) {};
		\node [style=flavour2] (90) at (2.75, -7) {};
		\node [style=flavour2] (91) at (3.75, -7) {};
		\node [style=none] (92) at (1.75, -8.5) {U};
		\node [style=none] (93) at (2.75, -8.5) {\textcolor{cyan}{U}};
		\node [style=none] (94) at (3.75, -8.5) {U};
		\node [style=none] (95) at (5, -8) {};
		\node [style=none] (96) at (7, -8) {};
		\node [style=gauge3] (97) at (9.75, -8) {};
		\node [style=none] (98) at (10.75, -8) {$\dots$};
		\node [style=none] (100) at (10.25, -8) {};
		\node [style=none] (101) at (9.75, -8.5) {U};
		\node [style=none] (102) at (8.5, -8) {$\dots$};
		\node [style=none] (103) at (9, -8) {};
		\node [style=none] (104) at (8, -8) {};
		\node [style=none] (105) at (2.75, -6.25) {Overbalanced U in the middle};
		\node [style=none] (110) at (9.525, -7.975) {};
		\node [style=none] (111) at (9.7, -7.8) {};
		\node [style=none] (112) at (10.65, -6.75) {};
		\node [style=none] (113) at (10.525, -6.575) {};
		\node [style=none] (114) at (9.925, -7.975) {};
		\node [style=none] (115) at (9.75, -7.8) {};
		\node [style=gauge3] (117) at (10.7, -6.625) {};
		\node [style=none] (118) at (10.725, -6.025) {U(1)};
		\node [style=gauge3] (119) at (11.7, -8) {};
		\node [style=none] (121) at (11.2, -8) {};
		\node [style=none] (122) at (11.7, -8.5) {U};
		\node [style=none] (123) at (11.925, -7.975) {};
		\node [style=none] (124) at (11.75, -7.8) {};
		\node [style=none] (125) at (10.75, -6.75) {};
		\node [style=none] (126) at (10.925, -6.575) {};
		\node [style=none] (127) at (11.525, -7.975) {};
		\node [style=none] (128) at (11.7, -7.8) {};
		\node [style=gauge3] (129) at (1.675, -11.75) {};
		\node [style=gauge3] (130) at (2.675, -11.75) {};
		\node [style=gauge3] (131) at (3.675, -11.75) {};
		\node [style=flavour2] (132) at (1.675, -10.75) {};
		\node [style=flavour2] (133) at (2.675, -10.75) {};
		\node [style=flavour2] (134) at (3.675, -10.75) {};
		\node [style=none] (135) at (1.675, -12.25) {U};
		\node [style=none] (136) at (2.675, -12.25) {\textcolor{cyan}{S}};
		\node [style=none] (137) at (3.675, -12.25) {U};
		\node [style=none] (138) at (4.925, -11.75) {};
		\node [style=none] (139) at (6.925, -11.75) {};
		\node [style=gauge3] (140) at (9.675, -11.75) {};
		\node [style=none] (141) at (10.675, -11.75) {$\dots$};
		\node [style=none] (142) at (10.175, -11.75) {};
		\node [style=none] (143) at (9.675, -12.25) {U};
		\node [style=none] (144) at (8.425, -11.75) {$\dots$};
		\node [style=none] (145) at (8.925, -11.75) {};
		\node [style=none] (146) at (7.925, -11.75) {};
		\node [style=none] (147) at (2.675, -10) {Overbalanced S  in the middle};
		\node [style=none] (155) at (9.65, -10.025) {U(1)};
		\node [style=gauge3] (156) at (11.625, -11.75) {};
		\node [style=none] (157) at (11.125, -11.75) {};
		\node [style=none] (158) at (11.625, -12.25) {U};
		\node [style=blacknode] (165) at (2.675, -11.75) {};
		\node [style=none] (166) at (12.925, -8) {$\dots$};
		\node [style=none] (167) at (13.425, -8) {};
		\node [style=none] (168) at (12.425, -8) {};
		\node [style=none] (169) at (12.925, -11.75) {$\dots$};
		\node [style=none] (170) at (13.425, -11.75) {};
		\node [style=none] (171) at (12.425, -11.75) {};
		\node [style=gauge3] (172) at (9.65, -10.5) {};
		\node [style=gauge3] (173) at (11.575, -10.5) {};
		\node [style=none] (174) at (9.525, -10.5) {};
		\node [style=none] (175) at (9.775, -10.5) {};
		\node [style=none] (176) at (9.525, -11.65) {};
		\node [style=none] (177) at (9.775, -11.65) {};
		\node [style=none] (178) at (11.475, -10.5) {};
		\node [style=none] (179) at (11.725, -10.5) {};
		\node [style=none] (180) at (11.475, -11.65) {};
		\node [style=none] (181) at (11.725, -11.65) {};
		\node [style=none] (182) at (11.625, -10.025) {U(1)};
	\end{pgfonlayer}
	\begin{pgfonlayer}{edgelayer}
		\draw (0) to (2);
		\draw (6) to (0);
		\draw (1) to (7);
		\draw (8) to (2);
		\draw [style=->] (15.center) to (16.center);
		\draw [style=->] (16.center) to (15.center);
		\draw (22.center) to (19);
		\draw (30.center) to (19);
		\draw (36.center) to (40.center);
		\draw (41.center) to (37.center);
		\draw (38.center) to (42.center);
		\draw (43.center) to (39.center);
		\draw (44) to (46);
		\draw (47) to (44);
		\draw (45) to (48);
		\draw (49) to (46);
		\draw [style=->] (53.center) to (54.center);
		\draw [style=->] (54.center) to (53.center);
		\draw (58.center) to (55);
		\draw (61.center) to (55);
		\draw (75.center) to (77.center);
		\draw (78.center) to (76.center);
		\draw (79.center) to (81.center);
		\draw (82.center) to (80.center);
		\draw (86) to (88);
		\draw (89) to (86);
		\draw (87) to (90);
		\draw (91) to (88);
		\draw [style=->] (95.center) to (96.center);
		\draw [style=->] (96.center) to (95.center);
		\draw (100.center) to (97);
		\draw (103.center) to (97);
		\draw (112.center) to (114.center);
		\draw (115.center) to (113.center);
		\draw (121.center) to (119);
		\draw (125.center) to (127.center);
		\draw (128.center) to (126.center);
		\draw (129) to (131);
		\draw (132) to (129);
		\draw (130) to (133);
		\draw (134) to (131);
		\draw [style=->] (138.center) to (139.center);
		\draw [style=->] (139.center) to (138.center);
		\draw (142.center) to (140);
		\draw (145.center) to (140);
		\draw (157.center) to (156);
		\draw (127.center) to (168.center);
		\draw (156) to (171.center);
		\draw (175.center) to (177.center);
		\draw (176.center) to (174.center);
		\draw (179.center) to (181.center);
		\draw (180.center) to (178.center);
	\end{pgfonlayer}
\end{tikzpicture}}
\label{fourcases}
\end{equation}
where we now use cyan to label overbalanced nodes. For the four examples, we will only focus on changing the central node of the \DynkinA whereas the two unitary groups on the side are always balanced. If the central group is balanced as well, then the mirror will have a single U(1) node with a multiplicity 4 link.  The exact length of gauge nodes in the 3d mirror depends on the ranks of the gauge groups in the \DynkinA, but nevertheless there will only be one gauge group with a U(1) bouquet (here, there is only a single U(1) in the bouquet). If the central node in the \DynkinA is SU and balanced, then there will be two U(1) nodes in the bouquet of the mirror, each having a link of multiplicity two. Now, if the central node is overbalanced, then the 3d mirror will have more than one node with bouquets. If the overbalanced node is U, then the U(1) in the mirror will be shared between two different gauge groups. If the overbalanced node is SU, then there will be two U(1)s in the mirror and they are each connected to their own gauge groups. To summarize: $k$ overbalanced gauge groups means $k+1$ nodes in the mirror will have U(1) bouquets. And $n$ SU nodes means the bouquets in the 3d mirror will have in total $n+1$ U(1) nodes. 

The above discussions means whenever you reach a node in the Dynkin quiver that is both SU and overbalanced, it will split the 3d mirror in such a way that the U(1) bouquets on one side of the quiver does not join up with the U(1) bouquets on the otherside. This means part of the \DynkinBCD quiver that is separated from the BCD edge by an overbalanced SU node can be treated as if it is an \DynkinA. Let us consider it with the following example:
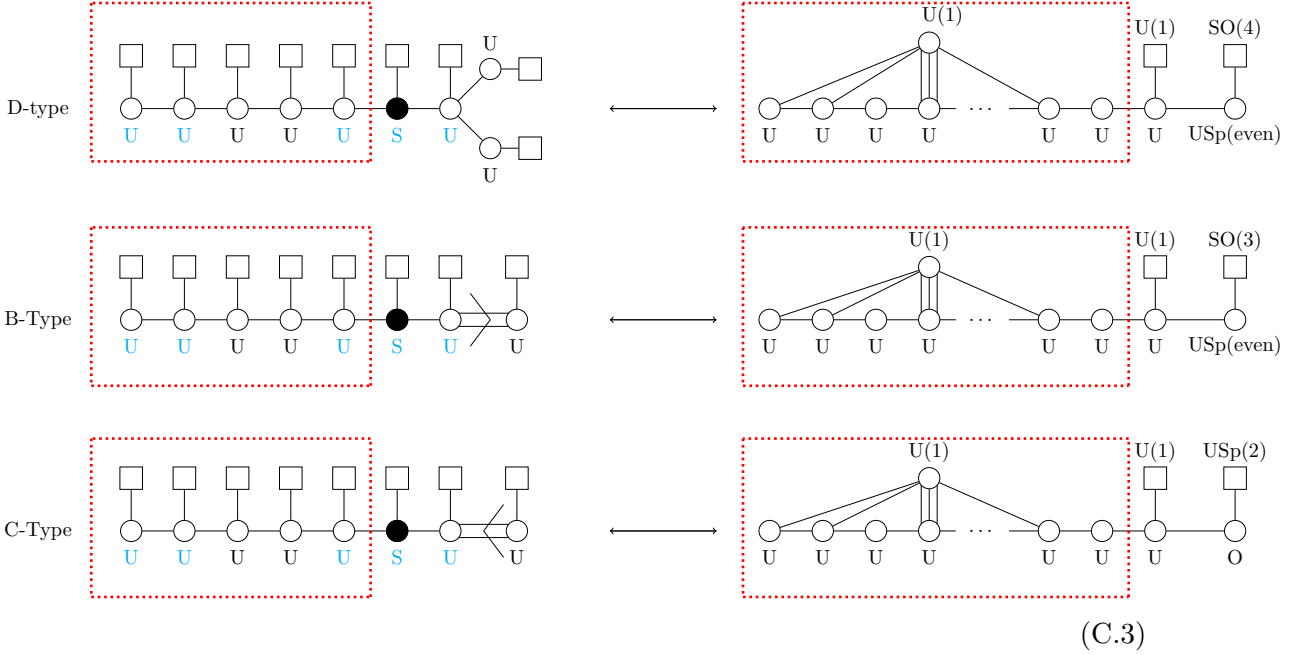
\begin{equation}
\scalebox{0.7}{\begin{tikzpicture}
	\begin{pgfonlayer}{nodelayer}
		\node [style=gauge3] (0) at (-8, 0) {};
		\node [style=gauge3] (1) at (-7, 0) {};
		\node [style=gauge3] (2) at (-6, 0) {};
		\node [style=gauge3] (4) at (-4, 0) {};
		\node [style=gauge3] (5) at (-3, 0) {};
		\node [style=gauge3] (6) at (-2.25, 0.75) {};
		\node [style=gauge3] (7) at (-2.25, -0.75) {};
		\node [style=gauge3] (8) at (-8, -8) {};
		\node [style=gauge3] (9) at (-7, -8) {};
		\node [style=gauge3] (10) at (-6, -8) {};
		\node [style=gauge3] (12) at (-4, -8) {};
		\node [style=gauge3] (13) at (-3, -8) {};
		\node [style=gauge3] (15) at (-1.75, -8) {};
		\node [style=none] (16) at (-1.75, -7.875) {};
		\node [style=none] (17) at (-1.75, -8.125) {};
		\node [style=none] (18) at (-3, -7.875) {};
		\node [style=none] (19) at (-3, -8.125) {};
		\node [style=none] (20) at (-2, -7.5) {};
		\node [style=none] (21) at (-2, -8.5) {};
		\node [style=none] (22) at (-2.375, -8) {};
		\node [style=gauge3] (23) at (-8, -4) {};
		\node [style=gauge3] (24) at (-7, -4) {};
		\node [style=gauge3] (25) at (-6, -4) {};
		\node [style=gauge3] (27) at (-4, -4) {};
		\node [style=gauge3] (28) at (-3, -4) {};
		\node [style=gauge3] (29) at (-1.75, -4) {};
		\node [style=none] (30) at (-1.75, -3.875) {};
		\node [style=none] (31) at (-1.75, -4.125) {};
		\node [style=none] (32) at (-3, -3.875) {};
		\node [style=none] (33) at (-3, -4.125) {};
		\node [style=none] (34) at (-2.625, -3.5) {};
		\node [style=none] (35) at (-2.625, -4.5) {};
		\node [style=none] (36) at (-2.25, -4) {};
		\node [style=flavour2] (43) at (-8, 1) {};
		\node [style=flavour2] (44) at (-7, 1) {};
		\node [style=flavour2] (45) at (-6, 1) {};
		\node [style=flavour2] (46) at (-4, 1) {};
		\node [style=flavour2] (47) at (-3, 1) {};
		\node [style=flavour2] (48) at (-1.5, 0.75) {};
		\node [style=flavour2] (49) at (-1.5, -0.75) {};
		\node [style=flavour2] (50) at (-3, -7) {};
		\node [style=flavour2] (51) at (-1.75, -7) {};
		\node [style=flavour2] (52) at (-4, -7) {};
		\node [style=flavour2] (53) at (-6, -7) {};
		\node [style=flavour2] (54) at (-7, -7) {};
		\node [style=flavour2] (55) at (-8, -7) {};
		\node [style=flavour2] (56) at (-8, -3) {};
		\node [style=flavour2] (57) at (-7, -3) {};
		\node [style=flavour2] (58) at (-6, -3) {};
		\node [style=flavour2] (59) at (-4, -3) {};
		\node [style=flavour2] (60) at (-3, -3) {};
		\node [style=flavour2] (61) at (-1.75, -3) {};
		\node [style=none] (63) at (-7, -0.5) {U};
		\node [style=none] (64) at (-6, -0.5) {U};
		\node [style=none] (65) at (-4, -0.5) {\textcolor{cyan}{S}};
		\node [style=none] (67) at (-2.25, 1.25) {U};
		\node [style=none] (68) at (-2.25, -1.25) {U};
		\node [style=none] (70) at (-7, -8.5) {U};
		\node [style=none] (71) at (-6, -8.5) {U};
		\node [style=none] (74) at (-1.75, -8.5) {U};
		\node [style=none] (76) at (-7, -4.5) {U};
		\node [style=none] (77) at (-6, -4.5) {U};
		\node [style=none] (80) at (-1.75, -4.5) {U};
		\node [style=none] (81) at (-10.75, 0) {D-type};
		\node [style=none] (82) at (-10.75, -4) {B-Type};
		\node [style=none] (83) at (-10.75, -8) {C-Type};
		\node [style=none] (84) at (0, 0) {};
		\node [style=none] (85) at (2, 0) {};
		\node [style=none] (86) at (0, -4) {};
		\node [style=none] (87) at (2, -4) {};
		\node [style=none] (88) at (0, -8) {};
		\node [style=none] (89) at (2, -8) {};
		\node [style=gauge3] (90) at (3, 0) {};
		\node [style=gauge3] (91) at (4, 0) {};
		\node [style=gauge3] (92) at (5, 0) {};
		\node [style=none] (93) at (7, 0) {$\dots$};
		\node [style=gauge3] (94) at (10.25, 0) {};
		\node [style=none] (95) at (7.5, 0) {};
		\node [style=none] (96) at (6.5, 0) {};
		\node [style=flavour2] (100) at (10.25, 1) {};
		\node [style=none] (101) at (3, -0.5) {U};
		\node [style=none] (102) at (4, -0.5) {U};
		\node [style=none] (103) at (5, -0.5) {U};
		\node [style=none] (104) at (10.25, -0.5) {U};
		\node [style=gauge3] (105) at (3, -4) {};
		\node [style=gauge3] (106) at (4, -4) {};
		\node [style=gauge3] (107) at (5, -4) {};
		\node [style=none] (108) at (7, -4) {$\dots$};
		\node [style=gauge3] (109) at (10.25, -4) {};
		\node [style=none] (110) at (7.5, -4) {};
		\node [style=none] (111) at (6.5, -4) {};
		\node [style=flavour2] (115) at (10.25, -3) {};
		\node [style=none] (116) at (3, -4.5) {U};
		\node [style=none] (117) at (4, -4.5) {U};
		\node [style=none] (118) at (5, -4.5) {U};
		\node [style=none] (119) at (10.25, -4.5) {U};
		\node [style=gauge3] (120) at (3, -8) {};
		\node [style=gauge3] (121) at (4, -8) {};
		\node [style=gauge3] (122) at (5, -8) {};
		\node [style=none] (123) at (7, -8) {$\dots$};
		\node [style=gauge3] (124) at (10.25, -8) {};
		\node [style=none] (125) at (7.5, -8) {};
		\node [style=none] (126) at (6.5, -8) {};
		\node [style=flavour2] (130) at (10.25, -7) {};
		\node [style=none] (131) at (3, -8.5) {U};
		\node [style=none] (132) at (4, -8.5) {U};
		\node [style=none] (133) at (5, -8.5) {U};
		\node [style=none] (134) at (10.25, -8.5) {U};
		\node [style=gauge3] (135) at (11.75, 0) {};
		\node [style=gauge3] (136) at (11.75, -4) {};
		\node [style=gauge3] (137) at (11.75, -8) {};
		\node [style=flavour2] (138) at (11.75, 1) {};
		\node [style=flavour2] (139) at (11.75, -3) {};
		\node [style=flavour2] (140) at (11.75, -7) {};
		\node [style=none] (141) at (11.75, -8.5) {O};
		\node [style=none] (142) at (11.75, -4.5) {USp(even)};
		\node [style=none] (143) at (11.75, -0.5) {USp(even)};
		\node [style=none] (144) at (11.75, 1.5) {SO(4)};
		\node [style=none] (145) at (11.75, -2.5) {SO(3)};
		\node [style=none] (146) at (11.75, -6.5) {USp(2)};
		\node [style=gauge3] (147) at (-5, 0) {};
		\node [style=gauge3] (148) at (-5, -4) {};
		\node [style=gauge3] (149) at (-5, -8) {};
		\node [style=none] (153) at (-4, -4.5) {\textcolor{cyan}{S}};
		\node [style=none] (154) at (-4, -8.5) {\textcolor{cyan}{S}};
		\node [style=flavour2] (155) at (-5, 1) {};
		\node [style=flavour2] (156) at (-5, -3) {};
		\node [style=flavour2] (157) at (-5, -7) {};
		\node [style=none] (161) at (-9.75, 2) {};
		\node [style=none] (162) at (-9.75, -1) {};
		\node [style=none] (163) at (-4.5, -1) {};
		\node [style=none] (164) at (-4.5, 2) {};
		\node [style=none] (165) at (-9.75, -2.25) {};
		\node [style=none] (166) at (-9.75, -5.25) {};
		\node [style=none] (167) at (-4.5, -5.25) {};
		\node [style=none] (168) at (-4.5, -2.25) {};
		\node [style=none] (169) at (-9.75, -6.25) {};
		\node [style=none] (170) at (-9.75, -9.25) {};
		\node [style=none] (171) at (-4.5, -9.25) {};
		\node [style=none] (172) at (-4.5, -6.25) {};
		\node [style=gauge3] (173) at (9.25, 0) {};
		\node [style=gauge3] (174) at (8.25, 0) {};
		\node [style=gauge3] (175) at (8.25, -4) {};
		\node [style=gauge3] (176) at (9.25, -4) {};
		\node [style=gauge3] (177) at (9.25, -8) {};
		\node [style=gauge3] (178) at (8.25, -8) {};
		\node [style=gauge3] (179) at (6, 0) {};
		\node [style=gauge3] (180) at (6, -4) {};
		\node [style=gauge3] (181) at (6, -8) {};
		\node [style=none] (182) at (6, -0.5) {U};
		\node [style=none] (183) at (6, -4.5) {U};
		\node [style=none] (184) at (6, -8.5) {U};
		\node [style=none] (188) at (9.25, -0.5) {U};
		\node [style=none] (189) at (8.25, -0.5) {U};
		\node [style=none] (190) at (8.25, -4.5) {U};
		\node [style=none] (191) at (9.25, -4.5) {U};
		\node [style=none] (192) at (8.25, -8.5) {U};
		\node [style=none] (193) at (9.25, -8.5) {U};
		\node [style=none] (200) at (2.5, 2) {};
		\node [style=none] (201) at (2.5, -1) {};
		\node [style=none] (202) at (9.75, -1) {};
		\node [style=none] (203) at (9.75, 2) {};
		\node [style=none] (204) at (2.5, -2.25) {};
		\node [style=none] (205) at (2.5, -5.25) {};
		\node [style=none] (206) at (9.75, -5.25) {};
		\node [style=none] (207) at (9.75, -2.25) {};
		\node [style=none] (208) at (2.5, -6.25) {};
		\node [style=none] (209) at (2.5, -9.25) {};
		\node [style=none] (210) at (9.75, -9.25) {};
		\node [style=none] (211) at (9.75, -6.25) {};
		\node [style=gauge3] (212) at (6, 1.25) {};
		\node [style=gauge3] (213) at (6, -3) {};
		\node [style=gauge3] (214) at (6, -7) {};
		\node [style=none] (215) at (6.25, 1.75) {U(1)};
		\node [style=none] (216) at (6, -2.5) {U(1)};
		\node [style=none] (217) at (6, -6.5) {U(1)};
		\node [style=none] (218) at (5.85, 1.25) {};
		\node [style=none] (219) at (6.15, 1.25) {};
		\node [style=none] (220) at (5.85, 0) {};
		\node [style=none] (221) at (6.15, 0) {};
		\node [style=none] (222) at (5.85, -3) {};
		\node [style=none] (223) at (6.15, -3) {};
		\node [style=none] (224) at (5.85, -4) {};
		\node [style=none] (225) at (6.15, -4) {};
		\node [style=none] (226) at (5.85, -7) {};
		\node [style=none] (227) at (6.15, -7) {};
		\node [style=none] (228) at (5.85, -8) {};
		\node [style=none] (229) at (6.15, -8) {};
		\node [style=gauge3] (230) at (-9, 0) {};
		\node [style=gauge3] (231) at (-9, -8) {};
		\node [style=gauge3] (232) at (-9, -4) {};
		\node [style=flavour2] (233) at (-9, 1) {};
		\node [style=flavour2] (234) at (-9, -7) {};
		\node [style=flavour2] (235) at (-9, -3) {};
		\node [style=none] (240) at (-8, -0.5) {\textcolor{cyan}{U}};
		\node [style=none] (241) at (-9, -4.5) {\textcolor{cyan}{U}};
		\node [style=none] (242) at (-9, -8.5) {\textcolor{cyan}{U}};
		\node [style=none] (243) at (-9, -0.5) {\textcolor{cyan}{U}};
		\node [style=none] (245) at (-5, -0.5) {\textcolor{cyan}{U}};
		\node [style=none] (246) at (-8, -4.5) {\textcolor{cyan}{U}};
		\node [style=none] (247) at (-5, -4.5) {\textcolor{cyan}{U}};
		\node [style=none] (248) at (-8, -8.5) {\textcolor{cyan}{U}};
		\node [style=none] (249) at (-5, -8.5) {\textcolor{cyan}{U}};
		\node [style=none] (250) at (-3, -0.5) {\textcolor{cyan}{U}};
		\node [style=none] (251) at (-3, -4.5) {\textcolor{cyan}{U}};
		\node [style=none] (253) at (-3, -8.5) {\textcolor{cyan}{U}};
		\node [style=none] (254) at (10.25, 1.5) {U(1)};
		\node [style=none] (255) at (10.25, -2.5) {U(1)};
		\node [style=none] (256) at (10.25, -6.5) {U(1)};
		\node [style=blacknode] (257) at (-4, 0) {};
		\node [style=blacknode] (258) at (-4, -4) {};
		\node [style=blacknode] (259) at (-4, -8) {};
	\end{pgfonlayer}
	\begin{pgfonlayer}{edgelayer}
		\draw (16.center) to (18.center);
		\draw (17.center) to (19.center);
		\draw (20.center) to (22.center);
		\draw (22.center) to (21.center);
		\draw (30.center) to (32.center);
		\draw (31.center) to (33.center);
		\draw (34.center) to (36.center);
		\draw (36.center) to (35.center);
		\draw (0) to (2);
		\draw (4) to (5);
		\draw (5) to (6);
		\draw (5) to (7);
		\draw (8) to (10);
		\draw (12) to (13);
		\draw (27) to (28);
		\draw (25) to (23);
		\draw (43) to (0);
		\draw (1) to (44);
		\draw (45) to (2);
		\draw (4) to (46);
		\draw (47) to (5);
		\draw (6) to (48);
		\draw (7) to (49);
		\draw (55) to (8);
		\draw (9) to (54);
		\draw (53) to (10);
		\draw (52) to (12);
		\draw (50) to (19.center);
		\draw (51) to (17.center);
		\draw (61) to (30.center);
		\draw (60) to (33.center);
		\draw (59) to (27);
		\draw (58) to (25);
		\draw (57) to (24);
		\draw (56) to (23);
		\draw [style=->] (84.center) to (85.center);
		\draw [style=->] (85.center) to (84.center);
		\draw [style=->] (86.center) to (87.center);
		\draw [style=->] (87.center) to (86.center);
		\draw [style=->] (88.center) to (89.center);
		\draw [style=->] (89.center) to (88.center);
		\draw (90) to (92);
		\draw (95.center) to (94);
		\draw (96.center) to (92);
		\draw (94) to (100);
		\draw (105) to (107);
		\draw (110.center) to (109);
		\draw (111.center) to (107);
		\draw (109) to (115);
		\draw (120) to (122);
		\draw (125.center) to (124);
		\draw (126.center) to (122);
		\draw (124) to (130);
		\draw (94) to (135);
		\draw (109) to (136);
		\draw (137) to (124);
		\draw (138) to (135);
		\draw (136) to (139);
		\draw (140) to (137);
		\draw (147) to (4);
		\draw (148) to (27);
		\draw (149) to (12);
		\draw (155) to (147);
		\draw (156) to (148);
		\draw (157) to (149);
		\draw [style=reddotted] (161.center) to (162.center);
		\draw [style=reddotted] (162.center) to (163.center);
		\draw [style=reddotted] (163.center) to (164.center);
		\draw [style=reddotted] (164.center) to (161.center);
		\draw [style=reddotted] (165.center) to (166.center);
		\draw [style=reddotted] (166.center) to (167.center);
		\draw [style=reddotted] (167.center) to (168.center);
		\draw [style=reddotted] (168.center) to (165.center);
		\draw [style=reddotted] (169.center) to (170.center);
		\draw [style=reddotted] (170.center) to (171.center);
		\draw [style=reddotted] (171.center) to (172.center);
		\draw [style=reddotted] (172.center) to (169.center);
		\draw [style=reddotted] (200.center) to (201.center);
		\draw [style=reddotted] (201.center) to (202.center);
		\draw [style=reddotted] (202.center) to (203.center);
		\draw [style=reddotted] (203.center) to (200.center);
		\draw [style=reddotted] (204.center) to (205.center);
		\draw [style=reddotted] (205.center) to (206.center);
		\draw [style=reddotted] (206.center) to (207.center);
		\draw [style=reddotted] (207.center) to (204.center);
		\draw [style=reddotted] (208.center) to (209.center);
		\draw [style=reddotted] (209.center) to (210.center);
		\draw [style=reddotted] (210.center) to (211.center);
		\draw [style=reddotted] (211.center) to (208.center);
		\draw (90) to (212);
		\draw (91) to (212);
		\draw (179) to (212);
		\draw (212) to (174);
		\draw (105) to (213);
		\draw (106) to (213);
		\draw (213) to (175);
		\draw (213) to (180);
		\draw (120) to (214);
		\draw (121) to (214);
		\draw (218.center) to (220.center);
		\draw (221.center) to (219.center);
		\draw (222.center) to (224.center);
		\draw (225.center) to (223.center);
		\draw (226.center) to (228.center);
		\draw (229.center) to (227.center);
		\draw (214) to (181);
		\draw (214) to (178);
		\draw (2) to (147);
		\draw (25) to (148);
		\draw (10) to (149);
		\draw (233) to (230);
		\draw (234) to (231);
		\draw (235) to (232);
		\draw (230) to (0);
		\draw (23) to (232);
		\draw (231) to (8);
	\end{pgfonlayer}
\end{tikzpicture}}
\label{superfamily1}
\end{equation}
Here, the red boxes indicate that changing U to SU nodes within the box for \DynkinBCD quiver will only affect the U(1) bouquets in the red boxes of the 3d mirror. On the other hand, the unitary nodes outside the red boxes in the \DynkinBCD quiver will only affect the bouquets (which are now flavor nodes) outside the red boxes in the 3d mirror. The split happens because of the overbalanced (cyan) special unitary node in the \DynkinBCD quiver. 

Even though the ranks of the gauge groups and flavor groups are arbitrary, certain features are fixed. Inside the red box of the Dynkin quiver, there are five unitary gauge nodes which means the red box of the mirror contains a single U(1) with $5+1=6$ links. Inside the box of the Dynkin quiver, if we turn any of the U to SU, the algorithm in determining the U(1) bouquets in the 3d mirror (which will all be inside the red box) is the same as for general DynkinAs. Details of such manipulations are outlined in \cite{USU}. Next, we turn some of the U to SU in the \DynkinBCD quivers and see the effect on the mirror:
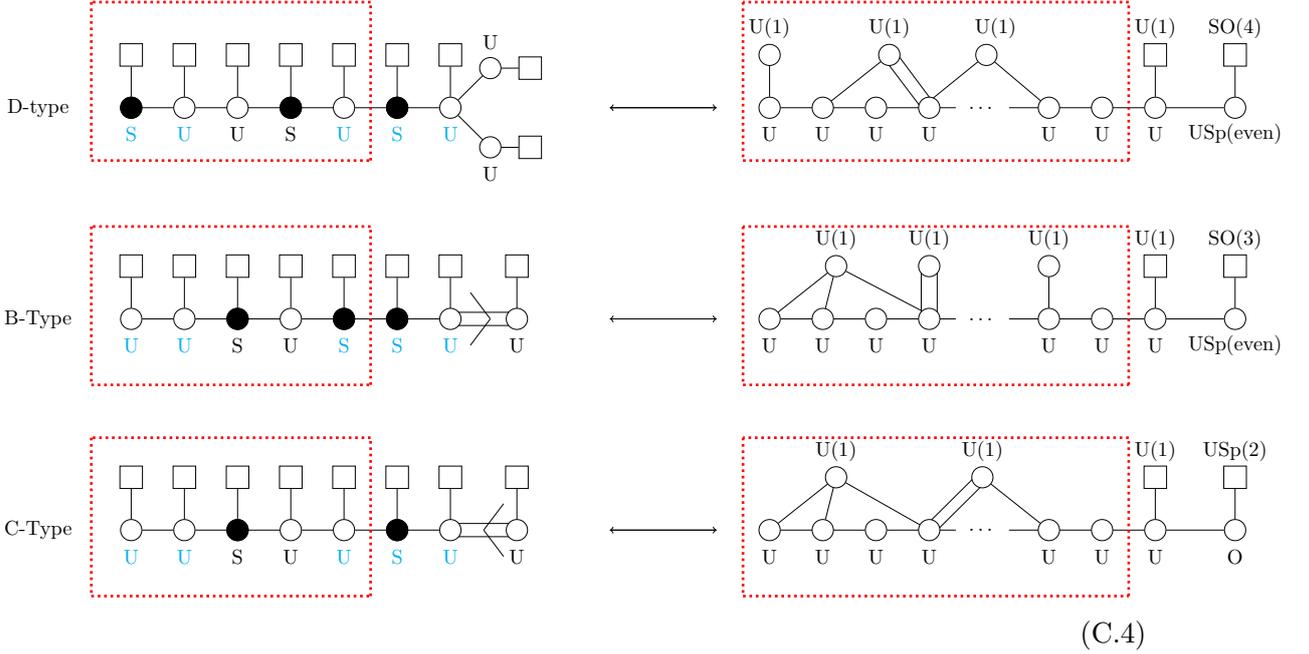
\begin{equation}
    \scalebox{0.7}{\begin{tikzpicture}
	\begin{pgfonlayer}{nodelayer}
		\node [style=gauge3] (0) at (-8, 0) {};
		\node [style=gauge3] (1) at (-7, 0) {};
		\node [style=gauge3] (2) at (-6, 0) {};
		\node [style=gauge3] (4) at (-4, 0) {};
		\node [style=gauge3] (5) at (-3, 0) {};
		\node [style=gauge3] (6) at (-2.25, 0.75) {};
		\node [style=gauge3] (7) at (-2.25, -0.75) {};
		\node [style=gauge3] (8) at (-8, -8) {};
		\node [style=gauge3] (9) at (-7, -8) {};
		\node [style=gauge3] (10) at (-6, -8) {};
		\node [style=gauge3] (12) at (-4, -8) {};
		\node [style=gauge3] (13) at (-3, -8) {};
		\node [style=gauge3] (15) at (-1.75, -8) {};
		\node [style=none] (16) at (-1.75, -7.875) {};
		\node [style=none] (17) at (-1.75, -8.125) {};
		\node [style=none] (18) at (-3, -7.875) {};
		\node [style=none] (19) at (-3, -8.125) {};
		\node [style=none] (20) at (-2, -7.5) {};
		\node [style=none] (21) at (-2, -8.5) {};
		\node [style=none] (22) at (-2.375, -8) {};
		\node [style=gauge3] (23) at (-8, -4) {};
		\node [style=gauge3] (24) at (-7, -4) {};
		\node [style=gauge3] (25) at (-6, -4) {};
		\node [style=gauge3] (27) at (-4, -4) {};
		\node [style=gauge3] (28) at (-3, -4) {};
		\node [style=gauge3] (29) at (-1.75, -4) {};
		\node [style=none] (30) at (-1.75, -3.875) {};
		\node [style=none] (31) at (-1.75, -4.125) {};
		\node [style=none] (32) at (-3, -3.875) {};
		\node [style=none] (33) at (-3, -4.125) {};
		\node [style=none] (34) at (-2.625, -3.5) {};
		\node [style=none] (35) at (-2.625, -4.5) {};
		\node [style=none] (36) at (-2.25, -4) {};
		\node [style=flavour2] (43) at (-8, 1) {};
		\node [style=flavour2] (44) at (-7, 1) {};
		\node [style=flavour2] (45) at (-6, 1) {};
		\node [style=flavour2] (46) at (-4, 1) {};
		\node [style=flavour2] (47) at (-3, 1) {};
		\node [style=flavour2] (48) at (-1.5, 0.75) {};
		\node [style=flavour2] (49) at (-1.5, -0.75) {};
		\node [style=flavour2] (50) at (-3, -7) {};
		\node [style=flavour2] (51) at (-1.75, -7) {};
		\node [style=flavour2] (52) at (-4, -7) {};
		\node [style=flavour2] (53) at (-6, -7) {};
		\node [style=flavour2] (54) at (-7, -7) {};
		\node [style=flavour2] (55) at (-8, -7) {};
		\node [style=flavour2] (56) at (-8, -3) {};
		\node [style=flavour2] (57) at (-7, -3) {};
		\node [style=flavour2] (58) at (-6, -3) {};
		\node [style=flavour2] (59) at (-4, -3) {};
		\node [style=flavour2] (60) at (-3, -3) {};
		\node [style=flavour2] (61) at (-1.75, -3) {};
		\node [style=none] (63) at (-7, -0.5) {U};
		\node [style=none] (64) at (-6, -0.5) {S};
		\node [style=none] (65) at (-4, -0.5) {\textcolor{cyan}{S}};
		\node [style=none] (67) at (-2.25, 1.25) {U};
		\node [style=none] (68) at (-2.25, -1.25) {U};
		\node [style=none] (70) at (-7, -8.5) {S};
		\node [style=none] (71) at (-6, -8.5) {U};
		\node [style=none] (74) at (-1.75, -8.5) {U};
		\node [style=none] (76) at (-7, -4.5) {S};
		\node [style=none] (77) at (-6, -4.5) {U};
		\node [style=none] (80) at (-1.75, -4.5) {U};
		\node [style=none] (81) at (-10.75, 0) {D-type};
		\node [style=none] (82) at (-10.75, -4) {B-Type};
		\node [style=none] (83) at (-10.75, -8) {C-Type};
		\node [style=none] (84) at (0, 0) {};
		\node [style=none] (85) at (2, 0) {};
		\node [style=none] (86) at (0, -4) {};
		\node [style=none] (87) at (2, -4) {};
		\node [style=none] (88) at (0, -8) {};
		\node [style=none] (89) at (2, -8) {};
		\node [style=gauge3] (90) at (3, 0) {};
		\node [style=gauge3] (91) at (4, 0) {};
		\node [style=gauge3] (92) at (5, 0) {};
		\node [style=none] (93) at (7, 0) {$\dots$};
		\node [style=gauge3] (94) at (10.25, 0) {};
		\node [style=none] (95) at (7.5, 0) {};
		\node [style=none] (96) at (6.5, 0) {};
		\node [style=flavour2] (100) at (10.25, 1) {};
		\node [style=none] (101) at (3, -0.5) {U};
		\node [style=none] (102) at (4, -0.5) {U};
		\node [style=none] (103) at (5, -0.5) {U};
		\node [style=none] (104) at (10.25, -0.5) {U};
		\node [style=gauge3] (105) at (3, -4) {};
		\node [style=gauge3] (106) at (4, -4) {};
		\node [style=gauge3] (107) at (5, -4) {};
		\node [style=none] (108) at (7, -4) {$\dots$};
		\node [style=gauge3] (109) at (10.25, -4) {};
		\node [style=none] (110) at (7.5, -4) {};
		\node [style=none] (111) at (6.5, -4) {};
		\node [style=flavour2] (115) at (10.25, -3) {};
		\node [style=none] (116) at (3, -4.5) {U};
		\node [style=none] (117) at (4, -4.5) {U};
		\node [style=none] (118) at (5, -4.5) {U};
		\node [style=none] (119) at (10.25, -4.5) {U};
		\node [style=gauge3] (120) at (3, -8) {};
		\node [style=gauge3] (121) at (4, -8) {};
		\node [style=gauge3] (122) at (5, -8) {};
		\node [style=none] (123) at (7, -8) {$\dots$};
		\node [style=gauge3] (124) at (10.25, -8) {};
		\node [style=none] (125) at (7.5, -8) {};
		\node [style=none] (126) at (6.5, -8) {};
		\node [style=flavour2] (130) at (10.25, -7) {};
		\node [style=none] (131) at (3, -8.5) {U};
		\node [style=none] (132) at (4, -8.5) {U};
		\node [style=none] (133) at (5, -8.5) {U};
		\node [style=none] (134) at (10.25, -8.5) {U};
		\node [style=gauge3] (135) at (11.75, 0) {};
		\node [style=gauge3] (136) at (11.75, -4) {};
		\node [style=gauge3] (137) at (11.75, -8) {};
		\node [style=flavour2] (138) at (11.75, 1) {};
		\node [style=flavour2] (139) at (11.75, -3) {};
		\node [style=flavour2] (140) at (11.75, -7) {};
		\node [style=none] (141) at (11.75, -8.5) {O};
		\node [style=none] (142) at (11.75, -4.5) {USp(even)};
		\node [style=none] (143) at (11.75, -0.5) {USp(even)};
		\node [style=none] (144) at (11.75, 1.5) {SO(4)};
		\node [style=none] (145) at (11.75, -2.5) {SO(3)};
		\node [style=none] (146) at (11.75, -6.5) {USp(2)};
		\node [style=gauge3] (147) at (-5, 0) {};
		\node [style=gauge3] (148) at (-5, -4) {};
		\node [style=gauge3] (149) at (-5, -8) {};
		\node [style=none] (153) at (-4, -4.5) {\textcolor{cyan}{S}};
		\node [style=none] (154) at (-4, -8.5) {\textcolor{cyan}{S}};
		\node [style=flavour2] (155) at (-5, 1) {};
		\node [style=flavour2] (156) at (-5, -3) {};
		\node [style=flavour2] (157) at (-5, -7) {};
		\node [style=none] (161) at (-9.75, 2) {};
		\node [style=none] (162) at (-9.75, -1) {};
		\node [style=none] (163) at (-4.5, -1) {};
		\node [style=none] (164) at (-4.5, 2) {};
		\node [style=none] (165) at (-9.75, -2.25) {};
		\node [style=none] (166) at (-9.75, -5.25) {};
		\node [style=none] (167) at (-4.5, -5.25) {};
		\node [style=none] (168) at (-4.5, -2.25) {};
		\node [style=none] (169) at (-9.75, -6.25) {};
		\node [style=none] (170) at (-9.75, -9.25) {};
		\node [style=none] (171) at (-4.5, -9.25) {};
		\node [style=none] (172) at (-4.5, -6.25) {};
		\node [style=gauge3] (173) at (9.25, 0) {};
		\node [style=gauge3] (174) at (8.25, 0) {};
		\node [style=gauge3] (175) at (8.25, -4) {};
		\node [style=gauge3] (176) at (9.25, -4) {};
		\node [style=gauge3] (177) at (9.25, -8) {};
		\node [style=gauge3] (178) at (8.25, -8) {};
		\node [style=gauge3] (179) at (6, 0) {};
		\node [style=gauge3] (180) at (6, -4) {};
		\node [style=gauge3] (181) at (6, -8) {};
		\node [style=none] (182) at (6, -0.5) {U};
		\node [style=none] (183) at (6, -4.5) {U};
		\node [style=none] (184) at (6, -8.5) {U};
		\node [style=none] (188) at (9.25, -0.5) {U};
		\node [style=none] (189) at (8.25, -0.5) {U};
		\node [style=none] (190) at (8.25, -4.5) {U};
		\node [style=none] (191) at (9.25, -4.5) {U};
		\node [style=none] (192) at (8.25, -8.5) {U};
		\node [style=none] (193) at (9.25, -8.5) {U};
		\node [style=none] (200) at (2.5, 2) {};
		\node [style=none] (201) at (2.5, -1) {};
		\node [style=none] (202) at (9.75, -1) {};
		\node [style=none] (203) at (9.75, 2) {};
		\node [style=none] (204) at (2.5, -2.25) {};
		\node [style=none] (205) at (2.5, -5.25) {};
		\node [style=none] (206) at (9.75, -5.25) {};
		\node [style=none] (207) at (9.75, -2.25) {};
		\node [style=none] (208) at (2.5, -6.25) {};
		\node [style=none] (209) at (2.5, -9.25) {};
		\node [style=none] (210) at (9.75, -9.25) {};
		\node [style=none] (211) at (9.75, -6.25) {};
		\node [style=gauge3] (213) at (4.25, -3) {};
		\node [style=gauge3] (214) at (4.25, -7) {};
		\node [style=none] (215) at (7.25, 1.5) {U(1)};
		\node [style=none] (216) at (6, -2.5) {U(1)};
		\node [style=none] (217) at (4.25, -6.5) {U(1)};
		\node [style=none] (218) at (5.1, 1) {};
		\node [style=none] (219) at (5.3, 1.15) {};
		\node [style=none] (220) at (5.85, 0) {};
		\node [style=none] (221) at (6.15, 0) {};
		\node [style=none] (222) at (5.85, -3) {};
		\node [style=none] (223) at (6.15, -3) {};
		\node [style=none] (224) at (5.85, -4) {};
		\node [style=none] (225) at (6.15, -4) {};
		\node [style=none] (226) at (6.85, -7) {};
		\node [style=none] (227) at (7.15, -7) {};
		\node [style=none] (228) at (5.85, -8) {};
		\node [style=none] (229) at (6.15, -8) {};
		\node [style=gauge3] (230) at (-9, 0) {};
		\node [style=gauge3] (231) at (-9, -8) {};
		\node [style=gauge3] (232) at (-9, -4) {};
		\node [style=flavour2] (233) at (-9, 1) {};
		\node [style=flavour2] (234) at (-9, -7) {};
		\node [style=flavour2] (235) at (-9, -3) {};
		\node [style=none] (240) at (-8, -0.5) {\textcolor{cyan}{U}};
		\node [style=none] (241) at (-9, -4.5) {\textcolor{cyan}{U}};
		\node [style=none] (242) at (-9, -8.5) {\textcolor{cyan}{U}};
		\node [style=none] (243) at (-9, -0.5) {\textcolor{cyan}{S}};
		\node [style=none] (245) at (-5, -0.5) {\textcolor{cyan}{U}};
		\node [style=none] (246) at (-8, -4.5) {\textcolor{cyan}{U}};
		\node [style=none] (247) at (-5, -4.5) {\textcolor{cyan}{S}};
		\node [style=none] (248) at (-8, -8.5) {\textcolor{cyan}{U}};
		\node [style=none] (249) at (-5, -8.5) {\textcolor{cyan}{U}};
		\node [style=none] (250) at (-3, -0.5) {\textcolor{cyan}{U}};
		\node [style=none] (251) at (-3, -4.5) {\textcolor{cyan}{U}};
		\node [style=none] (253) at (-3, -8.5) {\textcolor{cyan}{U}};
		\node [style=none] (254) at (10.25, 1.5) {U(1)};
		\node [style=none] (255) at (10.25, -2.5) {U(1)};
		\node [style=none] (256) at (10.25, -6.5) {U(1)};
		\node [style=gauge3] (257) at (3, 1) {};
		\node [style=gauge3] (258) at (7.075, 1) {};
		\node [style=gauge3] (259) at (5.25, 1) {};
		\node [style=gauge3] (260) at (8.25, -3) {};
		\node [style=gauge3] (261) at (6, -3) {};
		\node [style=gauge3] (262) at (7, -7) {};
		\node [style=none] (263) at (3, 1.5) {U(1)};
		\node [style=none] (264) at (5.25, 1.5) {U(1)};
		\node [style=none] (265) at (4.25, -2.5) {U(1)};
		\node [style=none] (267) at (8.25, -2.5) {U(1)};
		\node [style=none] (268) at (7, -6.5) {U(1)};
		\node [style=blacknode] (269) at (-9, 0) {};
		\node [style=blacknode] (270) at (-6, 0) {};
		\node [style=blacknode] (271) at (-4, 0) {};
		\node [style=blacknode] (272) at (-7, -4) {};
		\node [style=blacknode] (273) at (-5, -4) {};
		\node [style=blacknode] (274) at (-4, -4) {};
		\node [style=blacknode] (275) at (-7, -8) {};
		\node [style=blacknode] (276) at (-4, -8) {};
	\end{pgfonlayer}
	\begin{pgfonlayer}{edgelayer}
		\draw (16.center) to (18.center);
		\draw (17.center) to (19.center);
		\draw (20.center) to (22.center);
		\draw (22.center) to (21.center);
		\draw (30.center) to (32.center);
		\draw (31.center) to (33.center);
		\draw (34.center) to (36.center);
		\draw (36.center) to (35.center);
		\draw (0) to (2);
		\draw (4) to (5);
		\draw (5) to (6);
		\draw (5) to (7);
		\draw (8) to (10);
		\draw (12) to (13);
		\draw (27) to (28);
		\draw (25) to (23);
		\draw (43) to (0);
		\draw (1) to (44);
		\draw (45) to (2);
		\draw (4) to (46);
		\draw (47) to (5);
		\draw (6) to (48);
		\draw (7) to (49);
		\draw (55) to (8);
		\draw (9) to (54);
		\draw (53) to (10);
		\draw (52) to (12);
		\draw (50) to (19.center);
		\draw (51) to (17.center);
		\draw (61) to (30.center);
		\draw (60) to (33.center);
		\draw (59) to (27);
		\draw (58) to (25);
		\draw (57) to (24);
		\draw (56) to (23);
		\draw [style=->] (84.center) to (85.center);
		\draw [style=->] (85.center) to (84.center);
		\draw [style=->] (86.center) to (87.center);
		\draw [style=->] (87.center) to (86.center);
		\draw [style=->] (88.center) to (89.center);
		\draw [style=->] (89.center) to (88.center);
		\draw (90) to (92);
		\draw (95.center) to (94);
		\draw (96.center) to (92);
		\draw (94) to (100);
		\draw (105) to (107);
		\draw (110.center) to (109);
		\draw (111.center) to (107);
		\draw (109) to (115);
		\draw (120) to (122);
		\draw (125.center) to (124);
		\draw (126.center) to (122);
		\draw (124) to (130);
		\draw (94) to (135);
		\draw (109) to (136);
		\draw (137) to (124);
		\draw (138) to (135);
		\draw (136) to (139);
		\draw (140) to (137);
		\draw (147) to (4);
		\draw (148) to (27);
		\draw (149) to (12);
		\draw (155) to (147);
		\draw (156) to (148);
		\draw (157) to (149);
		\draw [style=reddotted] (161.center) to (162.center);
		\draw [style=reddotted] (162.center) to (163.center);
		\draw [style=reddotted] (163.center) to (164.center);
		\draw [style=reddotted] (164.center) to (161.center);
		\draw [style=reddotted] (165.center) to (166.center);
		\draw [style=reddotted] (166.center) to (167.center);
		\draw [style=reddotted] (167.center) to (168.center);
		\draw [style=reddotted] (168.center) to (165.center);
		\draw [style=reddotted] (169.center) to (170.center);
		\draw [style=reddotted] (170.center) to (171.center);
		\draw [style=reddotted] (171.center) to (172.center);
		\draw [style=reddotted] (172.center) to (169.center);
		\draw [style=reddotted] (200.center) to (201.center);
		\draw [style=reddotted] (201.center) to (202.center);
		\draw [style=reddotted] (202.center) to (203.center);
		\draw [style=reddotted] (203.center) to (200.center);
		\draw [style=reddotted] (204.center) to (205.center);
		\draw [style=reddotted] (205.center) to (206.center);
		\draw [style=reddotted] (206.center) to (207.center);
		\draw [style=reddotted] (207.center) to (204.center);
		\draw [style=reddotted] (208.center) to (209.center);
		\draw [style=reddotted] (209.center) to (210.center);
		\draw [style=reddotted] (210.center) to (211.center);
		\draw [style=reddotted] (211.center) to (208.center);
		\draw (105) to (213);
		\draw (106) to (213);
		\draw (120) to (214);
		\draw (121) to (214);
		\draw (218.center) to (220.center);
		\draw (221.center) to (219.center);
		\draw (222.center) to (224.center);
		\draw (225.center) to (223.center);
		\draw (226.center) to (228.center);
		\draw (229.center) to (227.center);
		\draw (214) to (181);
		\draw (2) to (147);
		\draw (25) to (148);
		\draw (10) to (149);
		\draw (233) to (230);
		\draw (234) to (231);
		\draw (235) to (232);
		\draw (230) to (0);
		\draw (23) to (232);
		\draw (231) to (8);
		\draw (257) to (90);
		\draw (258) to (174);
		\draw (91) to (259);
		\draw (220.center) to (258);
		\draw (213) to (225.center);
		\draw (260) to (175);
		\draw (262) to (178);
	\end{pgfonlayer}
\end{tikzpicture}}
\label{superduperfamily}
\end{equation}
As promised, the flavor nodes outside the red box in the mirror remains the same. 

Now, we turn our focus to the nodes outside of the red box which includes the BCD-type endings. As outlined in Section \S\ref{sec2}, if the gauge nodes in the BCD-type endings are all unitary, then the 3d mirror will have orthogonal/symplectic flavor nodes attached to the symplectic/orthogonal gauge groups. If BCD-type endings are special unitary, the result will be a bouquet of U(1)s rather than a flavor node. Here are some examples:
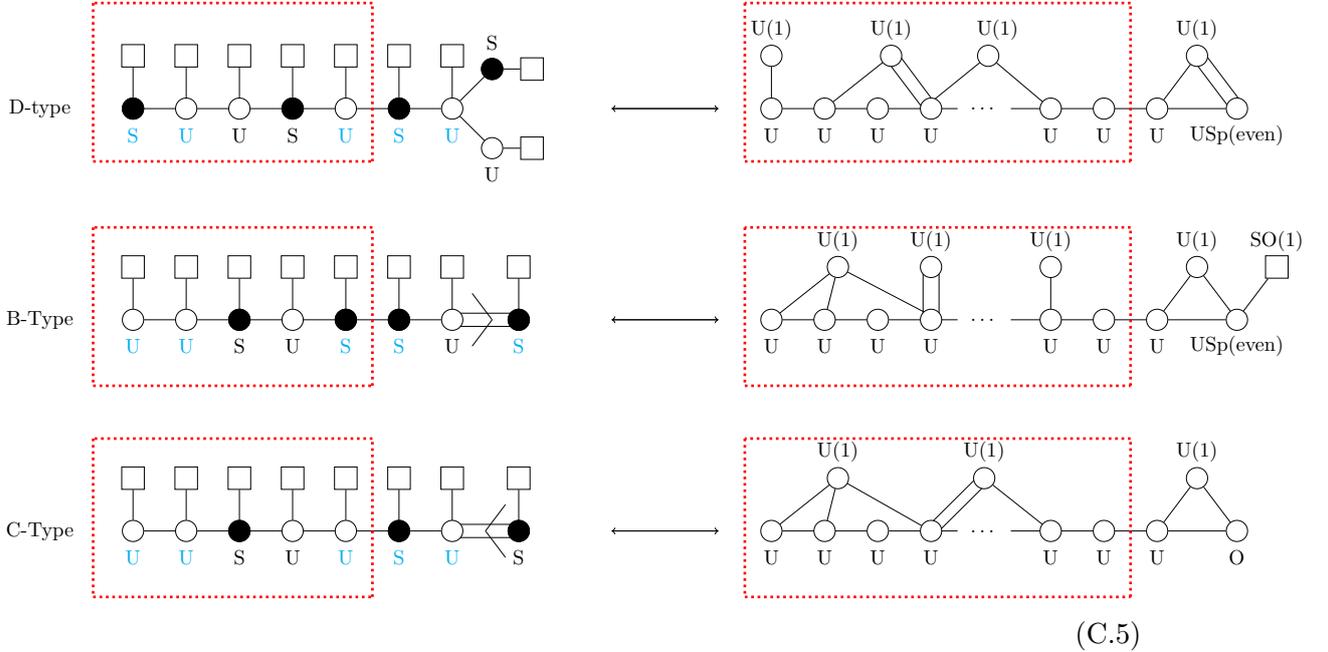
\begin{equation}
 \scalebox{0.7}{  \begin{tikzpicture}
	\begin{pgfonlayer}{nodelayer}
		\node [style=gauge3] (0) at (-8, 0) {};
		\node [style=gauge3] (1) at (-7, 0) {};
		\node [style=gauge3] (2) at (-6, 0) {};
		\node [style=gauge3] (4) at (-4, 0) {};
		\node [style=gauge3] (5) at (-3, 0) {};
		\node [style=gauge3] (6) at (-2.25, 0.75) {};
		\node [style=gauge3] (7) at (-2.25, -0.75) {};
		\node [style=gauge3] (8) at (-8, -8) {};
		\node [style=gauge3] (9) at (-7, -8) {};
		\node [style=gauge3] (10) at (-6, -8) {};
		\node [style=gauge3] (12) at (-4, -8) {};
		\node [style=gauge3] (13) at (-3, -8) {};
		\node [style=gauge3] (15) at (-1.75, -8) {};
		\node [style=none] (16) at (-1.75, -7.875) {};
		\node [style=none] (17) at (-1.75, -8.125) {};
		\node [style=none] (18) at (-3, -7.875) {};
		\node [style=none] (19) at (-3, -8.125) {};
		\node [style=none] (20) at (-2, -7.5) {};
		\node [style=none] (21) at (-2, -8.5) {};
		\node [style=none] (22) at (-2.375, -8) {};
		\node [style=gauge3] (23) at (-8, -4) {};
		\node [style=gauge3] (24) at (-7, -4) {};
		\node [style=gauge3] (25) at (-6, -4) {};
		\node [style=gauge3] (27) at (-4, -4) {};
		\node [style=gauge3] (28) at (-3, -4) {};
		\node [style=gauge3] (29) at (-1.75, -4) {};
		\node [style=none] (30) at (-1.75, -3.875) {};
		\node [style=none] (31) at (-1.75, -4.125) {};
		\node [style=none] (32) at (-3, -3.875) {};
		\node [style=none] (33) at (-3, -4.125) {};
		\node [style=none] (34) at (-2.625, -3.5) {};
		\node [style=none] (35) at (-2.625, -4.5) {};
		\node [style=none] (36) at (-2.25, -4) {};
		\node [style=flavour2] (43) at (-8, 1) {};
		\node [style=flavour2] (44) at (-7, 1) {};
		\node [style=flavour2] (45) at (-6, 1) {};
		\node [style=flavour2] (46) at (-4, 1) {};
		\node [style=flavour2] (47) at (-3, 1) {};
		\node [style=flavour2] (48) at (-1.5, 0.75) {};
		\node [style=flavour2] (49) at (-1.5, -0.75) {};
		\node [style=flavour2] (50) at (-3, -7) {};
		\node [style=flavour2] (51) at (-1.75, -7) {};
		\node [style=flavour2] (52) at (-4, -7) {};
		\node [style=flavour2] (53) at (-6, -7) {};
		\node [style=flavour2] (54) at (-7, -7) {};
		\node [style=flavour2] (55) at (-8, -7) {};
		\node [style=flavour2] (56) at (-8, -3) {};
		\node [style=flavour2] (57) at (-7, -3) {};
		\node [style=flavour2] (58) at (-6, -3) {};
		\node [style=flavour2] (59) at (-4, -3) {};
		\node [style=flavour2] (60) at (-3, -3) {};
		\node [style=flavour2] (61) at (-1.75, -3) {};
		\node [style=none] (63) at (-7, -0.5) {U};
		\node [style=none] (64) at (-6, -0.5) {S};
		\node [style=none] (65) at (-4, -0.5) {\textcolor{cyan}{S}};
		\node [style=none] (67) at (-2.25, 1.25) {S};
		\node [style=none] (68) at (-2.25, -1.25) {U};
		\node [style=none] (70) at (-7, -8.5) {S};
		\node [style=none] (71) at (-6, -8.5) {U};
		\node [style=none] (74) at (-1.75, -8.5) {S};
		\node [style=none] (76) at (-7, -4.5) {S};
		\node [style=none] (77) at (-6, -4.5) {U};
		\node [style=none] (80) at (-3, -4.5) {U};
		\node [style=none] (81) at (-10.75, 0) {D-type};
		\node [style=none] (82) at (-10.75, -4) {B-Type};
		\node [style=none] (83) at (-10.75, -8) {C-Type};
		\node [style=none] (84) at (0, 0) {};
		\node [style=none] (85) at (2, 0) {};
		\node [style=none] (86) at (0, -4) {};
		\node [style=none] (87) at (2, -4) {};
		\node [style=none] (88) at (0, -8) {};
		\node [style=none] (89) at (2, -8) {};
		\node [style=gauge3] (90) at (3, 0) {};
		\node [style=gauge3] (91) at (4, 0) {};
		\node [style=gauge3] (92) at (5, 0) {};
		\node [style=none] (93) at (7, 0) {$\dots$};
		\node [style=gauge3] (94) at (10.25, 0) {};
		\node [style=none] (95) at (7.5, 0) {};
		\node [style=none] (96) at (6.5, 0) {};
		\node [style=none] (101) at (3, -0.5) {U};
		\node [style=none] (102) at (4, -0.5) {U};
		\node [style=none] (103) at (5, -0.5) {U};
		\node [style=none] (104) at (10.25, -0.5) {U};
		\node [style=gauge3] (105) at (3, -4) {};
		\node [style=gauge3] (106) at (4, -4) {};
		\node [style=gauge3] (107) at (5, -4) {};
		\node [style=none] (108) at (7, -4) {$\dots$};
		\node [style=gauge3] (109) at (10.25, -4) {};
		\node [style=none] (110) at (7.5, -4) {};
		\node [style=none] (111) at (6.5, -4) {};
		\node [style=none] (116) at (3, -4.5) {U};
		\node [style=none] (117) at (4, -4.5) {U};
		\node [style=none] (118) at (5, -4.5) {U};
		\node [style=none] (119) at (10.25, -4.5) {U};
		\node [style=gauge3] (120) at (3, -8) {};
		\node [style=gauge3] (121) at (4, -8) {};
		\node [style=gauge3] (122) at (5, -8) {};
		\node [style=none] (123) at (7, -8) {$\dots$};
		\node [style=gauge3] (124) at (10.25, -8) {};
		\node [style=none] (125) at (7.5, -8) {};
		\node [style=none] (126) at (6.5, -8) {};
		\node [style=none] (131) at (3, -8.5) {U};
		\node [style=none] (132) at (4, -8.5) {U};
		\node [style=none] (133) at (5, -8.5) {U};
		\node [style=none] (134) at (10.25, -8.5) {U};
		\node [style=gauge3] (135) at (11.75, 0) {};
		\node [style=gauge3] (136) at (11.75, -4) {};
		\node [style=gauge3] (137) at (11.75, -8) {};
		\node [style=flavour2] (139) at (12.5, -3) {};
		\node [style=none] (141) at (11.75, -8.5) {O};
		\node [style=none] (142) at (11.75, -4.5) {USp(even)};
		\node [style=none] (143) at (11.75, -0.5) {USp(even)};
		\node [style=none] (145) at (12.5, -2.5) {SO(1)};
		\node [style=gauge3] (147) at (-5, 0) {};
		\node [style=gauge3] (148) at (-5, -4) {};
		\node [style=gauge3] (149) at (-5, -8) {};
		\node [style=none] (153) at (-4, -4.5) {\textcolor{cyan}{S}};
		\node [style=none] (154) at (-4, -8.5) {\textcolor{cyan}{S}};
		\node [style=flavour2] (155) at (-5, 1) {};
		\node [style=flavour2] (156) at (-5, -3) {};
		\node [style=flavour2] (157) at (-5, -7) {};
		\node [style=none] (161) at (-9.75, 2) {};
		\node [style=none] (162) at (-9.75, -1) {};
		\node [style=none] (163) at (-4.5, -1) {};
		\node [style=none] (164) at (-4.5, 2) {};
		\node [style=none] (165) at (-9.75, -2.25) {};
		\node [style=none] (166) at (-9.75, -5.25) {};
		\node [style=none] (167) at (-4.5, -5.25) {};
		\node [style=none] (168) at (-4.5, -2.25) {};
		\node [style=none] (169) at (-9.75, -6.25) {};
		\node [style=none] (170) at (-9.75, -9.25) {};
		\node [style=none] (171) at (-4.5, -9.25) {};
		\node [style=none] (172) at (-4.5, -6.25) {};
		\node [style=gauge3] (173) at (9.25, 0) {};
		\node [style=gauge3] (174) at (8.25, 0) {};
		\node [style=gauge3] (175) at (8.25, -4) {};
		\node [style=gauge3] (176) at (9.25, -4) {};
		\node [style=gauge3] (177) at (9.25, -8) {};
		\node [style=gauge3] (178) at (8.25, -8) {};
		\node [style=gauge3] (179) at (6, 0) {};
		\node [style=gauge3] (180) at (6, -4) {};
		\node [style=gauge3] (181) at (6, -8) {};
		\node [style=none] (182) at (6, -0.5) {U};
		\node [style=none] (183) at (6, -4.5) {U};
		\node [style=none] (184) at (6, -8.5) {U};
		\node [style=none] (188) at (9.25, -0.5) {U};
		\node [style=none] (189) at (8.25, -0.5) {U};
		\node [style=none] (190) at (8.25, -4.5) {U};
		\node [style=none] (191) at (9.25, -4.5) {U};
		\node [style=none] (192) at (8.25, -8.5) {U};
		\node [style=none] (193) at (9.25, -8.5) {U};
		\node [style=none] (200) at (2.5, 2) {};
		\node [style=none] (201) at (2.5, -1) {};
		\node [style=none] (202) at (9.75, -1) {};
		\node [style=none] (203) at (9.75, 2) {};
		\node [style=none] (204) at (2.5, -2.25) {};
		\node [style=none] (205) at (2.5, -5.25) {};
		\node [style=none] (206) at (9.75, -5.25) {};
		\node [style=none] (207) at (9.75, -2.25) {};
		\node [style=none] (208) at (2.5, -6.25) {};
		\node [style=none] (209) at (2.5, -9.25) {};
		\node [style=none] (210) at (9.75, -9.25) {};
		\node [style=none] (211) at (9.75, -6.25) {};
		\node [style=gauge3] (213) at (4.25, -3) {};
		\node [style=gauge3] (214) at (4.25, -7) {};
		\node [style=none] (215) at (7.25, 1.5) {U(1)};
		\node [style=none] (216) at (6, -2.5) {U(1)};
		\node [style=none] (217) at (4.25, -6.5) {U(1)};
		\node [style=none] (218) at (5.1, 1) {};
		\node [style=none] (219) at (5.3, 1.15) {};
		\node [style=none] (220) at (5.85, 0) {};
		\node [style=none] (221) at (6.15, 0) {};
		\node [style=none] (222) at (5.85, -3) {};
		\node [style=none] (223) at (6.15, -3) {};
		\node [style=none] (224) at (5.85, -4) {};
		\node [style=none] (225) at (6.15, -4) {};
		\node [style=none] (226) at (6.85, -7) {};
		\node [style=none] (227) at (7.15, -7) {};
		\node [style=none] (228) at (5.85, -8) {};
		\node [style=none] (229) at (6.15, -8) {};
		\node [style=gauge3] (230) at (-9, 0) {};
		\node [style=gauge3] (231) at (-9, -8) {};
		\node [style=gauge3] (232) at (-9, -4) {};
		\node [style=flavour2] (233) at (-9, 1) {};
		\node [style=flavour2] (234) at (-9, -7) {};
		\node [style=flavour2] (235) at (-9, -3) {};
		\node [style=none] (240) at (-8, -0.5) {\textcolor{cyan}{U}};
		\node [style=none] (241) at (-9, -4.5) {\textcolor{cyan}{U}};
		\node [style=none] (242) at (-9, -8.5) {\textcolor{cyan}{U}};
		\node [style=none] (243) at (-9, -0.5) {\textcolor{cyan}{S}};
		\node [style=none] (245) at (-5, -0.5) {\textcolor{cyan}{U}};
		\node [style=none] (246) at (-8, -4.5) {\textcolor{cyan}{U}};
		\node [style=none] (247) at (-5, -4.5) {\textcolor{cyan}{S}};
		\node [style=none] (248) at (-8, -8.5) {\textcolor{cyan}{U}};
		\node [style=none] (249) at (-5, -8.5) {\textcolor{cyan}{U}};
		\node [style=none] (250) at (-3, -0.5) {\textcolor{cyan}{U}};
		\node [style=none] (251) at (-1.75, -4.5) {\textcolor{cyan}{S}};
		\node [style=none] (253) at (-3, -8.5) {\textcolor{cyan}{U}};
		\node [style=none] (254) at (11, 1.5) {U(1)};
		\node [style=none] (255) at (11, -2.5) {U(1)};
		\node [style=none] (256) at (11, -6.5) {U(1)};
		\node [style=gauge3] (257) at (3, 1) {};
		\node [style=gauge3] (258) at (7.075, 1) {};
		\node [style=gauge3] (259) at (5.25, 1) {};
		\node [style=gauge3] (260) at (8.25, -3) {};
		\node [style=gauge3] (261) at (6, -3) {};
		\node [style=gauge3] (262) at (7, -7) {};
		\node [style=none] (263) at (3, 1.5) {U(1)};
		\node [style=none] (264) at (5.25, 1.5) {U(1)};
		\node [style=none] (265) at (4.25, -2.5) {U(1)};
		\node [style=none] (267) at (8.25, -2.5) {U(1)};
		\node [style=none] (268) at (7, -6.5) {U(1)};
		\node [style=gauge3] (269) at (11, -3) {};
		\node [style=gauge3] (270) at (11, 1) {};
		\node [style=none] (271) at (10.925, 0.925) {};
		\node [style=none] (272) at (11.1, 1.1) {};
		\node [style=none] (273) at (11.675, -0.075) {};
		\node [style=none] (274) at (11.85, 0.1) {};
		\node [style=gauge3] (275) at (11, -7) {};
		\node [style=blacknode] (276) at (-9, 0) {};
		\node [style=blacknode] (277) at (-6, 0) {};
		\node [style=blacknode] (278) at (-4, 0) {};
		\node [style=blacknode] (279) at (-2.25, 0.75) {};
		\node [style=blacknode] (280) at (-7, -4) {};
		\node [style=blacknode] (281) at (-5, -4) {};
		\node [style=blacknode] (282) at (-4, -4) {};
		\node [style=blacknode] (283) at (-1.75, -4) {};
		\node [style=blacknode] (284) at (-7, -8) {};
		\node [style=blacknode] (285) at (-4, -8) {};
		\node [style=blacknode] (286) at (-1.75, -8) {};
	\end{pgfonlayer}
	\begin{pgfonlayer}{edgelayer}
		\draw (16.center) to (18.center);
		\draw (17.center) to (19.center);
		\draw (20.center) to (22.center);
		\draw (22.center) to (21.center);
		\draw (30.center) to (32.center);
		\draw (31.center) to (33.center);
		\draw (34.center) to (36.center);
		\draw (36.center) to (35.center);
		\draw (0) to (2);
		\draw (4) to (5);
		\draw (5) to (6);
		\draw (5) to (7);
		\draw (8) to (10);
		\draw (12) to (13);
		\draw (27) to (28);
		\draw (25) to (23);
		\draw (43) to (0);
		\draw (1) to (44);
		\draw (45) to (2);
		\draw (4) to (46);
		\draw (47) to (5);
		\draw (6) to (48);
		\draw (7) to (49);
		\draw (55) to (8);
		\draw (9) to (54);
		\draw (53) to (10);
		\draw (52) to (12);
		\draw (50) to (19.center);
		\draw (51) to (17.center);
		\draw (61) to (30.center);
		\draw (60) to (33.center);
		\draw (59) to (27);
		\draw (58) to (25);
		\draw (57) to (24);
		\draw (56) to (23);
		\draw [style=->] (84.center) to (85.center);
		\draw [style=->] (85.center) to (84.center);
		\draw [style=->] (86.center) to (87.center);
		\draw [style=->] (87.center) to (86.center);
		\draw [style=->] (88.center) to (89.center);
		\draw [style=->] (89.center) to (88.center);
		\draw (90) to (92);
		\draw (95.center) to (94);
		\draw (96.center) to (92);
		\draw (105) to (107);
		\draw (110.center) to (109);
		\draw (111.center) to (107);
		\draw (120) to (122);
		\draw (125.center) to (124);
		\draw (126.center) to (122);
		\draw (94) to (135);
		\draw (109) to (136);
		\draw (137) to (124);
		\draw (136) to (139);
		\draw (147) to (4);
		\draw (148) to (27);
		\draw (149) to (12);
		\draw (155) to (147);
		\draw (156) to (148);
		\draw (157) to (149);
		\draw [style=reddotted] (161.center) to (162.center);
		\draw [style=reddotted] (162.center) to (163.center);
		\draw [style=reddotted] (163.center) to (164.center);
		\draw [style=reddotted] (164.center) to (161.center);
		\draw [style=reddotted] (165.center) to (166.center);
		\draw [style=reddotted] (166.center) to (167.center);
		\draw [style=reddotted] (167.center) to (168.center);
		\draw [style=reddotted] (168.center) to (165.center);
		\draw [style=reddotted] (169.center) to (170.center);
		\draw [style=reddotted] (170.center) to (171.center);
		\draw [style=reddotted] (171.center) to (172.center);
		\draw [style=reddotted] (172.center) to (169.center);
		\draw [style=reddotted] (200.center) to (201.center);
		\draw [style=reddotted] (201.center) to (202.center);
		\draw [style=reddotted] (202.center) to (203.center);
		\draw [style=reddotted] (203.center) to (200.center);
		\draw [style=reddotted] (204.center) to (205.center);
		\draw [style=reddotted] (205.center) to (206.center);
		\draw [style=reddotted] (206.center) to (207.center);
		\draw [style=reddotted] (207.center) to (204.center);
		\draw [style=reddotted] (208.center) to (209.center);
		\draw [style=reddotted] (209.center) to (210.center);
		\draw [style=reddotted] (210.center) to (211.center);
		\draw [style=reddotted] (211.center) to (208.center);
		\draw (105) to (213);
		\draw (106) to (213);
		\draw (120) to (214);
		\draw (121) to (214);
		\draw (218.center) to (220.center);
		\draw (221.center) to (219.center);
		\draw (222.center) to (224.center);
		\draw (225.center) to (223.center);
		\draw (226.center) to (228.center);
		\draw (229.center) to (227.center);
		\draw (214) to (181);
		\draw (2) to (147);
		\draw (25) to (148);
		\draw (10) to (149);
		\draw (233) to (230);
		\draw (234) to (231);
		\draw (235) to (232);
		\draw (230) to (0);
		\draw (23) to (232);
		\draw (231) to (8);
		\draw (257) to (90);
		\draw (258) to (174);
		\draw (91) to (259);
		\draw (220.center) to (258);
		\draw (213) to (225.center);
		\draw (260) to (175);
		\draw (262) to (178);
		\draw (109) to (269);
		\draw (269) to (136);
		\draw (94) to (270);
		\draw (271.center) to (273.center);
		\draw (274.center) to (272.center);
		\draw (275) to (124);
		\draw (275) to (137);
	\end{pgfonlayer}
\end{tikzpicture}}
\label{superfamily2}
\end{equation}
 Now, if both the BCD ending and the overbalanced gauge node attached to them are unitary, then we should expect something akin to the third diagram of \eqref{fourcases}. But rather than sharing a U(1) node, they will be sharing a flavor node instead. This is why in \eqref{superduperfamily} we have flavor groups attached to both the unitary and orthosymplectic gauge groups in the mirror.

To summarize, the procedure of finding the 3d mirror after replacing U with SU is nearly identical to the procedure for \DynkinA quivers. The only difference is that one needs to becareful with the BCD-ending and the gauge groups attached to it. With this in mind, the U(1) bouquets in the mirror can be easily determined. 

\subsection{D-type quiver with overbalanced node}
Let us  look at a DynkinD quiver  with overbalanced node and its 3d mirror in more detail:
\begin{equation}
\scalebox{0.8}{
}
\end{equation}
where the cyan node in the DynkinD quiver is overbalanced.
Turning all the nodes to special unitary gives:
\begin{equation}
 \scalebox{0.8}{   %
}
\end{equation}
The Coulomb branch phase gives:
\begin{equation}
\scalebox{0.8}{%
}
\end{equation}
The Higgs branch phase brane set up gives:
\begin{equation}
 \scalebox{0.7}{%
}
\end{equation}
Away from the overbalanced SU node, we can turn SU to U through the locking mechanism:
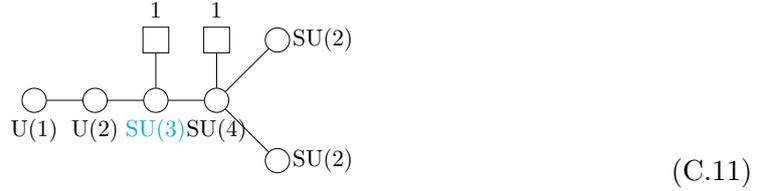
\begin{equation}
\scalebox{0.8}{  %
}
\end{equation}
and the Higgs phase brane set up with locking takes the form:
\begin{equation}
  \scalebox{0.8}{%
}
\end{equation}
On the other hand, turning the bifurcated nodes into unitary:
\begin{equation}
   \scalebox{0.8}{ %
}
\end{equation}
The Higgs branch phase is:
\begin{equation}
 \scalebox{0.7}{%
}
\end{equation}
Finally, turning the overbalanced node to unitary gives:
\begin{equation}
  \scalebox{0.8}{  %
}
\end{equation}
The Higgs phase gives:
\begin{equation}
   \scalebox{0.7}{%
}
\end{equation}

\bibliographystyle{JHEP}
\bibliography{bibli.bib}

\end{document}